\documentclass[11pt]{article}
\usepackage{eurosym}
\usepackage{amsfonts}
\usepackage{amssymb}
\usepackage{graphicx}
\usepackage{amsmath}
\usepackage{makeidx}
\usepackage{indentfirst}
\usepackage[T1]{fontenc}
\usepackage[utf8]{inputenc}

\newcounter{resultnum}[section]
\newcounter{conclusionnum}[section]
\newcounter{conditionnum}[section]
\newcounter{conjecturenum}[section]
\newcounter{examplenum}[section]
\newcounter{exercisenum}[section]
\newcounter{lemmanum}[section]
\newcounter{notationnum}[section]
\newcounter{theoremnum}[section]
\newcounter{definitionnum}[section]
\newcounter{corollarynum}[section]
\newcounter{remarknum}[section]
\newcounter{propositionnum}[section]
\newcounter{acknowledgementnum}[section]
\newcounter{algorithmnum}[section]
\newcounter{axiomnum}[section]
\newcounter{casenum}[section]
\newcounter{claimnum}[section]
\newcounter{summarynum}[section]
\newcounter{problemnum}[section]
\begin{document}

\title{Decoupling and integrability of nonassociative vacuum phase space
gravitational equations with star and R-flux parametric deformations}
\date{ March 10, 2021}
\author{ \vspace{.1 in} {\textbf{El\c{s}en Veli Veliev} } \thanks{%
email: elsen@kocaeli.edu.tr and elsenveli@hotmail.com} \\
%EndAName
{\small \textit{Department of Physics,\ Kocaeli University, 41380, Izmit,
Turkey}} \vspace{.2 in} \\
\vspace{.1 in} {\textbf{Lauren\c{t}iu Bubuianu}}\thanks{%
email: laurentiu.bubuianu@tvr.ro} \\
{\small \textit{SRTV - Studioul TVR Ia\c{s}i} and \textit{University
Appolonia}, 2 Muzicii street, Ia\c{s}i, 700399, Romania} \vspace{.2 in} \\
\vspace{.1 in} \textbf{Sergiu I. Vacaru} \thanks{
email: sergiu.vacaru@gmail.com \newline
\textit{Address for post correspondence in 2020-2021 as a visitor senior
researcher at YF CNU Ukraine is:\ } Yu. Gagarin street, 37-3, Chernivtsi,
Ukraine, 58008;\ authors' order reflects certain chronology of former and
futures research programs and involves equal contributions in obtaining new
results in this work} \\
{\small \textit{Physics Department, California State University at Fresno,
Fresno, CA 93740, USA; }}\\
{\small \textit{and Dep. Theoretical Physics and Computer Modelling,}}\\
{\small \textit{\ Yu. Fedkovych Chernivtsi National University}, 101
Storozhynetska street, Chernivtsi, 58029, Ukraine} }
\maketitle

\begin{abstract}
We prove that nonassociative star deformed vacuum Einstein equations can be
decoupled and integrated in certain general forms on phase spaces involving
real R-flux terms induced as parametric corrections on base Lorentz manifold
spacetimes. The geometric constructions are elaborated with parametric (on
respective Planck, $\hbar $, and string, $\kappa :=\mathit{\ell }%
_{s}^{3}/6\hbar $, constants) and nonholonomic dyadic decompositions of
fundamental geometric and physical objects. This is our second partner work
on elaborating nonassociative geometric and gravity theories with symmetric
and nonsymmetric metrics, (non) linear connections, star deformations
defined by generalized Moyal-Weyl products, endowed with quasi-Hopf algebra,
or other type algebraic and geometric structures, and all adapted to
nonholonomic distributions and frames. We construct exact and parametric
solutions for nonassociative vacuum configurations (with nontrivial or
effective cosmological constants) defined by star deformed generic
off-diagonal (non) symmetric metrics and (generalized) nonlinear and linear
connections. The coefficients of geometric objects defining such solutions
are determined by respective classes of generating and integration functions
and constants and may depend on all phase space coordinates [spacetime ones,
$(x^{i},t)$; and momentum like variables, $(p_{a},E$)]. Quasi-stationary
configurations are stated by solutions with spacetime Killing symmetry on a
time like vector $\partial _{t}$ but with possible dependencies on momentum
like coordinates on star deformed phase spaces. This geometric techniques of
decoupling and integrating nonlinear systems of physically important partial
differential equations (the anholonomic frame and connection deformation
method, AFCDM) is applied in our partner works for constructing
nonassociative and locally anisotropic generalizations of black hole and
cosmological solutions and elaborating geometric flow evolution and
classical and quantum information theories.

\vskip3pt

\end{abstract}

\tableofcontents

\section{Introduction}

There were formulated consistent approaches to nonassociative geometry and
gravity induced by the R-flux background of string theory \cite%
{blumenhagen16} and, for constructions with quasi-Hopf algebras, \cite%
{aschieri17} (see references therein). The formalism was developed up to a
level with nonassociative star deformed vacuum Einstein equations on phase
space $\mathcal{M}=T_{\shortparallel }^{\ast }V,$ with proofs that there are
nontrivial R-flux contributions resulting in real distortions of the Ricci
tensor for the Levi-Civita connection on Lorentz type spacetime manifold $V$.%
\footnote{\label{fncoord} See in next section details on index,
coordinatesand frames conventions. Here we note that we shall use local real
spacetime and complexified co-fiber, momentum like, coordinates $\
^{\shortparallel}u=(x,\ ^{\shortparallel }p)=\{\ ^{\shortparallel
}u^{\alpha}=(u^{k}=x^{k},\ ^{\shortparallel }p_{a}=(i\hbar ^{-1}p_{a})\},$
for complex unity $i^{2}=-1,$ where left up or low labels "$\
^{\shortparallel }$" will be used for geometric objects and spaces with dual
complexified momentum like coordinates; and coordinates $%
x=(x^{i})=(x^{1},x^{2},x^{3},x^{4}=t)$ on a four dimensional, 4-d, Lorentz
manifold with signature of metric $(+++-);$ for simplicity; in this paper,
we work with 4-d basic manifolds and 8-d (co) tangent bundles (i.e. phase
spaces); we can consider in similar forms spaces of higher dimension when
the base spacetime or phase spaces are of dimension 5, or 6, and respective
total dimension 10,11,12; if necessary, with certain imbedding and/or
compactification of coordinates.} That research was performed for
decompositions of (non)associative geometric and physical objects (such as
symmetric and nonsymmetric metrics, generalized (non) linear connections and
their torsion and curvature, etc.) up to linear orders on parameters $\hbar $
(Planck constant), $\kappa $ (string constant), which can be proportional to
the complex unity, and with nontrivial $\hbar \kappa $ real terms. It was
emphasized that for further developments it would be very important to study
various features of nonassociative models of gravity and matter field
interactions when star deformed Einstein equations are defined on phase
spaces and projected to spacetime manifolds, and to find explicit solutions
of such physically important nonlinear systems of nonassociative partial
differential equations, PDEs.

In our first partner work on nonassociative nonholonomic geometry and
gravity \cite{partner01}, we noted that because of generic nonlinearity and
tensor coupling of (non) associative / commutative gravitational equations,
even for vacuum configurations, the solutions of respective systems of PDEs
can not be constructed in general off-diagonal exact or parametric forms
depending on some phase and spacetime variables if we work only in holonomic
variables determined with respect to local coordinate frames and for
diagonalizable metrics. In such theories, R-flux star deformations extend
the geometric approach to constructions on phase spaces modelled as (co)
tangent bundles endowed with symmetric and nonsymmetric metrics and
generalized connection structures depending both on spacetime and momentum
like coordinates. Similar methods were studied in the geometry of algebraic
structures, metrics, and nonlinear and distinguished connections adapted to
nonholonomic distributions (in brief, N- and d-connections). Such a
formalism was developed in modified Finsler-Lagrange-Hamilton geometry, see
recent reviews and applications in physics in \cite{vacaru18,bubuianu18a},
and in generalized Einstein-Eisenhart-Moffat gravity extended for
nonholonomic Finsler like variables \cite%
{einstein25,einstein45,eisenhart51,eisenhart52,moffat79,
moffat95,moffat95a,moffat00,vacaru08aa,vacaru08bb,vacaru08cc}. To perform
rigorous mathematical formulations of nonassociative gravity and geometric
flow theories on total phase spaces and elaborate on methods of constructing
exact and parametric physically important solutions is not possible if we
work only in local coordinate bases and use star products stated only for
partial derivatives and adapting the geometric objects to respective
(linear) nonassociative / noncommutative /commutative algebraic structures.
It is important to apply a more advanced geometric techniques involving
nonassociative and noncommutative nonholonomic (i.e. non-integrable,
equivalently, anholonomic) distributions and frames, together with
nonholonomic deformations of (non) linear connections and respective
N-connection and parametric decompositions of (non) symmetric structures and
related distortions of fundamental geometric objects (for instance, of the
torsion and curvature tensors, distorted Ricci tensors etc.).

This work provides a nonholonomic dyadic formulation of nonassociative
geometry and vacuum gravity when a decomposition on so-called "shell by
shell" two dimensional variables determined by respective N-adapted frames
(in brief, s-structures and/or s-decompositions) allows to prove some
general decoupling and integrability properties of corresponding systems of
nonlinear PDEs. We generalize the approaches toward to the theory of
nonassociative gravity from proposed in \cite{blumenhagen16,aschieri17} to a
formalism when the geometric constructions are performed following the
method of star nonholonomic deformations elaborated in \cite{partner01}. Our
main goal is to write the nonassociative vacuum gravitational equations in
certain adapted nonholonomic variables with conventional 2+2+2+... splitting
of dimensions and in terms of auxiliary distinguished connections\footnote{%
d-connections, adapted to N-connection structures; for dyadic
decompositions, we use also the term s-connection} and, in result, to
generate new classes of exact and parametric physically important solutions.
The zero torsion Levi-Civita, LC, configurations can be extracted by
imposing additional nonholonomic constraints. Following the nonholonomic
dyadic approach, we can work on the entire phase space and formulate
nonassociative gravitational field equations using the geometric formalism
developed and outlined for commutative and noncommutative spaces in \cite%
{vacaru08bb,vacaru03,vacaru09a,vacaru18,bubuianu18a}.

For simplicity, in this paper, we restrict our considerations only to
nonassociative nonholonomic stationary vacuum configurations which can be
characterized also by nontrivial and/or effective cosmological constants
encoding or constraining nonholonomically the contributions of R-flux terms.
The stationary condition means that the metrics, connections, and related
fundamental geometric objects are with Killing symmetry on a vector $%
\partial _{4}=\partial _{t}$ for a time like spacetime coordinate $x^{4}=t$.
In our partner works, we elaborate on general methods for constructing exact
and parametric stationary and cosmological type solutions (when the
dependence on a time like variable is crucial) of nonassociative modified
Einstein equations with nontrivial and/or effective matter sources and for
generalized/ relativistic geometric flow evolution equations. Such a
geometric techniques is called in our works as the anholonomic frame and
connection deformation method, AFCDM,\footnote{%
in some previous works, we wrote in brief AFDM} of constructing exact
solutions in (non) commutative/ supersymmetric/ string / massive and other
type modified gravity and general relativity, GR, theories was developed in
\cite{bubuianu17b,bubuianu17, bubuianu19}, see also references therein, on
possible applications in modern cosmology and astrophysics, geometric
thermodynamics and classical and quantum information flow theories \cite%
{vacaru19,vacaru20}. Here we note that our nonholonomic dyadic\footnote{%
in mathematical and physical literature, there are used equivalently both
terms "dyad" and "diad" (the last one is from a corresponding Latin word);
in our works, such terms are introduced for conventional nonholonomic
decompostions of spaces and respective indices of geometric objects into
certain shells, s, of dimension 2; in this paper, we write "dyad" even in
some our previous articles it was used the term "diad".} approach to
generating exact solutions in (modified) gravity theories is more general
and different from other type methods with dyadic decompositions (for
instance, with the Newmann-Penrose formalism, see a review of geometric
methods and main results in \cite{misner,hawking73,wald82,kramer03}).
Applying the AFCDM, we work with distortions of connection structures which
allow to generate generic off-diagonal solutions depending on all spacetime
and phase space coordinates. This is not possible if we consider only the
LC--connection and other dyadic/ triad/ tetrad etc. methods.

It should be noted that algebras and geometries with nonassociative
structure arise from various developments on quantum mechanics with magnetic
monopole, string and M-theory with non-geometric fluxes etc., see \cite%
{kupriyanov15,kupriyanov18,szabo19} for reviews of such approaches. Here we
cite also a series of nonassociative theories with octonionic star products
and variables \cite{kurdgelaidze,okubo,castro1,castro2,castro3,gunaydin},
where the AFCDM can be applied for generating exact solutions and performing
deformation quantization, but we omit to study such models in this paper.

This work is structured as follows: Section \ref{sec2} provides a dyadic
formalism for the geometry of nonassociative nonholonomic star products and
deformations on phase spaces modelled as nonholonomic (co) tangent Lorentz
bundles. There are defined the N-connection and related frame and dual frame
structure with shell by shell decomposition (s-decomposition). The
nonholonomic dyadic constructions are considered for (non) associative phase
spaces and with projections to spacetime components involving R-flux
contributions.

Main features of star s-deformed nonsymmetric and symmetric metrics and
related nonassociative canonical s-connections and constraints for
torsionless LC-configurations are considered in section \ref{sec3}. It is
formulated a model of nonassociative dyadic geometry of phase spaces and
spacetime (non) symmetric metrics with canonical s-connection.

Section \ref{sec4} is devoted to elaborating a dyadic formalism with
nonholonomic (2+2)+(2+2) splitting for nonassociative s-decomposition of
vacuum Einstein equations encoding nontrivial and polarized cosomological
constants and/or sources induced by R-fluxes and parametric star
deformations of geometric objects. We follow the methods of nonassociative
differential geometry from section 3 of \cite{partner01} re-defined in
s-adapted variables at the end of this paper in appendix \ref{appendixa}.
There are considered s-variables for which the nonsymmetric components of
Ricci s-tensors are related to parametric and s-decompositions of the
nonsymmetric parts of s-metrics. Such nonassociative geometric constructions
involve complex (non) symmetric s-metric structures and canonical N- and
s-connections.

Section \ref{sec5} contains a further development of the AFCDM for
nonassociative vacuum gravity. There are stated and analyzed the conditions
on general decoupling and integrability of star nonholonomic deformed vacuum
Einstein equations. We show how corresponding systems of nonlinear PDEs spit
into coupled "shell by shell" systems of two plus two etc. equations with
coefficients depending recurrently on respective spacetime and phase
coordinates. Then we prove that such nonlinear s-systems are with a general
decoupling property and can be integrated in general form for general
off-diagonal ansatz for stationary and phase spaces Killing symmetries, for
instance, on $\partial _{t}$ and $\partial _{8}=\partial _{E}$. We study an
important class of nonassociative nonlinear symmetries relating generation
functions and generating sources connected to R-flux star deformations and
encoded into effective shell cosmological constants. The nonassociative
AFCDM is elaborated in such a form which is adapted to s-structures and
parametric decompositions involving symmetric and nonsymmetric components of
s-metrics and s-connections.

Conclusions are presented in section \ref{sec6}. In appendix \ref{appendixa}%
, there are provided main definitions and formulas for nonassociative
s-distinguished linear connections, s-connections, and introduced associated
covariant derivatives with formulas for computing s-components of
Riemannian, torsion and Ricci s-tensors. In appendix \ref{appendixb}, we
generalize for nonassociative vacuum gravity the procedure of generating
quasi-stationary solutions elaborated for associative and commutative phase
spaces in section 5 of \cite{bubuianu20}. We show how to construct
nonholonomic dyadic and parametric deformations of phase space metrics into
(non) associative / commutative phase space and spacetime configurations.

\section{Nonholonomic dyadic decompositions and nonassociative star products}

\label{sec2} In the first partner work \cite{partner01}, the nonassociative
vacuum Einstein equations from \cite{blumenhagen16,aschieri17} were
generalized and written in canonical nonholonomic variables on a phase space
modeled as a cotangent Lorentz bundle $\mathcal{M}=T_{\shortparallel }^{\ast}%
\mathbf{V}$ enabled with nonlinear connection, N-connection structure $\
^{\shortparallel }\mathbf{N}$ and a respectively defined N-adapted star
product using general frame transforms. We use boldface indices for spaces
and geometric objects enabled with (adapted to) N-connection structure as we
introduce below. In this section, we outline the N-connection geometry with
a splitting of type 4+4 on $TV$ and a $T^{\ast }V$ and further nonholonomic
dyadic decomposition into oriented shells $s=1,2,3,4$ with conventional
splitting of dimensions of type (2+2)+(2+2), in brief called a
s-decomposition. This is necessary for developing the anholonomic frame and
connection deformation method, AFCDM (on previous results and applications
to (non) commutative theories, see \cite%
{vacaru18,bubuianu17,bubuianu19,vacaru20}), for constructing exact and
parametric solutions in nonassociative gravity. Then, we study certain
important properties of the nonassociative star products introduced in
N-adapted form with s-splitting and quasi-Hopf structure, which generalize
the geometric constructions from \cite{aschieri17,partner01}.

\subsection{Nonlinear connections for dyadic splitting in commutative phase
spaces}

We summarize necessary concepts and definitions of geometric objects adapted
to N-connection structure and s-splitting of relativistic phase spaces with
real spacetime and complex velocity/momentum type variables.

\subsubsection{Conventions on dyadic local coordinates on phase spaces and
base spacetimes}

The geometric arena for elaborating on (non) associative / commutative
geometric and physical models in this work consists from (co) tangent
Lorentz bundles with conventional splitting of bases and (co) fiber like
coordinates and respective indices. We shall use such notations for local
real (spacetime and total phase space) and complex (co) fiber coordinates:%
\begin{eqnarray}
\mbox{ on }\mathbf{V} &\mbox{and}&\ _{s}\mathbf{V}:  \label{loccord} \\
x &=&\{x^{i}\}=~_{s}x=\{x^{i_{s}}\}=(x^{i_{1}},x^{a_{2}}\rightarrow
y^{a_{2}})=(x^{i_{2}}),\mbox{ with }x^{4}=t,  \notag \\
&&\mbox{ where }i,j,...=1,2,3,4;\mbox{ shells}:\ s=1:\mbox{ when }%
i_{1},j_{1},...=1,2;s=2,a_{2},b_{2},...=3,4;  \notag \\
\mbox{ on }T\mathbf{V} &\mbox{and}&T_{s}\mathbf{\mathbf{V}}:  \notag \\
\ u &=&(x,y)=\{u^{\alpha }=(u^{k}=x^{k},\ u^{a}=y^{a})\}=  \notag \\
\ _{s}u &=&(~_{s}x,~_{s}y)=\{u^{\alpha _{s}}=(u^{k_{s}}=x^{k_{s}},\
u^{a_{s}}=y^{a_{s}})\}=(x^{i_{1}},x^{i_{2}},x^{a_{3}}\rightarrow
y^{a_{3}},x^{a_{4}}\rightarrow y^{a_{4}}),\mbox{
shells }  \notag \\
&& s=1,2,3,4, \mbox{where }\alpha ,\beta
,...=1,2,...8;a,b,...=5,6,7,8;a_{3},b_{3},...=(5,6);\ a_{4},b_{4},..=(7,8);
\notag
\end{eqnarray}%
\begin{eqnarray}
\mbox{ on }T^{\ast }\mathbf{V} &\mbox{and}&T_{s}^{\ast }\mathbf{V}:  \notag
\\
\ \ \ ^{\shortmid }u=(x,\ \ \ \ ^{\shortmid }p) &=&\{\ \ u^{\alpha
}=(u^{k}=x^{k},\ \ ^{\shortmid }p_{a}=p_{a})\}  \notag \\
&=&(\ _{3}^{\shortmid }x,\ _{4}^{\shortmid }p)=\{\ ^{\shortmid }u^{\alpha
}=(\ ^{\shortmid }u^{k_{3}}=\ ^{\shortmid }x^{k_{3}},\ \ ^{\shortmid
}p_{a_{4}}=p_{a_{4}})\}=  \notag \\
\ _{s}^{\shortmid }u=(~_{s}x,\ _{s}^{\shortmid }p) &=&\{\ ^{\shortmid
}u^{\alpha _{s}}=(x^{k_{s}},\ \ ^{\shortmid
}p_{a_{s}}=p_{a_{s}})\}=(x^{i_{1}},x^{i_{2}},\ ^{\shortmid
}p_{a_{3}}=p_{a_{3}},\ \ ^{\shortmid }p_{a_{4}}=p_{a_{4}}),  \notag \\
&=&(~\ _{3}^{\shortmid }u~=\ _{3}^{\shortmid }x,\ _{4}^{\shortmid }p)=\{\ \
^{\shortmid }u^{\alpha _{3}}=(x^{i_{1}},x^{i_{2}},\ \ ^{\shortmid
}x^{i_{3}}\rightarrow ~\ ^{\shortmid }p_{a_{3}}),\ \ ^{\shortmid
}p_{a_{4}}\},  \notag \\
&&\mbox{ where }\ \ ^{\shortmid }x^{\alpha _{3}}=(x^{i_{1}},x^{a_{2}},\ \
^{\shortmid }p_{a_{3}}=p_{a_{3}}).  \notag
\end{eqnarray}%
\begin{eqnarray}
\mbox{ on }T_{\shortparallel }^{\ast }\mathbf{V} &\mbox{and}%
&T_{\shortparallel s}^{\ast }\mathbf{V}:  \notag \\
\ ^{\shortparallel }u=(x,\ ^{\shortparallel }p) &=&\{\ ^{\shortparallel
}u^{\alpha }=(u^{k}=x^{k},\ ^{\shortparallel }p_{a}=(i\hbar )^{-1}p_{a})\}
\notag \\
&=&(\ _{3}^{\shortparallel }x,\ _{4}^{\shortparallel }p)=\{\
^{\shortparallel }u^{\alpha }=(^{\shortparallel }u^{k_{3}}=\
^{\shortparallel }x^{k_{3}},\ ^{\shortparallel }p_{a_{4}}=(i\hbar
)^{-1}p_{a_{4}})\}=  \notag \\
\ _{s}^{\shortparallel }u=(~_{s}x,\ _{s}^{\shortparallel }p) &=&\{\
^{\shortparallel }u^{\alpha _{s}}=(x^{k_{s}},\ \ ^{\shortparallel
}p_{a_{s}}=(i\hbar )^{-1}p_{a_{s}})\}=(x^{i_{1}},x^{i_{2}},\
^{\shortparallel }p_{a_{3}}=(i\hbar )^{-1}p_{a_{3}},\ ^{\shortparallel
}p_{a_{4}}=(i\hbar )^{-1}p_{a_{4}}),  \notag \\
&=&(~\ _{3}^{\shortparallel }u~=\ _{3}^{\shortparallel }x,\
_{4}^{\shortparallel }p)=\{\ ^{\shortparallel }u^{\alpha
_{3}}=(x^{i_{1}},x^{i_{2}},\ ^{\shortparallel }x^{i_{3}}\rightarrow
~^{\shortparallel }p_{a_{3}}),\ ^{\shortparallel }p_{a_{4}}\},  \notag \\
&&\mbox{ where }\ ^{\shortparallel }x^{\alpha _{3}}=(x^{i_{1}},x^{i_{2}},\ \
^{\shortparallel }p_{a_{3}}=(i\hbar )^{-1}p_{a_{3}}).  \notag
\end{eqnarray}%
Such labels of local coordinates and respective splitting via oriented shell
by shell indices with $s=1,2,3,4$ can be considered on total spaces of
respective (co) tangent bundles and for a phase space model $%
T_{\shortparallel }^{\ast }\mathbf{V}$ with complex cofibers associated to a
real cotangent bundle $T^{\ast }\mathbf{V.}$ The notations from our works
are different from those introduced for formula (3.1) in \cite{blumenhagen16}%
, where $X_{K}=(\frac{p_{k}}{i\hbar },x^{k})$ and $\hbar =h/2\pi $ is the
Planck constant, is written instead of $\ ^{\shortparallel}u^{\alpha}$.
Similar coordinates with momentum like variables were used in \cite%
{aschieri17} but with a corresponding adapting for definition of quasi-Hopf
structures. In this work and \cite{partner01}, an up (or low, on
convenience), label "$\ ^{\shortparallel }$" is used for distinguishing
coordinates with "complexified momenta" from real phase coordinates $\
^{\shortmid }u^{\alpha }=(x^{k},p_{a})$ on $T^{\ast }\mathbf{V}$ (similar
conventions are considered in Finsler-Lagrange-Hamilton geometry \cite%
{vacaru18,bubuianu18a}). Such dyadic coordinates mix under general
coordinate transforms. Nevertheless, we can keep a shell oriented labelling
if we work with corresponding classes of local (co) frame structure with
corresponding adapting to nonholonomic distributions with dyadic
decomposition as we explain in next subsection. Dyadic structures, i.e.
nonholonomic shell decompositions, s-decompositions and/or s-structures, are
important for decoupling with respect to adapted frames various types of
modified Einstein equations. The formalism of N- and s-adapted labels and
respective abstract or frame coefficient notations used in this and partner
works is elaborated in such a way that allows to perform an unified
"symbolic" nonholonomic geometric calculus and provide proofs by analogy for
various types of nonassociative / noncommutative / commutative spaces with a
respective decoupling of physically important systems of PDEs.

\subsubsection{Nonlinear connections and dyadic splitting}

\paragraph{Nonlinear connection structures on commutative spacetimes and
phase spaces: \newline
}

To study R-flux nonassociative star deformations the phase spaces are
modelled as cotangent bundles $\ T^{\ast }V$ and $T_{\shortparallel}^{%
\ast}V. $ We shall elaborate on equivalent global and local (co) vector
bundles constructions keeping in mind that the complex unity $i$ before
momentum like coordinates is used for definition of star products. For
simplicity, we shall work with $\ ^{\shortparallel }u$ coordinates and
geometric objects on $T_{\shortparallel }^{\ast }V$ (if necessary, all
formulas can be written by analogy without complex unity). The conventional
horizontal, h, on base spacetime, and covertical, c, splitting of geometric
objects can be defined in global abstract forms by respective Whitney sums $%
\oplus $ of respective $h$- and $c$--distributions ($v$-distributions are
used for $TV$), when
\begin{equation}
\ ^{\shortmid }\mathbf{N}:\ T\mathbf{T}^{\ast }\mathbf{V}=hT^{\ast }V\oplus
cT^{\ast }V\mbox{ and  }\ \ ^{\shortparallel }\mathbf{N}:\ T\mathbf{T}%
_{\shortparallel }^{\ast }\mathbf{V}=hT_{\shortparallel }^{\ast }V\oplus
cT_{\shortparallel }^{\ast }V.  \label{ncon}
\end{equation}%
For a Lorentz base manifold, $\mathbf{V,}$ such a N--connection splitting is
with a nonholonomic $4+4$ decomposition which in local forms is written, for
instance, $\ ^{\shortparallel }\mathbf{N}=\ ^{\shortparallel }N_{ia}\frac{%
\partial } {\partial \ ^{\shortparallel }p_{a}}\otimes dx^{i}$, using
N-connection coefficients $\ ^{\shortparallel }\mathbf{N}=\{\
^{\shortparallel }N_{ia}\}.$ Here we note that in our works spacetime and
phase spaces and respective geometric objects are labeled by "bold face"
symbols if they are enabled with/ adapted to a N-connection structures and
written in so-called N-adapted form. We can always split a Lorentz manifold
into conventional h- and v-distributions considering a splitting $\mathbf{N}%
: \ T\mathbf{V}=hV\oplus vV$ using corresponding local dyads $\ ^{2}\mathbf{N%
} = N_{i_{1}}^{i_{2}}\frac{\partial }{\partial x^{i_{2}}}\otimes dx^{i_{1}}.$
This is a nonholonomic (2+2)-decomposition which is necessary for
constructing generic off-diagonal solutions on base manifolds.

\paragraph{N-adapted commutative coframe 4+4 structures: \newline
}

If a N-connection structure is introduced for a real Lorentz spacetime, we
can define respective classes of N-elongated (equivalently, N-adapted) bases
and dual bases (cobases) for nonholonomic dyadic splitting. Such $s=1$ and 2
dyads are linear on N-connection coefficients,
\begin{eqnarray}
\mathbf{e}_{i} &=&(\mathbf{e}_{i_{1}}=\frac{\partial }{\partial x^{i_{1}}}%
-N_{i_{1}}^{i_{2}}(x^{i})\frac{\partial }{\partial x^{i_{2}}}%
,e_{j_{2}}=\partial _{j_{2}}=\frac{\partial }{\partial x^{j_{2}}}),%
\mbox{ on
}\ T\mathbf{V;}  \label{nadap} \\
\mathbf{e}^{i} &=&(e^{i_{1}}=dx^{i_{1}},\mathbf{e}%
^{i_{2}}=dx^{i_{2}}+N_{i_{1}}^{i_{2}}(x^{k_{1}},x^{k_{2}})dx^{i_{1}}),%
\mbox{
on }T^{\ast }\mathbf{V.}  \notag
\end{eqnarray}%
Such a local basis $\mathbf{e}_{k}$ (\ref{nadap}) is nonholonomic if the
commutators
\begin{equation}
\mathbf{e}_{[k}\mathbf{e}_{j]}:=\mathbf{e}_{k}\mathbf{e}_{j}-\mathbf{e}_{j}%
\mathbf{e}_{k}=w_{kj}^{l}(u)\mathbf{e}_{l}  \label{anhrel}
\end{equation}%
contain nontrivial anholonomy coefficients $w_{kj}^{l}=%
\{w_{i_{1}j_{2}}^{k_{2}}=\partial
_{j_{2}}N_{i_{1}}^{k_{2}},w_{j_{1}i_{1}}^{k_{2}}=\mathbf{e}%
_{j_{1}}N_{i_{1}}^{k_{2}}- \mathbf{e}_{i_{1}}N_{j_{1}}^{k_{2}}\}.$ We say
that a basis is holonomic (integrable) (when it can be transformed via
coordinate transforms into a local coordinated basis, $\mathbf{e}%
_{k}\rightarrow \mathbf{\partial }_{k}=\partial /\partial x^{k}$) if $%
w_{kj}^{l}=0.$

Similarly, we can define N-elongated dual bases and cobases for cotangent
bundles, for instance, on a (complex) phase space,
\begin{eqnarray}
\ ^{\shortparallel }\mathbf{e}_{\alpha } &=&(\ ^{\shortparallel }\mathbf{e}%
_{i}=\frac{\partial }{\partial x^{i}}-\ ^{\shortparallel }N_{ia}(x,\
^{\shortparallel }p)\frac{\partial }{\partial \ \ ^{\shortparallel }p_{a}},\
\ ^{\shortparallel }e^{b}=\frac{\partial }{\partial \ \ ^{\shortparallel
}p_{b}}),\mbox{ on }\ T\mathbf{T}_{\shortparallel }^{\ast }\mathbf{V;}
\label{nadapdc} \\
\ ^{\shortparallel }\mathbf{e}^{\alpha } &=&(\ ^{\shortparallel
}e^{i}=dx^{i},\ ^{\shortparallel }\mathbf{e}_{a}=d\ \ ^{\shortparallel
}p_{a}+\ \ ^{\shortparallel }N_{ia}(x,\ \ ^{\shortparallel }p)dx^{i}),%
\mbox{on }T^{\ast }\mathbf{T}_{\shortparallel }^{\ast }\mathbf{V},  \notag
\end{eqnarray}%
where $\partial _{i}:=\partial /\partial x^{i},\ \ ^{\shortparallel
}\partial ^{b}:=\partial /\partial \ ^{\shortparallel }p_{b},\partial
_{\alpha }:=(\partial _{i},\partial _{a}),\ \ ^{\shortparallel }\partial
_{\alpha }:=(\partial _{i},\ \ ^{\shortparallel }\partial ^{a})$ etc.

\paragraph{Nonholonomic dyadic decomposition (s-decomposition): \newline
}

Total spaces' respective (2+2)+(2+2) splitting, see similar constructions in
\cite{vacaru18,bubuianu18a,bubuianu17,bubuianu19,vacaru20}, can be defined
in global forms by respective classes of N--connection structures:
\begin{eqnarray}
\ _{s}^{\shortmid }\mathbf{N}:\ \ _{s}T\mathbf{T}^{\ast }\mathbf{V} &=&\
^{1}hT^{\ast }V\oplus \ ^{2}vT^{\ast }V\oplus \ ^{3}cT^{\ast }V\oplus \
^{4}cT^{\ast }V\mbox{ and }  \notag \\
\ _{s}^{\shortparallel }\mathbf{N}:\ \ _{s}T\mathbf{T}_{\shortparallel
}^{\ast }\mathbf{V} &=&\ ^{1}hT_{\shortparallel }^{\ast }V\oplus \
^{2}vT_{\shortparallel }^{\ast }V\oplus \ ^{3}cT_{\shortparallel }^{\ast
}V\oplus \ ^{4}cT_{\shortparallel }^{\ast }V,  \label{ncon2}
\end{eqnarray}%
when respective decompositions into conventional 2-dim nonholonomic
distributions of $TTV,$ $TT^{\ast }V$ and $TT_{\shortparallel }^{\ast }V,$
\begin{eqnarray*}
\dim (\ ^{1}hT^{\ast }V) &=&\dim (\ ^{2}vT^{\ast }V)=\dim (\ ^{3}cT^{\ast
}V)=(\ ^{4}cT^{\ast }V)=2\mbox{
and } \\
\dim (\ ^{1}hT_{\shortparallel }^{\ast }V) &=&\dim (\ ^{2}vT_{\shortparallel
}^{\ast }V)=\dim (\ ^{3}cT_{\shortparallel }^{\ast }V)=(\
^{4}cT_{\shortparallel }^{\ast }V)=2.
\end{eqnarray*}%
In these formulas, left up labels like $\ ^{1}h,\ ^{3}c$ etc. state that
using nonholonomic (equivalently, anholonomic and/or non-integrable)
distributions we split respective 8-d total spaces into oriented 2-d shells
with numbers 1,2,3 and 4.

The nonholonomic dyadic splitting with N-connections (\ref{ncon2}) are
defined locally, for instance, by such coefficients
\begin{eqnarray}
\ _{s}^{\shortparallel }\mathbf{N} &=&%
\{N_{i_{1}}^{i_{2}}(x^{i_{1}},x^{a_{2}}),\ ^{\shortparallel
}N_{i_{1}a_{3}}(x^{i_{1}},x^{a_{2}},^{\shortparallel }p_{b_{3}}),\
^{\shortparallel }N_{i_{1}a_{4}}(x^{i_{1}},x^{a_{2}},^{\shortparallel
}p_{b_{3}},^{\shortparallel }p_{b_{4}}),  \label{ncon2coef} \\
&&\ ^{\shortparallel }N_{i_{2}a_{3}}(x^{i_{1}},x^{a_{2}},^{\shortparallel
}p_{b_{3}}),\ ^{\shortparallel
}N_{i_{2}a_{4}}(x^{i_{1}},x^{a_{2}},^{\shortparallel
}p_{b_{3}},^{\shortparallel }p_{b_{4}}),\ ^{\shortparallel }N_{\
a_{4}}^{a_{3}}(x^{i_{1}},x^{a_{2}},^{\shortparallel
}p_{b_{3}},^{\shortparallel }p_{b_{4}})\}.  \notag
\end{eqnarray}%
Such values with "shell by shell" decompositions allow us to define
N-adapted bases (s-adapted, or equivalently, s-bases),
\begin{equation*}
\ ^{\shortparallel }\mathbf{e}_{\alpha _{s}}=(\ \ ^{\shortparallel }\mathbf{e%
}_{i_{s}}=\ \frac{\partial }{\partial x^{i_{s}}}-\ ^{\shortparallel }N_{\
i_{s}a_{s}}\frac{\partial }{\partial \ ^{\shortparallel }p_{a_{s}}},\ \
^{\shortparallel }e^{b_{s}}=\frac{\partial }{\partial \ ^{\shortparallel
}p_{b_{s}}})\mbox{ on }\ \ _{s}T\mathbf{T}_{\shortparallel }^{\ast }\mathbf{%
V,}
\end{equation*}
\begin{eqnarray}
\mbox{ when }\ \ ^{\shortparallel }\mathbf{e}_{\alpha _{1}} &=&(\ \ \
^{\shortparallel }e_{i_{1}}=\frac{\partial }{\partial x^{i_{1}}}=\partial
_{i_{1}}),\mbox{ for }i_{1}=1,2;  \label{nadapbdsc} \\
\ \ ^{\shortparallel }\mathbf{e}_{\alpha _{2}} &=&(\ ^{\shortparallel }%
\mathbf{e}_{i_{1}}=\frac{\partial }{\partial x^{i_{1}}}-N_{i_{1}}^{a_{2}}%
\frac{\partial }{\partial x^{a_{2}}},\ ^{\shortparallel }e_{b_{2}}=\frac{%
\partial }{\partial x^{b_{2}}})  \notag \\
&=&(\ ^{\shortparallel }\mathbf{e}_{i_{1}}=\partial
_{i_{1}}-N_{i_{1}}^{a_{2}}\partial _{a_{2}},\ \ \ ^{\shortparallel
}e_{b_{2}}=\partial _{b_{2}}),\mbox{ for }b_{2}=3,4  \notag \\
\ \ \ ^{\shortparallel }\mathbf{e}_{\alpha _{3}} &=&(\ ^{\shortparallel }%
\mathbf{e}_{i_{2}}=\frac{\partial }{\partial x^{i_{2}}}-\ ^{\shortparallel
}N_{i_{2}a_{3}}\frac{\partial }{\partial \ ^{\shortparallel }p_{a_{3}}},\
^{\shortparallel }e^{b_{3}}=\frac{\partial }{\partial \ ^{\shortparallel
}p_{b_{3}}}),  \notag \\
&=&(\ ^{\shortparallel }\mathbf{e}_{i_{2}}=\partial _{i_{2}}-\
^{\shortparallel }N_{i_{2}a_{3}}\ \ ^{\shortparallel }\partial ^{a_{3}},\
^{\shortparallel }e^{b_{3}}=\ \ ^{\shortparallel }\partial ^{b_{3}}),%
\mbox{
for }i_{2}=(i_{1},a_{2});  \notag \\
\ \ ^{\shortparallel }\mathbf{e}_{\alpha _{4}} &=&(\ \ ^{\shortparallel }%
\mathbf{e}_{i_{3}}=\frac{\partial }{\partial x^{i_{3}}}-\ ^{\shortparallel
}N_{\ i_{3}a_{4}}\frac{\partial }{\partial \ ^{\shortparallel }p_{a_{4}}},\
^{\shortparallel }e^{b_{4}}=\frac{\partial }{\partial \ ^{\shortparallel
}p_{b_{4}}})  \notag \\
&=&(\ \ ^{\shortparallel }\mathbf{e}_{i_{3}}=\partial _{i_{3}}-\
^{\shortparallel }N_{\ i_{3}a_{4}}\ ^{\shortparallel }\partial ^{a_{4}},\ \
^{\shortparallel }e^{b_{4}}=\ ^{\shortparallel }\partial ^{b_{4}}),%
\mbox{
for }i_{3}=(i_{1},a_{2},a_{3}),  \notag
\end{eqnarray}%
where we follow the conventions for dyadic indices and coordinates (\ref%
{loccord}).

For dual dyadic shell by shell s-adapted bases to (\ref{nadapbdsc}),
s-cobases, we have%
\begin{equation}
\ \ ^{\shortparallel }\mathbf{e}^{\alpha _{s}}=(\ \ ^{\shortparallel }%
\mathbf{e}^{i_{s}}=dx^{i_{s}},\ ^{\shortparallel }\mathbf{e}_{a_{s}}=d\
^{\shortparallel }p_{a_{s}}+\ ^{\shortparallel }N_{\ i_{s}a_{s}}dx^{i_{s}})%
\mbox{
on }\ \ _{s}T^{\ast }\mathbf{T}_{\shortparallel }^{\ast }\mathbf{V,}%
\mbox{
when }  \label{nadapbdss}
\end{equation}
\begin{eqnarray*}
\ \ ^{\shortparallel }\mathbf{e}^{\alpha _{4}} &=&(\ \ ^{\shortparallel }%
\mathbf{e}^{i_{3}}=dx^{i_{3}},\ ^{\shortparallel }\mathbf{e}_{a_{4}}=d\
^{\shortparallel }p_{a_{4}}+\ ^{\shortparallel }N_{\ i_{3}a_{4}}dx^{i_{3}}),
\\
\ \ ^{\shortparallel }\mathbf{e}^{\alpha _{3}} &=&(\ \mathbf{e}%
^{i_{2}}=dx^{i_{2}},\ ^{\shortparallel }\mathbf{e}_{a_{3}}=d\
^{\shortparallel }p_{a_{3}}+\ ^{\shortparallel }N_{\ i_{2}a_{3}}dx^{i_{2}}),
\\
\ \ ^{\shortparallel }\mathbf{e}^{\alpha _{2}} &=&(\ \mathbf{e}%
^{i_{1}}=dx^{i_{1}},\ ^{\shortparallel }\mathbf{e}_{a_{2}}=d\
^{\shortparallel }p_{a_{3}}+\ ^{\shortparallel }N_{\ i_{2}a_{3}}dx^{i_{2}}),
\\
^{\shortparallel }\mathbf{e}^{\alpha _{1}} &=&(\mathbf{e}^{i_{1}}=dx^{i_{1}})
\end{eqnarray*}%
defined by the same N-connection coefficients (\ref{ncon2coef}) for
respective s-decomposition (in brief, we shall write also s-connection
defining corresponding nonholonomic s-structures) phase space.

\subsection{Nonassociativity, dyadic N-connections, and quasi-Hopf algebras}

We generalize for phase spaces with dyadic decomposition the concept of
nonassociative star product considered in \cite{mylonas12,mylonas13}. The
nonassociative geometric constructions were re-defined on nonholonomic phase
spaces and spacetime manifolds enabled with N-connection structure in our
partner work \cite{partner01}). The formulas provided is section 2 of \cite%
{blumenhagen16} and section 2 of \cite{aschieri17} for configuration spaces
can be obtained in the cases of trivial N-connection s-structures.

\subsubsection{Nonassociative star product and dyadic s-structures}

We begin with the definition of nonassociative star products in
h-coordinates for h-coefficients with decompositions in shells $s=1,2$ in
order to understand the geometric principles for dyadic and nonholonomic
modifications of coordinate frame constructions in \cite%
{blumenhagen16,mylonas12,mylonas13}. In next subsections, the formalism of
nonassociative star product is developed in s--adapted form on the full
phase space $T_{\shortparallel s}^{\ast }\mathbf{V,}$ involving four shell
by shell decompositions. The constructions will be extended to the case of
nontrivial quasi-Hopf structures with s-splitting in order to formulate a
corresponding self-consistent geometric approach with star deformations of
coordinate diffeomorphism in s-adapted forms.

Let us consider a full phase space $\mathcal{M}$ containing a spacetime
direction determined by a N-adapted direction $\ ^{\shortmid }\mathbf{e}%
_{i_{s}}$ and a momentum like cofiber directions $\ ^{\shortmid }e^{b_{s}}.$
For any two functions $\ z(x,p)$ and $\ q(x,p),$ we define a s-adapted star
product $\star _{s}$:
\begin{eqnarray}
z\star _{s}q&:=&\cdot \lbrack \mathcal{F}_{s}^{-1}(z,q)]  \label{starpn} \\
&=&\cdot \lbrack \exp \left( -\frac{1}{2}i\hbar (\ ^{\shortmid }\mathbf{e}%
_{i_{s}}\otimes \ ^{\shortmid }e^{i_{s}}-\ ^{\shortmid }e^{i_{s}}\otimes \
^{\shortmid }\mathbf{e}_{i_{s}})\right) +\frac{i\mathit{\ell }_{s}^{4}}{%
12\hbar }R^{i_{s}j_{s}a_{s}}(p_{a_{s}}\ ^{\shortmid }\mathbf{e}%
_{i_{s}}\otimes \ ^{\shortmid }\mathbf{e}_{j_{a}}-\ ^{\shortmid }\mathbf{e}%
_{j_{s}}\otimes p_{a_{s}}\ ^{\shortmid }\mathbf{e}_{i_{s}})]z\otimes q
\notag \\
&=&z\cdot q-\frac{i}{2}\hbar \lbrack (\ ^{\shortmid }\mathbf{e}_{i_{s}}z)(\
^{\shortmid }e^{i_{s}}q)-(\ ^{\shortmid }e^{i_{s}}z)(\ ^{\shortmid }\mathbf{e%
}_{i_{s}}q)]+\frac{i\mathit{\ell }_{s}^{4}}{6\hbar }%
R^{i_{s}j_{s}a_{s}}p_{a_{s}}(\ ^{\shortmid }\mathbf{e}_{i_{s}}z)(\
^{\shortmid }\mathbf{e}_{j_{s}}q)+\ldots ,  \notag
\end{eqnarray}%
where the constant $\mathit{\ell }$ defines R-flux contributions for a
antisymmetric $R^{i_{s}j_{s}a_{s}}$ background in string theory, with
s-indices. In these formulas, the tensor product $\otimes $ is used in a
form indicating on which factor of $z\otimes q$ the N-adapted derivatives
act with the dot form, when eventually the tensor products turn into usual
multiplications. In coordinate frames and for holonomic frames determined by
trivial N-connection structures, the star product (\ref{starpn}) coincides
with that considered in \cite{blumenhagen16}. In \cite{partner01}, we write $%
\star _{N}$ and extend the concept of nonholonomic star product on total
phase space (see next subsection for constructions with quasi-Hopf
structure). In coordinate base cases, we do not write boldface symbols and
do not use the labels "s" or "N" for star products and related geometric
objects.

\subsubsection{Quasi-Hopf s-structures}

\label{ssqhda}

We consider the universal enveloping Hopf algebra $U\emph{Vec}(\mathcal{M})$
of the Lie algebra of vector fields $\emph{Vec}(\mathcal{M})$ on a
nonholonomic manifold (it can be with s-structure) $\mathcal{M},$ see such
constructions for holonomic structures in \cite{aschieri17}. For non-trivial
N-connection s-structures of phase space $\mathcal{M}=T_{\shortparallel}^{%
\ast}\mathbf{V,}$ we elaborate on a nonholonomic algebraic and dyadic
formalism with d-vectors and d-tensors. A nonholonomic splitting $\emph{Vec}(%
\mathcal{M})=h\emph{Vec}(\mathcal{M})\oplus v\emph{Vec}(\mathcal{M})$
follows from the definition of N-connection structure (\ref{ncon}). This
defines a Hopf d-algebra $U\emph{Vec}(\mathcal{M},\ _{s}^{\shortparallel}N).
$ A N-adapted cochain twist element $\mathcal{F}_{N}$ with possible
s-decomposition $\mathcal{F}_{s}$ is parameterized
\begin{eqnarray}
\ ^{\shortparallel }\mathcal{F}_{N} &=&\ ^{\shortparallel }\mathcal{\check{F}%
}_{N}\ ^{\shortmid }\mathcal{\check{F}}_{RN}=\ ^{\shortmid }\mathcal{\check{F%
}}_{RN}\ ^{\shortparallel }\mathcal{\check{F}}_{N}=  \label{twisthopf} \\
\ ^{\shortparallel }\mathcal{F}_{s} &=&\ ^{\shortparallel }\mathcal{\check{F}%
}_{s}\ ^{\shortmid }\mathcal{\check{F}}_{Rs}=\ ^{\shortmid }\mathcal{\check{F%
}}_{Rs}\ ^{\shortparallel }\mathcal{\check{F}}_{s}  \notag
\end{eqnarray}%
for a nonholonomic Hopf 2-cocycle (determining a nonholonomic Moyal-Weyl
deformation of the phase space)%
\begin{equation*}
\ ^{\shortparallel }\mathcal{\check{F}}_{N}=\exp \left[ \left( \
^{\shortparallel }\mathbf{e}_{i}\otimes \ ^{\shortparallel }e^{n+i}-\
^{\shortparallel }e^{n+i}\otimes \ ^{\shortparallel }\mathbf{e}_{i}\right) %
\right] =\ ^{\shortparallel }\mathcal{\check{F}}_{s}=\exp \left[ \left( \
^{\shortparallel }\mathbf{e}_{i_{s}}\otimes \ ^{\shortparallel
}e^{n+i_{s}}-\ ^{\shortparallel }e^{n+i_{s}}\otimes \ ^{\shortparallel }%
\mathbf{e}_{i_{s}}\right) \right] ,
\end{equation*}%
where for 4-d base Lorentz manifolds $i=(i_{1},i_{2})$, or $i=(i_{1},a_{2})$%
, (with $i_1 =1,2, a_2 =3,4)$; such indices are used for contacting in
oriented form shall splitting of indices $a=(a_{3}=n+i_{1},a_{4}=n+i_{2})$
for a h1-v2 splitting of coefficients of geometric s-objects. Respective
nonholonomic 2-cocycles for a R-flux are
\begin{eqnarray*}
&&\ ^{\shortmid }\mathcal{\check{F}}_{RN}=\exp{[\frac{i\kappa }{2}%
R^{jka}\left( p_{a}\ ^{\shortmid }\mathbf{e}_{k}\otimes \ ^{\shortmid }%
\mathbf{e}_{j}-\ ^{\shortmid }\mathbf{e}_{j}\otimes p_{a}\ ^{\shortmid }%
\mathbf{e}_{k}\right)]}= \\
&&\ ^{\shortmid }\mathcal{\check{F}}_{Rs}= \exp{[\frac{i\kappa }{2}%
R^{j_{s}k_{s}a_{s}}\left( p_{a_{s}}\ ^{\shortmid }\mathbf{e}_{k_{s}}\otimes
\ ^{\shortmid }\mathbf{e}_{j_{s}}-\ ^{\shortmid }\mathbf{e}_{j_{s}}\otimes
p_{a_{s}}\ ^{\shortmid }\mathbf{e}_{k_{s}}\right)]},
\end{eqnarray*}%
where $\kappa :=\mathit{\ell }_{s}^{3}/6\hbar $ and $\hbar $ are treated as
independent small deformation parameters (in some formulas, we shall us as
parameters the values $\mathit{\ell }_{s}$ and $\hbar $). The twist
d-operators can be parameterized in terms of respective elements with dyadic
splitting of indices $\mathfrak{f}^{\alpha _{s}},\mathfrak{f}_{\beta
_{s}}\in $ $U\emph{Vec}(\mathcal{M})$ when
\begin{eqnarray*}
\ ^{\shortparallel }\mathcal{\check{F}}_{N}&:=&\mathfrak{f}^{\alpha }\otimes
\mathfrak{f}_{\alpha }=\ ^{\shortparallel }\mathcal{\check{F}}_{s}=\mathfrak{%
f}^{\alpha _{s}}\otimes \mathfrak{f}_{\alpha _{s}}=1\otimes 1+\mathit{O}%
(\hbar ,\mathit{\ell }_{s}^{3}), \\
&&\mbox{ with inverse twist }\ ^{\shortparallel }\mathcal{\check{F}}%
_{N}^{-1} :=\overline{\mathfrak{f}}^{\alpha }\otimes \overline{\mathfrak{f}}%
_{\alpha }=\ ^{\shortparallel }\mathcal{\check{F}}_{s}^{-1}:=\overline{%
\mathfrak{f}}^{\alpha _{s}}\otimes \overline{\mathfrak{f}}_{\alpha _{s}}; \\
\ ^{\shortparallel }\mathcal{\check{F}}_{RN}&:= &\mathfrak{f}_{RN}^{\alpha
}\otimes \mathfrak{f}_{\alpha }^{RN}=\ ^{\shortparallel }\mathcal{\check{F}}%
_{Rs}:=\mathfrak{f}_{Rs}^{\alpha _{s}}\otimes \mathfrak{f}_{\alpha
_{s}}^{Rs}=1\otimes 1+\mathit{O}(\kappa ),
\end{eqnarray*}%
where summation on low-up repeating indices and possible shall by shall
splitting is understood.

A Hopf s-algebra $U\emph{Vec}(\mathcal{M},\ _{s}^{\shortparallel }N)$ is
determined by such N-adapted structures:

\begin{description}
\item a co-product $\blacktriangle $ defined as $\blacktriangle (1)=1\otimes
1,\blacktriangle (\ ^{\shortparallel }\mathbf{e}_{\alpha _{s}})=1\otimes \
^{\shortparallel }\mathbf{e}_{\alpha _{s}}+\ ^{\shortparallel }\mathbf{e}%
_{\alpha _{s}}\otimes 1;$\

\item a co-unit $\epsilon $ defined as $\epsilon (1)=1,\epsilon (\
^{\shortparallel} \mathbf{e}_{\alpha _{s}})=0;$ and

\item and an antipode $\mathbf{S}$ defined as $\mathbf{S}(1)=1, \mathbf{S}(\
^{\shortparallel }\mathbf{e}_{\alpha _{s}})=-\ ^{\shortparallel }\mathbf{e}%
_{\alpha _{s}}$; such $\blacktriangle $ and $\epsilon $ d-operators are
extended to the total space $U\emph{Vec}(\mathcal{M},N)$ as d-algebra
homomorphisms and $\mathbf{S}$ extended as an d-algebra anti-homomorphism
(linear, anti-multiplicative and N-adapted). The N-connection s-splitting
states a s-decomposition $\blacktriangle =(h\blacktriangle ,\
^{s}v\blacktriangle ).$
\end{description}

A quasi-Hopf s-algebra $U\emph{Vec}^{\mathcal{F}}(\mathcal{M},\
_{s}^{\shortparallel }N)$ is generated as an extension in s-adapted form
with power series in $\hbar $ and $\kappa $ of a Hopf d-algebra $U\emph{Vec}(%
\mathcal{M},\ _{s}^{\shortparallel }N)$ using a twist $\ ^{\shortparallel }%
\mathcal{F}_{s}$ (\ref{twisthopf}). The algebraic Hopf s-structure is
preserved for respective $\mathcal{F}$-extended s-objects: co-product $%
\blacktriangle _{\mathcal{F}}=\mathcal{F}_{N}\blacktriangle \mathcal{F}%
_{N}^{-1} = \blacktriangle _{\mathcal{F}}=\mathcal{F}_{s}\blacktriangle
\mathcal{F}_{s}^{-1};$ quasi-antipode $\mathbf{S}_{\mathcal{F}}=\mathbf{S}$;
and co-unit $\epsilon _{\mathcal{F}}=\epsilon .$ Such s-operators satisfy
the properties:%
\begin{eqnarray*}
&&\blacktriangle _{\mathcal{F}}(\ ^{\shortparallel }\mathbf{e}_{i_{s}})
=1\otimes \ ^{\shortparallel }\mathbf{e}_{i_{s}}+\ ^{\shortparallel }\mathbf{%
e}_{i_{s}}\otimes 1,\blacktriangle _{\mathcal{F}}(\ ^{\shortparallel
}e^{a_{s}})=1\otimes \ ^{\shortparallel }e^{a_{s}}+\ ^{\shortparallel
}e^{a_{s}}\otimes 1+i\kappa R^{j_{s}k_{s}a_{s}}\ ^{\shortparallel }\mathbf{e}%
_{j_{s}}\otimes \ ^{\shortparallel }\mathbf{e}_{k_{s}}\mbox{ and } \\
&&\mathfrak{f}^{\alpha }\ \mathbf{S(}\mathfrak{f}_{\alpha })=\mathfrak{f}%
_{RN}^{\alpha }\ \mathbf{S}(\mathfrak{f}_{\alpha }^{RN})=\mathfrak{f}%
^{\alpha _{s}}\ \mathbf{S(}\mathfrak{f}_{\alpha _{s}})=\mathfrak{f}%
_{Rs}^{\alpha _{s}}\ \mathbf{S(}\mathfrak{f}_{\alpha _{s}}^{Rs})= \\
&&\overline{\mathfrak{f}}^{\alpha }\ \mathbf{S}(\overline{\mathfrak{f}}%
_{\alpha })=\overline{\mathfrak{f}}_{RN}^{\alpha }\ \mathbf{S}(\overline{%
\mathfrak{f}}_{\alpha }^{RN})=\overline{\mathfrak{f}}^{\alpha _{s}}\ \mathbf{%
S(}\overline{\mathfrak{f}}_{\alpha _{s}})=\overline{\mathfrak{f}}%
_{Rs}^{\alpha _{s}}\ \mathbf{S(}\overline{\mathfrak{f}}_{\alpha
_{s}}^{Rs})=1.
\end{eqnarray*}%
In holonomic and N- or s-adapted variants twist d-operators $\
^{\shortparallel }\mathcal{F}_{N}$ or $\ ^{\shortparallel }\mathcal{F}_{s}$
do not fulfill the 2-cocycle condition. In such cases, one considers
respective associator N- and s-operators. For Hopf d-algebraic structures,
we use respective associator operations $\Phi $ (defining a Hopf 3-cocycle;
and its inverse associator $\Phi ^{-1},$%
\begin{eqnarray*}
\Phi _{N} &=&\exp \left( \frac{\mathit{\ell }_{s}^{3}}{6}R^{n+j\ n+k\ n+i}\
^{\shortparallel }\mathbf{e}_{j}\otimes \ ^{\shortparallel }\mathbf{e}%
_{k}\otimes \ ^{\shortparallel }\mathbf{e}_{i}\right) = \\
\Phi _{s} &=&\exp \left( \frac{\mathit{\ell }_{s}^{3}}{6}R^{n+j_{s}\
n+k_{s}\ n+i_{s}}\ ^{\shortparallel }\mathbf{e}_{j_{s}}\otimes \
^{\shortparallel }\mathbf{e}_{k_{s}}\otimes \ ^{\shortparallel }\mathbf{e}%
_{i_{s}}\right) =\ _{1}\phi \otimes \ _{2}\phi \otimes \ _{3}\phi =1\otimes
1\otimes 1+O(\mathit{\ell }_{s}^{3}), \\
&&\mbox{ and }\Phi _{N}^{-1}=\Phi _{s}^{-1}=\ _{1}\overline{\phi }\otimes \
_{2}\overline{\phi }\otimes \ _{3}\overline{\phi }.
\end{eqnarray*}%
For instance, for s-operators, the co-product $\blacktriangle _{\mathcal{F}}$
is not co-associative because the 2-cocycle condition is not satisfied and
the property
\begin{equation*}
\Phi _{s}(\ ^{\shortparallel }\mathcal{F}_{s}\otimes 1)(\blacktriangle
id\otimes 1)\ ^{\shortparallel }\mathcal{F}_{s}=(1\otimes \ ^{\shortparallel}%
\mathcal{F}_{s}) (\blacktriangle id\otimes 1)\ ^{\shortparallel}\mathcal{F}%
_{s}
\end{equation*}%
results in a quasi-associativity condition
\begin{equation*}
\Phi _{s}(\blacktriangle _{\mathcal{F}}id\otimes 1)\blacktriangle _{\mathcal{%
F}}(\xi )=(id\otimes \blacktriangle _{\mathcal{F}})\blacktriangle _{\mathcal{%
F}}(\xi )\Phi _{s},\forall \xi \in U\emph{Vec}(\mathcal{M},\
_{s}^{\shortparallel }N).
\end{equation*}%
Considering dyadic decompositions, we use a sextuple $\left( U\emph{Vec}(%
\mathcal{M}, \ _{s}^{\shortparallel }N),\bullet ,\blacktriangle _{\mathcal{F}%
},\Phi _{s}, \mathbf{S,}\epsilon \right) $ and define on the d-vector space $%
U\emph{Vec}(\mathcal{M},\ _{s}^{\shortparallel }N),$ i.e. on $\ _{s}T%
\mathcal{M}=TT_{\shortparallel s}^{\ast }\mathbf{V,}$ a structure of
quasi-Hopf s-algebra $U\emph{Vec}^{\mathcal{F}}(\mathcal{M},\
_{s}^{\shortparallel }N).$ For our constructions, there are we considered
boldface d-operators in order to emphasize that we work with s-adapted
algebraic and geometric structures. We substitute $\ ^{\shortparallel }%
\mathbf{e}_{\alpha _{s}}\rightarrow \ ^{\shortparallel }\mathbf{\partial }%
_{\alpha _{s}}$ and changing respectively $\blacktriangle _{\mathcal{F}%
}\rightarrow \triangle _{\mathcal{F}},\Phi _{s}\rightarrow \Phi ,\mathbf{S}%
\rightarrow S$ etc., we obtain the definition of standard Hopf algebra $U%
\emph{Vec}(\mathcal{M})$ and coordinate base constructions as in \cite%
{drinf,aschieri17,mylonas13}. For N-connection splitting, such d-algebras
were introduced in \cite{partner01}.

In nonholonomic dyadic form, we can formulate such principles of s-adapted
twist deformation: We write $\mathcal{A}_{s}=(h\mathcal{A},~^{s}v\mathcal{A}%
) $ for a d-algebra with s-adapted representation of a Hopf d-algebra $U
\emph{Vec}(\mathcal{M},\ _{s}^{\shortparallel }N)$ and a twist deformation
induced by a R-flux into a quasi-Hopf d-algebra $U\emph{Vec}^{\mathcal{F}}(%
\mathcal{M},\ _{s}^{\shortparallel }N).$ Any d-vector $\vartheta
=(h\vartheta ,~^{s}v\vartheta )\in \emph{Vec}(\mathcal{M},\
_{s}^{\shortparallel }N)$ acts on $\mathcal{A}$ $_{s}.$ This action is
performed following such steps:

\begin{description}
\item 1) For every algebraic product $ab,$ we consider $\vartheta
(ab)=u(a)b+au(b)$ and we have a \ $U\emph{Vec}(\mathcal{M},\
_{s}^{\shortparallel }N)$-module d-algebra $\mathcal{A}_{s}.$

\item 2) This multiplication is deformed into a N- and s-adapted star
multiplication
\begin{eqnarray}
a\star _{N}b &=&\overline{\mathfrak{f}}^{\alpha }(a)\overline{\mathfrak{f}}%
_{\alpha }(b)=\overline{\mathfrak{f}}^{i}(a)\overline{\mathfrak{f}}_{i}(b)+\
^{\shortparallel }\overline{\mathfrak{f}}_{c}(a)\ ^{\shortparallel }%
\overline{\mathfrak{f}}^{c}(b)=  \label{starmult} \\
a\star _{s}b &=&\overline{\mathfrak{f}}^{\alpha _{s}}(a)\overline{\mathfrak{f%
}}_{\alpha _{s}}(b)=\overline{\mathfrak{f}}^{i_{s}}(a)\overline{\mathfrak{f}}%
_{i_{s}}(b)+\ ^{\shortparallel }\overline{\mathfrak{f}}_{c_{s}}(a)\
^{\shortparallel }\overline{\mathfrak{f}}^{c_{s}}(b)  \notag
\end{eqnarray}
resulting in a noncommmutative and nonassociative s-algebra $\mathcal{A}%
_{s}^{\star }=(h\mathcal{A}^{\star },~^{s}v\mathcal{A}^{\star }).$
\end{description}

Any $\mathcal{A}_{N}^{\star }$ and/or carries $\mathcal{A}_{s}^{\star }$ a
respective representation of the quasi-Hopf d-algebra $U\emph{Vec}^{\mathcal{%
F}}(\mathcal{M},\ _{s}^{\shortparallel }N)$ because for any element $\xi $
of this d-algebra we prove in N- and s-adapted form that
\begin{eqnarray*}
\xi (a\star _{N}b) &=&\xi \left( \overline{\mathfrak{f}}^{\alpha }(a)%
\overline{\mathfrak{f}}_{\alpha }(b)\right) =\xi _{(1_{0})}\left( \overline{%
\mathfrak{f}}^{\alpha }(a))\xi _{(2_{0})}(\overline{\mathfrak{f}}_{\alpha
}(b)\right) =\overline{\mathfrak{f}}^{\alpha }(\xi _{(1_{0})}(a))\xi
_{(2_{0})}(\overline{\mathfrak{f}}_{\alpha }(b))= \\
\xi (a\star _{s}b) &=&\xi \left( \overline{\mathfrak{f}}^{\alpha _{s}}(a)%
\overline{\mathfrak{f}}_{\alpha _{s}}(b)\right) =\xi _{(1_{0})}\left(
\overline{\mathfrak{f}}^{\alpha _{s}}(a))\xi _{(2_{0})}(\overline{\mathfrak{f%
}}_{\alpha _{s}}(b)\right) =\overline{\mathfrak{f}}^{\alpha _{s}}(\xi
_{(1_{0})}(a))\xi _{(2_{0})}(\overline{\mathfrak{f}}_{\alpha _{s}}(b)).
\end{eqnarray*}%
In such formulas, there are used un-deformed N- and s-adapted co-products:%
\begin{eqnarray*}
\blacktriangle (\xi ) &=&\xi _{(1_{0})}\otimes _{N}\xi _{(2_{0})}%
\mbox{ when
}\blacktriangle (\xi )\mathcal{F}_{N}^{-1}=\mathcal{F}_{N}^{-1}%
\blacktriangle _{\mathcal{F}}(\xi ),\mbox{ and } \\
\blacktriangle (\xi ) &=&\xi _{(1_{0})}\otimes _{s}\xi _{(2_{0})}%
\mbox{ when
}\blacktriangle (\xi )\mathcal{F}_{s}^{-1}=\mathcal{F}_{s}^{-1}%
\blacktriangle _{\mathcal{F}}(\xi ).
\end{eqnarray*}%
Such constructions can be used for N- and s-adapting other type of algebraic
structures (for instance, Clifford algebras, octonionic algebras etc.).

\paragraph{Hopf s-algebras, dyadic N-adapted Lie derivatives, and $\mathcal{R%
}$ action on functions on $T_{\shortparallel }^{\ast }\mathbf{V}$: \newline
}

Hereafter, we provide only the formulas for s-structures. For a commutative
d-algebra $\mathcal{A}_{s}$ and it s-adapted star deformation $\mathcal{A}%
_{s}^{\star }$, we can control noncommutativity via shall by shall action of
$\mathcal{R}$-matrix, $\mathcal{R}_{s}=\mathcal{F}_{s}^{-2},$ when for the
star product (\ref{starmult}) of two elements $a$ and $b$ is computed
\begin{eqnarray*}
a\star _{s}b &=&\overline{R}^{\gamma }(b)\star _{s}\overline{R}_{\gamma
}(a):=b^{\overline{\gamma }}\star _{s}a_{\overline{\gamma }}=b^{\overline{k}%
}\star _{s}a_{\overline{k}}+\ ^{\shortparallel }b_{_{\overline{c}}}\star
_{N}\ ^{\shortparallel }a^{_{\overline{c}}}, \\
&=&\overline{\mathfrak{f}}^{\alpha _{s}}(a)\overline{\mathfrak{f}}_{\alpha
_{s}}(b)=\overline{\mathfrak{f}}_{\alpha _{s}}(b)\overline{\mathfrak{f}}%
^{\alpha _{s}}(a),\mbox{ where }b^{\overline{\gamma }_{s}}:=\overline{R}%
^{\gamma _{s}}(b)\mbox{ and }a_{\overline{\gamma }_{s}}:=\overline{R}%
_{\gamma _{s}}(a).
\end{eqnarray*}%
For an associative d-algebra $\mathcal{A}_{s}$ is associative, we can
consider the associator d-operator%
\begin{equation}
\Phi :(a\star _{s}b)\star _{s}c=\ ^{\phi _{1}}a\star _{s}(\ ^{\phi
_{2}}b\star _{s}\ ^{\phi _{3}}c),  \label{aux13}
\end{equation}%
where, for instance, $\ ^{\phi _{1}}a=\phi _{1}(a).$

To understand how the quasi-Hopf and dyadic structures can be adapted to act
in a distinguished self-consistent form we can consider a simplest case of
algebraic s-operations for a function $q\in C^{\infty }(\mathcal{M},\
_{s}^{\shortparallel }N).$ This can be for the space of functions is $%
C^{\infty }(\mathcal{M)}$ when the usual partial derivatives can be
transformed into N-adapted and/or s-adapted ones. The Lie d-derivative $%
\mathcal{L}_{\xi }(q):=\xi (q)$ can be defined in s-adapted form when any
d-vector field $\xi =\xi ^{\beta _{s}} \ ^{\shortparallel }\mathbf{e}_{\beta
_{s}}$ is subjected to a s-adapted dyadic decomposition$.$ The action of the
Lie s-algebra $\emph{Vec}(\mathcal{M},\ _{s}^{\shortparallel }N)$ on such
functions can be extended to an action of the universal enveloping d-algebra
$U\emph{Vec}(\mathcal{M},\ _{s}^{\shortparallel }N)$ if we define the Lie
d-derivative (we may write equivalently s-derivative, respectively,
s-vectors) on products of s-vectors $\mathcal{L}_{\ ^{1}\xi \ ^{2}\xi ...\
^{m}\xi }:=\mathcal{L}_{\ ^{1}\xi \ }\circ \mathcal{L}_{\ ^{2}\xi }\circ
...\circ \mathcal{L}_{\ ^{m}\xi }$ and considering decompositions by
linearity.

At the next step, we deform\ with power series on $\hbar $ and $\kappa $ the
$U\emph{Vec}(\mathcal{M},\ _{s}^{\shortparallel }N)$-module d-algebra $%
C^{\infty }(\mathcal{M},\ _{s}^{\shortparallel }N)$ into $U\emph{Vec}^{%
\mathcal{F}}(\mathcal{M},\ _{s}^{\shortparallel }N)$-module d-algebra $%
\mathcal{A}_{s}^{\star }:=C_{\star }^{\infty }(\mathcal{M},\
_{s}^{\shortparallel }N)$ for which the d-vector space is considered the
same as $C^{\infty }(\mathcal{M},\ _{s}^{\shortparallel }N)$ but with
multiplication defined by a s-adapted star product (\ref{starpn}). We state
that for two functions $q,z\in C^{\infty }(\mathcal{M},\
_{s}^{\shortparallel }N),$ the s-adapted star product possess such dyadic
properties:%
\begin{eqnarray}
q\star _{s}z &=&\overline{\mathfrak{f}}^{\alpha _{s}}(q)\cdot \overline{%
\mathfrak{f}}_{\alpha _{s}}(z)  \label{starpnh} \\
&=&\overline{R}^{\alpha _{s}}(z)\star _{s}\overline{R}_{\alpha _{s}}(q):=z^{%
\overline{\gamma }_{s}}\star _{s}q_{\overline{\gamma }_{s}}%
\mbox{ controls
noncommutativity };  \notag \\
\mbox{ and }\Phi &:&(z\star _{s}q)\star _{s}f=\ ^{\phi _{1}}z\star _{s}(\
^{\phi _{2}}q\star _{s}\ ^{\phi _{3}}f)\mbox{ controls nonassociativity}.
\notag
\end{eqnarray}%
In these formulas, the constant function $1$ on $\mathcal{M}$ is also the
unit of the star s-algebra $\mathcal{A}_{s}^{\star }$ which is stated by the
property that $q\star _{s}1=q=1\star _{s}q.$

We can consider a s-adapted star commutator of functions, $\left[ q,z\right]
_{\star s}= q\star _{s}z-z\star _{s}q,$ and define on nonholonomic dyadic
decompositions on phase space $(\mathcal{M},\ _{s}^{\shortparallel }N)$ a
quasi-Poisson coordinate s-algebra:
\begin{equation*}
\left[ x^{j_{s}},x^{k_{s}}\right] _{\star s}=2i\kappa R^{n+j_{s}\ n+k_{s}\
a_{s}}p_{a_{s}},\left[ x^{i_{s}},p_{n+j_{s}}\right] _{\star s}=i\hbar \delta
_{\ j_{s}}^{i_{s}}\mbox{ and }\left[ p_{n+i_{s}},p_{n+j_{s}}\right] _{\star
s}=0,
\end{equation*}%
for a parabolic R-flux nonholonomic background, and a nontrivial Jacobiator%
\begin{equation*}
\left[ x^{i_{s}},x^{j_{s}},x^{k_{s}}\right] _{\star _{s}}=\mathit{\ell }%
_{s}^{3}R^{n+i_{s}\ n+j_{s}\ n+k_{s}}.
\end{equation*}%
We note that such formal coordinate functions on $T_{\shortparallel s}^{\ast}%
\mathbf{V}$ are computed with indices determined with respect to $\
^{\shortparallel }\mathbf{e}_{\beta _{s}}$ (\ref{nadapbdsc}) and following
the conventions on s-coordinates (\ref{loccord}).

\subsection{Forms and d-tensors for quasi-Hopf s-adapted structures}

\subsubsection{Nonassociative dyadic deformations of star exterior products}

We denote the exterior algebra of differential forms with dyadic
decomposition $\Omega ^{\natural }(\mathcal{M},\ _{s}^{\shortparallel }N)$
and study N-adapted nonassociative nonholonomic s-deformations of star
exterior products, which is written $\Omega _{\star }^{\natural }(\mathcal{M}%
,\ _{s}^{\shortparallel }N).$ The constructions start with zero-forms (which
are functions with star s-adapted quasi-Hopf product (\ref{starpnh}), $%
\Omega _{\star }^{0}(\mathcal{M},\ _{s}^{\shortparallel }N)=\mathcal{A}%
_{s}^{\star }.$ Then, the algebra of differential forms is defined in a
standard nonholonomic form but with nonassociative product $\wedge _{\star
_{s}}.$ We state that for any couple of 1-forms $\ ^{\shortparallel }\omega
=\{\ _{s}^{\shortparallel }\omega \},\ ^{\shortparallel }\gamma =\{\
_{s}^{\shortparallel }\gamma \}\in $ $\Omega _{\star }^{1}(\mathcal{M},\
_{s}^{\shortparallel }N)$
\begin{equation*}
\ ^{\shortparallel }\omega \wedge _{\star _{N}}\ ^{\shortparallel }\gamma =\
^{\shortparallel }\omega \wedge _{\star _{s}}\ ^{\shortparallel }\gamma =\
_{s}^{\shortparallel }\omega \wedge _{\star s}\ _{s}^{\shortparallel }\gamma
=\overline{\mathfrak{f}}^{\alpha }(\ ^{\shortparallel }\omega )\cdot
\overline{\mathfrak{f}}_{\alpha }(\ ^{\shortparallel }\gamma )=\overline{%
\mathfrak{f}}^{\alpha _{s}}(\ _{s}^{\shortparallel }\omega )\cdot \overline{%
\mathfrak{f}}_{\alpha _{s}}(\ _{s}^{\shortparallel }\gamma ).
\end{equation*}%
The exterior derivative can be written in s-adapted form,%
\begin{equation*}
\ ^{\shortparallel }\mathbf{d=\{\ _{s}^{\shortparallel }\mathbf{d}\}:\ }%
\Omega _{\star }^{\natural }(\mathcal{M},\ _{s}^{\shortparallel
}N)\rightarrow \Omega _{\star }^{\natural +1}(\mathcal{M},\
_{s}^{\shortparallel }N).
\end{equation*}%
This allows us to work with the un-deformed Leibniz rule,%
\begin{eqnarray*}
\ ^{\shortparallel }\mathbf{d}(\ ^{\shortparallel }\omega \wedge _{\star
_{N}}\ ^{\shortparallel }\gamma ) &=&\ ^{\shortparallel }\mathbf{d}\
^{\shortparallel }\omega \wedge _{\star _{N}}\ ^{\shortparallel }\gamma
+(-1)^{\mid \omega \mid }\ ^{\shortparallel }\omega \wedge _{\star _{N}}\
^{\shortparallel }\mathbf{d}\ ^{\shortparallel }\gamma = \\
\ ^{\shortparallel }\mathbf{d}(\ ^{\shortparallel }\omega \wedge _{\star
_{s}}\ ^{\shortparallel }\gamma ) &=&\ _{s}^{\shortparallel }\mathbf{d}(\
_{s}^{\shortparallel }\omega \wedge _{\star _{s}}\ _{s}^{\shortparallel
}\gamma )=\ _{s}^{\shortparallel }\mathbf{d}\ _{s}^{\shortparallel }\omega
\wedge _{\star s}\ _{s}^{\shortparallel }\gamma +(-1)^{\mid \omega \mid }\
_{s}^{\shortparallel }\omega \wedge _{\star _{N}}\ _{s}^{\shortparallel }%
\mathbf{d}\ _{s}^{\shortparallel }\gamma ,
\end{eqnarray*}%
where $\ _{s}^{\shortparallel }\omega $ is a homogeneous form of degree $%
|\omega |.$ We consider that $\ _{s}^{\shortparallel }\mathbf{d}\overline{%
\mathfrak{f}}^{\alpha _{s}}(\ _{s}^{\shortparallel }\omega )=\overline{%
\mathfrak{f}}^{\alpha _{s}}(\ _{s}^{\shortparallel }\mathbf{d}\
_{s}^{\shortparallel }\omega )$ and $\ _{s}^{\shortparallel }\mathbf{d}%
\overline{\mathfrak{f}}_{\alpha _{s}}(\ _{s}^{\shortparallel }\omega )=%
\overline{\mathfrak{f}}_{\alpha _{s}}(\ _{s}^{\shortparallel }\mathbf{d}\
_{s}^{\shortparallel }\omega ).$ Here, we emphasize that we shall use
symbols $\ ^{\shortparallel }\mathbf{d}$ or $\ _{s}^{\shortparallel }\mathbf{%
\mathbf{d}}$ depending on the fact if it is necessary to state explicitly,
or not, a s-decomposition of corresponding d- / s-operators.

A corresponding s-adapted exterior differential calculus can be elaborated
via exterior products between functions, i.e. 0-forms and 1-forms, and
respective s-adapted star deformations. This way, we generate the $%
C^{\infty}(\mathcal{M},\ _{s}^{\shortparallel }N)$-bimodule and resulting $%
\mathcal{A}_{s}^{\star }$--bimodule s-structure of spaces of 1-forms, which
for dyadic decompositions we denote $\Omega _{s}^{1}(\mathcal{M)}=\Omega
^{1}(\mathcal{M},\ _{s}^{\shortparallel }N)$ and $\Omega _{\star s}^{1}(%
\mathcal{M)=}\Omega _{\star }^{1}(\mathcal{M},\ _{s}^{\shortparallel }N).$

For a s-adapted differential form nonassociative calculus, we present some
important formulas when the s-components
\begin{eqnarray*}
f\star _{N}\ ^{\shortparallel }e^{j} &=&f\cdot \ ^{\shortparallel }e^{j}-%
\frac{\kappa }{2}R^{jkb}\mathbf{e}_{k}f\cdot \ ^{\shortparallel }\mathbf{e}%
_{b}=\ ^{\shortparallel }\mathbf{e}^{j}\star _{N}f-\ ^{\shortparallel }%
\mathbf{e}_{b}\star _{N}\kappa R^{jkb}\mathbf{e}_{k}f \\
f\star _{s}\ ^{\shortparallel }e^{j_{s}} &=&f\cdot \ ^{\shortparallel
}e^{j_{s}}-\frac{\kappa }{2}R^{j_{s}k_{s}b}\mathbf{e}_{k_{s}}f\cdot \
^{\shortparallel }\mathbf{e}_{b_{s}}=\ ^{\shortparallel }\mathbf{e}%
^{j_{s}}\star _{s}f-\ ^{\shortparallel }\mathbf{e}_{b_{s}}\star _{s}\kappa
R^{j_{s}k_{s}b_{s}}\mathbf{e}_{k_{s}}f, \\
f\star _{N}\ ^{\shortparallel }\mathbf{e}_{a} &=&f\cdot \ ^{\shortparallel }%
\mathbf{e}_{a}=\ ^{\shortparallel }\mathbf{e}_{a}\star _{N}f\rightarrow
f\star _{s}\ ^{\shortparallel }\mathbf{e}_{a_{s}}=f\cdot \ ^{\shortparallel }%
\mathbf{e}_{a_{s}}=\ ^{\shortparallel }\mathbf{e}_{a_{s}}\star _{s}f
\end{eqnarray*}%
are written with s-indices. We have also
\begin{equation*}
f\star _{N}\ ^{\shortparallel }\mathbf{e}^{\alpha }=\ ^{\shortparallel }%
\mathbf{e}^{\gamma }\star _{N}(f\delta _{\ \gamma }^{\alpha }-i\kappa
\mathcal{R}_{\quad \gamma }^{\alpha \beta }\ ^{\shortparallel }\mathbf{e}%
_{\beta }f)\rightarrow f\star _{N}\ ^{\shortparallel }\mathbf{e}^{\alpha
_{s}}=\ ^{\shortparallel }\mathbf{e}^{\gamma _{s}}\star _{s}(f\delta _{\
\gamma _{s}}^{\alpha _{s}}-i\kappa \mathcal{R}_{\quad \gamma _{s}}^{\alpha
_{s}\beta _{s}}\ ^{\shortparallel }\mathbf{e}_{\beta _{s}}f),
\end{equation*}%
when the absolute differential is star deformed in s-adapted form%
\begin{eqnarray*}
\ ^{\shortparallel }\mathbf{d}f &=&\ ^{\shortparallel }\mathbf{e}_{\beta }f\
^{\shortparallel }\mathbf{e}^{\beta }=\ ^{\shortparallel }\mathbf{e}_{\beta
}f\star _{N}\ ^{\shortparallel }\mathbf{e}^{\beta }=\ ^{\shortparallel }%
\mathbf{e}^{\beta }\star _{N}\ ^{\shortparallel }\mathbf{e}_{\beta }f= \\
\ _{s}^{\shortparallel }\mathbf{d}f &=&\ ^{\shortparallel }\mathbf{e}_{\beta
_{s}}f\ ^{\shortparallel }\mathbf{e}^{\beta _{s}}=\ ^{\shortparallel }%
\mathbf{e}_{\beta _{s}}f\star _{s}\ ^{\shortparallel }\mathbf{e}^{\beta
_{s}}=\ ^{\shortparallel }\mathbf{e}^{\beta _{s}}\star _{s}\
^{\shortparallel }\mathbf{e}_{\beta _{s}}f.
\end{eqnarray*}%
Such formulas can be re-defined in local coordinate bases still keeping, or
not, a shell s-splitting of indices,
\begin{eqnarray*}
f\star _{N}\ dx^{j} &=&f\cdot \ dx^{j}-\frac{i\kappa }{2}R^{jkb}\frac{%
\partial f}{\partial x^{k}}\cdot dp_{b}=\ dx^{j}\star _{N}f-\ dp_{b}\star
_{N}i\kappa R^{jkb}\frac{\partial f}{\partial x^{k}}\rightarrow \\
f\star _{s}\ dx^{j_{s}} &=&f\cdot \ dx^{j_{s}}-\frac{i\kappa }{2}%
R^{j_{s}k_{s}b_{s}}\frac{\partial f}{\partial x^{k_{s}}}\cdot dp_{b_{s}}=\
dx^{j_{s}}\star _{s}f-\ dp_{b_{s}}\star _{s}i\kappa R^{j_{s}k_{s}b_{s}}\frac{%
\partial f}{\partial x^{k_{s}}}, \\
f\star _{N}dp_{a} &=&f\cdot \ dp_{a}=dp_{a}\star _{N}f\rightarrow f\star
_{s}dp_{a_{s}}=f\cdot \ dp_{a_{s}}=dp_{a_{s}}\star _{s}f.
\end{eqnarray*}

Above presented formalism can be extended to d-tensors involving N- and
s-connections and quasi-Hopf s-structures. All such formulas are similar to
those for N-adapted constructions in if $\ ^{\shortparallel }\mathbf{e}%
_{\beta _{s}}\rightarrow \ ^{\shortparallel }\mathbf{e}_{\beta }$ and $\star
_{s}\rightarrow \star _{N},$ etc. In this work, we shall omit proofs and
details if certain (non) associative formulas are s-analogs of respective
N-adapted ones.

\subsubsection{s-tensor and dyadic quasi-Hopf star deformations}

\label{ssdtqh}We can deform always a s-adapted tensor product $\otimes
_{C^{\infty }(\mathcal{M},\ _{s}^{\shortparallel }N)}$ over $C^{\infty }(%
\mathcal{M},\ _{s}^{\shortparallel }N)$ into a s-tensor product $%
\otimes_{\star _{s}}$ over s-algebra $\mathcal{A}_{s}^{\star }.$ We follow
the convention
\begin{eqnarray*}
\ ^{\shortparallel }\mathbf{B}\otimes _{\star _{N}}\ ^{\shortparallel }%
\mathbf{Q} &=&\overline{\mathfrak{f}}^{\alpha }(\ ^{\shortparallel }\mathbf{B%
})\otimes _{C^{\infty }(\mathcal{M},N)}\overline{\mathfrak{f}}_{\alpha }(\
^{\shortparallel }\mathbf{Q})= \\
\ ^{\shortparallel }\mathbf{B}\otimes _{\star _{s}}\ ^{\shortparallel }%
\mathbf{Q} &=&\overline{\mathfrak{f}}^{\alpha _{s}}(\ ^{\shortparallel }%
\mathbf{B})\otimes _{C^{\infty }(\mathcal{M},\ _{s}^{\shortparallel }N)}%
\overline{\mathfrak{f}}_{\alpha _{s}}(\ ^{\shortparallel }\mathbf{Q}),
\end{eqnarray*}%
where the s-adapted twist on the s-tensors $\ ^{\shortparallel }\mathbf{B}$
and $\ ^{\shortparallel }\mathbf{Q}$ is taken via the Lie s-derivative and
the nonassociativity for any $a\in \mathcal{A}_{N}^{\star },$ is involved as
\begin{eqnarray*}
(\ ^{\shortparallel }\mathbf{B}\star _{N}a)\otimes _{\star _{N}}\
^{\shortparallel }\mathbf{Q} &\mathbf{=}&\ _{\phi _{1}}^{\shortparallel }%
\mathbf{B}\otimes _{\star _{N}}(\ _{\phi _{2}}a\star _{N}\ _{\phi
_{3}}^{\shortparallel }\mathbf{Q)=} \\
(\ ^{\shortparallel }\mathbf{B}\star _{s}a)\otimes _{\star _{s}}\
^{\shortparallel }\mathbf{Q} &\mathbf{=}&\ _{\phi _{1}}^{\shortparallel }%
\mathbf{B}\otimes _{\star _{s}}(\ _{\phi _{2}}a\star _{s}\ _{\phi
_{3}}^{\shortparallel }\mathbf{Q).}
\end{eqnarray*}%
We cite for details on N-adapted actions of associators $\phi _{1},\phi
_{2}, $ and $\phi _{3}$ the section 2.6.2 in \cite{partner01} having a
straightforward s-structure re-formulation.

The N- and s-adapted star product of a function $f$ and s-components of
partial derivatives
\begin{eqnarray*}
f\star _{N}\ ^{\shortparallel }\mathbf{e}_{i} &=&f\cdot \ ^{\shortparallel }%
\mathbf{e}_{i}=\ ^{\shortparallel }\mathbf{e}_{i}\star _{N}f\rightarrow
f\star _{s}\ ^{\shortparallel }\mathbf{e}_{i_{s}}=f\cdot \ ^{\shortparallel }%
\mathbf{e}_{i_{s}}=\ ^{\shortparallel }\mathbf{e}_{i_{s}}\star _{s}f, \\
f\star _{N}\ ^{\shortparallel }e^{a} &=&f\cdot \ ^{\shortparallel }e^{a}-%
\frac{i\kappa }{2}R^{jka}\mathbf{e}_{j}f\cdot \ ^{\shortparallel }\mathbf{e}%
_{k}=\ ^{\shortparallel }e^{a}\star _{N}f-\ ^{\shortparallel }\mathbf{e}%
_{j}\star _{N}i\kappa R^{ajk}\mathbf{e}_{k}f\rightarrow \\
f\star _{s}\ ^{\shortparallel }e^{a_{s}} &=&f\cdot \ ^{\shortparallel
}e^{a_{s}}-\frac{i\kappa }{2}R^{j_{s}k_{s}a_{s}}\mathbf{e}_{j_{s}}f\cdot \
^{\shortparallel }\mathbf{e}_{k_{s}}=\ ^{\shortparallel }e^{a_{s}}\star
_{s}f-\ ^{\shortparallel }\mathbf{e}_{j_{s}}\star _{s}i\kappa
R^{a_{s}j_{s}k_{s}}\mathbf{e}_{k_{s}}f.
\end{eqnarray*}%
Here we nota that, for instance, $\ ^{\shortparallel }\mathbf{e}%
_{i_{s}}\star _{s}f$ denotes the right s-adapted \ $\mathcal{A}_{s}^{\star }$%
--action on $\emph{Vec}_{\star _{s}}$ (this is not on action of $\
^{\shortparallel }\mathbf{e}_{i_{s}}$ on the function $f$). For dyadic
indices,
\begin{equation*}
f\star _{s}\ ^{\shortparallel }\mathbf{e}_{\alpha _{s}}=\ ^{\shortparallel }%
\mathbf{e}_{\gamma _{s}}\star _{s}(f\delta _{\ \alpha _{s}}^{\gamma
_{s}}+i\kappa \mathcal{R}_{\quad \alpha _{s}}^{\gamma _{s}\beta _{s}}\
^{\shortparallel }\mathbf{e}_{\beta _{s}}f).
\end{equation*}%
In similar forms, using s-adapted star tensor products, we can extend the
dyadic $\mathcal{A}_{s}^{\star }$ --bimodule $\emph{Vec}_{\star _{s}}$ of
s-vector fields to the $\Omega _{\star }^{\natural }$--bimodule $\emph{Vec}%
_{\star _{s}}^{\natural }=\emph{Vec}_{\star _{s}}$ $\otimes _{\star _{s}}\
\Omega _{\star }^{\natural }(\mathcal{M},\ _{s}^{\shortparallel}N).$ This
way, we re-express the d-tensor and d-form $\star _{N}$-calculus from
section 2 of \cite{partner01} into a respective s-adapted one which allows
2+2+... decoupling of fundamental geometric and physical important formulas
(see next sections).

\section{Dyadic geometry of star deformed (non) symmetric s-metrics}

\label{sec3}Nonassociative geometric models with star R-flux deformed
symmetric and nonsymmetric metric structures and respective modified vacuum
Einstein equations for the LC-connection are elaborated in Refs. \cite%
{blumenhagen16,aschieri17} in abstract and coordinate frames. The goal of
this section is to re-define the generalizations with N-connection structure
from section 3.3 of \cite{partner01} in s-adapted frames which will be
applied in next section for proofs of decoupling and integrability of
nonassociative vacuum gravitational equations. The key constructions will be
based on performing nonholonomic dyadic decompositions of (non) symmetric
metric s-structures and respective canonical s-connection and nonholonomic
constraints to LC-connection structures. Nonassociative canonical Riemann
and Ricci s-tensors adapted to quasi-Hopf s-structure are defined and
computed following methods from appendix \ref{appendixa}.

\subsection{The geometry of phase space s-metrics and canonical s-connections%
}

The goal of this subsection is to study main properties of (non) symmetric
metric and related linear connection structures which are generated by N-
and s-adapted star deformations to quasi-Hopf structures on phase spaces $%
\mathcal{M}=\mathbf{T}_{\shortparallel }^{\ast }\mathbf{V.}$

\subsubsection{Symmetric and nonsymmetric s-metric structures and star
deformations}

A (pseudo) Riemannian symmetric metric on cotangent Lorentz bundle $T^{\ast
}V$ is defined by a metric $\ ^{\shortparallel }g=\{\ ^{\shortparallel
}g_{\alpha \beta }\}\in $ $TT^{\ast }V\otimes TT^{\ast }V,$ which for
nonassociative generalizations of Einstein gravity is chosen to be of local
signature $(+,+,+,-;+,+,+,-).$ Such a metric can be expressed in a
nonholonomic dyadic form for shells $s=1,2,3,4$, i.e. as a s-metric on phase
space $\mathcal{M}=\mathbf{T}_{\shortparallel }^{\ast }\mathbf{V}$, with
respect to symmetric tensor products of s-bases $\ ^{\shortparallel }\mathbf{%
\ e}^{\alpha _{s}}\in T_{s}^{\ast }\mathbf{T}_{\shortparallel }^{\ast }%
\mathbf{V}$ (\ref{nadapbdsc}),
\begin{eqnarray}
g &\rightarrow &\ _{s}^{\shortparallel }\mathbf{g}=(h_{1}\ ^{\shortparallel }%
\mathbf{g},v_{2}\ ^{\shortparallel }\mathbf{g,}c_{3}\ ^{\shortparallel }%
\mathbf{g,}c_{4}\ ^{\shortparallel }\mathbf{g})\in T\mathbf{T}%
_{\shortparallel }^{\ast }\mathbf{V}\otimes _{\star N}T\mathbf{T}%
_{\shortparallel }^{\ast }\mathbf{V}  \label{sdm} \\
&=&\ ^{\shortparallel }\mathbf{g}_{\alpha _{s}\beta _{s}}(\
_{s}^{\shortparallel }u)\ ^{\shortparallel }\mathbf{e}^{\alpha _{s}}\otimes
_{\star s}\ ^{\shortparallel }\mathbf{e}^{\beta _{s}}=\{\ ^{\shortparallel }%
\mathbf{g}_{\alpha _{s}\beta _{s}}=(\ ^{\shortparallel }\mathbf{g}%
_{i_{1}j_{1}},\ ^{\shortparallel }\mathbf{g}_{a_{2}b_{2}},\ ^{\shortparallel
}\mathbf{g}^{a_{3}b_{3}},\ ^{\shortparallel }\mathbf{g}^{a_{4}b_{4}})\}.
\notag
\end{eqnarray}%
In the case of R-flux deformations to nonassociative geometry, symmetric
metric transforms into symmetric and nonsymmetric ones \cite%
{blumenhagen16,aschieri17} (for nonholonomic N-adapted constructions, see
section 3.3 of \cite{partner01}). We use such s-adapted parameterizations
(in brief, we shall write s-metrics)
\begin{eqnarray}
&&\mbox{ symmetric: }\ _{\star s}^{\shortparallel }\mathbf{g}=(h_{1}\
_{\star s}^{\shortparallel }\mathbf{g},v_{2}\ _{\star s}^{\shortparallel }%
\mathbf{g,}c_{3}\ _{\star s}^{\shortparallel }\mathbf{g,}c_{4}\ _{\star
s}^{\shortparallel }\mathbf{g})  \label{ssdm} \\
&=&\{\ _{\star }^{\shortparallel }\mathbf{g}_{\alpha _{s}\beta _{s}}=\
_{\star }^{\shortparallel }\mathbf{g}_{\beta _{s}\alpha _{s}}=(\ _{\star
}^{\shortparallel }\mathbf{g}_{i_{1}j_{1}}=\ _{\star }^{\shortparallel }%
\mathbf{g}_{j_{1}i_{1}},\ _{\star }^{\shortparallel }\mathbf{g}%
_{a_{2}b_{2}}=\ _{\star }^{\shortparallel }\mathbf{g}_{b_{2}a_{2}},\ _{\star
}^{\shortparallel }\mathbf{g}^{a_{3}b_{3}}=\ \ _{\star }^{\shortparallel }%
\mathbf{g}^{b_{3}a_{3}},\ _{\star }^{\shortparallel }\mathbf{g}%
^{a_{4}b_{4}}=\ _{\star }^{\shortparallel }\mathbf{g}^{b_{4}a_{4}})\},
\notag \\
&&\mbox{ and }  \notag \\
&&\mbox{ nonsymmetric: }\ _{\star s}^{\shortparallel }\mathfrak{g}=(h_{1}\
_{\star s}^{\shortparallel }\mathfrak{g},v_{2}\ _{\star s}^{\shortparallel }%
\mathfrak{g},c_{3}\ _{\star s}^{\shortparallel }\mathfrak{g,}c_{4}\ _{\star
s}^{\shortparallel }\mathfrak{g})  \label{nssdm} \\
&=&\{\ _{\star }^{\shortparallel }\mathfrak{g}_{\alpha _{s}\beta _{s}}=(\
_{\star }^{\shortparallel }\mathfrak{g}_{i_{1}j_{1}}\neq \ _{\star
}^{\shortparallel }\mathfrak{g}_{j_{1}i_{1}},\ \ _{\star }^{\shortparallel }%
\mathfrak{g}_{a_{2}b_{2}}\neq \ _{\star }^{\shortparallel }\mathfrak{g}%
_{b_{2}a_{2}}\ _{\star }^{\shortparallel }\mathfrak{g}^{a_{3}b_{3}}\neq \
_{\star }^{\shortparallel }\mathfrak{g}^{b_{3}a_{3}},\ _{\star
}^{\shortparallel }\mathfrak{g}^{a_{4}b_{4}}\neq \ _{\star }^{\shortparallel
}\mathfrak{g}^{b_{4}a_{4}})\neq \ _{\star }^{\shortparallel }\mathfrak{g}%
_{\beta _{s}\alpha _{s}}\}.  \notag
\end{eqnarray}%
In coordinate bases, when $\ ^{\shortparallel }\mathbf{e}^{\alpha
_{s}}\rightarrow \ ^{\shortparallel }e^{\alpha }=d\ ^{\shortparallel
}u^{\alpha }\in T^{\ast }T_{\shortparallel }^{\ast }V$ and we can omit the
shell/dyadic label $s$, such star deformed metric structures can be
considered generic in off-diagonal forms,
\begin{equation}
\mbox{ symmetric },\ _{\star }^{\shortparallel }g=\{\ _{\star
}^{\shortparallel }g_{\alpha \beta }\neq \ _{\star }^{\shortparallel
}g_{\beta \alpha }\},\mbox{ and nonsymmetric },\ _{\star }^{\shortparallel }%
\mathsf{G}=\{\ _{\star }^{\shortparallel }\mathsf{G}_{\alpha \beta }\neq \
_{\star }^{\shortparallel }\mathsf{G}_{\beta \alpha }\},  \label{offdns}
\end{equation}%
Here we note that for fixed N-/ s-connection and respective coordinate
structures, we have $\ _{\star }^{\shortparallel }\mathbf{g}=\ _{\star
s}^{\shortparallel }\mathbf{g}=\ _{\star }^{\shortparallel }g$ and $\
_{\star }^{\shortparallel }\mathfrak{g}=\ _{\star s}^{\shortparallel }%
\mathfrak{g}=\ _{\star }^{\shortparallel }\mathsf{G,}$ but with different
block $(4\times 4)+(4\times 4),[(2\times 2)+(2\times 2)]+[(2\times
2)+(2\times 2)],$ and $8\times 8$ matrices of coefficients when N-adapted,
s-adapted or local coordinate (co) frame decompositions are considered. In
this paper, we follow the approach with quasi-Hopf s-algebras because it
allows to elaborate on s-adapted diffeomorphisms and star nonholonomic
deformations. This can be used for straightforward generalizations to
classical and quantum physical and information theory models. We write $%
\mathsf{G}$ for nonsymmetric metrics as in Section 5 of \cite{aschieri17}
but revise the notations and the formulas in s-adapted form and with
boldface symbols and labels $N,s,\ _{\star }^{\shortparallel },$ etc.

A s--metric $\ _{s}^{\shortparallel }\mathbf{g}$ on a associative and
commutative tangent Lorentz co-bundle with shell by shell dyadic
decomposition, $\mathbf{T}_{s\shortparallel }^{\ast }\mathbf{V,}$ is a
symmetric tensor which can be described by respective coefficients of
s-adapted (\ref{nadapbdsc}) and/or local coordinate co-bases,
\begin{equation}
\ _{s}^{\shortparallel }\mathbf{g}=\ ^{\shortparallel }\mathbf{g}_{\alpha
_{s}\beta _{s}}(~_{s}x,\ _{s}^{\shortparallel }p)\ ^{\shortparallel }\mathbf{%
e}^{\alpha _{s}}\mathbf{\otimes \ ^{\shortparallel }e}^{\beta _{s}}=\
^{\shortparallel }g_{\underline{\alpha }_{s}\underline{\beta }_{s}}(x,\
^{\shortparallel }p)d\ ^{\shortparallel }u^{\underline{\alpha }_{s}}\mathbf{%
\otimes }d\ ^{\shortparallel }u^{\underline{\beta }_{s}}=\ ^{\shortparallel
}g_{\underline{\alpha }\underline{\beta }}(x,\ ^{\shortparallel }p)d\
^{\shortparallel }u^{\underline{\alpha }}\mathbf{\otimes }d\
^{\shortparallel }u^{\underline{\beta }}.  \label{commetr}
\end{equation}%
We use labels for local coordinates in the form (\ref{loccord}) when certain
indices are underlined in order to state that they are for some coefficients
with respect to local coordinate (co) bases. On base manifold and total
space we can consider general coordinate transforms which mix shell's
coordinates, we can omit, for simplicity, s-parametrization for such
coordinates. The coefficients from (\ref{loccord}) are defined via frame
transforms,
\begin{equation*}
\ ^{\shortparallel }\mathbf{g}_{\alpha _{s}\beta _{s}}=\ \ ^{\shortparallel
}e_{\ \alpha _{s}}^{\underline{\alpha }_{s}}\ \ ^{\shortparallel }e_{\ \beta
_{s}}^{\underline{\beta }_{s}}\ \ ^{\shortparallel }g_{\underline{\alpha }%
_{s}\underline{\beta }_{s}}=\ \ ^{\shortparallel }e_{\ \alpha _{s}}^{%
\underline{\alpha }}\ \ ^{\shortparallel }e_{\ \beta _{s}}^{\underline{\beta
}}\ \ ^{\shortparallel }g_{\underline{\alpha }\underline{\beta }},
\end{equation*}
relating respective block dyadic decompositions with off-diagonal matrices.
For instance, with respect to local coordinate dual basis $d\
^{\shortparallel }u^{\underline{\alpha }},$ get such recurrent off-diagonal
parameterizations:
\begin{eqnarray}
\ ^{\shortparallel }g_{\underline{\alpha }_{1}\underline{\beta }%
_{1}}(x^{k_{1}}) &=&\ ^{\shortparallel }g_{\underline{i}_{1}\underline{i}%
_{1}}=\ ^{\shortparallel }g_{i_{1}i_{1}}=\ g_{i_{1}j_{1}}(x^{k_{1}}),%
\mbox{
with identification of indices for }s=1,  \notag \\
\ \ ^{\shortparallel }g_{\underline{\alpha }_{2}\underline{\beta }%
_{2}}(x^{k_{1}},x^{c_{2}}) &=&\left[
\begin{array}{cc}
\ g_{i_{1}j_{1}}+g_{a_{2}b_{2}}N_{i_{1}}^{a_{2}}N_{j_{1}}^{b_{2}} &
g_{a_{2}e_{2}}N_{i_{1}}^{e_{2}} \\
\ g_{a_{2}e_{2}}N_{i_{1}}^{e_{2}} & \ \ ^{\shortparallel }g_{a_{2}b_{2}}\
\end{array}%
\right] s=2,\mbox{ for }  \label{offd} \\
g_{a_{2}b_{2}} &=&g_{a_{2}b_{2}}(x^{k_{1}},x^{c_{2}}),N_{i_{1}}^{a_{2}}=
N_{i_{1}}^{a_{2}}(x^{k_{1}},x^{c_{2}});  \notag \\
\ \ ^{\shortparallel }g_{\underline{\alpha _{3}}\underline{\beta }%
_{3}}(x^{k_{1}},x^{c_{2}},~^{\shortparallel }p_{f_{3}}) &=&\left[
\begin{array}{cc}
\ \ ^{\shortparallel }g_{i_{2}j_{2}}+\ ^{\shortparallel }g^{a_{3}b_{3}}\
^{\shortparallel }N_{i_{2}a_{3}}\ ^{\shortparallel }N_{j_{2}b_{3}} & \ \
^{\shortparallel }g^{a_{3}e_{3}}\ \ ^{\shortparallel }N_{j_{2}e_{3}} \\
\ \ ^{\shortparallel }g^{b_{3}e_{3}}\ \ ^{\shortparallel }N_{i_{2}e_{3}} & \
\ ^{\shortparallel }g^{a_{3}b_{3}}%
\end{array}%
\right] ,s=3,\mbox{ for }  \notag \\
\ ^{\shortparallel }g_{i_{2}j_{2}} &=&\ ^{\shortparallel
}g_{i_{2}j_{2}}(x^{k_{1}},x^{c_{2}}),\ ^{\shortparallel }g^{a_{3}b_{3}}=\
^{\shortparallel }g^{a_{3}b_{3}}(x^{k_{1}},x^{c_{2}},~^{\shortparallel
}p_{f_{3}}),  \notag \\
\ ^{\shortparallel }N_{j_{2}e_{3}} &=&\
^{\shortparallel}N_{j_{2}e_{3}}(x^{k_{1}},x^{c_{2}}, \ ^{\shortparallel
}p_{f_{3}});  \notag \\
\ ^{\shortparallel }g_{\underline{\alpha }_{4} \underline{\beta _{4}}%
}(x^{k_{1}},x^{c_{2}}, \ ^{\shortparallel }p_{f_{3}},\
^{\shortparallel}p_{f_{4}}) &=&\left[
\begin{array}{cc}
\ \ ^{\shortparallel }g_{i_{3}j_{3}}\ +\ ^{\shortparallel }g^{a_{4}b_{4}}\
^{\shortparallel }N_{i_{3}a_{4}}\ \ ^{\shortparallel }N_{j_{3}b_{4}} & \ \
^{\shortparallel }g^{a_{3}e_{3}}\ ^{\shortparallel }N_{j_{3}e_{4}} \\
\ \ ^{\shortparallel }g^{b_{4}e_{4}}\ ^{\shortparallel }N_{i_{3}e_{4}} & \ \
^{\shortparallel }g^{a_{4}b_{4}}%
\end{array}%
\right] ,s=4,\mbox{ for }  \notag \\
\ ^{\shortparallel }g_{i_{3}j_{3}} &=&\ ^{\shortparallel
}g_{i_{3}j_{3}}(x^{k_{1}},x^{c_{2}},~^{\shortparallel }p_{f_{3}}),\
^{\shortparallel }g^{a_{4}b_{4}}=\ ^{\shortparallel
}g^{a_{4}b_{4}}(x^{k_{1}},x^{c_{2}},~^{\shortparallel
}p_{f_{3}},~^{\shortparallel }p_{f_{4}}),  \notag \\
\ ^{\shortparallel }N_{i_{3}a_{4}} &=&\ ^{\shortparallel
}N_{i_{3}a_{4}}(x^{k_{1}},x^{c_{2}},~^{\shortparallel
}p_{f_{3}},~^{\shortparallel }p_{f_{4}}).\   \notag
\end{eqnarray}%
Equivalently, shell by shell, above formulas define s-adapted metric and
N-connection coefficients are parameterized shell by shell in the form (\ref%
{sdm}),
\begin{eqnarray*}
\ ^{\shortparallel }\mathbf{g}_{\alpha _{s}\beta _{s}} &=&[\
^{\shortparallel }\mathbf{g}_{i_{1}j_{1}}=g_{i_{1}j_{1}}(x^{k_{1}}),\
^{\shortparallel }\mathbf{g}%
_{a_{2}b_{2}}=g_{a_{2}b_{2}}(x^{k_{1}},x^{c_{2}}), \\
&& \ ^{\shortparallel }\mathbf{g}^{a_{3}b_{3}} =\ ^{\shortparallel
}g^{a_{3}b_{3}}(x^{k_{1}},x^{c_{2}},\ ^{\shortparallel }p_{f_{3}}),\
^{\shortparallel }\mathbf{g}^{a_{4}b_{4}}=\ ^{\shortparallel
}g^{a_{4}b_{4}}(x^{k_{1}},x^{c_{2}},~^{\shortparallel
}p_{f_{3}},~^{\shortparallel }p_{f_{4}})]; \\
\ ^{\shortparallel }N_{i_{s-1}a_{s}}
&=&[N_{i_{1}}^{a_{2}}=N_{i_{1}}^{a_{2}}(x^{k_{1}},x^{c_{2}}),\
^{\shortparallel }N_{j_{2}e_{3}}=\ ^{\shortparallel
}N_{j_{2}e_{3}}(x^{k_{1}},x^{c_{2}},~^{\shortparallel }p_{f_{3}}),\
^{\shortparallel }N_{i_{3}a_{4}}=\ ^{\shortparallel
}N_{i_{3}a_{4}}(x^{k_{1}},x^{c_{2}},~^{\shortparallel
}p_{f_{3}},~^{\shortparallel }p_{f_{4}})],
\end{eqnarray*}%
where the N-connection s-coordinate coefficients are chosen in the form (\ref%
{ncon2coef}). Such s-parameterizations can be used for proofs of a general
decoupling and integration of nonholonomic gravitational and geometric flow
systems of PDEs in modified gravity (non) commutative theories, see reviews
\cite{vacaru18,bubuianu18a,bubuianu17,bubuianu19,vacaru20}.

\subsubsection{Canonical s-connections as distortions of dyadic
LC-connections}

For nonholonomic distributions on phase space, we can define different
d-connection structures following different geometric principles of adapting
to respective N-connection structures. Using a s-metric $\
_{s}^{\shortparallel }\mathbf{g}$ (\ref{sdm}), we can work in equivalent
form with two different linear connections which are adapted or not to
respective nonholonomic distributions:
\begin{equation}
(\ _{s}^{\shortparallel}\mathbf{g,\ _{s}^{\shortparallel}N})\rightarrow
\left\{
\begin{array}{cc}
\ ^{\shortparallel }\mathbf{\nabla :} & \ ^{\shortparallel }\mathbf{\nabla }%
\ \ _{s}^{\shortparallel }\mathbf{g}=0;\ _{\nabla }^{\shortparallel }
\mathcal{T}=0, \mbox{\  LC--connection }; \\
\ _{s}^{\shortparallel }\widehat{\mathbf{D}}: &
\begin{array}{c}
\ _{s}^{\shortparallel}\widehat{\mathbf{D}}\ _{s}^{\shortparallel}\mathbf{g}%
=0;\ h_{1}\ ^{\shortparallel}\widehat{\mathcal{T}}=0,v_{2}\ ^{\shortparallel}%
\widehat{\mathcal{T}}=0,c_{3}\ ^{\shortparallel}\widehat{\mathcal{T}}%
=0,c_{4}\ ^{\shortparallel}\widehat{\mathcal{T}}=0, \\
h_{1}v_{2}\ ^{\shortparallel}\widehat{\mathcal{T}}\neq 0,h_{1}c_{s} \
^{\shortparallel}\widehat{\mathcal{T}}\neq 0,v_{2}c_{s}\ ^{\shortparallel}
\widehat{\mathcal{T}}\neq 0,c_{3}c_{4}\ ^{\shortparallel}\widehat{\mathcal{T}%
}\neq 0,\mbox{ canonical s--connection},%
\end{array}%
\end{array}%
\right.  \label{twocon}
\end{equation}%
where $\ _{s}^{\shortparallel }\widehat{\mathbf{D}}= (h_{1}\
^{\shortparallel} \widehat{\mathbf{D}},\ v_{2}\ ^{\shortparallel }\widehat{%
\mathbf{D}},\ c_{3}\ ^{\shortparallel }\widehat{\mathbf{D}}, \ c_{4}\
^{\shortparallel }\widehat{\mathbf{D}}),$ with dyadic (co) vertical
splitting, is a s-connection adapted to a N--connection structure $\
_{s}^{\shortparallel }\mathbf{N.}$ Such linear (affine) and nonlinear
connections and their s-torsion are nonholonomic but commutative variants of
nonassociative values introduced in appendix, see formulas (\ref{sconhopf})
and (\ref{dtorshform}). We use "hat" values in order to emphasize that such
s-adapted values are determined by a s-metric structure following certain
nonholonomic constraints involving (partial) zero torsion and metric
compatibility conditions. The LC-connection $\ ^{\shortparallel }\mathbf{%
\nabla }$ is not a d-/ or s-connection because it does not preserve the
N-connection splitting and/or dyadic decompositions under parallel
transports.

We provide here the s-adapted coefficients of the (associative and
commutative) canonical s-connection $\ _{s}^{\shortparallel }\widehat{%
\mathbf{D}}$ on $T\mathbf{T}_{\shortparallel }^{\ast }\mathbf{V},$ which is
an example with trivial R-fluxes of nonassociative s-connection $\
_{s}^{\shortparallel }\mathbf{D}^{\star }$ with respective dyadic
decomposition (\ref{irevndecomdc}) (see labeling of s-indices in those
formulas):
\begin{eqnarray*}
\ _{s}^{\shortparallel }\widehat{\mathbf{D}} &=&\{\ ^{\shortparallel}%
\widehat{\mathbf{\Gamma }}_{\alpha _{s}\beta _{s}}^{\gamma _{s}}=( \widehat{L%
}_{j_{1}k_{1}}^{i_{1}},\widehat{L}_{b_{2}\ k_{1}}^{a_{2}},\widehat{C}_{\
j_{1}c_{2}}^{i_{1}\ },\widehat{C}_{b_{2}c_{2}}^{a_{2}};\ ^{\shortparallel }%
\widehat{L}_{j_{2}k_{2}}^{i_{2}},\ ^{\shortparallel }\widehat{L}%
_{a_{3}k_{2}}^{\ b_{3}},\ ^{\shortparallel }\widehat{C}_{\ j_{2}}^{i_{2}\
c_{3}}, \\
&&\ ^{\shortparallel }\widehat{C}_{a_{3}b_{3}}^{\quad c_{3}};\
^{\shortparallel }\widehat{L}_{j_{3}k_{3}}^{i_{3}},^{\shortparallel }%
\widehat{L}_{a_{4}\ k_{3}}^{\ b_{4}},\ ^{\shortparallel }C_{\ j_{3}}^{i_{3}\
c_{4}},\ ^{\shortparallel }C_{a_{4}b_{4}}^{\quad c_{4}})\},
\end{eqnarray*}
\begin{eqnarray}
\mbox{ where }\widehat{L}_{j_{1}k_{1}}^{i_{1}} &=&\frac{1}{2}\ \
^{\shortparallel }g^{i_{1}r_{1}}(\ \ ^{\shortparallel }\mathbf{e}_{k_{1}}\
^{\shortparallel }g_{j_{1}r_{1}}+\ \ ^{\shortparallel }\mathbf{e}_{j_{1}}\
^{\shortparallel }g_{k_{1}r_{1}}-\ ^{\shortparallel }\mathbf{e}_{r_{1}}\
^{\shortparallel }g_{j_{1}k_{1}}),\   \label{canhcs} \\
\widehat{L}_{b_{2}\ k_{1}}^{a_{2}} &=&\ ^{\shortparallel }e_{b_{2}}(\
^{\shortparallel }N_{k_{1}}^{a_{2}})+\frac{1}{2}\ ^{\shortparallel
}g^{b_{2}c_{2}}(\ ^{\shortparallel }e_{k_{1}}\ ^{\shortparallel
}g_{b_{2}c_{2}}-\ ^{\shortparallel }g_{d_{2}c_{2}}\ ^{\shortparallel
}e_{b_{2}}\ ^{\shortparallel }N_{k_{1}}^{d_{2}}-\ ^{\shortparallel
}g_{d_{2}b_{2}}\ ^{\shortparallel }e_{c_{2}}\ ^{\shortparallel
}N_{k_{1}}^{d_{2}}),  \notag \\
\widehat{C}_{\ j_{1}c_{2}}^{i_{1}\ } &=&\frac{1}{2}\ \ ^{\shortparallel
}g^{ik}\ \ ^{\shortparallel }e_{c}\ \ ^{\shortparallel }g_{jk},\ \ \
\widehat{C}_{b_{2}c_{2}}^{a_{2}}=\frac{1}{2}\ \ ^{\shortparallel
}g^{a_{2}d_{2}}(\ \ ^{\shortparallel }e_{c_{2}}\ ^{\shortparallel
}g_{b_{2}d_{2}}+\ ^{\shortparallel }e_{b_{2}}\ ^{\shortparallel
}g_{c_{2}d_{2}}-\ ^{\shortparallel }e_{d_{2}}\ ^{\shortparallel
}g_{b_{2}c_{2}}),  \notag \\
\ \ ^{\shortparallel }\widehat{L}_{a_{3}\ k_{2}}^{\ b_{3}} &=&\
^{\shortparallel }e^{b_{3}}(\ ^{\shortparallel }N_{a_{3}k_{2}})+\frac{1}{2}\
^{\shortparallel }g_{a_{3}c_{3}}(\ ^{\shortparallel }e_{k_{2}}\
^{\shortparallel }g^{b_{3}c_{3}}-\ ^{\shortparallel }g^{d_{3}c_{3}}\ \
^{\shortparallel }e^{b_{3}}\ ^{\shortparallel }N_{d_{3}k_{2}}-\
^{\shortparallel }g^{d_{3}b_{3}}\ ^{\shortparallel }e^{c}\ ^{\shortparallel
}N_{d_{3}k_{2}}),  \notag \\
\ \ ^{\shortparallel }\widehat{C}_{\ j_{2}}^{i_{2}\ c_{3}} &=&\frac{1}{2}\ \
^{\shortparallel }g^{i_{2}k_{2}}\ ^{\shortparallel }e^{c_{3}}\
^{\shortparallel }g_{j_{2}k_{2}},\ \ \ ^{\shortparallel }\widehat{C}_{\
a_{3}}^{b_{3}\ c_{3}}=\frac{1}{2}\ ^{\shortparallel }g_{a_{3}d_{3}}(\
^{\shortparallel }e^{c_{3}}\ ^{\shortparallel
}g^{b_{3}d_{3}}+~^{\shortparallel }e^{b_{3}}\ \ ^{\shortparallel
}g^{c_{3}d_{3}}-\ ^{\shortparallel }e^{d_{3}}\ ^{\shortparallel
}g^{b_{3}c_{3}}),  \notag \\
\ \ ^{\shortparallel }\widehat{L}_{a_{4}\ k_{3}}^{\ b_{4}} &=&\ \
^{\shortparallel }e^{b_{4}}(\ \ ^{\shortparallel }N_{a_{4}k_{3}})+\frac{1}{2}%
\ ^{\shortparallel }g_{a_{4}c_{4}}(\ ^{\shortparallel }e_{k_{3}}\
^{\shortparallel }g^{b_{4}c_{4}}-\ ^{\shortparallel }g^{d_{4}c_{4}}\
^{\shortparallel }e^{b_{4}}\ ^{\shortparallel }N_{d_{4}k_{3}}-\
^{\shortparallel }g^{d_{4}b_{4}}\ ^{\shortparallel }e^{c_{4}}\
^{\shortparallel }N_{d_{4}k_{3}}),  \notag \\
\ \ ^{\shortparallel }\widehat{C}_{\ j_{3}}^{i_{3}\ c_{4}} &=&\frac{1}{2}\
^{\shortparallel }g^{i_{3}k_{3}}\ ^{\shortparallel }e^{c_{4}}\
^{\shortparallel }g_{j_{3}k_{3}},\ \ \ ^{\shortparallel }\widehat{C}_{\
a_{4}}^{b_{4}\ c_{4}}=\frac{1}{2}\ \ ^{\shortparallel }g_{a_{4}d_{4}}(\
^{\shortparallel }e^{c_{4}}\ ^{\shortparallel }g^{b_{4}d_{4}}+\
^{\shortparallel }e^{b_{4}}\ ^{\shortparallel }g^{c_{4}d_{4}}-\
^{\shortparallel }e^{d_{4}}\ ^{\shortparallel }g^{b_{4}c_{4}}).  \notag
\end{eqnarray}%
By straightforward computations with such coefficients (see similar details
and proofs for $\mathbf{T}_{\shortmid }^{\ast }\mathbf{V}$ in \cite%
{vacaru18,bubuianu18a}), we can check that there are satisfied all necessary
conditions from the definition of the canonical s-connection in (\ref{twocon}%
). Introducing s-coefficients (\ref{canhcs}) in formulas (\ref{candricci})
and (\ref{candriccidist}), we can compute respective s-adapted coefficients
for the canonical Ricci s-tensor, corresponding Ricci scalars and
distortions from respective values determined by the LC-connection.

A LC-connection $\ ^{\shortparallel }\mathbf{\nabla }$ is not a d-connection
and can not be expressed as a s-connection. Nevertheless, we can compute a
canonical distortion relation to canonical s-connection,
\begin{equation}
\ _{s}^{\shortparallel }\widehat{\mathbf{D}}=\ ^{\shortparallel }\nabla +\
_{s}^{\shortparallel }\widehat{\mathbf{Z}}.  \label{candistr}
\end{equation}%
The distortion s-tensor, $\ _{s}^{\shortparallel }\widehat{\mathbf{Z}}= \{\
^{\shortparallel }\widehat{\mathbf{Z}}_{\ \beta _{s}\gamma _{s}}^{\alpha
_{s}}[\ ^{\shortparallel }\widehat{\mathbf{T}}_{\ \beta _{s}\gamma
_{s}}^{\alpha _{s}}]\},$ is an algebraic combination of the coefficients the
canonical torsion s-tensor $\ _{s}^{\shortparallel}\widehat{\mathcal{T}}=
\{\ ^{\shortparallel }\widehat{\mathbf{T}}_{\ \beta _{s}\gamma _{s}}^{\alpha
_{s}}\}$ of $\ _{s}^{\shortparallel }\widehat{\mathbf{D}}.$ The canonical
s-torsion is a commutative variant of the nonassociative d-torsion
parameterized as in formulas (\ref{storsnonassoc}), without star labels but
with "hats" for canonical values. We shall provide explicit formulas for
(non) associative canonical s--torsions in next sections. This distortion is
determined by the s-connection coefficients and anholonomy coefficients $\
^{\shortparallel}w_{\alpha _{s}\beta _{s}}^{\gamma _{s}},$ see similar
formulas (\ref{anhrel}) computed for s-frames $\ ^{\shortparallel }\mathbf{e}%
_{\alpha _{s}}$ (\ref{nadapbdsc}).

\subsubsection{Canonical nonholonomic Ricci and Einstein s-tensors}

For both type linear connections $\ ^{\shortparallel }\mathbf{\nabla }$ and $%
\ _{s}^{\shortparallel }\widehat{\mathbf{D}}$ (\ref{twocon}), we can define
and compute in standard forms (as in metric-affine geometry) respective
torsions, $\ _{\nabla }^{\shortparallel }\mathcal{T}=0$ and $\
_{s}^{\shortparallel }\widehat{\mathcal{T}},$ and curvatures, $\
_{\nabla}^{\shortparallel }\mathcal{R}=\{\ _{\nabla }^{\shortparallel }R_{\
\beta \gamma \delta }^{\alpha }\}$ and $\ _{s}^{\shortparallel }\widehat{%
\mathcal{R}}= \{\ ^{\shortparallel }\widehat{\mathbf{R}}_{\ \beta _{s}\gamma
_{s}\delta _{s}}^{\alpha _{s}}\}.$ Such general frame/coordinate and s-frame
can defined and computed in with respect to arbitrary, coordinate, and/or
N-/ s-adapted frames.

The canonical Ricci s-tensor $\ _{s}^{\shortparallel }\widehat{\mathcal{R}}%
ic= \{\ ^{\shortparallel }\widehat{\mathbf{R}}_{\ \beta _{s}\gamma _{s}}:= \
^{\shortparallel }\widehat{\mathbf{R}}_{\ \alpha _{s}\beta _{s}\gamma
_{s}}^{\gamma _{s}}\}$ and, for the LC-connection, $\ ^{\shortparallel}Ric=
\{\ ^{\shortparallel }R_{\ \beta \gamma }:= \ ^{\shortparallel }R_{\ \alpha
\beta \gamma }^{\gamma }\}$ (we omit the label $\nabla $ writing formulas in
not "boldface" forms and without shell labels if such formulas do not result
in ambiguities) are defined and computed by a respective contracting the
first and fours indices. The canonical s-tensor $\ _{s}^{\shortparallel }%
\widehat{\mathcal{R}}ic$ is characterized by s-adapted coefficients,%
\begin{eqnarray}
\ ^{\shortparallel }\widehat{\mathbf{R}}_{\ \beta _{s}\gamma _{s}} &=&\{\
^{\shortparallel }\widehat{R}_{\ h_{1}j_{1}}=\ \ ^{\shortparallel }\widehat{%
\mathbf{R}}_{~h_{1}j_{1}i_{1}}^{i_{1}},\ \ ^{\shortparallel }\widehat{P}%
_{j_{1}a_{2}}^{\ }=-\ ^{\shortparallel }\widehat{\mathbf{R}}_{~\
j_{1}i_{1}a_{2}}^{i_{1}\quad }\ ,\ ^{\shortparallel }\widehat{P}_{\star \
b_{2}k_{1}}=\ ^{\shortparallel }\widehat{\mathbf{R}}%
_{~b_{2}k_{1}c_{2}}^{c_{2}\ },\ ^{\shortparallel }\widehat{S}_{\star
b_{2}c_{2}\ }=\ ^{\shortparallel }\widehat{\mathbf{R}}_{~b_{2}c_{2}a_{2}\
}^{a_{2}\ },  \notag \\
&&\ ^{\shortparallel }\widehat{P}_{~j_{2}}^{\ a_{3}}=-\ \ ^{\shortparallel }%
\widehat{\mathbf{R}}_{~\ j_{2}i_{2}}^{i_{2}\quad a_{3}}\ ,\ ^{\shortparallel
}\widehat{P}_{~\ k_{2}}^{b_{3}\quad }=\ ^{\shortparallel }\widehat{\mathbf{R}%
}_{~c_{3}\ k_{2}}^{\ b_{3}\ c_{3}},\ ^{\shortparallel }\widehat{S}_{\
}^{b_{3}c_{3}\quad }=\ \ ^{\shortparallel }\widehat{\mathbf{R}}_{a_{3}\
}^{\quad b_{3}c_{3}a_{3}}\ ,  \label{candricci} \\
&&\ ^{\shortparallel }\widehat{P}_{~j_{3}}^{\ a_{4}}=-\ ^{\shortparallel }%
\widehat{\mathbf{R}}_{~\ j_{3}i_{3}}^{i_{3}\quad a_{4}}\ ,\ ^{\shortparallel
}\widehat{P}_{~\ k_{3}}^{b_{4}\quad }=\ ^{\shortparallel }\widehat{\mathbf{R}%
}_{~c_{4}\ k_{3}}^{\ b_{4}\ c_{4}},\ ^{\shortparallel }\widehat{S}_{\
}^{b_{4}c_{4}\quad }=\ ^{\shortparallel }\widehat{\mathbf{R}}_{a_{4}\
}^{\quad b_{4}c_{4}a_{4}}\},  \notag
\end{eqnarray}%
are defined as the associative commutative part (the component of order ($%
i\hbar )^{0}$ and $(i\kappa )^{0}$ of (\ref{driccinahc}) with s-coefficients
(\ref{nadriemhopf1c}). Introducing in such formulas the canonical
s-connection coefficients (\ref{canhcs}), we can express the values $\
^{\shortparallel }\widehat{\mathbf{R}}_{\ \beta _{s}\gamma _{s}}$ (\ref%
{candricci}) in terms of the coefficients of s-metric and N-/ s-connection.
We note that in our nonholonomic dyadic approach $\ _{s}^{\shortparallel }
\widehat{\mathcal{R}}ic$ is an associative and commutative variant of the
nonassociative Ricci s-tensor $\ _{s}^{\shortparallel }\mathcal{\Re }%
ic^{\star }= \{\ ^{\shortparallel }\mathbf{R}ic_{\alpha _{s}\beta
_{s}}^{\star }\}$ (\ref{driccinahc}) with s-adapted distributions
parameterized in such forms that
\begin{equation*}
\begin{array}{cc}
\ _{s}^{\shortparallel }\mathcal{\Re }^{\star }(\ref{stardcurvh})\implies \
_{s}^{\shortparallel }\mathcal{\Re }ic^{\star }(\ref{driccina})\quad
\begin{array}{c}
\longrightarrow \\
(i\hbar )^{0},(i\kappa )^{0}%
\end{array}%
\quad \ _{s}^{\shortparallel }\mathcal{R}ic\quad
\begin{array}{c}
\implies \\
(\ _{s}^{\shortparallel }\mathbf{g,}\ _{s}^{\shortparallel }\mathbf{\widehat{%
\mathbf{D}}=\ ^{\shortparallel }\mathbf{\nabla }+}\ _{s}^{\shortparallel }%
\widehat{\mathbf{Z}})%
\end{array}
&
\begin{array}{c}
\ _{s}^{\shortparallel }\widehat{\mathcal{R}}ic=\{\ ^{\shortparallel }%
\widehat{\mathbf{R}}_{\alpha _{s}\beta _{s}}\} \\
\mbox{ on }\mathbf{T}_{\shortparallel }^{\ast }\mathbf{V}%
\end{array}
\\
&
\begin{array}{c}
\mbox{ nonholonomic } \\
\mbox{ s-transforms }\&\mbox{ lifts}%
\end{array}
\\
&
\begin{array}{c}
\Uparrow \\
R_{ij}\mbox{ on }V%
\end{array}%
\
\end{array}%
\end{equation*}

Contracting the coefficients of inverse (s-) metric with respective
coefficients of the canonical Ricci (s-) tensor, we can define two different
scalar curvatures,
\begin{eqnarray}
\ _{s}^{\shortparallel }\widehat{\mathbf{R}}sc:= &&\ ^{\shortparallel }%
\mathbf{g}^{\alpha _{s}\beta _{s}}\ ^{\shortparallel }\widehat{\mathbf{R}}%
_{\alpha _{s}\beta _{s}}=\ ^{\shortparallel }g^{i_{1}j_{1}}\
^{\shortparallel }\widehat{R}_{i_{1}j_{1}}+\ ^{\shortparallel
}g^{a_{2}b_{2}}\ ^{\shortparallel }\widehat{R}_{a_{2}b_{2}}+\
^{\shortparallel }g_{a_{3}b_{3}}\ ^{\shortparallel }\widehat{R}%
^{a_{3}b_{3}}+\ ^{\shortparallel }g_{a_{4}b_{4}}\ ^{\shortparallel }\widehat{%
R}^{a_{4}b_{4}}  \notag \\
&&\mbox{ and }\ ^{\shortparallel }R:=\ ^{\shortparallel }\mathbf{g}^{\alpha
\beta }\ ^{\shortparallel }R_{\alpha \beta }.  \label{canricscal1}
\end{eqnarray}%
The modified Einstein equations for $(\ _{s}^{\shortparallel }\mathbf{g,}\
_{s}^{\shortparallel }\widehat{\mathbf{D}})$ on $\mathbf{T}_{\shortparallel
}^{\ast }\mathbf{V}$ with a nontrivial cosmological constant $\
^{\shortparallel }\lambda $ can be postulated using the same geometric
principles as in GR (see, for instance, \cite{misner}) but extended to
cotangent Lorentz bundles, see an axiomatic approach and details in \cite%
{vacaru18,bubuianu18a}, in the form
\begin{equation}
\ _{s}^{\shortparallel }\widehat{E}n:=\ _{s}^{\shortparallel }\widehat{%
\mathcal{R}}ic-\frac{1}{2}\ _{s}^{\shortparallel }\widehat{\mathbf{R}}sc=\
^{\shortparallel }\lambda \ _{s}^{\shortparallel }\mathbf{g},
\label{cnsveinst1}
\end{equation}%
where $\ _{s}^{\shortparallel }\widehat{E}n=\{\ _{s}^{\shortparallel}%
\widehat{E}_{_{\alpha _{s}\beta _{s}}}\}$ is by definition the canonical
Einstein s-tensor. Such forms of commutative nonholonomic modifications of
vacuum gravitational Einstein equations are constructed via lifts of metrics
and related linear connection $\nabla ,$ or possible other type
d-connections, from Lorentz spacetime to it cotangent bundle and canonical
deformation of the Levi-Civita connection, $\ ^{\shortparallel }\mathbf{%
\nabla }$ $\rightarrow $ $\ _{s}^{\shortparallel }\widehat{\mathbf{D}}.$
Here we note that $\ _{s}^{\shortparallel }\widehat{\mathbf{D}}(\
_{s}^{\shortparallel }\widehat{E}n)\neq 0,$ which is typical for theories
with nonholonomic constraints.\footnote{%
For instance, the conservation laws in nonholonomic Lagrange mechanics
should be revised by introducing additional multiples associated to
nonholonomic constraints of different type.} Using distortions of
d-connections of type (\ref{candistr}), we can always nonholonomically
deform such systems of nonlinear PDEs into equivalent ones with $\
_{s}^{\shortparallel}\widehat{E}n\rightarrow \ _{\nabla }^{\shortparallel
}En,$ when $\ ^{\shortparallel }\nabla (\ _{\nabla }^{\shortparallel }En)=0.$
Such canonical s-distortions of geometric objects and physically important
gravitational and geometric flows equations, are determined by respective
distortions of the curvature and Ricci tensors,
\begin{equation}
\ _{s}^{\shortparallel }\widehat{\mathcal{R}}=\ \ _{\nabla }^{\shortparallel
}\mathcal{R+}\ \ \ _{s}^{\shortparallel }\widehat{\mathcal{Z}},\
_{s}^{\shortparallel }\widehat{\mathcal{R}}ic=\ \ _{\nabla }^{\shortparallel
}Ric+\ \ _{s}^{\shortparallel }\widehat{\mathcal{Z}}ic,\
_{s}^{\shortparallel }\widehat{\mathbf{R}}sc=\ \ _{\nabla }^{\shortparallel
}Rsc+\ \ _{s}^{\shortparallel }\widehat{\mathcal{Z}}sc,\mbox{
and }\ _{s}^{\shortparallel }\widehat{E}n=\ _{\nabla }^{\shortparallel }En+\
\ _{s}^{\shortparallel }\widehat{\mathcal{Z}}n,  \label{candriccidist}
\end{equation}%
with corresponding distortion s-tensors $\ _{\nabla }^{\shortparallel}%
\widehat{\mathcal{Z}}$ and $\ _{\nabla }^{\shortparallel }\widehat{\mathcal{Z%
}}ic$ determined by $(\ _{s}^{\shortparallel }\widehat{\mathbf{Z}}, \
_{s}^{\shortparallel }\mathbf{g)}$ on a (pseudo) Riemannian phase background
with $\ ^{\shortparallel }\nabla .$ We omit abstract and cumbersome
s-adapted/ coordinate formulas for such geometric objects.

To construct generic off-diagonal solutions and study their physical
properties in certain models of phase gravity, we can consider effective
cosmological constants on any shell $s=1,2,3,4$ and/or for a h-
/c-decomposition. We shall use for star parametric (on $\hbar ,\kappa ,$ and
$\hbar \kappa )$ deformations the solutions of associative/ commutative
nonholonomic vacuum gravitational equations
\begin{equation}
\ ^{\shortparallel }\widehat{\mathbf{R}}_{\ \beta _{s}\gamma _{s}}=\
_{s}^{\shortparallel }\lambda \ ^{\shortparallel }\mathbf{g}_{\ \beta
_{s}\gamma _{s}}.  \label{cnsveinst2}
\end{equation}%
These equations transform into (\ref{cnsveinst1}) if $\
_{s}^{\shortparallel}\lambda =\ ^{\shortparallel }\lambda $ for all values
of $s.$ The goal of this work is to prove that nonassociative vacuum
equations constructed as R-flux and star canonical s-deformations of (\ref%
{cnsveinst2}) can be decoupled and integrated in certain general form for
quasi-stationary configurations (when in corresponding s-adapted variables
the metric's s-coefficients and other fundamental geometric objects do not
depend on time like coordinate $x^{4}=t).$

We conclude that in the associative and commutative case (with zero R-flux),
the geometry of phase space $\mathbf{T}_{\shortparallel }^{\ast }\mathbf{V}$
can be described by a standard (pseudo) Riemannian geometry (we can consider
momentum coordinates multiplied to complex unity). Equivalently, we can
consider models with nonholonomic induced torsion by the N-connection
structure on $\mathbf{T}_{\shortparallel }^{\ast }\mathbf{V}$ as we
explained in details in \cite{partner01}. In this work, we use nonholonomic
dyadic decompositions when (from the same metric structure) we can derive
both a canonical s-connection and the LC-connection, and can apply the AFCDM
for constructing exact solutions. We note that all geometric constructions
can be performed equivalently working with different geometric data $\left(\
^{\shortparallel}\mathbf{g}, \ ^{\shortparallel }\nabla \right)$ and $(\
_{s}^{\shortparallel}\mathbf{g}, \ _{s}^{\shortparallel}\mathbf{N}, \
_{s}^{\shortparallel}\widehat{\mathbf{D}}).$

\subsection{Nonassociative canonical s-connections and LC-configurations}

To provide physical motivations for star deformations (they can result both
in noncommutative and noncommutative geometries) of commutative models of
gravity and geometric flow theories and their physically important classes
of solutions, we elaborate a formalism of nonassiative R-flux determined
star s-adapted deformations of (non) symmetric metrics and (non) linear
connections for phase spaces with quasi-Hopf s-structure.

\subsubsection{Conventions on parameterizing nonassociative R-flux star
deformations}

We extend for nonholonomic dyadic decompositions the Conventions 1 and 2
analyzed in section 2.2.1 of \cite{partner01} on constructing nonassociative
geometries:

\begin{itemize}
\item \textbf{Convention 1:\ } R-flux star deformations to nonassocitative
geometric models are elaborating in coordinate frames and beginning with a
flat (co) fiber metric $\mathbf{\ ^{\shortparallel }}\eta .$ In \cite%
{blumenhagen16,aschieri17}, star products $\star $ are defined in terms of
coordinate bases $\mathbf{\ ^{\shortparallel }\partial }$ and respective
nonassociative generalizations of (pseudo) Riemann geometry are constructed
in using R-flux deformations to symmetric, $\ _{\star }^{\shortparallel }g,$
and nonsymmetric, $\ _{\star }^{\shortparallel }\mathsf{G,}$ star-metric
structures and a related nonassociative variant of LC-connection $\mathbf{\
^{\shortparallel }}\nabla ^{\star }.$ There are considered such
star-deformations of geometric structures:
\begin{equation*}
(\ ^{\shortparallel}\eta ,\ ^{\shortparallel }\partial ,\ ^{\shortparallel
}\nabla )\rightarrow (\star \ ,\ \mathcal{A}^{\star },\ _{\star
}^{\shortparallel }g, \ _{\star }^{\shortparallel}\mathsf{G},\ \
^{\shortparallel }\partial \ ^{\shortparallel}\nabla ^{\star }),
\end{equation*}%
with possible adapting of constructions to quasi-Hopf algebras $\ \mathcal{A}%
^{\star }$, or other type algebraic and geometric structures. The
nonassocitative gravity theories were elaborated to a level of vacuum
gravitational equations $\ _{\nabla }^{\shortparallel }Ric^{\star}=0$ on a
target star deformed phase space. There were computed real (proportional to
parameters products $\hbar \kappa )$ R-flux contributions for projections on
pseudo-Riemannian spacetime $V.$ Such fundamental geometric and
gravitational equations are formulated in terms of tensor and linear
connection objects with coefficients computed with respect to coordinate
(co) bases $d\ ^{\shortparallel }u^{\alpha }$ and $\
^{\shortparallel}\partial _{\alpha }$ and their star deformations.
Technically, it is very difficult to decouple and solve in certain explicit
exact/ parametric forms such systems of nonlinear PDEs encoding
nonassociative R-flux contributions etc., and to elaborate on quantum
deforming of exact/parametric generic off-diagonal solutions in such
theories.

\item \textbf{Convention 2: }Our main goal is to elaborate a nonassociative
generalization of the AFCDM method which allows to construct exact solutions
defined by generic off-diagonal symmetric and nonsymmetric metrics and
generalized (non) linear connections. The coefficients of geometric objects
for such solutions depend, in general, on all possible phase space and
spacetime coordinates and various classes of off-diagonal solutions can be
constructed in different modified noncommutative and commutative gravity
theories, in particular, in GR. Such a geometric techniques can be
formulated using respective nonholonomic frame and connection deformations
to certain nonholonomic and dyadic N- /s-adapted structures resulting in
systems of nonlinear PDEs which can be decoupled and integrated in certain
general forms. In a series of recent our works (the first partner one is
\cite{partner01}), we elaborate and apply geometric methods when commutative
theories are formulated in certain general nonholonomic frame forms and then
deformed by star nonholonomic products constructed for $\ ^{\shortparallel}%
\mathbf{e}_{\alpha }$ adapted to N-connection structure $\ ^{\shortparallel}%
\mathbf{N}$. We extend the conditions of Convention 2 in \cite{partner01} to
the case of star products (\ref{starpn}) defined with nonholonomic dyadic
decompositions on $\ ^{\shortparallel}\mathbf{e}_{\alpha _{s}}$ with R-flux
terms. There are computed respective star deformations of canonical
s-adapted geometric objects into nonassociative ones, with symmetric, $\
_{\star s}^{\shortparallel }\mathbf{g}$ (\ref{ssdm}), and nonsymmetric, $\
_{\star s}^{\shortparallel}\mathbf{\mathfrak{g}}$ (\ref{nssdm}), star
s-metrics and canonical star s-connection $\ _{s}^{\shortparallel }\mathbf{D}%
^{\star }$. The nonholonomic dyadic star deformations of geometric canonical
s-structures are defined by such star transforms of geometric data:%
\begin{equation}
\begin{array}{ccc}
(\star _{N},\ \ \mathcal{A}_{N}^{\star },\ _{\star }^{\shortparallel }%
\mathbf{g,\ _{\star }^{\shortparallel }\mathfrak{g,}\ \ ^{\shortparallel }N},%
\mathbf{\ \ ^{\shortparallel }e}_{\alpha }\mathbf{,\ \mathbf{\mathbf{\mathbf{%
\ ^{\shortparallel }}}}D}^{\star }) & \Leftrightarrow & (\star _{s},\ \
\mathcal{A}_{s}^{\star },\ _{\star s}^{\shortparallel }\mathbf{g,\ _{\star
s}^{\shortparallel }\mathfrak{g,}}\ \ _{s}^{\shortparallel }\mathbf{N},%
\mathbf{\ \ ^{\shortparallel }e}_{\alpha _{s}}\mathbf{,\ \mathbf{\mathbf{%
\mathbf{\ _{s}^{\shortparallel }}}}D}^{\star }) \\
& \Uparrow &  \\
(\ \ ^{\shortparallel }\mathbf{g,\ \ ^{\shortparallel }N},\mathbf{\ \
^{\shortparallel }e}_{\alpha }\mathbf{,}\ ^{\shortparallel }\widehat{\mathbf{%
D}}) & \Leftrightarrow & (\ \ _{s}^{\shortparallel }\mathbf{g,\ \
_{s}^{\shortparallel }N},\mathbf{\ \ ^{\shortparallel }e}_{\alpha _{s}}%
\mathbf{,}\ _{s}^{\shortparallel }\widehat{\mathbf{D}}),%
\end{array}
\label{conv2s}
\end{equation}%
for $\ ^{\shortparallel}\mathbf{D}^{\star }=\ ^{\shortparallel}\nabla
^{\star}+ \ ^{\shortparallel }\widehat{\mathbf{Z}}^{\star }$ and $\
_{s}^{\shortparallel }\mathbf{D}^{\star }=\ ^{\shortparallel }\nabla
^{\star}+ \ _{s}^{\shortparallel }\widehat{\mathbf{Z}}^{\star }.$ This way,
we prove in section \ref{sec5} that there are certain general decoupling and
integration properties of nonassociative vacuum Einstein equations in
canonical nonholonomic s-variables, written in the form (\ref{cnsveinst2}),
for stationary configurations and parametric decompositions on $\hbar
,\kappa ,$ and $\hbar \kappa .$
\end{itemize}

\subsubsection{Nonsymmetric metrics and s-metrics and their inverses}

We study s-adapted metric structures in nonassociative nonholonomic dyadic
differential geometry and related star deformations of the LC-connections
and canonical s-connections on phase spaces endowed with quasi-Hopf
s-structures. A star metric symmetric s-tensor (\ref{ssdm}) can be
represented in the form
\begin{equation*}
\ _{\star s}^{\shortparallel }\mathbf{g=\ }_{\star }^{\shortparallel }%
\mathbf{g}_{\alpha _{s}\beta _{s}}\star _{s}(\ ^{\shortparallel }\mathbf{e}%
^{\alpha _{s}}\otimes _{\star s}\ ^{\shortparallel }\mathbf{e}^{\beta
_{s}})\in \Omega _{\star }^{1}\otimes _{\star s}\Omega _{\star }^{1}.
\end{equation*}%
There are considered real-valued s-adapted coefficients $\
_{\star}^{\shortparallel }\mathbf{g(\ ^{\shortparallel }\mathbf{e}}_{\alpha
_{s}}\mathbf{,\ ^{\shortparallel }\mathbf{e}}_{\beta _{s}})=$ $\
_{\star}^{\shortparallel }\mathbf{g}_{\alpha _{s}\beta _{s}}= \
_{\star}^{\shortparallel }\mathbf{g}_{\beta _{s}\alpha _{s}}$ $\in \mathcal{A%
}_{s}^{\star }$, which follows from the property $\ _{\star
s}^{\shortparallel}\mathbf{g(\ ^{\shortparallel }z,\ ^{\shortparallel }v)}=
\ _{\star s}^{\shortparallel }\mathbf{g}(\mathbf{\ _{\intercal
}^{\shortparallel }v},\mathbf{\ _{\intercal }^{\shortparallel }z})$ for all $%
\mathbf{\ ^{\shortparallel }z=\{\ _{s}^{\shortparallel }z\},\
^{\shortparallel }v=\{\ _{s}^{\shortparallel }v\}\in }\emph{Vec}_{\star
_{s}} $ when the braiding s-operator "$\mathbf{\ _{\intercal }^{{}}}$%
"\thinspace\ is defined as in formula (\ref{braidop}). Such a s-metric is
compatible with a star s-connection $\ _{s}^{\shortparallel }\mathbf{D}%
^{\star }$ (\ref{sconhopf}) if the condition
\begin{equation}
\ _{s}^{\shortparallel}\mathbf{D}^{\star }\ _{\star s}^{\shortparallel }%
\mathbf{g}=0  \label{mcompnas}
\end{equation}
is satisfied.

R-flux deformations result also in nonsymmetric metric structures \cite%
{blumenhagen16,aschieri17} and such geometric d- / s-objects have to be
considered additionally to the symmetric metric. A nonsymmetric metric (\ref%
{offdns}) on phase space $\mathbf{T}_{\shortparallel }^{\ast }\mathbf{V}$
can be defined in generic off-diagonal form with respect to local coordinate
bases when%
\begin{equation}
\ _{\star }^{\shortparallel }\mathsf{G}_{\alpha \beta }=\ _{\star
}^{\shortparallel }g_{\alpha \beta }-i\kappa \mathcal{R}_{\quad \alpha
}^{\tau \xi }\ \mathbf{^{\shortparallel }}\partial _{\xi }\ _{\star
}^{\shortparallel }g_{\beta \tau }.  \label{offdns1}
\end{equation}%
In general, such coefficients define a nonsymmetric $8\times 8$ matrix. With
respect to a dyadic s-adapted basis $\ ^{\shortparallel }\mathbf{e}_{\xi
_{s}}$ and tensor products of their dual, a nonsymmetric s-metric structure (%
\ref{nssdm}) can be parameterized in a \newline
$[(2\times 2)+(2\times 2)]+$ $[(2\times 2)+(2\times 2)]$ block form,
\begin{equation}
\ _{\star }^{\shortparallel }\mathfrak{g}_{\alpha _{s}\beta _{s}}=\ _{\star
}^{\shortparallel }\mathbf{g}_{\alpha _{s}\beta _{s}}-i\kappa \overline{%
\mathcal{R}}_{\quad \alpha _{s}}^{\tau _{s}\xi _{s}}\ \mathbf{%
^{\shortparallel }e}_{\xi _{s}}\ _{\star }^{\shortparallel }\mathbf{g}%
_{\beta _{s}\tau _{s}}.  \label{dmss1}
\end{equation}%
In these formulas, the R-flux coefficients re-defined in s-adapted form $%
\overline{\mathcal{R}}_{\quad \alpha _{s}}^{\tau _{s}\xi _{s}}$ determine a
star nonsymmetric generalization of the commutative s-metric (\ref{commetr})
with respective frame transforms from coordinate cobases and inversely, then
the metric structure can be represented equivalently in s-adapted and/or
(not adapted) coordinate forms,
\begin{equation*}
\ _{\star s}^{\shortparallel }\mathfrak{g}=\ _{\star }^{\shortparallel }%
\mathfrak{g}_{\alpha _{s}\beta _{s}}\star _{s}(\ ^{\shortparallel }\mathbf{e}%
^{\alpha _{s}}\otimes _{\star s}\ ^{\shortparallel }\mathbf{e}^{\beta
_{s}})=\ _{\star }^{\shortparallel }\mathsf{G}_{\alpha \beta }\star (\ d\
^{\shortparallel }u^{\alpha }\otimes _{\star }d\ ^{\shortparallel }u^{\beta
}),
\end{equation*}%
where $\ _{\star }^{\shortparallel }\mathfrak{g}_{\alpha _{s}\beta _{s}}\neq
\ _{\star }^{\shortparallel }\mathfrak{g}_{\beta _{s}\alpha _{s}}$ and $\
_{\star }^{\shortparallel }\mathsf{G}_{\alpha \beta }\neq \ _{\star
}^{\shortparallel }\mathsf{G}_{\beta \alpha }.$

For quasi-Hopf configurations, a holonomic formalism of constructing
inversions of matrices in $\mathcal{A}^{\star }$ is elaborated in section
5.2 of \cite{aschieri17}. It allows us to compute the inverse matrix $\
^{\shortparallel }\overline{\mathsf{G}}^{-1}=\{\ _{\star }^{\shortparallel }%
\mathsf{G}^{\alpha \beta }\}$ of a matrix $\ ^{\shortparallel }\overline{%
\mathsf{G}}=\{\ _{\star }^{\shortparallel }\mathsf{G}_{\alpha \beta }\}$ as
a solution of algebraic equations $\ _{\star }^{\shortparallel }\mathsf{G}%
^{\alpha \beta }\cdot \ _{\star }^{\shortparallel }\mathsf{G}_{\beta \gamma
}=\ _{\star }^{\shortparallel }\mathsf{G}_{\gamma \beta }\cdot $ $\ _{\star
}^{\shortparallel }\mathsf{G}^{\beta \alpha }=\delta _{\beta }^{\alpha }.$
Such matrix formulas are derived using geometric series, when
\begin{equation}
\ _{\star }^{\shortparallel }\mathsf{G}^{\alpha \beta }=\ ^{\shortparallel
}g^{\alpha \beta }-i\kappa \ ^{\shortparallel }g^{\alpha \tau }\mathcal{R}%
_{\quad \tau }^{\mu \nu }(\partial _{\mu }\ ^{\shortparallel }g_{\nu
\varepsilon })\ ^{\shortparallel }g^{\varepsilon \beta }+O(\kappa ^{2}),
\label{offdns1inv}
\end{equation}%
which $\mathcal{R}_{\quad \tau }^{\mu \nu }$ constructing via nonholonomic
transforms. Similar formulas can be written in block d-forms for matrices
constructed from s-adapted coefficients of a nonsymmetric star s-metric $\
_{\star }^{\shortparallel }\mathfrak{g}_{\alpha _{s}\beta _{s}}$ and,
respectively, of a symmetric star d-metric $\ _{\star }^{\shortparallel}
\mathbf{g}_{\alpha _{s}\beta _{s}}.$

\subsubsection{Star deformed LC-connections and canonical s-connections}

We define star deformations of the two linear connections structure (\ref%
{twocon}) using a star d-metric $\ _{\star }^{\shortparallel }\mathbf{g}$
and work in equivalent forms with respective two different linear
connections: {\small
\begin{equation}
(\ _{\star s}^{\shortparallel }\mathbf{g,\ _{s}^{\shortparallel }N}%
)\rightarrow \left\{
\begin{array}{cc}
\ _{\star }^{\shortparallel }\mathbf{\nabla :} &
\begin{array}{c}
\ _{\star }^{\shortparallel }\mathbf{\nabla }\ _{\star s}^{\shortparallel }%
\mathbf{g}=0;\ _{\nabla }^{\shortparallel }\mathcal{T}^{\star }=0,%
\mbox{\
star LC-connection}; \\
\end{array}
\\
\ _{s}^{\shortparallel }\widehat{\mathbf{D}}^{\star }: &
\begin{array}{c}
\ _{s}^{\shortparallel }\widehat{\mathbf{D}}^{\star }\ _{\star
s}^{\shortparallel }\mathbf{g}=0;\ h_{1}\ ^{\shortparallel }\widehat{%
\mathcal{T}}^{\star }=0,v_{2}\ ^{\shortparallel }\widehat{\mathcal{T}}%
^{\star }=0,c_{3}\ ^{\shortparallel }\widehat{\mathcal{T}}^{\star }=0,c_{4}\
^{\shortparallel }\widehat{\mathcal{T}}^{\star }=0, \\
h_{1}v_{2}\ ^{\shortparallel }\widehat{\mathcal{T}}^{\star }\neq
0,h_{1}c_{s}\ ^{\shortparallel }\widehat{\mathcal{T}}^{\star }\neq
0,v_{2}c_{s}\ ^{\shortparallel }\widehat{\mathcal{T}}^{\star }\neq
0,c_{3}c_{4}\ ^{\shortparallel }\widehat{\mathcal{T}}^{\star }\neq 0,%
\mbox{canonical s-connection},%
\end{array}%
\end{array}%
\right.  \label{twoconsstar}
\end{equation}%
} where $\ _{s}^{\shortparallel }\widehat{\mathbf{D}}^{\star }=(h_{1}\
^{\shortparallel }\widehat{\mathbf{D}}^{\star },\ v_{2}\ ^{\shortparallel }%
\widehat{\mathbf{D}}^{\star },\ c_{3}\ ^{\shortparallel }\widehat{\mathbf{D}}%
^{\star },\ c_{4}\ ^{\shortparallel }\widehat{\mathbf{D}}^{\star })$ encodes
a nonholonomic dyadic horizontal and (co) vertical splitting, is a
s-connection adapted to a nonlinear s--connection structure $\
_{s}^{\shortparallel }\mathbf{N.}$\footnote{%
We shall write in brief for such formulas that they are with a h1-v2-c3-c4
decomposition.} Such a canonical s-connection satisfies the metricity
conditions (\ref{mcompnas}) but contains certain "mixing shells" nontrivial
torsion coefficients of $\ ^{\shortparallel }\widehat{\mathcal{T}}^{\star }$
even "pure" shell torsions vanish in s-adapted (co) frames $\
^{\shortparallel }\mathbf{e}_{\alpha _{s}}$ (\ref{nadapbdsc}) and $\
^{\shortparallel }\mathbf{e}^{\alpha _{s}}$ (\ref{nadapbdss}).

In above formulas, the nonassociative metric compatibility of both linear
connections is stated as in (\ref{mcompnas}) and the star canonical
s-torsion components are parameterized following decompositions(\ref%
{storsnonassoc}), when
\begin{eqnarray}
\mathbf{\mathbf{\mathbf{\mathbf{\ \ \ }}}}_{s}^{\shortparallel }\mathcal{T}%
^{\star } &=&\{\mathbf{\mathbf{\mathbf{\mathbf{\ ^{\shortparallel }}}}}%
\widehat{\mathbf{T}}_{\star \alpha _{s}\beta _{s}}^{\gamma _{s}}=(\mathbf{%
\mathbf{\mathbf{\mathbf{\ ^{\shortparallel }}}}}\widehat{T}_{\star \
j_{1}k_{1}}^{i_{1}},\mathbf{\mathbf{\mathbf{\mathbf{\ ^{\shortparallel }}}}}%
\widehat{T}_{\star a_{2}j_{1}}^{i_{1}\ },\mathbf{\mathbf{\mathbf{\mathbf{\ \
^{\shortparallel }}}}}\widehat{T}_{\star j_{1}c_{2}}^{i_{1}\ },\mathbf{%
\mathbf{\mathbf{\mathbf{\ ^{\shortparallel }}}}}\widehat{T}_{\star
j_{1}i_{1}}^{c_{2}},\mathbf{\mathbf{\mathbf{\mathbf{\ ^{\shortparallel }}}}}%
\widehat{T}_{\star a_{2}\ j_{1}}^{\ c_{2}},\mathbf{\mathbf{\mathbf{\mathbf{\
\ ^{\shortparallel }}}}}\widehat{T}_{\star b_{2}c_{2}\ }^{\ a_{2}},\mathbf{%
\mathbf{\mathbf{\mathbf{\ }}}}  \label{nacanscoef} \\
&&\mathbf{\mathbf{\mathbf{\mathbf{^{\shortparallel }}}}}\widehat{T}_{\star \
j_{2}}^{i_{2}\ a_{3}},\ \mathbf{\mathbf{\mathbf{\mathbf{\ ^{\shortparallel }}%
}}}\widehat{T}_{\star a_{3}j_{2}i_{2}},\mathbf{\mathbf{\mathbf{\mathbf{\ \
^{\shortparallel }}}}}\widehat{T}_{\star c_{3}\ j_{2}}^{\ a_{3}},\mathbf{%
\mathbf{\mathbf{\mathbf{\ ^{\shortparallel }}}}}\widehat{T}_{\star a_{3}\
}^{\ b_{3}c_{3}},\mathbf{\mathbf{\mathbf{\mathbf{\ ^{\shortparallel }}}}}%
\widehat{T}_{\star \ j_{3}}^{i_{3}\ a_{4}},\mathbf{\mathbf{\mathbf{\mathbf{\
^{\shortparallel }}}}}\widehat{T}_{\star a_{4}j_{3}i_{3}},\mathbf{\mathbf{%
\mathbf{\mathbf{\ ^{\shortparallel }}}}}\widehat{T}_{\star c_{4}\ j_{3}}^{\
a_{4}},\mathbf{\mathbf{\mathbf{\mathbf{\ \ ^{\shortparallel }}}}}\widehat{T}%
_{\star a_{4}\ }^{\ b_{4}c_{4}})\}.  \notag
\end{eqnarray}%
The coefficients for such canonical s-torsion (\ref{storsnonassoc}) are
computed
\begin{eqnarray}
\mathbf{\mathbf{\mathbf{\mathbf{\ \ ^{\shortparallel }}}}}\widehat{T}_{\star
\ j_{1}k_{1}}^{i_{1}} &=&\mathbf{\mathbf{\mathbf{\mathbf{\ \
^{\shortparallel }}}}}\widehat{L}_{\star j_{1}k_{1}}^{i_{1}}-\mathbf{\mathbf{%
\mathbf{\mathbf{\ \ ^{\shortparallel }}}}}\widehat{L}_{\star
k_{1}j_{1}}^{i_{1}},\mathbf{\mathbf{\mathbf{\mathbf{\ ^{\shortparallel }}}}}%
\widehat{T}_{\star a_{2}j_{1}}^{i_{1}\ }=\mathbf{\mathbf{\mathbf{\mathbf{\
^{\shortparallel }}}}}\widehat{C}_{\star a_{2}j_{1}}^{i_{1}},\ \mathbf{%
\mathbf{\mathbf{\mathbf{\ \ ^{\shortparallel }}}}}\widehat{T}_{\star
j_{1}c_{2}}^{i_{1}\ }=\mathbf{\mathbf{\mathbf{\mathbf{\ ^{\shortparallel }}}}%
}\widehat{C}_{\star j_{1}c_{2}}^{i_{1}},\mathbf{\mathbf{\mathbf{\mathbf{\
^{\shortparallel }}}}}\widehat{T}_{\star j_{1}i_{1}}^{c_{2}}=-\ \mathbf{%
\mathbf{\mathbf{\mathbf{\ \ ^{\shortparallel }}}}}\Omega _{\star
j_{1}i_{1}}^{c_{2}},  \notag \\
\mathbf{\mathbf{\mathbf{\mathbf{\ \ ^{\shortparallel }}}}}\widehat{T}_{\star
a_{2}\ j_{1}}^{\ c_{2}} &=&\mathbf{\mathbf{\mathbf{\mathbf{\ \
^{\shortparallel }}}}}\widehat{L}_{\star a_{2}\ j_{1}}^{c_{2}}-\mathbf{%
\mathbf{\mathbf{\mathbf{\ ^{\shortparallel }}}}}e_{a_{2}}(\mathbf{\mathbf{%
\mathbf{\mathbf{\ ^{\shortparallel }}}}}N_{\star j_{1}}^{c_{2}}),\mathbf{%
\mathbf{\mathbf{\mathbf{\ \ ^{\shortparallel }}}}}\widehat{T}_{\star
b_{2}c_{2}\ }^{\ a_{2}}=\mathbf{\mathbf{\mathbf{\mathbf{\ ^{\shortparallel }}%
}}}\widehat{C}_{\star b_{2}c_{2}\ }^{a_{2}}-\mathbf{\mathbf{\mathbf{\mathbf{%
\ \ ^{\shortparallel }}}}}\widehat{C}_{\star c_{2}b_{2}\ }^{a_{2}};
\label{nacanscoef1} \\
\mathbf{\mathbf{\mathbf{\mathbf{\ \ ^{\shortparallel }}}}}\widehat{T}_{\star
\ j_{2}}^{i_{2}\ a_{3}} &=&\mathbf{\mathbf{\mathbf{\mathbf{\
^{\shortparallel }}}}}\widehat{C}_{\star j_{2}}^{i_{2}a_{3}},\ \mathbf{%
\mathbf{\mathbf{\mathbf{\ ^{\shortparallel }}}}}\widehat{T}_{\star
a_{3}j_{2}i_{2}}=-\ \mathbf{\mathbf{\mathbf{\mathbf{\ \ ^{\shortparallel }}}}%
}\Omega _{\star a_{3}j_{2}i_{2}},\mathbf{\mathbf{\mathbf{\mathbf{\ \
^{\shortparallel }}}}}\widehat{T}_{\star c_{3}\ j_{2}}^{\ a_{3}}=\mathbf{%
\mathbf{\mathbf{\mathbf{\ \ ^{\shortparallel }}}}}\widehat{L}_{\star c_{3}\
j_{2}}^{a_{3}}-\mathbf{\mathbf{\mathbf{\mathbf{\ ^{\shortparallel }}}}}%
e^{a_{3}}(\mathbf{\mathbf{\mathbf{\mathbf{\ ^{\shortparallel }}}}}N_{\star
c_{3}j_{2}}),  \notag \\
\mathbf{\mathbf{\mathbf{\mathbf{\ \ ^{\shortparallel }}}}}\widehat{T}_{\star
a_{3}\ }^{\ b_{3}c_{3}} &=&\mathbf{\mathbf{\mathbf{\mathbf{\
^{\shortparallel }}}}}\widehat{C}_{\star a_{3}}^{\ b_{3}c_{3}}-\mathbf{%
\mathbf{\mathbf{\mathbf{\ \ ^{\shortparallel }}}}}\widehat{C}_{\star
a_{3}}^{\ c_{3}b_{3}},  \notag \\
\mathbf{\mathbf{\mathbf{\mathbf{\ ^{\shortparallel }}}}}\widehat{T}_{\star \
j_{3}}^{i_{3}\ a_{4}} &=&\mathbf{\mathbf{\mathbf{\mathbf{\ ^{\shortparallel }%
}}}}\widehat{C}_{\star j_{3}}^{i_{3}a_{4}},\ \mathbf{\mathbf{\mathbf{\mathbf{%
\ ^{\shortparallel }}}}}\widehat{T}_{\star a_{4}j_{3}i_{3}}=-\ \mathbf{%
\mathbf{\mathbf{\mathbf{\ \ ^{\shortparallel }}}}}\Omega _{\star
a_{4}j_{3}i_{3}},\mathbf{\mathbf{\mathbf{\mathbf{\ \ ^{\shortparallel }}}}}%
\widehat{T}_{\star c_{4}\ j_{3}}^{\ a_{4}}=\mathbf{\mathbf{\mathbf{\mathbf{\
\ ^{\shortparallel }}}}}\widehat{L}_{\star c_{4}\ j_{3}}^{~a_{4}}-\mathbf{%
\mathbf{\mathbf{\mathbf{\ ^{\shortparallel }}}}}e^{a_{4}}(\mathbf{\mathbf{%
\mathbf{\mathbf{\ ^{\shortparallel }}}}}N_{\star c_{4}j_{3}}),  \notag \\
\mathbf{\mathbf{\mathbf{\mathbf{\ \ ^{\shortparallel }}}}}\widehat{T}_{\star
a_{4}\ }^{\ b_{4}c_{4}} &=&\mathbf{\mathbf{\mathbf{\mathbf{\
^{\shortparallel }}}}}\widehat{C}_{\star a_{4}}^{\ b_{4}c_{4}}-\mathbf{%
\mathbf{\mathbf{\mathbf{\ \ ^{\shortparallel }}}}}\widehat{C}_{\star
a_{4}}^{\ c_{4}b_{4}}.  \notag
\end{eqnarray}%
The next step is to find the dependence of canonical torsion s-coefficients (%
\ref{nacanscoef1}) on s-metric and s-connection coefficients by introducing
a set of s-adapted Christoffel symbols
\begin{equation}
\ _{[0]}^{\shortparallel }\widehat{\mathbf{\Gamma }}_{\star \gamma
_{s}\alpha _{s}\beta _{s}}=\ _{\star }^{\shortparallel }\mathbf{g}_{\gamma
_{s}\tau _{s}}\ _{[0]}^{\shortparallel }\widehat{\mathbf{\Gamma }}_{\star
\beta _{s}\alpha _{s}}^{\tau _{s}},  \label{0canconnonas}
\end{equation}%
for instance, with $\ _{[0]}^{\shortparallel }\widehat{L}_{\star
i_{1}j_{1}k_{1}}=\ _{\star }^{\shortparallel }g_{i_{1}m_{1}}\ \
_{[0]}^{\shortparallel }\widehat{L}_{\star j_{1}k_{1}}^{m_{1}},$ ...,$\
_{[0]}^{\shortparallel }\widehat{L}_{\star \ k_{3}}^{\ a_{4}b_{4}}=\ _{\star
}^{\shortparallel }g^{a_{4}c_{4}}\ \ _{[0]}^{\shortparallel }\widehat{L}%
_{\star c_{4}\ k_{3}}^{\ b_{4}},$ $\ _{[0]}^{\shortparallel }\widehat{C}%
_{\star i_{3}j_{3}}^{\quad c_{4}}=\ _{\star }^{\shortparallel
}g_{i_{3}m_{3}}\ _{[0]}^{\shortparallel }\widehat{C}_{\star \ j_{3}}^{m_{3}\
c_{4}},$ $\ _{[0]}^{\shortparallel }\widehat{C}_{\star }^{b_{4}e_{4}c_{4}}=\
_{\star }^{\shortparallel }g^{a_{4}e_{4}}\ _{[0]}^{\shortparallel }\widehat{C%
}_{\star \ e_{4}}^{b_{4}\ c_{4}},$ where s-components of a star deformation
of canonical s-connection (\ref{canhcs}) are computed with respect to
nonholonomic s-bases, using dyadic shell N-connection coefficients $\
_{s}^{\shortparallel }\mathbf{N}=\{\ ^{\shortparallel }N_{a_{s}i_{1}}\mathbf{%
\}}$ and the symmetric star s-metric for $\ _{\star}^{\shortparallel }%
\mathbf{g}_{\alpha _{s}\beta _{s}}$ (\ref{ssdm}),
\begin{eqnarray}
\ _{[0]}^{\shortparallel }\widehat{L}_{\star j_{1}k_{1}}^{i_{1}} &=&\frac{1}{%
2}\ \ \ _{\star }^{\shortparallel }g^{i_{1}r_{1}}(\ \ ^{\shortparallel }%
\mathbf{e}_{k_{1}}\ \ _{\star }^{\shortparallel }g_{j_{1}r_{1}}+\ \
^{\shortparallel }\mathbf{e}_{j_{1}}\ \ _{\star }^{\shortparallel
}g_{k_{1}r_{1}}-\ ^{\shortparallel }\mathbf{e}_{r_{1}}\ \ _{\star
}^{\shortparallel }g_{j_{1}k_{1}}),\   \label{0canscnas} \\
\ _{[0]}^{\shortparallel }\widehat{L}_{\star b_{2}\ k_{1}}^{a_{2}} &=&\
^{\shortparallel }e_{b_{2}}(\ ^{\shortparallel }N_{k_{1}}^{a_{2}})+\frac{1}{2%
}\ \ _{\star }^{\shortparallel }g^{b_{2}c_{2}}(\ ^{\shortparallel
}e_{k_{1}}\ \ _{\star }^{\shortparallel }g_{b_{2}c_{2}}-\ _{\star
}^{\shortparallel }g_{d_{2}c_{2}}\ ^{\shortparallel }e_{b_{2}}\
^{\shortparallel }N_{k_{1}}^{d_{2}}-\ \ _{\star }^{\shortparallel
}g_{d_{2}b_{2}}\ ^{\shortparallel }e_{c_{2}}\ ^{\shortparallel
}N_{k_{1}}^{d_{2}}),  \notag \\
\ _{[0]}^{\shortparallel }\widehat{C}_{\ \star j_{1}c_{2}}^{i_{1}\ } &=&%
\frac{1}{2}\ _{\star }^{\shortparallel }g^{ik}\ \ ^{\shortparallel }e_{c}\
_{\star }^{\shortparallel }g_{jk},\ \ \ \ _{[0]}^{\shortparallel }\widehat{C}%
_{\star b_{2}c_{2}}^{a_{2}}=\frac{1}{2}\ _{\star }^{\shortparallel
}g^{a_{2}d_{2}}(\ \ ^{\shortparallel }e_{c_{2}}\ _{\star }^{\shortparallel
}g_{b_{2}d_{2}}+\ ^{\shortparallel }e_{b_{2}}\ _{\star }^{\shortparallel
}g_{c_{2}d_{2}}-\ ^{\shortparallel }e_{d_{2}}\ _{\star }^{\shortparallel
}g_{b_{2}c_{2}}),  \notag \\
\ \ _{[0]}^{\shortparallel }\widehat{L}_{\star a_{3}\ k_{2}}^{\ b_{3}} &=&\
^{\shortparallel }e^{b_{3}}(\ ^{\shortparallel }N_{a_{3}k_{2}})+\frac{1}{2}\
\ _{\star }^{\shortparallel }g_{a_{3}c_{3}}(\ ^{\shortparallel }e_{k_{2}}\ \
_{\star }^{\shortparallel }g^{b_{3}c_{3}}-\ _{\star }^{\shortparallel
}g^{d_{3}c_{3}}\ \ ^{\shortparallel }e^{b_{3}}\ ^{\shortparallel
}N_{d_{3}k_{2}}-\ \ _{\star }^{\shortparallel }g^{d_{3}b_{3}}\
^{\shortparallel }e^{c}\ ^{\shortparallel }N_{d_{3}k_{2}}),  \notag \\
\ _{[0]}^{\shortparallel }\widehat{C}_{\star \ j_{2}}^{i_{2}\ c_{3}} &=&%
\frac{1}{2}\ _{\star }^{\shortparallel }g^{i_{2}k_{2}}\ ^{\shortparallel
}e^{c_{3}}\ \ _{\star }^{\shortparallel }g_{j_{2}k_{2}},\ \
_{[0]}^{\shortparallel }\widehat{C}_{\star \ a_{3}}^{b_{3}\ c_{3}}=\frac{1}{2%
}\ _{\star }^{\shortparallel }g_{a_{3}d_{3}}(\ ^{\shortparallel }e^{c_{3}}\
\ _{\star }^{\shortparallel }g^{b_{3}d_{3}}+~^{\shortparallel }e^{b_{3}}\
_{\star }^{\shortparallel }g^{c_{3}d_{3}}-\ ^{\shortparallel }e^{d_{3}}\
_{\star }^{\shortparallel }g^{b_{3}c_{3}}),  \notag \\
\ \ \ _{[0]}^{\shortparallel }\widehat{L}_{\star a_{4}\ k_{3}}^{\ b_{4}}
&=&\ \ ^{\shortparallel }e^{b_{4}}(\ ^{\shortparallel }N_{a_{4}k_{3}})+\frac{%
1}{2}\ \ _{\star }^{\shortparallel }g_{a_{4}c_{4}}(\ ^{\shortparallel
}e_{k_{3}}\ \ _{\star }^{\shortparallel }g^{b_{4}c_{4}}-\ \ _{\star
}^{\shortparallel }g^{d_{4}c_{4}}\ ^{\shortparallel }e^{b_{4}}\
^{\shortparallel }N_{d_{4}k_{3}}-\ _{\star }^{\shortparallel
}g^{d_{4}b_{4}}\ ^{\shortparallel }e^{c_{4}}\ ^{\shortparallel
}N_{d_{4}k_{3}}),  \notag \\
\ \ \ _{[0]}^{\shortparallel }\widehat{C}_{\star \ j_{3}}^{i_{3}\ c_{4}} &=&%
\frac{1}{2}\ _{\star }^{\shortparallel }g^{i_{3}k_{3}}\ ^{\shortparallel
}e^{c_{4}}\ _{\star }^{\shortparallel }g_{j_{3}k_{3}},\ \ \ \
_{[0]}^{\shortparallel }\widehat{C}_{\ \star a_{4}}^{b_{4}\ c_{4}}=\frac{1}{2%
}\ \ _{\star }^{\shortparallel }g_{a_{4}d_{4}}(\ ^{\shortparallel
}e^{c_{4}}\ _{\star }^{\shortparallel }g^{b_{4}d_{4}}+\ ^{\shortparallel
}e^{b_{4}}\ _{\star }^{\shortparallel }g^{c_{4}d_{4}}-\ ^{\shortparallel
}e^{d_{4}}\ _{\star }^{\shortparallel }g^{b_{4}c_{4}}).  \notag
\end{eqnarray}%
We parameterize R-flux s-adapted contributions (being proportional to $%
\kappa $) using s-adapted frames and symmetric s-metric coefficients,
\begin{equation}
\ _{[1]}^{\shortparallel }\widehat{\mathbf{\Gamma }}_{\star \alpha _{s}\beta
_{s}\mu _{s}}=\frac{1}{2}\overline{\mathcal{R}}_{\quad \mu _{s}}^{\xi
_{s}\tau _{s}}\ (\ \mathbf{^{\shortparallel }e}_{\xi _{s}}\ \mathbf{%
^{\shortparallel }e}_{\alpha _{s}}\ _{\star }^{\shortparallel }\mathbf{g}%
_{\beta _{s}\tau _{s}}+\ \mathbf{^{\shortparallel }e}_{\xi _{s}}\mathbf{%
^{\shortparallel }e}_{\beta _{s}}\ _{\star }^{\shortparallel }\mathbf{g}%
_{\alpha _{s}\tau _{s}}).  \label{aux311}
\end{equation}%
In result, we define in complete parametric form the N-adapted coefficients
of the star canonical d-connection $\ ^{\shortparallel}\widehat{\mathbf{D}}%
^{\star }=\{\ ^{\shortparallel }\widehat{\mathbf{\Gamma }}_{\star \alpha
\beta }^{\gamma }\}$ considering a s-decomposition used torsion s-tensors (%
\ref{irevndecomdc}),
\begin{eqnarray*}
\ \ ^{\shortparallel }\widehat{\mathbf{\Gamma }}_{\star \alpha _{s}\beta
_{s}}^{\gamma _{s}} &=&\ \ _{[0]}^{\shortparallel }\widehat{\mathbf{\Gamma }}%
_{\star \alpha _{s}\beta _{s}}^{\gamma _{s}}+i\kappa \
_{[1]}^{\shortparallel }\widehat{\mathbf{\Gamma }}_{\star \alpha _{s}\beta
_{s}}^{\gamma _{s}}=(\ ^{\shortparallel }\widehat{L}_{\star
j_{1}k_{1}}^{i_{1}},\ \ ^{\shortparallel }\widehat{L}_{\star b_{2}\
k_{1}}^{a_{2}},\ \ ^{\shortparallel }\widehat{C}_{\star \
j_{1}c_{2}}^{i_{1}\ },\ \ ^{\shortparallel }\widehat{C}_{\star
b_{2}c_{2}}^{a_{2}};\ \  \\
&&\ \ ^{\shortparallel }\widehat{L}_{\star j_{2}k_{2}}^{i_{2}},\ \
^{\shortparallel }\widehat{L}_{\star a_{3}k_{2}}^{\ b_{3}},\ \
^{\shortparallel }\widehat{C}_{\star \ j_{2}}^{i_{2}\ c_{3}},\ \
^{\shortparallel }\widehat{C}_{\star a_{3}b_{3}}^{\quad c_{3}};\ \
^{\shortparallel }\widehat{L}_{\star j_{3}k_{3}}^{i_{3}},\ \
^{\shortparallel }\widehat{L}_{\star a_{4}\ k_{3}}^{\ b_{4}},\ \
^{\shortparallel }\widehat{C}_{\star \ j_{3}}^{i_{3}\ c_{4}},\ \
^{\shortparallel }\widehat{C}_{\star a_{4}b_{4}}^{\quad c_{4}}).
\end{eqnarray*}%
The nonsymmetric s-metric $\ _{\star }^{\shortparallel }\mathfrak{g}_{\alpha
_{s}\beta _{s}}$ (\ref{dmss1}) is involved following the condition%
\begin{equation}
\ \mathbf{^{\shortparallel }e}^{\tau _{s}}\star _{s}\ _{\star
}^{\shortparallel }\mathfrak{g}_{\tau _{s}\gamma _{s}}\star _{s}\
^{\shortparallel }\widehat{\mathbf{\Gamma }}_{\star \alpha _{s}\beta
_{s}}^{\gamma _{s}}=\ \mathbf{^{\shortparallel }e}^{\tau _{s}}\star _{s}(\
_{[0]}^{\shortparallel }\widehat{\mathbf{\Gamma }}_{\star \tau _{s}\alpha
_{s}\beta _{s}}+i\kappa \ _{[1]}^{\shortparallel }\widehat{\mathbf{\Gamma }}%
_{\star \tau _{s}\alpha _{s}\beta _{s}}).  \label{eqnasdmdc}
\end{equation}%
In these formulas, the s-components of $\ _{[0]}^{\shortparallel }\widehat{%
\mathbf{\Gamma }}_{\star \tau _{s}\alpha _{s}\beta _{s}}$ are determined by
formulas (\ref{0canconnonas}), when $\ _{[1]}^{\shortparallel }\widehat{%
\mathbf{\Gamma }}_{\star \alpha _{s}\beta _{s}\mu _{s}}$(\ref{aux311})
involve s-compositions of $\ _{\star }^{\shortparallel }\mathbf{g}_{\beta
_{s}\tau _{s}}.$

For nonassociative LC-configurations, formulas similar to (\ref{eqnasdmdc})
were considered in \cite{aschieri17}. Following our system of notations, we
can extract the coefficients $\ ^{\shortparallel }\nabla ^{\star }= \{\
^{\shortparallel}\Gamma _{\star \beta \alpha }^{\gamma }\}$ with respect to
local coordinate dual frames $\ ^{\shortparallel }\partial _{\tau },$
\begin{equation*}
\ _{\star }^{\shortparallel }\mathsf{G}_{\gamma \nu }\star \mathbf{\mathbf{\
^{\shortparallel }}}\Gamma _{\star \alpha \beta }^{\nu }=\frac{1}{2}\left( \
\mathbf{^{\shortparallel }\partial }_{\beta }\ _{\star }^{\shortparallel
}g_{\alpha \gamma }+\ \mathbf{^{\shortparallel }\partial }_{\alpha }\
_{\star }^{\shortparallel }g_{\beta \gamma }-\ \mathbf{^{\shortparallel
}\partial }_{\gamma }\ _{\star }^{\shortparallel }g_{\alpha \beta }+i\kappa
\mathcal{R}_{\ \gamma }^{\xi \tau }\ (\ \mathbf{^{\shortparallel }}\partial
_{\xi }\ \mathbf{^{\shortparallel }}\partial _{\beta }\ _{\star
}^{\shortparallel }g_{\alpha \tau }+\ \mathbf{^{\shortparallel }}\partial
_{\xi }\ \mathbf{^{\shortparallel }}\partial _{\alpha }\ _{\star
}^{\shortparallel }g_{\beta \tau })\right) .
\end{equation*}%
In such formulas, it is used the off-diagonal nonsymmetric metric $\ _{\star
}^{\shortparallel }\mathsf{G}_{\alpha \beta }$ computed as in (\ref{offdns1}%
). All formulas can be rewritten in s-adapted form if $\
^{\shortparallel}\partial _{\alpha }\rightarrow \ ^{\shortparallel}\mathbf{e}%
_{\alpha _{s}}.$ We can compute also s-adapted coefficients of
nonassociative star deformed $\nabla$ and star deformation of the
associative and commutative canonical distortion s-relation (\ref{candistr}%
),
\begin{equation}
\ _{s}^{\shortparallel}\widehat{\mathbf{D}}^{\star }=\
^{\shortparallel}\nabla ^{\star }+\ _{\star s}^{\shortparallel }\widehat{%
\mathbf{Z}}.  \label{candistrnas}
\end{equation}%
The s-adapted coefficients of distortion s-tensor, $\
_{s\star}^{\shortparallel}\widehat{\mathbf{Z}}= \{\ ^{\shortparallel }
\widehat{\mathbf{Z}}_{\ \star \beta _{s}\gamma _{s}}^{\alpha _{s}}[\
^{\shortparallel}\widehat{\mathbf{T}}_{\ \star \beta _{s}\gamma
_{s}}^{\alpha _{s}}]\}$ can be computed as an algebraic combination of the
coefficients of the nonassociative canonical s-tensor $\
_{s}^{\shortparallel }\mathcal{T}^{\star }= \{\ ^{\shortparallel }\widehat{%
\mathbf{T}}_{\ \star \beta _{s}\gamma _{s}}^{\alpha _{s}}\}$ for
s-components (\ref{nacanscoef1}) computed using formulas (\ref{0canconnonas}%
) and (\ref{0canscnas}).

Using nonholonomic dyadic decompositions, we can describe any model of
nonassociative phase geometry determined by a nonsymmetric s-metric
structure $\ _{\star }^{\shortparallel }\mathfrak{g}_{\alpha _{s}\beta _{s}}$%
. For any given N-connection and R-flux s-coefficients data, we can consider
only the symmetric star s-metric $\ _{\star }^{\shortparallel }\mathbf{g}%
_{\alpha _{s}\beta _{s}}$, see formulas (\ref{dmss1}). Such a nonassociative
geometry with symmetric and nonsymmetric s-metrics can be described
equivalently both in terms of the star LC-connection $\
^{\shortparallel}\nabla ^{\star }$ and/or the star canonical s-connection $\
_{s}^{\shortparallel }\widehat{\mathbf{D}}^{\star }.$ To construct solutions
of physically important systems of PDEs, it is convenient to work with
nonholonomic dyadic canonical geometric data $(\ _{\star s}^{\shortparallel }%
\mathfrak{g}, \ _{s}^{\shortparallel }\widehat{\mathbf{D}}^{\star }).$ Then
we can redefine the constructions in terms of star deformed
LC-configurations using s-distortions (\ref{candistrnas}). In certain
important cases, we can extract such nonassociative LC-configurations
imposing some zero s-torsion conditions%
\begin{equation}
\ _{\star s}^{\shortparallel }\widehat{\mathbf{Z}}=0,%
\mbox{ which is
equivalent to }\ _{s}^{\shortparallel }\widehat{\mathbf{D}}_{\mid \
_{s}^{\shortparallel }\widehat{\mathbf{T}}=0}^{\star }=\ ^{\shortparallel
}\nabla ^{\star }.  \label{lccondnonass}
\end{equation}%
Even we can find nontrivial solutions which result in zero coefficients of (%
\ref{nacanscoef1}), when $\ ^{\shortparallel}\widehat{\mathbf{T}}_{\
\star\beta _{s}\gamma _{s}}^{\alpha _{s}}=0,$ which is equivalent to (\ref%
{lccondnonass}), the anholonomy coefficients $\ ^{\shortparallel }w_{\alpha
_{s}\beta _{s}}^{\gamma _{s}}$ may be not zero (formulas (\ref{anhrel}) can
be written for any (dual) s-shell). So, in general, we work with
nonholonomic s-frames and involved (non) symmetric s-metrics can be written
in local coordinate forms as generic off-diagonal matrices. Such matrices
can not be diagonalized by coordinate transforms on a finite spacetime or
phase space region if the anholonomy coefficients do not vanish. This
criteria works in all types of nonassociative / noncommutative geometric
models and their constraints to commutative nonholonomic configurations.

Nonassociative fundamental geometric s-objects from (\ref{conv2s}) can be
decomposed on parameters $\hbar ,\kappa ,$ and $\hbar \kappa $ which is
important to compute R-flux deformations of associative and commutative
geometric models. In appendix \ref{assparamet}, we provide respective
formulas for such decompositions of nonassociative symmetric and
nonsymmetric metrics and canonical s-connections.

\subsection{Nonassociative parametric decompositions of canonical Riemann
and Ricci s-tensors}

We show how nonholonomic dyadic star deformations of important s-objects can
be performed in parametric form but omit cumbersome formulas for further
h1-v2-c3-c4 decompositions.

\subsubsection{Star and parametric deformed canonical curvature s-tenors}

The nonassociative canonical Riemann s-tensor can be computed by introducing
the star canonical s-connection coefficients $\ ^{\shortparallel }\widehat{%
\mathbf{\Gamma }}_{\star \alpha _{s}\beta _{s}}^{\gamma _{s}}$ (\ref%
{eqnasdmdc}) in formulas (\ref{nadriemhopf}). In "hat" s-adapted variables,
we have
\begin{eqnarray}
\mathbf{\mathbf{\mathbf{\mathbf{\ ^{\shortparallel }}}}}\widehat{\mathcal{%
\Re }}_{\quad \alpha _{s}\beta _{s}\gamma _{s}}^{\star \mu _{s}} &=&\mathbf{%
\mathbf{\mathbf{\mathbf{\ _{1}^{\shortparallel }}}}}\widehat{\mathcal{\Re }}%
_{\quad \alpha _{s}\beta _{s}\gamma _{s}}^{\star \mu _{s}}+\mathbf{\mathbf{%
\mathbf{\mathbf{\ _{2}^{\shortparallel }}}}}\widehat{\mathcal{\Re }}_{\quad
\alpha _{s}\beta _{s}\gamma _{s}}^{\star \mu _{s}},\mbox{ where }
\label{nadriemhopfcan} \\
\mathbf{\mathbf{\mathbf{\mathbf{\ _{1}^{\shortparallel }}}}}\widehat{%
\mathcal{\Re }}_{\quad \alpha _{s}\beta _{s}\gamma _{s}}^{\star \mu _{s}}
&=&\ \mathbf{^{\shortparallel }e}_{\gamma _{s}}\mathbf{\ ^{\shortparallel }}%
\widehat{\Gamma }_{\star \alpha _{s}\beta _{s}}^{\mu _{s}}-\ \mathbf{%
^{\shortparallel }e}_{\beta _{s}}\mathbf{\ ^{\shortparallel }}\widehat{%
\Gamma }_{\star \alpha _{s}\gamma _{s}}^{\mu }+\mathbf{\ ^{\shortparallel }}%
\widehat{\Gamma }_{\star \nu _{s}\tau _{s}}^{\mu _{s}}\star _{s}(\delta _{\
\gamma _{s}}^{\tau _{s}}\mathbf{\ ^{\shortparallel }}\widehat{\Gamma }%
_{\star \alpha _{s}\beta _{s}}^{\nu _{s}}-\delta _{\ \beta _{s}}^{\tau _{s}}%
\mathbf{\ ^{\shortparallel }}\widehat{\Gamma }_{\star \alpha _{s}\gamma
_{s}}^{\nu _{s}})+\mathbf{\ ^{\shortparallel }}w_{\beta _{s}\gamma
_{s}}^{\tau _{s}}\star _{s}\mathbf{\ ^{\shortparallel }}\widehat{\Gamma }%
_{\star \alpha _{s}\tau _{s}}^{\mu _{s}},  \notag \\
\ _{2}^{\shortparallel }\widehat{\mathcal{\Re }}_{\quad \alpha _{s}\beta
_{s}\gamma _{s}}^{\star \mu _{s}} &=&i\kappa \ ^{\shortparallel }\widehat{%
\Gamma }_{\star \nu _{s}\tau _{s}}^{\mu _{s}}\star _{s}(\mathcal{R}_{\quad
\gamma _{s}}^{\tau _{s}\xi _{s}}\ \mathbf{^{\shortparallel }e}_{\xi _{s}}%
\mathbf{\ ^{\shortparallel }}\widehat{\Gamma }_{\star \alpha _{s}\beta
_{s}}^{\nu _{s}}-\mathcal{R}_{\quad \beta _{s}}^{\tau _{s}\xi _{s}}\ \mathbf{%
^{\shortparallel }e}_{\xi _{s}}\mathbf{\ ^{\shortparallel }}\widehat{\Gamma }%
_{\star \alpha _{s}\gamma _{s}}^{\nu _{s}}).  \notag
\end{eqnarray}%
Further h1-v2-c3-c4 decompositions in s-adapted coefficients ("hat" variants
of (\ref{nadriemhopf1c})), can be performed using formulas (\ref{0canscnas})
and (\ref{aux311}).

Parametric decompositions of the star canonical s-connections, see (\ref%
{aux39}) and (\ref{aux51}), can be computed in explicit s-form
\begin{equation*}
\ ^{\shortparallel }\widehat{\mathbf{\Gamma }}_{\star \alpha _{s}\beta
_{s}}^{\gamma _{s}}=\ _{[0]}^{\shortparallel }\widehat{\mathbf{\Gamma }}%
_{\star \alpha _{s}\beta _{s}}^{\nu _{s}}+i\kappa \ _{[1]}^{\shortparallel }%
\widehat{\mathbf{\Gamma }}_{\star \alpha _{s}\beta _{s}}^{\nu _{s}}=\
_{[00]}^{\shortparallel }\widehat{\Gamma }_{\ast \alpha _{s}\beta _{s}}^{\nu
_{s}}+\ _{[01]}^{\shortparallel }\widehat{\Gamma }_{\ast \alpha _{s}\beta
_{s}}^{\nu _{s}}(\hbar )+\ _{[10]}^{\shortparallel }\widehat{\Gamma }_{\ast
\alpha _{s}\beta _{s}}^{\nu _{s}}(\kappa )+\ _{[11]}^{\shortparallel }%
\widehat{\Gamma }_{\ast \alpha _{s}\beta _{s}}^{\nu _{s}}(\hbar \kappa
)+O(\hbar ^{2},\kappa ^{2},...).
\end{equation*}%
Then, introducing such parametric formulas in (\ref{nadriemhopfcan}), we
compute%
\begin{equation*}
\mathbf{\mathbf{\mathbf{\mathbf{\ ^{\shortparallel }}}}}\widehat{\mathcal{%
\Re }}_{\quad \alpha _{s}\beta _{s}\gamma _{s}}^{\star \mu _{s}}=\mathbf{%
\mathbf{\mathbf{\mathbf{\ }}}}\ _{[00]}^{\shortparallel }\widehat{\mathcal{%
\Re }}_{\quad \alpha _{s}\beta _{s}\gamma _{s}}^{\star \mu _{s}}+\mathbf{%
\mathbf{\mathbf{\mathbf{\ }}}}\ _{[01]}^{\shortparallel }\widehat{\mathcal{%
\Re }}_{\quad \alpha _{s}\beta _{s}\gamma _{s}}^{\star \mu _{s}}(\hbar )+%
\mathbf{\mathbf{\mathbf{\mathbf{\ }}}}\ _{[10]}^{\shortparallel }\widehat{%
\mathcal{\Re }}_{\quad \alpha _{s}\beta _{s}\gamma _{s}}^{\star \mu
_{s}}(\kappa )+\ _{[11]}^{\shortparallel }\widehat{\mathcal{\Re }}_{\quad
\alpha _{s}\beta _{s}\gamma _{s}}^{\star \mu _{s}}(\hbar \kappa )+O(\hbar
^{2},\kappa ^{2},...).
\end{equation*}%
To include in such a scheme nonholonomic dyadic decompositions of (non)
commutative canonical Riemann s-tensors and their star deformations into
nonassociative ones we can fix such nonholonomic s-distributions on $%
\mathcal{M}=\mathbf{T}_{s\shortparallel }^{\ast }\mathbf{V}$ when the values
$\ _{[00]}^{\shortparallel }\widehat{\mathcal{\Re }}_{\quad \alpha _{s}\beta
_{s}\gamma _{s}}^{\star \mu _{s}}$ define a "hat" variant of "not star
deformed" curvature s-tensor for the canonical s-connection $\
^{\shortparallel}\widehat{\Gamma }_{\star \alpha _{s}\tau _{s}}^{\mu _{s}}$ (%
\ref{twocon}). We omit cumbersome explicit formulas for the s-coefficients
up till order $\hbar ,\kappa $ and $\hbar \kappa $, for above geometric
s-objects, because such values are not used in this paper.

\subsubsection{Parametric decomposition of star canonical Ricci s-tensors
and their distortions}

Contracting on the fist and forth indices in formulas (\ref{nadriemhopfcan})
and above formulas with parametric decompositions for the star canonical
s-connection $\ ^{\shortparallel }\widehat{\Gamma }_{\star \alpha _{s}\gamma
_{s}}^{\mu _{s}},$ we define and compute the s-components the nonassociative
canonical Ricci d-tensor as a "hat" variant of (\ref{driccina}),
\begin{eqnarray}
\mathbf{\mathbf{\mathbf{\mathbf{\ _{s}^{\shortparallel }}}}}\widehat{%
\mathcal{\Re }}ic^{\star } &=&\mathbf{\mathbf{\mathbf{\mathbf{\
^{\shortparallel }}}}}\widehat{\mathbf{\mathbf{\mathbf{\mathbf{R}}}}}%
ic_{\alpha _{s}\beta _{s}}^{\star }\star _{s}(\ \mathbf{^{\shortparallel }e}%
^{\alpha _{s}}\otimes _{\star s}\ \mathbf{^{\shortparallel }e}^{\beta _{s}}),%
\mbox{ where }  \label{driccicanonstar} \\
&&\mathbf{\mathbf{\mathbf{\mathbf{\ ^{\shortparallel }}}}}\widehat{\mathbf{%
\mathbf{\mathbf{\mathbf{R}}}}}ic_{\alpha _{s}\beta _{s}}^{\star }:=\
_{s}^{\shortparallel }\widehat{\mathcal{\Re }}ic^{\star }(\mathbf{\ }\
^{\shortparallel }\mathbf{e}_{\alpha _{s}},\ ^{\shortparallel }\mathbf{e}%
_{\beta _{s}})=\mathbf{\langle }\ \mathbf{\mathbf{\mathbf{\mathbf{\
^{\shortparallel }}}}}\widehat{\mathbf{\mathbf{\mathbf{\mathbf{R}}}}}ic_{\mu
_{s}\nu _{s}}^{\star }\star _{s}(\ \mathbf{^{\shortparallel }e}^{\mu
_{s}}\otimes _{\star _{s}}\ \mathbf{^{\shortparallel }e}^{\nu _{s}}),\mathbf{%
\mathbf{\ }\ ^{\shortparallel }\mathbf{e}}_{\alpha _{s}}\mathbf{\otimes
_{\star s}\ ^{\shortparallel }\mathbf{e}}_{\beta _{s}}\mathbf{\rangle }%
_{\star _{s}}.  \notag
\end{eqnarray}%
The s-adapted coefficients are
\begin{eqnarray}
\mathbf{\mathbf{\mathbf{\mathbf{\ ^{\shortparallel }}}}}\widehat{\mathbf{%
\mathbf{\mathbf{\mathbf{R}}}}}ic_{\alpha _{s}\beta _{s}}^{\star } &=&\mathbf{%
\mathbf{\mathbf{\mathbf{\ ^{\shortparallel }}}}}\widehat{\mathcal{\Re }}%
_{\quad \alpha _{s}\beta _{s}\mu _{s}}^{\star \mu _{s}}=\mathbf{\mathbf{%
\mathbf{\mathbf{\ \mathbf{\mathbf{\mathbf{\mathbf{\ }}}}}}}}\
_{[00]}^{\shortparallel }\widehat{\mathbf{\mathbf{\mathbf{\mathbf{R}}}}}%
ic_{\alpha _{s}\beta _{s}}^{\star }+\mathbf{\mathbf{\mathbf{\mathbf{\ \ }}}}%
_{[01]}^{\shortparallel }\widehat{\mathbf{\mathbf{\mathbf{\mathbf{R}}}}}%
ic_{\alpha _{s}\beta _{s}}^{\star }(\hbar )+\mathbf{\mathbf{\mathbf{\mathbf{%
\ }}}}_{[10]}^{\shortparallel }\widehat{\mathbf{\mathbf{\mathbf{\mathbf{R}}}}%
}ic_{\alpha _{s}\beta _{s}}^{\star }(\kappa )  \notag \\
&&+\mathbf{\mathbf{\mathbf{\mathbf{\ }}}}_{[11]}^{\shortparallel }\widehat{%
\mathbf{\mathbf{\mathbf{\mathbf{R}}}}}ic_{\alpha _{s}\beta _{s}}^{\star
}(\hbar \kappa )+O(\hbar ^{2},\kappa ^{2},...),\mbox{where}  \notag \\
&&\ _{[00]}^{\shortparallel }\widehat{\mathbf{R}}ic_{\alpha _{s}\beta
_{s}}^{\star }=\ _{[00]}^{\shortparallel }\widehat{\mathcal{\Re }}_{\quad
\alpha _{s}\beta _{s}\mu _{s}}^{\star \mu _{s}}\mathbf{\mathbf{\mathbf{%
\mathbf{\ ,}}}}\ _{[01]}^{\shortparallel }\mathbf{\mathbf{\mathbf{\mathbf{%
\widehat{\mathbf{\mathbf{\mathbf{\mathbf{R}}}}}}}}}ic_{\alpha _{s}\beta
_{s}}^{\star }=\ _{[01]}^{\shortparallel }\mathbf{\mathbf{\mathbf{\mathbf{%
\widehat{\mathcal{\Re }}}}}}_{\quad \alpha _{s}\beta _{s}\mu _{s}}^{\star
\mu _{s}},  \label{driccicanonstar1} \\
&&\ _{[10]}^{\shortparallel }\mathbf{\mathbf{\mathbf{\mathbf{\widehat{%
\mathbf{\mathbf{\mathbf{\mathbf{R}}}}}}}}}ic_{\alpha _{s}\beta _{s}}^{\star
}=\ _{[10]}^{\shortparallel }\mathbf{\mathbf{\mathbf{\mathbf{\widehat{%
\mathcal{\Re }}}}}}_{\quad \alpha _{s}\beta _{s}\mu _{s}}^{\star \mu _{s}},\
_{[11]}^{\shortparallel }\widehat{\mathbf{\mathbf{\mathbf{\mathbf{R}}}}}%
ic_{\alpha _{s}\beta _{s}}^{\star }=\ _{[11]}^{\shortparallel }\mathbf{%
\mathbf{\mathbf{\mathbf{\widehat{\mathcal{\Re }}}}}}_{\quad \alpha _{s}\beta
_{s}\mu _{s}}^{\star \mu _{s}}.  \notag
\end{eqnarray}%
In appendix \ref{assnaricci}, we show how to compute the parametric
[00],[01],[10], and [11] s-coefficients from (\ref{driccicanonstar1}).

The h1-v2-c3-c4 splitting of a canonical Ricci s-connection is given by a
corresponding "hat" variant of (\ref{driccinahc}),
\begin{eqnarray}
\ \mathbf{\mathbf{\mathbf{\mathbf{^{\shortparallel }}}}}\widehat{\mathbf{%
\mathbf{\mathbf{\mathbf{R}}}}}ic_{\alpha _{s}\beta _{s}}^{\star } &=&\{\
^{\shortparallel }\widehat{R}_{\ \star h_{1}j_{1}}=\ \ ^{\shortparallel }%
\widehat{\mathcal{\Re }}_{\ \star h_{1}j_{1}i_{1}}^{i_{1}},\ \
^{\shortparallel }\widehat{P}_{\star j_{1}a_{2}}^{\ }=-\ ^{\shortparallel }%
\widehat{\mathcal{\Re }}_{\star \ j_{1}i_{1}a_{2}}^{i_{1}\quad }\ ,\
^{\shortparallel }\widehat{P}_{\star \ b_{2}k_{1}}=\ ^{\shortparallel }%
\widehat{\mathcal{\Re }}_{\star b_{2}k_{1}c_{2}}^{c_{2}\ },\
^{\shortparallel }\widehat{S}_{\star b_{2}c_{2}\ }=\ ^{\shortparallel }%
\widehat{\mathcal{\Re }}_{\star b_{2}c_{2}a_{2}\ }^{a_{2}\ },  \notag \\
\ ^{\shortparallel }\widehat{P}_{\star j_{2}}^{\ a_{3}} &=&-\
^{\shortparallel }\widehat{\mathcal{\Re }}_{\star \ j_{2}i_{2}}^{i_{2}\quad
a_{3}}\ ,\ ^{\shortparallel }\widehat{P}_{\star \ k_{2}}^{b_{3}\quad }=\
^{\shortparallel }\widehat{\mathcal{\Re }}_{\star c_{3}\ k_{2}}^{\ b_{3}\
c_{3}},\ ^{\shortparallel }\widehat{S}_{\star \ }^{b_{3}c_{3}\quad }=\
^{\shortparallel }\widehat{\mathcal{\Re }}_{\star a_{3}\ }^{\quad
b_{3}c_{3}a_{3}}\ ,  \notag \\
\ ^{\shortparallel }\widehat{P}_{\star j_{3}}^{\ a_{4}} &=&-\
^{\shortparallel }\widehat{\mathcal{\Re }}_{\star \ j_{3}i_{3}}^{i_{3}\quad
a_{4}}\ ,\ ^{\shortparallel }\widehat{P}_{\star \ k_{3}}^{b_{4}\quad }=\
^{\shortparallel }\widehat{\mathcal{\Re }}_{\star c_{4}\ k_{3}}^{\ b_{4}\
c_{4}},\ ^{\shortparallel }\widehat{S}_{\star \ }^{b_{4}c_{4}\quad }=\
^{\shortparallel }\widehat{\mathcal{\Re }}_{\star a_{4}\ }^{\quad
b_{4}c_{4}a_{4}}\ \}.  \label{hcnonassocrcaa}
\end{eqnarray}%
Each component of (\ref{hcnonassocrcaa}) can be decomposed in respective $%
[00],[01],[10]$ and $[11]$ terms, for instance,%
\begin{eqnarray}
\ _{[00]}^{\shortparallel }\widehat{\mathbf{\mathbf{\mathbf{\mathbf{R}}}}}%
ic_{\alpha _{s}\beta _{s}}^{\star } &=&\mathbf{\{}\ \
_{[00]}^{\shortparallel }\widehat{R}_{\ \star h_{1}j_{1}}=\ \
_{[00]}^{\shortparallel }\widehat{\mathcal{\Re }}_{\ \star
h_{1}j_{1}i_{1}}^{i_{1}},\ _{[00]}^{\shortparallel }\widehat{P}_{\star
j_{1}a_{2}}^{\ }=-\ \ _{[00]}^{\shortparallel }\widehat{\mathcal{\Re }}%
_{\star \ j_{1}i_{1}a_{2}}^{i_{1}\quad }\ ,\ \ _{[00]}^{\shortparallel }%
\widehat{P}_{\star \ b_{2}k_{1}}=\ _{[00]}^{\shortparallel }\widehat{%
\mathcal{\Re }}_{\star b_{2}k_{1}c_{2}}^{c_{2}\ },...\},  \notag \\
\ _{[01]}^{\shortparallel }\widehat{\mathbf{\mathbf{\mathbf{\mathbf{R}}}}}%
ic_{\alpha _{s}\beta _{s}}^{\star } &=&\{...,-\ \ _{[01]}^{\shortparallel }%
\widehat{\mathcal{\Re }}_{\star \ j_{2}i_{2}}^{i_{2}\quad a_{3}}\ ,\ \
_{[01]}^{\shortparallel }\widehat{P}_{\star \ k_{2}}^{b_{3}\quad }=\ \
_{[01]}^{\shortparallel }\widehat{\mathcal{\Re }}_{\star c_{3}\ k_{2}}^{\
b_{3}\ c_{3}},\ \ _{[01]}^{\shortparallel }\widehat{S}_{\star \
}^{b_{3}c_{3}\quad }=\ _{[01]}^{\shortparallel }\widehat{\mathcal{\Re }}%
_{\star a_{3}\ }^{\quad b_{3}c_{3}a_{3}},...\}  \label{hcnonassocrcana} \\
\ _{[10]}^{\shortparallel }\widehat{\mathbf{\mathbf{\mathbf{\mathbf{R}}}}}%
ic_{\alpha _{s}\beta _{s}}^{\star } &=&\{...,\ \ _{[10]}^{\shortparallel }%
\widehat{P}_{\star j_{3}}^{\ a_{4}}=-\ \ _{[10]}^{\shortparallel }\widehat{%
\mathcal{\Re }}_{\star \ j_{3}i_{3}}^{i_{3}\quad a_{4}}\ ,\ \
_{[10]}^{\shortparallel }\widehat{P}_{\star \ k_{3}}^{b_{4}\quad }=\ \
_{[10]}^{\shortparallel }\widehat{\mathcal{\Re }}_{\star c_{4}\ k_{3}}^{\
b_{4}\ c_{4}},\ \ _{[10]}^{\shortparallel }\widehat{S}_{\star \
}^{b_{4}c_{4}\quad }=\ \ _{[10]}^{\shortparallel }\widehat{\mathcal{\Re }}%
_{\star a_{4}\ }^{\quad b_{4}c_{4}a_{4}}\}  \notag \\
\ _{[11]}^{\shortparallel }\widehat{\mathbf{\mathbf{\mathbf{\mathbf{R}}}}}%
ic_{\alpha _{s}\beta _{s}}^{\star } &=&\mathbf{\{}\ \
_{[11]}^{\shortparallel }\widehat{R}_{\ \star h_{1}j_{1}}=\ \
_{[11]}^{\shortparallel }\widehat{\mathcal{\Re }}_{\ \star
h_{1}j_{1}i_{1}}^{i_{1}},\ _{[11]}^{\shortparallel }\widehat{P}_{\star
j_{1}a_{2}}^{\ }=-\ \ _{[11]}^{\shortparallel }\widehat{\mathcal{\Re }}%
_{\star \ j_{1}i_{1}a_{2}}^{i_{1}\quad }\ ,\ \ _{[11]}^{\shortparallel }%
\widehat{P}_{\star \ b_{2}k_{1}}=\ _{[11]}^{\shortparallel }\widehat{%
\mathcal{\Re }}_{\star b_{2}k_{1}c_{2}}^{c_{2}\ },...\}.  \notag
\end{eqnarray}

We can arrange the nonholonomic distributions on phase space that $\
_{[00]}^{\shortparallel }\widehat{\mathbf{R}}ic_{\alpha _{s}\beta
_{s}}^{\star }= \ ^{\shortparallel }\widehat{\mathbf{R}}_{\ \alpha _{s}\beta
_{s}}$ (\ref{candricci}) are determined by associative and commutative
s-adapted canonical s-connection (\ref{candricci}) but other $%
[01,10,11]:=\left\lceil \hbar ,\kappa \right\rceil $ components contain
nonassociative and noncommutative contributions from star product
deformations, which can be real or complex ones. In parametric form, we can
express the star s-deformed Ricci s-tensor in the form
\begin{equation}
\ _{s}^{\shortparallel }\widehat{\mathcal{R}}ic^{\star }=\{\
^{\shortparallel }\widehat{\mathbf{R}}_{\ \star \beta _{s}\gamma _{s}}\}=\
_{s}^{\shortparallel }\widehat{\mathcal{R}}ic+\ _{s}^{\shortparallel }%
\widehat{\mathcal{K}}ic\left\lceil \hbar ,\kappa \right\rceil =\{\
^{\shortparallel }\widehat{\mathbf{R}}_{\ \beta _{s}\gamma _{s}}+\
^{\shortparallel }\widehat{\mathbf{K}}_{\ \beta _{s}\gamma _{s}}\left\lceil
\hbar ,\kappa \right\rceil \},  \label{paramsricci}
\end{equation}%
where $\ _{s}^{\shortparallel }\widehat{\mathcal{K}}ic=\{\ ^{\shortparallel }%
\widehat{\mathbf{K}}_{\ \beta _{s}\gamma _{s}}\left\lceil \hbar ,\kappa
\right\rceil \}\}$ encode nonassociative parametric deformations of the
canonical Ricci s-tenor. If we take an associative and commutative solution $%
\ _{\star }^{\shortparallel }\mathbf{g}_{\alpha _{s}\beta _{s}}$ of
nonholonomic dyadic vacuum Einstein equations (\ref{cnsveinst2}), we can
compute certain as R-flux and star canonical s-deformations to $\ _{\star
}^{\shortparallel }\mathfrak{g}_{\tau _{s}\gamma _{s}}.$ In section \ref%
{sec6}, we shall generate parametric solutions for some classes of
stationary nonassociative s-metrics $\ _{\star }^{\shortparallel }\mathfrak{g%
}_{\tau _{s}\gamma _{s}}$ when nonassocitaive distortions of Ricci s-tensors
of type $\ ^{\shortparallel }\widehat{\mathbf{K}}_{\ \beta _{s}\gamma
_{s}}\left\lceil \hbar ,\kappa \right\rceil $ in (\ref{paramsricci}) are
encoded in certain effective sources and/or cosmological constants, see
subsection \ref{ssnonassocsourc}. Using respective formulas for parametric
splitting of the nonassociative canonical Ricci s-tensor presented in
appendix \ref{assnaricci} (see respective formulas (\ref{ric51}), (\ref%
{ric52}) and (\ref{ric53})),
\begin{equation}
\ ^{\shortparallel }\widehat{\mathbf{K}}_{\ \beta _{s}\gamma _{s}}=\mathbf{%
\mathbf{\mathbf{\mathbf{\ \ }}}}_{[01]}^{\shortparallel }\widehat{\mathbf{%
\mathbf{\mathbf{\mathbf{R}}}}}ic_{\beta _{s}\gamma _{s}}^{\star }+\mathbf{%
\mathbf{\mathbf{\mathbf{\ \ }}}}_{[10]}^{\shortparallel }\widehat{\mathbf{%
\mathbf{\mathbf{\mathbf{R}}}}}ic_{\beta _{s}\gamma _{s}}^{\star }+\mathbf{%
\mathbf{\mathbf{\mathbf{\ \ }}}}_{[11]}^{\shortparallel }\widehat{\mathbf{%
\mathbf{\mathbf{\mathbf{R}}}}}ic_{\beta _{s}\gamma _{s}}^{\star },
\label{paramsriccins}
\end{equation}%
where $\ ^{\shortparallel }\widehat{\mathbf{R}}_{\ \beta _{s}\gamma _{s}}=%
\mathbf{\mathbf{\mathbf{\mathbf{\ \ }}}}_{[00]}^{\shortparallel }\widehat{%
\mathbf{\mathbf{\mathbf{\mathbf{R}}}}}ic_{\beta _{s}\gamma _{s}}^{\star }$ (%
\ref{ric50}) can be computted using formulas for commutative canonical Ricci
s-tensors presented in our previous works \cite%
{bubuianu20,bubuianu19,bubuianu17,bubuianu17b}.

In our nonholonomic frame approach, we can elaborate (non) commutative and
(non) associative theories in equivalent forms using different classes of
linear connection structures defined by corresponding adapting to N- /s-
structures. The canonical s-connection $\ _{s}^{\shortparallel }\widehat{%
\mathbf{D}}^{\star }$ allows to decouple and solve in some general
off-diagonal forms (exactly, or with dependence on certain parameters) star
generalizations of (modified) Einstein equations as we prove in sections \ref%
{sec5} and \ref{sec6}. In another turn, the LC-connection $\
^{\shortparallel }\nabla ^{\star }$ is a standard one used in GR and various
MGTs. If both type of geometric constructions are determined by the same
(non) associative metric s-structure ( $\ _{\star }^{\shortparallel}%
\mathfrak{g}_{\tau _{s}\gamma _{s}})$ $\ _{\star }^{\shortparallel }\mathbf{g%
}_{\alpha _{s}\beta _{s}}$, we can re-define mutually the geometric/
physical theories via corresponding frame transforms and connections
deformations.

Introducing star distortions of the canonical s-connection $,\
_{s}^{\shortparallel }\widehat{\mathbf{D}}^{\star }=\
^{\shortparallel}\nabla ^{\star }+\ _{\star s}^{\shortparallel }\widehat{%
\mathbf{Z}}$ (\ref{candistrnas}) into formulas for $\ ^{\shortparallel }%
\widehat{\mathcal{\Re }}ic^{\star }$ (\ref{driccicanonstar}), we can compute
the s-distortions of the nonassocitative canonical Ricci scalar and Ricci
s-tensors (see formulas (\ref{candriccidist}) for commutative
s-configurations). In nonholonomic dyadic form, we obtain
\begin{equation}
\ _{s}^{\shortparallel }\widehat{\mathcal{R}}sc^{\star}= \
_{\nabla}^{\shortparallel }\mathcal{R}sc^{\star }+ \ _{\nabla
}^{\shortparallel s}\widehat{\mathcal{Z}}sc^{\star } \mbox{ and }\ \
_{s}^{\shortparallel }\widehat{\mathcal{\Re }}ic^{\star }= \ _{\nabla
}^{\shortparallel }\mathcal{\Re }ic^{\star }+\ \ _{\nabla }^{\shortparallel
s}\widehat{\mathcal{Z}}ic^{\star },  \label{candriccidistna}
\end{equation}%
with corresponding distortion tensors $\ _{\nabla }^{\shortparallel s}%
\widehat{\mathcal{Z}}^{\star }$ and $\ _{\nabla }^{\shortparallel s}\widehat{%
\mathcal{Z}}ic^{\star },$ where left labels state that we consider canonical
s-deformations from the nonassociative LC-connection. We can extract also
LC-configurations if impose additionally the zero torsion conditions (\ref%
{lccondnonass}).

\section{Nonassociative nonholonomic dyadic vacuum Einstein equations}

\label{sec4} The goal of this section is to reformulate in nonholonomic
dyadic s-adapted canonical variables the nonassociative vacuum Einstein
equations formulated for quasi-Hopf structures \cite{aschieri17}. In a
similar form, we can consider s-adapted constructions for nonassociative
geometric objects and equations from \cite{blumenhagen16} but we omit such
details in this work. In \cite{partner01}, we generalized that
nonassociative geometric approach for N-adapted configurations with R-flux
and star modifications of vacuum phase space gravitational equations with a
nontrivial cosmological constant. We develop a nonholonomic geometric
approach for deriving fundamental physical equations following geometric
principles (in our case, with nonsymmetric metrics and generalized
connections) as in \cite{misner,vacaru18,bubuianu18a}. In this paper, such
associative and commutative vacuum gravitational are formulated using the
canonical s-connection (\ref{cnsveinst2}). Nonassociative modifications are
encoded in parametric distortions of the canonical Ricci s-tensor (\ref%
{paramsricci}).

\subsection{Nonassociative vacuum Einstein equations with (non) symmetric
s-metrics}

Following Convention 2 (\ref{conv2s}) and using formulas for fundamental
geometric s-objects (for (non) associative s-metrics, see (\ref{sdm}), (\ref%
{ssdm}) and (\ref{nssdm}); for s-connections, see (\ref{candistr}) and (\ref%
{candistrnas}); and distortions of s-connections, see (\ref{candriccidist})
and (\ref{candriccidistna})) and respective geometric s-distortions, we can
work with such equivalent geometric data: {\small
\begin{equation}
\begin{array}{ccc}
(\ ^{\shortparallel}{\partial}_{\alpha},\ ^{\shortparallel}{g}_{\alpha
\beta},\ ^{\shortparallel}{\nabla }_{\beta},\ _{\nabla}^{\shortparallel}{R}%
_{\mu \nu}) &
\begin{array}{c}
\mbox{ commutative phase } \\
\mbox{canonic N-/s-deforms}%
\end{array}%
{\ \Longrightarrow} & \left\{
\begin{array}{c}
(\mathbf{\mathbf{\mathbf{\mathbf{\ ^{\shortparallel }}}}e}_{\alpha },\mathbf{%
\mathbf{\mathbf{\mathbf{\ ^{\shortparallel }}}}g}_{\alpha \beta },\mathbf{%
\mathbf{\mathbf{\mathbf{\ ^{\shortparallel }}}}}\widehat{\mathbf{D}}_{\beta
}=\ ^{\shortparallel }\nabla _{\beta }+\ ^{\shortparallel }\widehat{\mathbf{Z%
}}_{\beta },\mathbf{\mathbf{\mathbf{\mathbf{\ ^{\shortparallel }}}}}\widehat{%
\mathbf{R}}_{\mu \nu }) \\
(\mathbf{\mathbf{\mathbf{\mathbf{\ ^{\shortparallel }}}}e}_{\alpha _{s}},%
\mathbf{\mathbf{\mathbf{\mathbf{\ ^{\shortparallel }}}}g}_{\alpha _{s}\beta
_{s}},\mathbf{\mathbf{\mathbf{\mathbf{\ ^{\shortparallel }}}}}\widehat{%
\mathbf{D}}_{\beta _{s}}=\ ^{\shortparallel }\nabla _{\beta _{s}}+\
^{\shortparallel }\widehat{\mathbf{Z}}_{\beta _{s}},\mathbf{\mathbf{\mathbf{%
\mathbf{\ ^{\shortparallel }}}}}\widehat{\mathbf{R}}_{\mu _{s}\nu _{s}})%
\end{array}%
\right. \\
&
\begin{array}{c}
\mbox{nonassoc. canonic} \\
\mbox{star N-/s-deforms}%
\end{array}%
{\ \Longrightarrow } & \left\{
\begin{array}{c}
(\mathbf{\mathbf{\mathbf{\mathbf{\ ^{\shortparallel }}}}e}_{\alpha },\
_{\star }^{\shortparallel }\mathfrak{g}_{\alpha \beta },\ ^{\shortparallel }%
\widehat{\mathbf{D}}_{\beta }^{\star }=\ ^{\shortparallel }\nabla _{\beta
}^{\star }+\ _{\star }^{\shortparallel }\widehat{\mathbf{Z}}_{\beta },%
\mathbf{\mathbf{\mathbf{\mathbf{\ ^{\shortparallel }}}}}\widehat{\mathbf{%
\mathbf{\mathbf{\mathbf{R}}}}}ic_{\mu \nu }^{\star }) \\
(\mathbf{\mathbf{\mathbf{\mathbf{\ ^{\shortparallel }}}}e}_{\alpha _{s}},\
_{\star }^{\shortparallel }\mathfrak{g}_{\alpha _{s}\beta _{s}},\
^{\shortparallel }\widehat{\mathbf{D}}_{\beta _{s}}^{\star }=\
^{\shortparallel }\nabla _{\beta _{s}}^{\star }+\ _{\star }^{\shortparallel }%
\widehat{\mathbf{Z}}_{\beta _{s}},\mathbf{\mathbf{\mathbf{\mathbf{\
^{\shortparallel }}}}}\widehat{\mathbf{\mathbf{\mathbf{\mathbf{R}}}}}ic_{\mu
_{s}\nu _{s}}^{\star })%
\end{array}%
\right.%
\end{array}
\label{nonholstardef}
\end{equation}
} Using the canonical s-connection, we shall be able to decouple
nonassociative vacuum gravitational equations in a parametric form which is
similar to the case of associative (non) commutative gravity, see section %
\ref{sec5}.

\subsubsection{Parametric decomposition of nonsymmetric star deformed
s-metrics}

For any nonsymmetric s-metric $\ _{\star }^{\shortparallel }\mathfrak{g}%
_{\alpha _{s}\beta _{s}}= \ _{\star }^{\shortparallel }\mathfrak{g}_{\alpha
_{s}\beta _{s}}^{[0]}+ \ _{\star }^{\shortparallel }\mathfrak{g}_{\alpha
_{s}\beta _{s}}^{[1]}(\kappa )$ (see formulas (\ref{nssdm}) and (\ref{aux37}%
), we can define and compute the symmetric part,
\begin{eqnarray}
\ _{\star }^{\shortparallel }\mathfrak{\check{g}}_{\alpha _{s}\beta _{s}} &=&%
\frac{1}{2}(\ _{\star }^{\shortparallel }\mathfrak{g}_{\alpha _{s}\beta
_{s}}+\ _{\star }^{\shortparallel }\mathfrak{g}_{\beta _{s}\alpha _{s}})=\
_{\star }^{\shortparallel }\mathbf{g}_{\alpha _{s}\beta _{s}}-\frac{i\kappa
}{2}\left( \overline{\mathcal{R}}_{\quad \beta _{s}}^{\tau _{s}\xi _{s}}\
\mathbf{^{\shortparallel }e}_{\xi _{s}}\ _{\star }^{\shortparallel }\mathbf{g%
}_{\tau _{s}\alpha _{s}}+\overline{\mathcal{R}}_{\quad \alpha _{s}}^{\tau
_{s}\xi _{s}}\ \mathbf{^{\shortparallel }e}_{\xi _{s}}\ _{\star
}^{\shortparallel }\mathbf{g}_{\beta _{s}\tau _{s}}\right)  \label{aux40b} \\
&=&\ _{\star }^{\shortparallel }\mathfrak{\check{g}}_{\alpha _{s}\beta
_{s}}^{[0]}+\ _{\star }^{\shortparallel }\mathfrak{\check{g}}_{\alpha
_{s}\beta _{s}}^{[1]}(\kappa ),  \notag \\
&&\mbox{ for }\ _{\star }^{\shortparallel }\mathfrak{\check{g}}_{\alpha
_{s}\beta _{s}}^{[0]}=\ _{\star }^{\shortparallel }\mathbf{g}_{\alpha
_{s}\beta _{s}}\mbox{ and }\ _{\star }^{\shortparallel }\mathfrak{\check{g}}%
_{\alpha _{s}\beta _{s}}^{[1]}(\kappa )=-\frac{i\kappa }{2}\left( \overline{%
\mathcal{R}}_{\quad \beta _{s}}^{\tau _{s}\xi _{s}}\ \mathbf{%
^{\shortparallel }e}_{\xi _{s}}\ _{\star }^{\shortparallel }\mathbf{g}_{\tau
_{s}\alpha _{s}}+\overline{\mathcal{R}}_{\quad \alpha _{s}}^{\tau _{s}\xi
_{s}}\ \mathbf{^{\shortparallel }e}_{\xi _{s}}\ _{\star }^{\shortparallel }%
\mathbf{g}_{\beta _{s}\tau _{s}}\right) .  \notag
\end{eqnarray}%
For nonholonomic dyadic decompositions, see formulas (\ref{ssdm}) and (\ref%
{dmss1}). The anti-symmetric part of a s-metric can be written with respect
to s-frames when it is defined by R-fluxes,
\begin{eqnarray}
\ _{\star }^{\shortparallel }\mathfrak{a}_{\alpha _{s}\beta _{s}}:= &&\frac{1%
}{2}(\ _{\star }^{\shortparallel }\mathfrak{g}_{\alpha _{s}\beta _{s}}-\
_{\star }^{\shortparallel }\mathfrak{g}_{\beta _{s}\alpha _{s}})=\frac{%
i\kappa }{2}\left( \overline{\mathcal{R}}_{\quad \beta _{s}}^{\tau _{s}\xi
_{s}}\ \mathbf{^{\shortparallel }e}_{\xi _{s}}\ _{\star }^{\shortparallel }%
\mathbf{g}_{\tau _{s}\alpha _{s}}-\overline{\mathcal{R}}_{\quad \alpha
_{s}}^{\tau _{s}\xi _{s}}\ \mathbf{^{\shortparallel }e}_{\xi _{s}}\ _{\star
}^{\shortparallel }\mathbf{g}_{\beta _{s}\tau _{s}}\right)  \notag \\
&=&\ _{\star }^{\shortparallel }\mathfrak{a}_{\alpha _{s}\beta
_{s}}^{[1]}(\kappa )=\frac{1}{2}(\ _{\star }^{\shortparallel }\mathfrak{g}%
_{\alpha _{s}\beta _{s}}^{[1]}(\kappa )-\ _{\star }^{\shortparallel }%
\mathfrak{g}_{\beta _{s}\alpha _{s}}^{[1]}(\kappa )).  \label{aux40a}
\end{eqnarray}%
We state $\ _{\star }^{\shortparallel }\mathfrak{a}_{\alpha \beta }^{[0]}=0$
if we consider nonassociative star deformations of commutative theories with
symmetric metrics,%
\begin{equation}
\ _{\star }^{\shortparallel }\mathfrak{g}_{\alpha _{s}\beta _{s}}=\ _{\star
}^{\shortparallel }\mathfrak{\check{g}}_{\alpha _{s}\beta _{s}}+\ _{\star
}^{\shortparallel }\mathfrak{a}_{\alpha _{s}\beta _{s}}.  \label{splitdmetr}
\end{equation}

In geometric models with star deformation, we have to apply a more
sophisticate procedure for computing inverse metrics. In Appendix \ref%
{assinvsm}, we outline most important formulas for the case nonholonomic
dyadic decompositions, see formulas (\ref{aux38a}), etc. The symmetric and
nonsymmetric parts of a nonsymmetric s-metric with parametric decompositions
can be written similarly to
\begin{equation*}
\ _{\star }^{\shortparallel }\mathfrak{g}^{\alpha _{s}\beta _{s}}=\ _{\star
}^{\shortparallel }\mathfrak{\check{g}}^{\alpha _{s}\beta _{s}}+\ _{\star
}^{\shortparallel }\mathfrak{a}^{\alpha _{s}\beta _{s}},
\end{equation*}%
when, in general, $\ _{\star }^{\shortparallel }\mathfrak{\check{g}}^{\alpha
_{s}\beta _{s}}$ is not the inverse to $\ _{\star }^{\shortparallel }%
\mathfrak{\check{g}}_{\alpha _{s}\beta _{s}}$ and $\ _{\star
}^{\shortparallel }\mathfrak{a}^{\alpha _{s}\beta _{s}}$ is not inverse to $%
\ _{\star }^{\shortparallel }\mathfrak{a}_{\alpha _{s}\beta _{s}}.$ The
inverse symmetric and nonsymmetric values are computed respectively:
\begin{eqnarray}
\ _{\star }^{\shortparallel }\mathfrak{\check{g}}^{\alpha _{s}\beta _{s}} &=&%
\frac{1}{2}(\ _{\star }^{\shortparallel }\mathfrak{\check{g}}^{\alpha
_{s}\beta _{s}}+\ _{\star }^{\shortparallel }\mathfrak{\check{g}}^{\beta
_{s}\alpha _{s}})  \notag \\
&=&\frac{1}{2}(\ _{\star }^{\shortparallel }\mathbf{g}^{\alpha _{s}\beta
_{s}}+\ _{\star }^{\shortparallel }\mathbf{g}^{\beta _{s}\alpha _{s}})
\notag \\
&&-\frac{i\kappa }{2}(\ _{\star }^{\shortparallel }\mathbf{g}^{\alpha
_{s}\tau _{s}}\overline{\mathcal{R}}_{\quad \tau _{s}}^{\mu _{s}\nu _{s}}(\
\mathbf{^{\shortparallel }e}_{\mu _{s}}\ \ _{\star }^{\shortparallel }%
\mathbf{g}_{\nu _{s}\varepsilon _{s}})\ \ _{\star }^{\shortparallel }\mathbf{%
g}^{\varepsilon _{s}\beta _{s}}+\ _{\star }^{\shortparallel }\mathbf{g}%
^{\beta _{s}\tau _{s}}\overline{\mathcal{R}}_{\quad \tau _{s}}^{\mu _{s}\nu
_{s}}(\ \mathbf{^{\shortparallel }e}_{\mu _{s}}\ \ _{\star }^{\shortparallel
}\mathbf{g}_{\nu _{s}\varepsilon _{s}})\ \ _{\star }^{\shortparallel }%
\mathbf{g}^{\varepsilon _{s}\alpha _{s}})+O(\kappa ^{2})\   \notag \\
&=&\frac{1}{2}(\ _{\ast }^{\shortparallel }\mathfrak{g}_{[0]}^{\alpha
_{s}\beta _{s}}+\ _{\star }^{\shortparallel }\mathfrak{g}_{[1]}^{\alpha
_{s}\beta _{s}}(\kappa )+\ _{\ast }^{\shortparallel }\mathfrak{g}%
_{[0]}^{\beta _{s}\alpha _{s}}+\ _{\star }^{\shortparallel }\mathfrak{g}%
_{[1]}^{\beta _{s}\alpha _{s}}(\kappa ))+O(\kappa ^{2})  \label{aux40}
\end{eqnarray}%
\begin{eqnarray*}
&=&\ _{\ast }^{\shortparallel }\mathfrak{g}_{[0]}^{\alpha _{s}\beta _{s}}+%
\frac{1}{2}(\ _{\star }^{\shortparallel }\mathfrak{g}_{[1]}^{\alpha
_{s}\beta _{s}}(\kappa )+\ _{\star }^{\shortparallel }\mathfrak{g}%
_{[1]}^{\beta _{s}\alpha _{s}}(\kappa ))+O(\kappa ^{2})=\ _{\ast
}^{\shortparallel }\mathfrak{g}_{[0]}^{\alpha _{s}\beta _{s}}+\ _{\ast
}^{\shortparallel }\mathfrak{\check{g}}_{[1]}^{\alpha _{s}\beta _{s}}(\kappa
)+O(\kappa ^{2}) \\
&=&\ _{\ast }^{\shortparallel }\mathfrak{g}_{[00]}^{\alpha _{s}\beta _{s}}+\
_{\ast }^{\shortparallel }\mathfrak{g}_{[01]}^{\alpha _{s}\beta _{s}}(\hbar
)+\ _{\ast }^{\shortparallel }\mathfrak{\check{g}}_{[10]}^{\alpha _{s}\beta
_{s}}(\kappa )+\ _{\ast }^{\shortparallel }\mathfrak{\check{g}}%
_{[11]}^{\alpha _{s}\beta _{s}}(\hbar \kappa )+O(\hbar ^{2},\kappa ^{2}),%
\mbox{ for } \\
\ _{\ast }^{\shortparallel }\mathfrak{\check{g}}_{[0]}^{\alpha _{s}\beta
_{s}} &=&\ _{\ast }^{\shortparallel }\mathfrak{g}_{[0]}^{\alpha _{s}\beta
_{s}},\ _{\ast }^{\shortparallel }\mathfrak{\check{g}}_{[1]}^{\alpha
_{s}\beta _{s}}(\kappa )=\ _{\ast }^{\shortparallel }\mathfrak{g}%
_{[1]}^{\alpha _{s}\beta _{s}}(\kappa ),\ _{\ast }^{\shortparallel }%
\mathfrak{g}_{[0]}^{\alpha _{s}\beta _{s}}=\ _{\ast }^{\shortparallel }%
\mathfrak{g}_{[0]}^{\beta _{s}\alpha _{s}},
\end{eqnarray*}%
\begin{eqnarray*}
\mbox{and }\ _{\star }^{\shortparallel }\mathfrak{a}^{\alpha _{s}\beta _{s}}
&=&\frac{1}{2}(\ _{\star }^{\shortparallel }\mathfrak{\check{g}}^{\alpha
_{s}\beta _{s}}-\ _{\star }^{\shortparallel }\mathfrak{\check{g}}^{\beta
_{s}\alpha _{s}}) \\
&=&\frac{1}{2}(\ _{\star }^{\shortparallel }\mathbf{g}^{\alpha _{s}\beta
_{s}}-\ _{\star }^{\shortparallel }\mathbf{g}^{\beta _{s}\alpha _{s}}) \\
&&-\frac{i\kappa }{2}(\ _{\star }^{\shortparallel }\mathbf{g}^{\alpha
_{s}\tau _{s}}\overline{\mathcal{R}}_{\quad \tau _{s}}^{\mu _{s}\nu _{s}}(\
\mathbf{^{\shortparallel }e}_{\mu _{s}}\ \ _{\star }^{\shortparallel }%
\mathbf{g}_{\nu _{s}\varepsilon _{s}})\ \ _{\star }^{\shortparallel }\mathbf{%
g}^{\varepsilon _{s}\beta _{s}}-\ _{\star }^{\shortparallel }\mathbf{g}%
^{\beta _{s}\tau _{s}}\overline{\mathcal{R}}_{\quad \tau _{s}}^{\mu _{s}\nu
_{s}}(\ \mathbf{^{\shortparallel }e}_{\mu _{s}}\ _{\star }^{\shortparallel }%
\mathbf{g}_{\nu _{s}\varepsilon _{s}})\ \ _{\star }^{\shortparallel }\mathbf{%
g}^{\varepsilon _{s}\alpha _{s}})+O(\kappa ^{2})\  \\
&=&\frac{1}{2}(\ _{\ast }^{\shortparallel }\mathfrak{g}_{[0]}^{\alpha
_{s}\beta _{s}}+\ _{\star }^{\shortparallel }\mathfrak{g}_{[1]}^{\alpha
_{s}\beta _{s}}(\kappa )-\ _{\ast }^{\shortparallel }\mathfrak{g}%
_{[0]}^{\beta _{s}\alpha _{s}}-\ _{\star }^{\shortparallel }\mathfrak{g}%
_{[1]}^{\beta _{s}\alpha _{s}}(\kappa ))+O(\kappa ^{2})
\end{eqnarray*}%
\begin{eqnarray*}
&=&\frac{1}{2}(\ _{\star }^{\shortparallel }\mathfrak{g}_{[1]}^{\alpha
_{s}\beta _{s}}(\kappa )-\ _{\star }^{\shortparallel }\mathfrak{g}%
_{[1]}^{\beta _{s}\alpha _{s}}(\kappa ))+O(\kappa ^{2})=\ _{\star
}^{\shortparallel }\mathfrak{a}_{[1]}^{\alpha _{s}\beta _{s}}(\kappa
)+O(\kappa ^{2}) \\
&=&\ _{\ast }^{\shortparallel }\mathfrak{a}_{[10]}^{\alpha _{s}\beta
_{s}}(\kappa )+\ _{\ast }^{\shortparallel }\mathfrak{a}_{[11]}^{\alpha
_{s}\beta _{s}}(\hbar \kappa )+O(\hbar ^{2},\kappa ^{2}),\mbox{ for } \\
\ _{\ast }^{\shortparallel }\mathfrak{g}_{[0]}^{\alpha _{s}\beta _{s}} &=&\
_{\ast }^{\shortparallel }\mathfrak{g}_{[0]}^{\beta _{s}\alpha _{s}},\
_{\star }^{\shortparallel }\mathfrak{a}_{[1]}^{\alpha _{s}\beta _{s}}(\kappa
)=\frac{1}{2}(\ _{\star }^{\shortparallel }\mathfrak{g}_{[1]}^{\alpha
_{s}\beta _{s}}(\kappa )-\ _{\star }^{\shortparallel }\mathfrak{g}%
_{[1]}^{\beta _{s}\alpha _{s}}(\kappa )),\ _{\ast }^{\shortparallel }%
\mathfrak{a}_{[00]}^{\alpha _{s}\beta _{s}}=0,\quad _{\ast }^{\shortparallel
}\mathfrak{a}_{[01]}^{\alpha _{s}\beta _{s}}(\hbar )=0.
\end{eqnarray*}
Any coefficient of above geometric s-objects can be also subjected to
h1-v2-c3-c4 decompositions.

\subsubsection{Star deformed canonical s-adapted nonassociative vacuum
gravitational equations}

In general, the canonical Ricci s-tensors are not symmetric for all (non)
commutative and nonassociative cases. Contracting respectively with the
symmetric and nonsymmetric components of a (star deformed) inverse s-metric,
we define and compute the nonassociative nonholonomic canonical Ricci scalar
curvature:%
\begin{eqnarray}
\ _{s}^{\shortparallel }\widehat{\mathbf{R}}sc^{\star }&:=&\ _{\star
}^{\shortparallel }\mathfrak{g}^{\mu _{s}\nu _{s}}\mathbf{\mathbf{\mathbf{%
\mathbf{\ ^{\shortparallel }}}}}\widehat{\mathbf{\mathbf{\mathbf{\mathbf{R}}}%
}}ic_{\mu _{s}\nu _{s}}^{\star }=\left( \ _{\star }^{\shortparallel }%
\mathfrak{\check{g}}^{\mu _{s}\nu _{s}}+\ _{\star }^{\shortparallel }%
\mathfrak{a}^{\mu _{s}\nu _{s}}\right) \left( \mathbf{\mathbf{\mathbf{%
\mathbf{\ ^{\shortparallel }}}}}\widehat{\mathbf{\mathbf{\mathbf{\mathbf{R}}}%
}}ic_{(\mu _{s}\nu _{s})}^{\star }+\mathbf{\mathbf{\mathbf{\mathbf{\
^{\shortparallel }}}}}\widehat{\mathbf{\mathbf{\mathbf{\mathbf{R}}}}}%
ic_{[\mu _{s}\nu _{s}]}^{\star }\right) =\ _{s}^{\shortparallel }\widehat{%
\mathbf{\mathbf{\mathbf{\mathbf{R}}}}}ss^{\star }+\ _{s}^{\shortparallel }%
\widehat{\mathbf{\mathbf{\mathbf{\mathbf{R}}}}}sa^{\star },  \notag \\
&&\mbox{ where }\ _{s}^{\shortparallel }\widehat{\mathbf{\mathbf{\mathbf{%
\mathbf{R}}}}}ss^{\star }=:\ _{\star }^{\shortparallel }\mathfrak{\check{g}}%
^{\mu _{s}\nu _{s}}\mathbf{\mathbf{\mathbf{\mathbf{\ ^{\shortparallel }}}}}%
\widehat{\mathbf{\mathbf{\mathbf{\mathbf{R}}}}}ic_{(\mu _{s}\nu
_{s})}^{\star }\mbox{ and }\ _{s}^{\shortparallel }\widehat{\mathbf{\mathbf{%
\mathbf{\mathbf{R}}}}}sa^{\star }:=\ _{\star }^{\shortparallel }\mathfrak{a}%
^{\mu _{s}\nu _{s}}\mathbf{\mathbf{\mathbf{\mathbf{\ ^{\shortparallel }}}}}%
\widehat{\mathbf{\mathbf{\mathbf{\mathbf{R}}}}}ic_{[\mu _{s}\nu
_{s}]}^{\star }.  \label{ricciscsymnonsym}
\end{eqnarray}%
In this paper, respective symmetric $\left( ...\right) $ and anti-symmetric $%
\left[ ...\right] $ decompositions are performed using respective
symmetrization and anti-symmetrization with multiple $1/2.$ For instance,
for the second rank s-tensors, $\mathbf{\mathbf{\mathbf{\mathbf{\
^{\shortparallel }}}}}\widehat{\mathbf{\mathbf{\mathbf{\mathbf{R}}}}}ic_{\mu
_{s}\nu _{s}}^{\star }=\mathbf{\mathbf{\mathbf{\mathbf{\ ^{\shortparallel }}}%
}}\widehat{\mathbf{\mathbf{\mathbf{\mathbf{R}}}}}ic_{(\mu _{s}\nu
_{s})}^{\star }+\mathbf{\mathbf{\mathbf{\mathbf{\ ^{\shortparallel }}}}}%
\widehat{\mathbf{\mathbf{\mathbf{\mathbf{R}}}}}ic_{[\mu _{s}\nu
_{s}]}^{\star }.$

Applying star nonholonomic s-deformations (\ref{nonholstardef}) to (\ref%
{cnsveinst1}), we can define and compute s-adapted components of
nonassociative vacuum gravitational equations,
\begin{equation}
\ ^{\shortparallel }\widehat{\mathbf{\mathbf{\mathbf{\mathbf{R}}}}}%
ic_{\alpha _{s}\beta _{s}}^{\star }-\frac{1}{2}\ _{\star }^{\shortparallel }%
\mathfrak{g}_{\alpha _{s}\beta _{s}}\ _{s}^{\shortparallel }\widehat{\mathbf{%
\mathbf{\mathbf{\mathbf{R}}}}}sc^{\star }=\ _{s}^{\shortparallel }\lambda
\mathbf{\mathbf{\mathbf{\mathbf{\ }}}}\ _{\star }^{\shortparallel }\mathfrak{%
g}_{\alpha _{s}\beta _{s}}.  \label{nonassocdeinst1}
\end{equation}%
It is convenient to write down such systems of nonlinear PDEs in a form
emphasizing explicitly the symmetric and nonsymmetric components of s-metric
(\ref{splitdmetr}),%
\begin{eqnarray}
\ _{\star }^{\shortparallel }\mathfrak{\check{g}}_{\mu _{s}\nu _{s}}(\
_{s}^{\shortparallel }\lambda +\frac{1}{2}\ _{s}^{\shortparallel }\widehat{%
\mathbf{\mathbf{\mathbf{\mathbf{R}}}}}sc^{\star }) &=&\mathbf{\mathbf{%
\mathbf{\mathbf{\ ^{\shortparallel }}}}}\widehat{\mathbf{\mathbf{\mathbf{%
\mathbf{R}}}}}ic_{(\mu _{s}\nu _{s})}^{\star }\mbox{ and }
\label{nonassocdeinst2a} \\
\ _{\star }^{\shortparallel }\mathfrak{a}_{\mu _{s}\nu _{s}}(\
_{s}^{\shortparallel }\lambda +\frac{1}{2}\ _{s}^{\shortparallel }\widehat{%
\mathbf{\mathbf{\mathbf{\mathbf{R}}}}}sc^{\star }) &=&\mathbf{\mathbf{%
\mathbf{\mathbf{\ ^{\shortparallel }}}}}\widehat{\mathbf{\mathbf{\mathbf{%
\mathbf{R}}}}}ic_{[\mu _{s}\nu _{s}]}^{\star }.  \label{nonassocdeinst2b}
\end{eqnarray}%
Form the second system of equations, we observe that if we impose the
nonholonomic conditions that
\begin{equation}
(\ _{s}^{\shortparallel }\lambda +\frac{1}{2}\ _{s}^{\shortparallel }%
\widehat{\mathbf{\mathbf{\mathbf{\mathbf{R}}}}}sc^{\star })=0\mbox{ and }%
\mathbf{\mathbf{\mathbf{\mathbf{\ ^{\shortparallel }}}}}\widehat{\mathbf{%
\mathbf{\mathbf{\mathbf{R}}}}}ic_{[\mu _{s}\nu _{s}]}^{\star }=0
\label{zeroeffsource}
\end{equation}%
the s-adapted R-fluxes generate nonsymmetric components of s--metrics $\
_{\star }^{\shortparallel }\mathfrak{a}_{\mu _{s}\nu _{s}}[\overline{%
\mathcal{R}}_{\quad \beta _{s}}^{\tau _{s}\xi _{s}}\ \mathbf{%
^{\shortparallel }e}_{\xi _{s}}\ _{\star }^{\shortparallel }\mathbf{g}_{\tau
_{s}\alpha _{s}}]$ (\ref{aux40a}) for arbitrary symmetric solutions $\
_{\star }^{\shortparallel }\mathfrak{\check{g}}_{\mu _{s}\nu _{s}}=\ _{\star
}^{\shortparallel }\mathfrak{g}_{\mu _{s}\nu _{s}}[\ _{\star
}^{\shortparallel }\mathbf{g}_{\beta _{s}\tau _{s}},\overline{\mathcal{R}}%
_{\quad \beta _{s}}^{\tau _{s}\xi _{s}}\ \mathbf{^{\shortparallel }e}_{\xi
_{s}}\ _{\star }^{\shortparallel }\mathbf{g}_{\tau _{s}\alpha _{s}}]$ of (%
\ref{aux40b}) where coefficients are constructed as functionals. The
symmetric part can taken to be a solution of
\begin{equation}
\mathbf{\mathbf{\mathbf{\mathbf{\ ^{\shortparallel }}}}}\widehat{\mathbf{%
\mathbf{\mathbf{\mathbf{R}}}}}ic_{\mu _{s}\nu _{s}}^{\star }=\
_{s}^{\shortparallel }\lambda \mathbf{\mathbf{\mathbf{\mathbf{\ }}}}\
_{\star }^{\shortparallel }\mathfrak{g}_{\alpha _{s}\beta _{s}},
\label{nonassocdeinst2aa}
\end{equation}%
which is a star s-adapted deformation of (\ref{cnsveinst2}). The symmetric
solutions of such PDEs of zero order on parameters $\hbar $ and $\kappa $
can be used as prime ones for generating general (non) symmetric solutions
with parametric dependence in nonassociative gravity and geometric flow
theories, see next section.

For both type symmetric and nonsymmetric solutions of above systems of
nonlinear PDEs for nonassociative vacuum gravity, we can extract
LC-configurations imposing at the end certain nonholonomic conditions (\ref%
{lccondnonass}) when $\ _{s}^{\shortparallel} \widehat{\mathbf{D}}_{\mid \
_{s}^{\shortparallel }\widehat{\mathbf{T}}=0}^{\star }=\
^{\shortparallel}\nabla ^{\star }.$ The s-adapted and decoupled equations (%
\ref{nonassocdeinst2a}) and/or (\ref{nonassocdeinst2b})(they are equivalent
to (\ref{nonassocdeinst2aa}) ) can be re-written equivalently for $\
^{\shortparallel }\nabla ^{\star }$ using canonical star distortions (\ref%
{candistrnas}). For the LC-configurations, the Ricci tensor is always
symmetric but in such holonomic variables we obtain vacuum gravitational
equations written in "more coupled" nonlinear forms which can not be solved
in some exact and well-defined parametric forms. Here we also emphasize that
the systems of nonlinear PDEs with nonholonomic dyadic splitting of type (%
\ref{nonassocdeinst1}) and, respectively, (\ref{nonassocdeinst2a}) and (\ref%
{nonassocdeinst2b}) etc. are self-consistent. Really, we can always start
with a (off-) diagonal solution of commutative vacuum Einstein equations and
compute \ s-adapted nonholonomic deforms and/or star deformations with
respective term in nonassociative analogous of s-adapted star deformations,
for a symmetric $\ ^{\shortparallel }\mathbf{g}_{\alpha _{s}\beta _{s}}.$ Of
course, we can work with more general (non) associative configurations
prescribing some commutative ansatz for $\ _{\star }^{\shortparallel }%
\mathfrak{\check{g}}_{\alpha _{s}\beta _{s}}$ and $\ _{\star
}^{\shortparallel }\mathfrak{a}_{\mu _{s}\nu _{s}}$ (we can consider
nonsymmetric canonical Ricci s-tensors) in a self-consistent form for a
chosen s-adapted frame $\ ^{\shortparallel }\mathbf{e}_{\mu _{s}}$ and then
try to generate parametric solutions of vacuum nonassociative gravitational
field equations with nontrivial R-flux contributions. The main issue in such
cases will be to keep such a s-adapted nonassociative gravitational
dynamical / geometric evolution scenarios when certain decoupling of star
deformed systems of PDEs will be possible.

There are nontrivial solutions of (\ref{nonassocdeinst2a}) and/or (\ref%
{nonassocdeinst2b}) for arbitrary parametric decomposition of $\
_{\star}^{\shortparallel }\mathfrak{a}_{\mu \nu }$ if
\begin{eqnarray}
\mathbf{\mathbf{\mathbf{\mathbf{\ ^{\shortparallel }}}}}\widehat{\mathbf{%
\mathbf{\mathbf{\mathbf{R}}}}}ic_{[\mu _{s}\nu _{s}]}^{\star } &=&0,\mathbf{%
\mathbf{\mathbf{\mathbf{\ ^{\shortparallel }}}}}\widehat{\mathbf{\mathbf{%
\mathbf{\mathbf{R}}}}}ic_{(\mu _{s}\nu _{s})}^{\star }=\mathbf{\mathbf{%
\mathbf{\mathbf{\ ^{\shortparallel }}}}}\widehat{\mathbf{\mathbf{\mathbf{%
\mathbf{R}}}}}ic_{\mu _{s}\nu _{s}}^{\star }=\mathbf{\mathbf{\mathbf{\mathbf{%
\ ^{\shortparallel }}}}}\widehat{\mathbf{\mathbf{\mathbf{\mathbf{R}}}}}%
ic_{\nu _{s}\mu _{s}}^{\star }=0,\mbox{ when }  \label{purestarvacuum} \\
\ _{s}^{\shortparallel }\widehat{\mathbf{R}}sc^{\star } &=&\
_{s}^{\shortparallel }\widehat{\mathbf{\mathbf{\mathbf{\mathbf{R}}}}}%
ss^{\star }=\ _{\star }^{\shortparallel }\mathfrak{\check{g}}^{\mu _{s}\nu
_{s}}\mathbf{\mathbf{\mathbf{\mathbf{\ ^{\shortparallel }}}}}\widehat{%
\mathbf{\mathbf{\mathbf{\mathbf{R}}}}}ic_{\mu _{s}\nu _{s}}^{\star }=\
_{\star }^{\shortparallel }\mathfrak{\check{g}}^{i_{1}j_{1}}\mathbf{\mathbf{%
\mathbf{\mathbf{\ ^{\shortparallel }}}}}\widehat{\mathbf{\mathbf{\mathbf{%
\mathbf{R}}}}}ic_{i_{1}j_{1}}^{\star }+\ _{\star }^{\shortparallel }%
\mathfrak{\check{g}}^{a_{2}b_{2}}\mathbf{\mathbf{\mathbf{\mathbf{\
^{\shortparallel }}}}}\widehat{\mathbf{\mathbf{\mathbf{\mathbf{R}}}}}%
ic_{a_{2}b_{2}}^{\star }  \notag \\
&& +\ _{\star }^{\shortparallel }\mathfrak{\check{g}}_{a_{3}b_{3}}\mathbf{%
\mathbf{\mathbf{\mathbf{\ ^{\shortparallel }}}}}\widehat{\mathbf{\mathbf{%
\mathbf{\mathbf{R}}}}}ic^{\star a_{3}b_{3}}+\ _{\star }^{\shortparallel }%
\mathfrak{\check{g}}_{a_{4}b_{4}}\mathbf{\mathbf{\mathbf{\mathbf{\
^{\shortparallel }}}}}\widehat{\mathbf{\mathbf{\mathbf{\mathbf{R}}}}}%
ic^{\star a_{4}b_{4}} =0.  \notag
\end{eqnarray}%
Such s-adapted star deformations with terms proportional up to $\hbar ,
\kappa $ and $\hbar \kappa $ can be considered for generating nonassociative
generalizations of black hole solutions from GR. Such solutions will be
constructed in partner works but some important and characteristic
properties for stationary configurations will be analyzed in section \ref%
{ssnonassocvsol},

In a more general context, we can compute nonzero $\ ^{\shortparallel}%
\widehat{\mathbf{R}}sc^{\star }= \ ^{1}\Lambda +\ ^{2}\Lambda +\ ^{3}\Lambda
+\ ^{4}\Lambda $ for certain effective shell polarized cosmological
constants $\ ^{s}\Lambda (\ _{s}^{\shortparallel }u)$ with certain
dependence on shell coordinates. Such polarizations of a nonassociative
gravitational vacuum determined by determined by nonlinear interactions in a
nonassociative gravitational "eather" which is a star nonholonomic
deformation of a vacuum configuration in GR. The nonassociative vacuum
gravitational equations with polarized cosmological shell constants,
\begin{equation}
~\mathbf{\mathbf{\mathbf{\mathbf{^{\shortparallel }}}}}\widehat{\mathbf{%
\mathbf{\mathbf{\mathbf{R}}}}}ic_{\alpha _{s}\beta _{s}}^{\star }(\
^{\shortparallel }u^{\gamma _{s}})=\ ^{s}\Lambda (\ ^{\shortparallel
}u^{\gamma _{s}})\ _{\star }^{\shortparallel }\mathfrak{\check{g}}_{\alpha
_{s}\beta _{s}}(\ ^{\shortparallel }u^{\gamma _{s}}),
\label{cannonsymparamc1}
\end{equation}%
split in dyadic form for respective shells:
\begin{eqnarray}
~\mathbf{\mathbf{\mathbf{\mathbf{^{\shortparallel }}}}}\widehat{\mathbf{%
\mathbf{\mathbf{\mathbf{R}}}}}ic_{i_{1}j_{1}}^{\star }(x^{k_{1}})
&=&~^{1}\Lambda (x^{k_{1}})\ _{\star }^{\shortparallel }\mathfrak{\check{g}}%
_{i_{1}j_{1}}(x^{k_{1}}),  \label{cannonsymparamcosm} \\
~\mathbf{\mathbf{\mathbf{\mathbf{^{\shortparallel }}}}}\widehat{\mathbf{%
\mathbf{\mathbf{\mathbf{R}}}}}ic_{a_{2}b_{2}}^{\star }(x^{k_{1}},x^{c_{2}})
&=&~^{2}\Lambda (x^{k_{1}},x^{c_{2}})\ _{\star }^{\shortparallel }\mathfrak{%
\check{g}}_{a_{2}b_{2}}(x^{k_{1}},x^{c_{2}}),  \notag \\
\mathbf{\mathbf{\mathbf{\mathbf{\ ^{\shortparallel }}}}}\widehat{\mathbf{%
\mathbf{\mathbf{\mathbf{R}}}}}ic^{\star a_{3}b_{3}}(x^{k_{1}},x^{c_{2}},%
\mathbf{\mathbf{\mathbf{\mathbf{\ ^{\shortparallel }}}}}p_{b_{3}})
&=&~^{3}\Lambda (x^{k_{1}},x^{c_{2}},\mathbf{\mathbf{\mathbf{\mathbf{\
^{\shortparallel }}}}}p_{b_{3}})\ _{\star }^{\shortparallel }\mathfrak{%
\check{g}}^{a_{3}b_{3}}(x^{k_{1}},x^{c_{2}},\mathbf{\mathbf{\mathbf{\mathbf{%
\ ^{\shortparallel }}}}}p_{b_{3}}),  \notag \\
\mathbf{\mathbf{\mathbf{\mathbf{\ ^{\shortparallel }}}}}\widehat{\mathbf{%
\mathbf{\mathbf{\mathbf{R}}}}}ic^{\star a_{4}b_{4}}(x^{k_{1}},x^{c_{2}},%
\mathbf{\mathbf{\mathbf{\mathbf{\ ^{\shortparallel }}}}}p_{b_{3}},\mathbf{%
\mathbf{\mathbf{\mathbf{\ ^{\shortparallel }}}}}p_{b_{4}}) &=&~^{4}\Lambda
(x^{k_{1}},x^{c_{2}},\mathbf{\mathbf{\mathbf{\mathbf{\ ^{\shortparallel }}}}}%
p_{b_{3}},\mathbf{\mathbf{\mathbf{\mathbf{\ ^{\shortparallel }}}}}%
p_{b_{4}})\ _{\star }^{\shortparallel }\mathfrak{\check{g}}%
^{a_{4}b_{4}}(x^{k_{1}},x^{c_{2}},\mathbf{\mathbf{\mathbf{\mathbf{\
^{\shortparallel }}}}}p_{b_{3}},\mathbf{\mathbf{\mathbf{\mathbf{\
^{\shortparallel }}}}}p_{b_{4}}).  \notag
\end{eqnarray}%
We can impose certain nonholonomic constraints%
\begin{equation}
\mathbf{\mathbf{\mathbf{\mathbf{\ ^{\shortparallel }}}}}\lambda +\frac{1}{2}%
\mathbf{\mathbf{\mathbf{\mathbf{\ ^{\shortparallel }}}}}\widehat{\mathbf{%
\mathbf{\mathbf{\mathbf{R}}}}}sc^{\star }=\mathbf{\mathbf{\mathbf{\mathbf{\
^{\shortparallel }}}}}\lambda +~^{1}\Lambda +~^{2}\Lambda +~^{3}\Lambda
+~^{4}\Lambda =0,  \label{nonhconstr1}
\end{equation}%
which allow to generate arbitrary nonsymmetric $\ _{\star }^{\shortparallel}
\mathfrak{a}_{\mu _{s}\nu _{s}}$ driven by a general off-diagonal symmetric
and associative/ commutative s-metric$\ ^{\shortparallel }\mathbf{g}%
_{\alpha_{s}\beta _{s}}$ and respective parametric R-flux deformations as in
(\ref{aux40b}). Such nonholonomic constraints impose a nonholonomic dynamic/
geometric evolution in phase space for (non) symmetric metrics. In section %
\ref{sec6}, we shall integrate in general forms, recurrently, systems of
nonlinear PDEs (\ref{cannonsymparamcosm}) with parametric s-decompositions
of $\ _{\star }^{\shortparallel }\mathfrak{g}_{\mu _{s}\nu _{s}}$ and $\
_{\star }^{\shortparallel }\mathfrak{a}_{\mu _{s}\nu _{s}}.$ We shall also
define certain generalized nonassociative nonlinear symmetries which will
allow to modify the generating functions and effective sources in such
solutions to those determined with certain constant relations $\
^{s}\Lambda(\ _{s}^{\shortparallel }u)\rightarrow $ $~_{0}^{s}\Lambda
=const, $ in particular, with $\ _{0}^{s}\Lambda =\ _{0}^{s}\Lambda =\
^{\shortparallel} \lambda /4.$

\subsection{Vacuum gravitational equations with effective nonassociative
s-sources}

\label{ssnonassocsourc}Using parametric s-decompositions with $\
_{\star}^{\shortparallel}\mathfrak{\check{g}}_{\alpha _{s}\beta
_{s}}^{[0]}=\ _{\star }^{\shortparallel }\mathbf{g}_{\alpha _{s}\beta _{s}}$
(\ref{aux40b}) for the canonical Ricci s-tensor, we express the
nonassociative vacuum gravitational field equations (\ref{cannonsymparamc1})
in the form
\begin{eqnarray}
\ ^{\shortparallel }\widehat{\mathbf{R}}_{\ \beta _{s}\gamma _{s}}~ &=&\
^{\shortparallel }\mathbf{K}_{_{\beta _{s}\gamma _{s}}},%
\mbox{ for effective
nonassociative sources }  \label{cannonsymparamc2} \\
\ ^{\shortparallel }\mathbf{K}_{_{\beta _{s}\gamma _{s}}} &=&\
_{[0]}^{\shortparallel }\Upsilon _{_{\beta _{s}\gamma _{s}}}+\
_{[1]}^{\shortparallel }\mathbf{K}_{_{\beta _{s}\gamma _{s}}}\left\lceil
\hbar ,\kappa \right\rceil ,\mbox{ where }  \notag \\
&&\ _{[0]}^{\shortparallel }\Upsilon _{_{\beta _{s}\gamma _{s}}}=\
^{s}\Lambda (\ ^{\shortparallel }u^{\gamma _{s}})\ _{\star }^{\shortparallel
}\mathbf{g}_{\beta _{s}\gamma _{s}}\ \mbox{and }  \label{assocsourc2} \\
&&\ _{[1]}^{\shortparallel }\mathbf{K}_{_{\beta _{s}\gamma _{s}}}\left\lceil
\hbar ,\kappa \right\rceil =\ ^{s}\Lambda (\ ^{\shortparallel }u^{\gamma
_{s}})\ _{\star }^{\shortparallel }\mathfrak{\check{g}}_{\beta _{s}\gamma
_{s}}^{[1]}(\kappa )-\ ^{\shortparallel }\widehat{\mathbf{K}}_{\ \beta
_{s}\gamma _{s}}\left\lceil \hbar ,\kappa \right\rceil  \label{nassocsourc2}
\end{eqnarray}%
is an effective parametric source with coefficients proportional to $\hbar
,\kappa $ and $\hbar \kappa .$

In this work, we shall apply a procedure for decoupling and integrating
nonassociative vacuum gravitational equations in parametric form when
additional conditions of type (\ref{zeroeffsource}) on (effective)
cosmological constants and certain scalars can be always imposed. For such
conditions, we shall be able to decouple the equations for the symmetric and
nonsymmetric components of s-metrics. We shall begin with decoupling of the
associative and commutative part of (\ref{cannonsymparamc2}) with certain
general parametrization of generating sources and finding certain general
integrals for $\ _{\star }^{\shortparallel }\mathbf{g}_{\beta _{s}\gamma
_{s}\mid \hbar ,\kappa =0}=\ ^{\shortparallel }\mathbf{g}_{\beta _{s}\gamma
_{s}}$ following the results of our recent work \cite{bubuianu20}, see also
similar results in \cite{bubuianu19,bubuianu17,bubuianu17b}.

\subsubsection{Dyadic parametrization of effective nonassociatvie s-sources}

We are not able to decouple and integrate in certain general/ explicit forms
modified (non) associative equations for general sources. There are
necessary also a series of additional assumptions on parameterizations of
nonassociative R-flux induced effective sources which allow to generate
generic off-diagonal solutions in explicit forms. In this work, we shall
parameterize (effective) sources (\ref{assocsourc2}) and (\ref{nassocsourc2}%
) are parameterized in s-adapted diagonal form%
\begin{eqnarray}
\ ^{\shortparallel }\mathbf{K}_{\ \beta _{s}}^{\alpha _{s}} &=&\{[\ \
_{1}^{\shortparallel }\Upsilon (x^{k_{1}})+\ _{1}^{\shortparallel }\mathbf{K}%
(\kappa ,x^{k_{1}})]\delta _{j_{1}}^{i_{1}},[\ \ _{2}^{\shortparallel
}\Upsilon (x^{k_{1}},x^{3})+\ _{2}^{\shortparallel }\mathbf{K}(\kappa
,x^{k_{1}},x^{3})]\delta _{b_{2}}^{a_{2}},  \label{ansatzsourc} \\
&&[\ \ _{3}^{\shortparallel }\Upsilon (x^{k_{2}},\ ^{\shortparallel
}p_{6})+\ _{3}^{\shortparallel }\mathbf{K}(\kappa ,x^{k_{2}},\
^{\shortparallel }p_{6})]\ \delta _{b_{3}}^{a_{3}},[\ \ _{4}^{\shortparallel
}\Upsilon (x^{k_{3}},\ ^{\shortparallel }p_{8})+\ _{4}^{\shortparallel }%
\mathbf{K}(\kappa ,x^{k_{3}},\ ^{\shortparallel }p_{8})]\delta
_{b_{4}}^{a_{4}}\}  \notag \\
&=&\{\ _{1}^{\shortparallel }\mathcal{K}(\kappa ,x^{k_{1}}),\
_{2}^{\shortparallel }\mathcal{K}(\kappa ,x^{k_{1}},x^{3}),\
_{3}^{\shortparallel }\mathcal{K}(\kappa ,x^{k_{2}},\ ^{\shortparallel
}p_{6}),\ _{4}^{\shortparallel }\mathcal{K}(\kappa ,x^{k_{3}},\
^{\shortparallel }p_{8})\},\mbox{ where } \ _{s}^{\shortparallel }\mathcal{K=%
}\ \ _{s}^{\shortparallel }\Upsilon +\ _{s}^{\shortparallel }\mathbf{K}.
\notag
\end{eqnarray}%
In these formulas, we consider a commutative spacetime background determined
by $\ ^{\shortparallel }\mathbf{g}_{\beta _{s}\gamma _{s}}$ (which can be
used for transforming up/low indices into low/up ones up till orders $\kappa
^{2}$ and $\hbar ^{2}$) when the total spacetime geometry is defined by a
parametric (non) symmetric s-metric. This will allow in next section to
prove a parametric decoupling property of (\ref{cannonsymparamc2}) for
quasi-stationary (non) symmetric star metrics. In well defined dyadic frames
the s-coefficients of such s-metrics do not depend on time like coordinated $%
x^{4}=t$ for the shells $s=1$ and 2. We shall consider for phase s-metrics
explicit dependencies on energy type coordinate $\ ^{\shortparallel
}p_{8}\sim E.$

A s-metric (and respective solutions of a (non) associative/ commutative )
gravitational equations is \textsf{quasi-stationary} if the corresponding
(non) associative phase spacetime geometric s-objects possess a Killing
symmetry on $\partial _{4}=\partial _{t}$ on shell $s=2$ and on $\
^{\shortparallel }\partial _{7},$ or $\ ^{\shortparallel }\partial _{8},$
for all shells up to $s=4.$ The AFCDM allows to construct very general
classes of exact and parametric solutions for other conditions, for
instance, with Killing symmetry $\partial _{3}$ on shell $s=2$ and on $\
^{\shortparallel }\partial _{8},$ or on $\ ^{\shortparallel }\partial _{7},$
up to all shells $s=4.$ The existence of such Killing vectors allows to
integrate and find off-diagonal solutions in very general forms. In
principle, we can construct more general classes of solutions without
assumptions on existence of some Killing symmetries, see review of results
and references in \cite{vacaru18} but we do not present such cumbersome
formulas in this paper.\footnote{%
One could be constructed also various examples of exact solutions, for
instance, corresponding to triadic splitting $3+3+3+...,$ or a tetradic $%
4+4+...$one. Such higher dimension, $\geq 3,$ shell by shell, decompositions
can encode certain topology of respective dimension 3,4, etc. which may be
different from the case of nonholonomic dyadic splitting. It is important to
apply more advanced geometric and topologic methods to study properties of
such physically important systems of nonlinear PDEs. One of the priorities
of dyadic decompositions is that for 2-d shells we can always conventionally
diagonalize block matrices which allow to decouple nonlinear equations in
certain "convenient" systems of references/ coordinates. For such
assumptions, we can construct solutions which depend, in principle, on all
spacetime and phase space coordinates. Finally, we note that $\
^{\shortparallel }\mathcal{K}_{\ \beta _{s}}^{\alpha _{s}}$ (\ref%
{ansatzsourc}) can be related to a general non-diagonal source $\
^{\shortparallel }\widehat{\Upsilon }_{\alpha _{s}^{\prime }\beta
_{s}^{\prime }}$ via frame transforms $\ \ ^{\shortparallel }\widehat{%
\Upsilon }_{\alpha _{s}^{\prime }\beta _{s}^{\prime }}=e_{\ \alpha
_{s}^{\prime }}^{\alpha _{s}}e_{\ \beta _{s}^{\prime }}^{\beta _{s}}\ \
^{\shortparallel }\mathcal{K}_{\alpha _{s}\beta _{s}}.$}

The nonassociative nonholonomic and parametric vacuum gravitational
equations (\ref{cannonsymparamc2}) with R-flux modified sources (\ref%
{ansatzsourc}) can be written in dyadic form as
\begin{eqnarray}
\ ^{\shortparallel }\widehat{\mathbf{R}}_{\ \beta _{s}}^{\alpha _{s}}~ &=&\
^{\shortparallel }\mathbf{K}_{\ \beta _{s}}^{\alpha _{s}}[\hbar \kappa
\overline{\mathcal{R}}]~,\mbox{ where }  \label{cannonsymparamc2a} \\
\ ^{\shortparallel }\mathbf{K}_{\ \beta _{s}}^{\alpha _{s}}~
&=&[~_{1}^{\shortparallel }\mathcal{K}(\kappa ,x^{k_{1}})\delta
_{i_{1}}^{j_{1}},~_{2}^{\shortparallel }\mathcal{K}(\kappa
,x^{k_{1}},x^{3})\delta _{b_{2}}^{a_{2}},~_{3}^{\shortparallel }\mathcal{K}%
(\kappa ,x^{k_{2}},\ ^{\shortparallel }p_{6})\delta
_{a_{3}}^{b_{3}},~_{4}^{\shortparallel }\mathcal{K}(\kappa ,x^{k_{3}},\
^{\shortparallel }p_{8})\delta _{a_{4}}^{b_{4}}].  \notag
\end{eqnarray}%
Such systems of nonlinear PDEs can be decoupled and solved in exact and
parametric forms on 8-d phase spaces applying the AFCDM \cite%
{bubuianu20,bubuianu19} for nonholonomic 2+2+2+2 splitting. In this work, we
shall adapt the constructions to the case of momentum variables with complex
unity and R-flux induced nonsymmetric and symmetric star deformed metrics.

Prescribing four couples of effective sources for certain diagonal data $\
_{s}^{\shortparallel }\mathcal{K}=\left[ ~_{1}^{\shortparallel }\mathcal{K},
\ _{2}^{\shortparallel }\mathcal{K},\ _{3}^{\shortparallel }\mathcal{K},
_{4}^{\shortparallel }\mathcal{K}\right] $ in (\ref{cannonsymparamc2a}), we
constrain nonholonomically shell by shell in dyadic adapted form the
gravitational dynamics. For further constructions, we shall consider that we
can prescribe any $\ _{s}^{\shortparallel }\mathcal{K}$ as \textbf{%
generating sources} encoding nonassociative geometric data. This will allow
to express certain generic off-diagonal classes of exact/parametric
solutions using some explicit formulas. We shall prove that such values and
generating functions for sources are related to conventional cosmological
constants via nonlinear symmetries, when the nonsymmetric parts of the
s-metrics and canonical Ricci s-tensors can be computed as R-flux
deformations of some off-diagonal symmetric metric configurations.

\subsubsection{Computing parametric R-flux induced (non) symmetric metrics
and effective source}

Considering an associative background solution $\ _{\star }^{\shortparallel }%
\mathbf{g}_{\beta _{s}\gamma _{s}}\simeq \ ^{\shortparallel }\mathbf{g}%
_{\beta _{s}\gamma _{s}}$ of $^{\shortparallel }\widehat{\mathbf{R}}_{\
\beta _{s}\gamma _{s}}~=\ _{[0]}^{\shortparallel }\Upsilon _{_{\beta
_{s}\gamma _{s}}}$ from (\ref{cannonsymparamc2}), we can compute the
nonsymmetric s-adapted coefficients for s-metric (\ref{aux40a}) as induced
ones by R-fluxes. The dyadic splitting is chosen with zero values of the
commutative h1 and v2 components,
\begin{eqnarray}
\ _{\star }^{\shortparallel }\mathfrak{a}_{\alpha _{s}\beta _{s}}:= &&[\
_{\star }^{\shortparallel }\mathfrak{a}_{i_{1}j_{1}}=0,\ _{\star
}^{\shortparallel }\mathfrak{a}_{a_{2}b_{2}}=0,  \label{nsm1} \\
\ _{\star }^{\shortparallel }\mathfrak{a}^{a_{3}b_{3}} &=&\frac{i\kappa }{2}(%
\overline{\mathcal{R}}_{c_{3}\quad }^{\ n+k_{s}a_{3}}\ \mathbf{%
^{\shortparallel }e}_{k_{s}}\ ^{\shortparallel }\mathbf{g}^{c_{3}b_{3}}-%
\overline{\mathcal{R}}_{c_{3}\quad }^{\ n+k_{s}b_{3}}\ \mathbf{%
^{\shortparallel }e}_{k_{s}}\ ^{\shortparallel }\mathbf{g}^{c_{3}a_{3}})=\
_{\star }^{\shortparallel }\mathfrak{a}_{[1]}^{a_{3}b_{3}}(\kappa )=\frac{1}{%
2}(\ _{\star }^{\shortparallel }\mathfrak{g}_{[1]}^{a_{3}b_{3}}-\ _{\star
}^{\shortparallel }\mathfrak{g}_{[1]}^{b_{3}a_{3}})(\kappa ),  \notag \\
\ _{\star }^{\shortparallel }\mathfrak{a}^{a_{4}b_{4}} &=&\frac{i\kappa }{2}(%
\overline{\mathcal{R}}_{c_{4}\quad }^{\ n+k_{s}a_{4}}\ \mathbf{%
^{\shortparallel }e}_{k_{s}}\ ^{\shortparallel }\mathbf{g}^{c_{4}b_{4}}-%
\overline{\mathcal{R}}_{c_{4}\quad }^{\ n+k_{s}b_{4}}\ \mathbf{%
^{\shortparallel }e}_{k_{s}}\ ^{\shortparallel }\mathbf{g}^{c_{4}a_{4}})=\
_{\star }^{\shortparallel }\mathfrak{a}_{[1]}^{a_{4}b_{4}}(\kappa )=\frac{1}{%
2}(\ _{\star }^{\shortparallel }\mathfrak{g}_{[1]}^{a_{4}b_{4}}-\ _{\star
}^{\shortparallel }\mathfrak{g}_{[1]}^{b_{4}a_{4}})(\kappa )],  \notag
\end{eqnarray}%
where $i_{2}=(i_{1},a_{2}),$ for $s=1,2,$ following convention on s-adapted
indices and coordinates. The symmetric part of s-metric is computed
following parametric decompositions (\ref{aux40b}),
\begin{eqnarray}
\ ^{\shortparallel }\mathfrak{\check{g}}_{\alpha _{s}\beta _{s}} &=&\
^{\shortparallel }\mathbf{g}_{\alpha _{s}\beta _{s}}+\ _{\star
}^{\shortparallel }\mathfrak{\check{g}}_{\alpha _{s}\beta _{s}}^{[1]}(\kappa
),\mbox{ for }  \label{ssm1} \\
\ _{\star }^{\shortparallel }\mathfrak{\check{g}}_{\alpha _{s}\beta
_{s}}^{[1]}(\kappa ) &=&-\frac{i\kappa }{2}\left( \overline{\mathcal{R}}%
_{\quad \beta _{s}}^{\tau _{s}\xi _{s}}\ \mathbf{^{\shortparallel }e}_{\xi
_{s}}\ ^{\shortparallel }\mathbf{g}_{\tau _{s}\alpha _{s}}+\overline{%
\mathcal{R}}_{\quad \alpha _{s}}^{\tau _{s}\xi _{s}}\ \mathbf{%
^{\shortparallel }e}_{\xi _{s}}\ ^{\shortparallel }\mathbf{g}_{\beta
_{s}\tau _{s}}\right),  \notag
\end{eqnarray}%
where $\ ^{\shortparallel }\mathbf{g}_{\beta _{s}\tau _{s}}$ encode
associative contributions from all shells $s=1,2,3,4$ but also result in
symmetric R-flux deformations to star metrics.

So, prescribing any generating sources $\ ^{\shortparallel }\mathbf{K}_{\
\beta _{s}}^{\alpha _{s}}$ (\ref{ansatzsourc}) with nonassociative
parametric deformations and using block $[(2\times 2)+(2\times 2)]+[(2\times
2)+(2\times 2)]$ parametrization of symmetric and nonsymmetric parts of star
deformed s-metrics, $\ _{\star }^{\shortparallel }\mathfrak{g}_{\mu _{s}\nu
_{s}}=(\ _{\star }^{\shortparallel}\mathfrak{g}_{i_{1}j_{1}},\
_{\star}^{\shortparallel} \mathfrak{g}_{a_{2}b_{2}},\ _{\star
}^{\shortparallel}\mathfrak{g}^{a_{3}b_{3}}, \ _{\star }^{\shortparallel }%
\mathfrak{g}^{a_{4}b_{4}})$ and $\ _{\star }^{\shortparallel }\mathfrak{a}%
_{\mu _{s}\nu _{s}}=(0,0,\ _{\star }^{\shortparallel }\mathfrak{a}%
_{c_{3}b_{3}}, \ _{\star }^{\shortparallel }\mathfrak{a}_{c_{4}b_{4}}),$ we
can compute such R-flux deformations for any stated associative (off-)
diagonal solution $\ ^{\shortparallel }\mathbf{g}_{\alpha _{s}\beta _{s}}.$
In a similar form, there are computed complex and real R-flux deformations
of the canonical Ricci s-tensor (\ref{ric51}), (\ref{ric52}) and (\ref{ric53}%
) subjected to nonholonomic constraints by generating sources in (\ref%
{paramsriccins}).

The main goal of this work is to extend the AFDM to nonassociative vacuum
gravity described by dyadic equations (\ref{cannonsymparamc2}) with
effective sources (\ref{ansatzsourc}) and prove a general decoupling
property of such nonlinear systems of PDEs.

\subsection{Adapted lifts and horizontal projections of nonassociative
canonical Ricci s-tensors}

On a Lorentz spacetime manifold $V$ and it (co) tangent bundle (and, for
instance, on phase spaces $\mathcal{M}=T^{\ast }\mathbf{V}$ and/or $\
^{\shortparallel }\mathcal{M}$), we can define nonholonomic dyadic
structures with shells $s=1,2,3,4$. Then, the geometric constructions can
subjected to s-adapted nonassociative star deformations. Such deformations
can be reordered using the associator and projected back to $\mathbf{V}$ in
s-adapted form. Functions and s-forms can be lifted from $\mathbf{V}$ to $%
\mathcal{M}$ as pullbacks of s-forms using a respective s-adapted canonical
projection $\pi :\ T^{\ast }\mathbf{V}\rightarrow \mathbf{V}.$ To work in
inverse direction, with an embedding $\sigma _{s}:\ \mathbf{V}\rightarrow
\mathcal{M}$, which is given by zero section $x\rightarrow \sigma _{s}(\
^{1}x,\ ^{2}x)=(\ ^{1}x,~^{2}x,0,0),$ we pull back s-forms on $\mathcal{M}$
to s-forms on $\mathbf{V}.$ Lifts of s-vectors are constructed using
s-adapted foliations of $\mathcal{M}$ with constant momentum leaves with 2+2
splitting. Such leafs are chosen to be diffeomorphic to $\mathbf{V}$ when
for $\partial _{i_{2}}$ on $\mathbf{V}$ are transformed into a s-vector on $%
\mathcal{M}$, when
\begin{eqnarray*}
b^{i_{2}}(~^{1}x,~^{2}x)\partial _{i_{2}} &\rightarrow &\pi ^{\ast
}(b^{i_{2}})(~^{1}x,~^{2}x,~_{3}^{\shortparallel }p,~~_{4}^{\shortparallel
}p)\partial _{i_{2}},\mbox{ where } \\
\pi ^{\ast }(b^{i_{2}})(x,p)\partial _{i_{2}} &=&(b^{i_{2}})(\pi
(~^{1}x,~^{2}x,~_{3}^{\shortparallel }p,~~_{4}^{\shortparallel
}p))=b^{i_{2}}(~^{1}x,~^{2}x).
\end{eqnarray*}
Acting in such directions, we project s-vector fields on $\mathcal{M}$ into
s-vector fields on $\mathbf{V}$ via zero section of $\sigma _{s},$ which can
be written in s-adapted coordinate form,
\begin{eqnarray*}
&&b^{i_{1}}(~^{1}x,~^{2}x,~_{3}^{\shortparallel }p,~~_{4}^{\shortparallel
}p)\partial _{i_{1}}+b^{a_{2}}(~^{1}x,~^{2}x,~_{3}^{\shortparallel
}p,~~_{4}^{\shortparallel }p)\partial
_{a_{2}}+b_{a_{3}}(~^{1}x,~^{2}x,~_{3}^{\shortparallel
}p,~~_{4}^{\shortparallel }p)\partial
^{a_{3}}+b_{a_{4}}(~^{1}x,~^{2}x,~_{3}^{\shortparallel
}p,~~_{4}^{\shortparallel }p)\partial ^{a_{4}} \\
&&\rightarrow b^{i_{1}}(~^{1}x,~^{2}x,0,~0)\partial
_{i_{1}}+b^{a_{2}}(~^{1}x,~^{2}x,0,~0)\partial _{a_{2}}.
\end{eqnarray*}

To apply the AFCDM, we have to work on nonholonomic manifolds and (co)
tangent Lorentz bundles with dyadic s-splitting and perform the
constructions in s-adapted form, with formulas of type
\begin{eqnarray*}
&&\ ^{\shortparallel }b^{i_{1}}(~^{1}x)\ ^{\shortparallel }\mathbf{e}%
_{i_{1}}+\ ^{\shortparallel }b^{a_{2}}(~^{1}x,~^{2}x)\ ^{\shortparallel }%
\mathbf{e}_{a_{2}}+\ ^{\shortparallel
}b_{a_{3}}(~^{1}x,~^{2}x,~_{3}^{\shortparallel }p)\ ^{\shortparallel }%
\mathbf{e}^{a_{3}}+\ ^{\shortparallel
}b_{a_{4}}(~^{1}x,~^{2}x,~_{3}^{\shortparallel }p,~~_{4}^{\shortparallel
}p)\ ^{\shortparallel }\partial ^{a_{4}} \\
&\rightarrow &\ ^{\shortparallel }b^{i_{1}}(~^{1}x)\ ^{\shortparallel }%
\mathbf{e}_{i_{1}}+\ ^{\shortparallel }b^{a_{2}}(~^{1}x,~^{2}x)\
^{\shortparallel }\mathbf{e}_{a_{2}},
\end{eqnarray*}%
see s-adapted bases (\ref{nadapdc}). Any metric tensor $\widehat{g}
_{j_{2}k_{2}}(\ ^{1}x,\ ^{2}x)dx^{j_{2}}\otimes dx^{k_{2}}$ on $\mathbf{V}$,
when $j_{2}=(j_{1},b_{2})$, can be lifted into an off-diagonal metric and/or
equivalent s--metric $\ _{s}^{\shortparallel }\mathbf{g}$ (\ref{commetr}) on
$\mathbf{T}_{s\shortparallel }^{\ast }\mathbf{V}$. In s-adapted form, we can
consider such shell by shell parameterizations using matrices (for symmetric
s-tensors):%
\begin{eqnarray*}
\mbox{ for }s &=&1,2;\mbox{ for }\underline{\alpha }_{2}\underline{\beta }%
_{2},...,\mbox{ or }i_{2},j_{2},...=1,2,3,4;=(i_{1},a_{2}),(j_{1},b_{2}),%
\mbox{ for }i_{1},j_{1}=1,2;a_{2,}b_{2},...=3,4; \\
\ \widehat{g}_{j_{1}k_{1}}(\ ^{1}x) &\rightarrow &\left\vert
\begin{array}{c}
\ ^{\shortparallel }\widehat{g}_{\underline{\alpha }_{2}\underline{\beta }%
_{2}}(u^{\gamma _{2}}),\mbox{ or }\ ^{\shortparallel }\widehat{g}_{%
\underline{i}_{2}\underline{j}_{2}}(x^{k_{2}}), \\
\mbox{ equivalent index } \\
\mbox{ convention } \\
\ ^{\shortparallel }\widehat{g}_{\underline{i}_{2}\underline{j}%
_{2}}(x^{k_{1}},x^{c_{2}})%
\end{array}%
\right\vert = \\
&&\left[
\begin{array}{cc}
\ ^{\shortparallel }\widehat{g}%
_{i_{1}j_{1}}(x^{k_{1}})+g_{a_{2}b_{2}}(x^{k_{2}})N_{i_{1}}^{a_{2}}(x^{k_{2}}) N_{j_{1}}^{b_{2}}(x^{k_{2}})
& g_{a_{2}e_{2}}(x^{k_{2}})\ N_{j_{1}}^{e_{2}}(x^{k_{2}}) \\
\ \ g_{b_{2}e_{2}}(x^{k_{2}})N_{i_{1}}^{e_{2}}(x^{k_{2}}) & \
g_{a_{2}b_{2}}(x^{k_{2}})\
\end{array}%
\right] \\
&\simeq &\left[ \ \widehat{\mathbf{g}}_{j_{1}k_{1}}(x^{k_{1}}),\mathbf{g}%
_{a_{2}b_{2}}(x^{k_{1}},x^{c_{2}})\right] =\ ^{\shortparallel }\mathbf{g}%
_{\mu _{2}\nu _{2}}(x^{k_{1}},x^{c_{2}})=\left[
\begin{array}{cc}
\mathbf{g}_{j_{1}k_{1}}(x^{k_{1}},x^{c_{2}}) & 0 \\
0 & \ \mathbf{g}_{a_{2}b_{2}}(x^{k_{1}},x^{c_{2}})%
\end{array}%
\right] \\
&\longrightarrow &\widehat{\mathbf{g}}_{j_{2}k_{2}}(x^{i_{2}})=\left[
\begin{array}{cc}
\mathbf{g}_{j_{1}k_{1}}(x^{k_{1}},x^{c_{2}}) & 0 \\
0 & \ ^{\shortparallel }\mathbf{g}_{a_{2}b_{2}}(x^{k_{1}},x^{c_{2}})%
\end{array}%
\right] ,\mbox{ general s-coordinates };
\end{eqnarray*}

\begin{eqnarray*}
\mbox{ for }s &=&3;\mbox{ for }\underline{\alpha }_{3}\underline{\beta }%
_{3},...,\mbox{ or }i_{3},j_{3}...=1,2,3,4,5,6;=(i_{2},a_{3}),(j_{2},b_{3}),%
\mbox{ for }a_{3,}b_{3},...=5,6; \\
\ \widehat{g}_{j_{2}k_{2}}(x^{i_{2}}) &\rightarrow &\left\vert
\begin{array}{c}
\ ^{\shortparallel }\widehat{g}_{\underline{\alpha }_{3}\underline{\beta }%
_{3}}(\ ^{\shortparallel }u^{\gamma _{3}}),\mbox{ or }\ ^{\shortparallel }%
\widehat{g}_{\underline{i}_{2}\underline{j}_{2}}(\ ^{\shortparallel
}x^{k_{3}}), \\
\mbox{ equivalent index } \\
\mbox{ convention } \\
\ ^{\shortparallel }\widehat{g}_{\underline{i}_{3}\underline{j}%
_{3}}(x^{i_{2}},\ ^{\shortparallel }p_{a_{3}})%
\end{array}%
\right\vert = \\
&&\left[
\begin{array}{cc}
\ \widehat{g}_{i_{2}j_{2}}(x^{k_{2}})+\ ^{\shortparallel }\widehat{g}%
^{a_{3}b_{3}}(\ ^{\shortparallel }x^{k_{3}})\ ^{\shortparallel
}N_{i_{2}a_{3}}(\ ^{\shortparallel }x^{k_{3}})\ \ ^{\shortparallel
}N_{j_{2}b_{3}}(\ ^{\shortparallel }x^{k_{3}}) & \ \ ^{\shortparallel }%
\widehat{g}^{a_{3}e_{3}}(\ ^{\shortparallel }x^{k_{3}})\ ^{\shortparallel
}N_{j_{2}e_{3}}(\ ^{\shortparallel }x^{k_{3}}) \\
\ \ ^{\shortparallel }\widehat{g}^{b_{3}e_{3}}(\ ^{\shortparallel
}x^{k_{3}})\ ^{\shortparallel }N_{i_{2}e_{3}}(\ ^{\shortparallel }x^{k_{3}})
& \ \ ^{\shortparallel }\widehat{g}^{a_{3}b_{3}}(\ ^{\shortparallel
}x^{k_{3}})\
\end{array}%
\right] \\
&\simeq &\left[ \ \ ^{\shortparallel }\mathbf{g}_{j_{2}k_{2}}(x^{k_{2}},\
^{\shortparallel }p_{c_{3}}),\ ^{\shortparallel }\widehat{\mathbf{g}}%
^{a_{3}b_{3}}(x^{k_{2}},\ ^{\shortparallel }p_{c_{3}})\right] =\
^{\shortparallel }\widehat{\mathbf{g}}_{\mu _{3}\nu _{3}}(\ ^{\shortparallel
}u^{\gamma _{3}})=\left[
\begin{array}{cc}
\ \widehat{\mathbf{g}}_{j_{2}k_{2}}(\ ^{\shortparallel }u^{\gamma _{3}}) & 0
\\
0 & \ ^{\shortparallel }\widehat{\mathbf{g}}^{a_{3}b_{3}}(\ ^{\shortparallel
}u^{\gamma _{3}})%
\end{array}%
\right] \\
&\longrightarrow &\ ^{\shortparallel }\widehat{\mathbf{g}}_{j_{3}k_{3}}(\
^{\shortparallel }x^{i_{3}})=\left[
\begin{array}{cc}
\ ^{\shortparallel }\mathbf{g}_{j_{2}k_{2}}(\ ^{\shortparallel }x^{i_{3}}) &
0 \\
0 & \ ^{\shortparallel }\mathbf{g}^{a_{3}b_{3}}(\ ^{\shortparallel
}x^{i_{3}})%
\end{array}%
\right] ;
\end{eqnarray*}%
\begin{eqnarray}
\mbox{ for }s &=&4;\mbox{ for }\underline{\alpha }_{4}\underline{\beta }%
_{4},...=1,2,3,...7,8;=(i_{3},a_{4}),(j_{3},b_{4}),\mbox{ for }%
a_{4,}b_{4},...=7,8;  \notag \\
\ ^{\shortparallel }\widehat{\mathbf{g}}_{j_{3}k_{3}}(\ ^{\shortparallel
}x^{i_{3}}) &\rightarrow &\ ^{\shortparallel }\widehat{g}_{\underline{\alpha
}_{4}\underline{\beta }_{4}}(\ _{4}^{\shortparallel }u)=\ ^{\shortparallel }%
\widehat{g}_{\underline{\alpha }_{4}\underline{\beta }_{4}}(\
^{\shortparallel }u^{\gamma _{4}})=\ ^{\shortparallel }\widehat{g}_{%
\underline{\alpha }_{4}\underline{\beta }_{4}}(\ _{3}^{\shortparallel }x,\
_{4}^{\shortparallel }p)  \notag \\
&=&\left[
\begin{array}{cc}
\ \ ^{\shortparallel }\widehat{g}_{i_{3}j_{3}}(\ ^{\shortparallel
}x^{k_{3}})+\ ^{\shortparallel }\widehat{g}^{a_{4}b_{4}}(\
_{4}^{\shortparallel }u)\ ^{\shortparallel }N_{i_{3}a_{4}}(\
_{4}^{\shortparallel }u)\ \ ^{\shortparallel }N_{j_{3}b_{4}}(\
_{4}^{\shortparallel }u) & \ \ ^{\shortparallel }\widehat{g}^{a_{4}e_{4}}(\
_{4}^{\shortparallel }u)\ ^{\shortparallel }N_{j_{3}e_{4}}(\
_{4}^{\shortparallel }u) \\
\ \ ^{\shortparallel }\widehat{g}^{b_{4}e_{4}}(\ _{4}^{\shortparallel }u)\
^{\shortparallel }N_{i_{3}e_{4}}(\ _{4}^{\shortparallel }u) & \ \
^{\shortparallel }\widehat{g}^{a_{4}b_{4}}(\ _{4}^{\shortparallel }u)\
\end{array}%
\right]  \notag \\
&\simeq &\left[ \mathbf{g}_{j_{3}k_{3}}(\ ^{\shortparallel }x^{i_{3}},\
^{\shortparallel }p_{c_{4}}),\ ^{\shortparallel }\mathbf{g}^{a_{3}b_{3}}(\
^{\shortparallel }x^{i_{3}},\ ^{\shortparallel }p_{c_{4}})\right] =\
^{\shortparallel }\mathbf{g}_{\mu _{4}\nu _{4}}(\ _{4}^{\shortparallel }u)=%
\left[
\begin{array}{cc}
\ ^{\shortparallel }\mathbf{g}_{j_{3}k_{3}}(\ _{4}^{\shortparallel }u) & 0
\\
0 & \ ^{\shortparallel }\widehat{\mathbf{g}}^{a_{4}b_{4}}(\
_{4}^{\shortparallel }u)%
\end{array}%
\right]  \notag \\
&\longrightarrow &\ ^{\shortparallel }\mathbf{g}_{\mu _{4}\nu _{4}}=\left[
\begin{array}{cc}
\mathbf{g}_{j_{3}k_{3}}(\ ^{\shortparallel }x^{i_{3}},\ ^{\shortparallel
}p_{c_{4}}) & 0 \\
0 & \ ^{\shortparallel }\mathbf{g}^{a_{4}b_{4}}(\ ^{\shortparallel
}x^{i_{3}},\ ^{\shortparallel }p_{c_{4}})%
\end{array}%
\right] .  \label{smetr1assoc}
\end{eqnarray}%
In above formulas, there are used different variants of s-adapted notations
in order to familiarize readers with such a symbolic nonholonomic and dyadic
geometric calculus.

We can use an associative s-metric adapted in the form $\ ^{\shortparallel }%
\mathbf{g}_{\alpha _{s}\beta _{s}}$ (\ref{smetr1assoc}) $\ $instead of $\
_{\star }^{\shortparallel }\mathbf{g}_{\alpha _{s}\beta _{s}}$ and study
s-adapted R-flux deformations to nonsymmetric s-metrics $\ ^{\shortparallel }%
\mathbf{g}_{\alpha \beta }\rightarrow \ _{\star }^{\shortparallel }\mathfrak{%
g}_{\alpha \beta }$ (\ref{dmss1}), when
\begin{equation*}
\ _{s}^{\shortparallel }\mathbf{g=}\ ^{\shortparallel }\mathbf{g}_{\alpha
_{s}\beta _{s}}\ ^{\shortparallel }\mathbf{e}^{\alpha _{s}}\otimes _{\star
s}\ ^{\shortparallel }\mathbf{e}^{\beta _{s}}\rightarrow \ _{\star
}^{\shortparallel }\mathfrak{g}=\ _{\star }^{\shortparallel }\mathfrak{g}%
_{\alpha _{s}\beta _{s}}\star _{s}(\ ^{\shortparallel }\mathbf{e}^{\alpha
_{s}}\otimes _{\star s}\ ^{\shortparallel }\mathbf{e}^{\beta _{s}}).
\end{equation*}%
An important physical case is to consider R-deformations of s-metrics
involving h1-v2-configurations determined by spacetime metrics, $\mathbf{g}
_{i_{2}j_{2}}(x^{k_{2}})\subset \ _{\star }^{\shortparallel }\mathfrak{g}%
_{\alpha _{s}\beta _{s}}.$ For such nonassociative metric configurations,
the linear R-flux corrections on shells $s=3,4$ can be parameterized in
off-diagonal and/or dyadic block terms,
\begin{eqnarray}
\ ^{\shortparallel }\mathbf{g}_{\alpha _{3}\beta _{3}} &=&\left[
\begin{array}{cc}
\mathbf{g}_{j_{2}k_{2}}(x^{i_{2}},\ ^{\shortparallel }p_{c_{3}}) & 0 \\
0 & \ ^{\shortparallel }\mathbf{g}^{a_{3}b_{3}}(x^{i_{2}},\ ^{\shortparallel
}p_{c_{3}})%
\end{array}%
\right]  \label{auxm61} \\
&\rightarrow &\ _{\star }^{\shortparallel }\mathfrak{g}_{\alpha _{3}\beta
_{3}}=\left[
\begin{array}{cc}
\mathbf{g}_{j_{2}k_{2}} & -\frac{i\kappa }{2}\overline{\mathcal{R}}_{\quad
\quad \quad n+k_{1}}^{n+i_{1}\ n+l_{2}}\ ^{\shortparallel }\mathbf{e}_{i_{1}}%
\mathbf{g}_{j_{2}l_{2}} \\
-\frac{i\kappa }{2}\overline{\mathcal{R}}_{\quad \quad \quad
n+j_{1}}^{n+i_{1}\ n+l_{2}}\ ^{\shortparallel }\mathbf{e}_{i_{1}}\mathbf{g}%
_{k_{2}l_{2}} & \ ^{\shortparallel }\mathbf{g}^{a_{3}b_{3}}%
\end{array}%
\right] ,  \notag \\
\mbox{ if }\mathbf{g}_{j_{2}k_{2}} &=&\mathbf{g}_{j_{2}k_{2}}(x^{i_{2}})%
\rightarrow \left[
\begin{array}{cc}
\mathbf{g}_{j_{2}k_{2}}(x^{i_{2}}) & -\frac{i\kappa }{2}\overline{\mathcal{R}%
}_{\quad \quad \quad n+k_{1}}^{n+i_{1}\ n+l_{2}}\mathbf{\partial }_{i_{1}}%
\mathbf{g}_{j_{2}l_{2}} \\
-\frac{i\kappa }{2}\overline{\mathcal{R}}_{\quad \quad \quad
n+j_{1}}^{n+i_{1}\ n+l_{2}}\mathbf{\partial }_{i_{1}}\mathbf{g}_{k_{2}l_{2}}
& \ ^{\shortparallel }\mathbf{g}^{a_{3}b_{3}}%
\end{array}%
\right] ;  \notag
\end{eqnarray}
\begin{eqnarray*}
\ \ ^{\shortparallel }\mathbf{g}_{\alpha _{4}\beta _{4}} &=&\left[
\begin{array}{cc}
\mathbf{g}_{j_{3}k_{3}}(x^{i_{3}},\ ^{\shortparallel }p_{c_{4}}) & 0 \\
0 & \ ^{\shortparallel }\mathbf{g}^{a_{4}b_{4}}(x^{i_{3}},\ ^{\shortparallel
}p_{c_{4}})%
\end{array}%
\right] \\
&\rightarrow &\ _{\star }^{\shortparallel }\mathfrak{g}_{\alpha _{4}\beta
_{4}}=\left[
\begin{array}{cc}
\mathbf{g}_{j_{2}k_{2}} & -\frac{i\kappa }{2}\overline{\mathcal{R}}_{\quad
\quad \quad n+k_{2}}^{n+i_{2}\ n+l_{2}}\ ^{\shortparallel }\mathbf{e}_{i_{2}}%
\mathbf{g}_{j_{2}l_{2}} \\
-\frac{i\kappa }{2}\overline{\mathcal{R}}_{\quad \quad \quad
n+j_{2}}^{n+i_{2}\ n+l_{2}}\ ^{\shortparallel }\mathbf{e}_{i_{2}}\mathbf{g}%
_{k_{2}l_{2}} & \ ^{\shortparallel }\mathbf{g}^{a_{4}b_{4}}%
\end{array}%
\right] , \\
\mbox{ if }\mathbf{g}_{j_{2}k_{2}} &=&\mathbf{g}_{j_{2}k_{2}}(x^{i_{2}})%
\rightarrow \left[
\begin{array}{cc}
\mathbf{g}_{j_{2}k_{2}}(x^{i_{2}}) & -\frac{i\kappa }{2}\overline{\mathcal{R}%
}_{\quad \quad \quad n+k_{2}}^{n+i_{2}\ n+l_{2}}\mathbf{\partial }_{i_{2}}%
\mathbf{g}_{j_{2}l_{2}} \\
-\frac{i\kappa }{2}\overline{\mathcal{R}}_{\quad \quad \quad
n+j_{2}}^{n+i_{2}\ n+l_{2}}\mathbf{\partial }_{i_{2}}\mathbf{g}_{k_{2}l_{2}}
& \ ^{\shortparallel }\mathbf{g}^{a_{4}b_{4}}%
\end{array}%
\right] .
\end{eqnarray*}%
We note that in above formulas with nonassociative dyadic decompositions the
R-flux terms mix and interrelate shells for a background dyadic splitting.

The h1-v2-projection of a Ricci s-tensor $\ ^{\shortparallel}\widehat{%
\mathbf{\mathbf{\mathbf{\mathbf{R}}}}}ic_{\alpha _{2}\beta _{2}}^{\star
}=\{\ ^{\shortparallel }\widehat{R}_{\ \star h_{2}j_{2}}\},$ see formulas (%
\ref{hcnonassocrcana}), is a canonical s-connection analog of the
nonassociative Ricci tensor for LC-configuration and projections on $V.$ If
we impose the conditions of zero distortions, $\ _{\star }^{\shortparallel s}%
\widehat{\mathbf{Z}}=0,$ i. e.$\ _{s}^{\shortparallel }\widehat{\mathbf{D}}%
_{\mid \ _{s}^{\shortparallel }\widehat{\mathbf{T}}=0}^{\star }=\
^{\shortparallel }\nabla ^{\star },$ see formulas (\ref{lccondnonass}) and (%
\ref{candistrnas}), and then performing the constructions \ a corresponding
local basis with dyadic splitting (then restricting the components of the
Ricci tensors and s-tensors, $Ric_{\alpha _{s}\beta s},$ to spacetime
directions and setting the momentum dependence to zero), we can construct
geometric configurations with
\begin{equation*}
\ \ ^{\shortparallel }\widehat{R}_{\ \star h_{2}j_{2}}(~_{s}x,\
_{s}^{\shortparallel }p)\rightarrow \ \ _{\nabla }^{\shortparallel }\mathcal{%
\Re }ic_{h_{s}j_{s}}^{\star }(~_{s}x)=Ric_{h_{2}j_{2}}^{\circ
}(~_{s}x)=\sigma ^{\ast }(\ \ _{\nabla }^{\shortparallel }\mathcal{\Re }%
ic_{h_{2}j_{12}}^{\star }(~_{s}x,\ _{s}^{\shortparallel }p))=R_{hj}(~_{s}x,\
.).
\end{equation*}%
In N-adapted form such nonassociative nonholonomic configurations and their
spacetime projections were defined in section 4.4 of \cite{partner01}. In
this work, the dyadic formalism allows to generate such geometric models
using s-metrics and star R-flux deformations via formulas (\ref{smetr1assoc}%
) and (\ref{auxm61}) in order to show how such configurations can be
extracted from solutions of nonassociative vacuum gravitational equations
with parametric decompositions $\hbar $ and $\kappa $ (see, for instance, (%
\ref{purestarvacuum})). Considering nonholonomic constraint to the case of
LC-connection we obtain the nonassociative vacuum gravitational equations on
phase space postulated in \cite{aschieri17},
\begin{equation}
\ \ _{\nabla }^{\shortparallel }\mathcal{\Re }ic_{\alpha _{s}\beta
_{s}}^{\star }(~_{s}x,\ _{s}^{\shortparallel }p)=0.  \label{veinstlc}
\end{equation}%
For certain additional assumptions, every solution of such a system of
nonlinear PDEs can be transformed as a solution of
\begin{equation}
Ric_{h_{2}j_{2}}^{\circ }(x^{k_{2}})=0  \label{veinsthlc}
\end{equation}%
on the base spacteime $V.$ In holonomic form, not all solutions of (\ref%
{veinsthlc}) can be lifted to solutions of (\ref{veinstlc}) on the phase
space because the "non-geometric" R-fluxes define nontrivial nonholonomic
distributions both on the phase space and the spacetime manifold.

In a series of partner works (see \cite{partner01} as the first one), we
elaborate on geometrization of "non-geometric" string theories with R-fluxes
using nonholonomic phase geometry. Following geometric principles similar to
\cite{misner,vacaru18,bubuianu18a,vacaru16} but for respective nonholonomic
dyadic (non) symmetric and canonical s-connection structures, we can derive
nonassociative dynamical gravitational and matter field equations and
develop geometric flow evolution models completely determined on phase
spaces. Such constructions can be completed with a geometric method of
constructing exact and parametric solutions for nonassociative parametric
physically important nonlinear systems of PDEs.

\subsection{First order parametric nonassociative s-adapted spacetime
corrections}

In this subsection, we analyze some important physical properties of h1-v2
components of nonassociative nonholonomic dyadic and parametric vacuum
gravitational equations (\ref{cannonsymparamc2}). Projecting such formulas
on a base spacetime $V,$ we can compute nontrivial real R-flux
contributions. Such s-adapted configurations can be generated by effective
sources prescribed only on shells $s=1$ and $2,$ when $\ ^{\shortparallel }%
\mathbf{K} _{\ \beta _{s}}^{\alpha _{s}}(x^{k_{2}})$ (\ref{ansatzsourc})
impose respective classes of nonholonomic constraints on nontrivial star
distortions of the canonical Ricci s-tensors depending, in general, on all
phase space coordinates but parameterized with generating sources defined
below as in formulas (\ref{cannonsymparamc2hv}), when $\
_{s}^{\shortparallel }\mathcal{K}=\left[\ _{1}^{\shortparallel }\mathcal{K}%
(x^{k_{1}}),\ _{2}^{\shortparallel }\mathcal{K}(x^{k_{2}})\right] ,$ for $%
s=1,2,$ .

We consider a nontrivial dependence on momentum coordinates for the $s=1$
and $2$ components of a s-metric $\mathbf{g}_{j_{2}k_{2}}(~_{s}x,\
_{s}^{\shortparallel }p)$ (in particular, one can be fixed configurations
with $\mathbf{g}_{j_{2}k_{2}}(\ _{s}x)$) for an arbitrary value of a
c-component $\ ^{\shortparallel }\mathbf{g}^{a_{s}b_{s}}(\ _{s}x,\
_{s}^{\shortparallel }p)$ in (\ref{auxm61}). Nonholonomic dyadic s-frame (%
\ref{nadapbdsc}) and (\ref{nadapbdss}) can be prescribed with "small"
canonical distortions (\ref{candistrnas}) and (\ref{canhcs}) (we can
consider general frame and coordinate transforms after the computations have
been performed for such special nonholonomic configruations) when%
\begin{eqnarray*}
\ _{[00]}^{\shortparallel }\widehat{\mathbf{\Gamma }}_{\ j_{2}k_{2}}^{i_{2}}
&=&\ ^{\shortparallel }\widehat{\mathbf{\Gamma }}_{\ j_{2}k_{2}}^{i_{2}}=(\
\widehat{L}_{j_{1}k_{1}}^{i_{1}},\widehat{L}_{b_{2}\ k_{1}}^{a_{2}},\widehat{%
C}_{\ j_{1}c_{2}}^{i_{1}\ },\widehat{C}_{b_{2}c_{2}}^{a_{2}}); \\
\ \ _{[01]}^{\shortparallel }\widehat{\mathbf{\Gamma }}_{\
j_{2}k_{2}}^{i_{2}} &=&0;
\end{eqnarray*}%
\begin{eqnarray*}
\ _{[10]}^{\shortparallel }\widehat{\mathbf{\Gamma }}_{\ j_{2}k_{2}}^{i_{2}}
&=&-i\kappa \overline{\mathcal{R}}_{\quad \quad \quad }^{n+o_{2}\
a_{s}~n+l_{2}}\ p_{a_{s}}\ ^{\shortparallel }\mathbf{g}^{i_{2}q_{2}}(\
^{\shortparallel }\mathbf{e}_{l_{2}}\ ^{\shortparallel }\mathbf{g}%
_{q_{2}r_{2}})(\ ^{\shortparallel }\mathbf{e}_{o_{2}}\ ^{\shortparallel }%
\widehat{\mathbf{\Gamma }}_{\ j_{2}k_{2}}^{r_{2}}), \\
\ _{[10]}^{\shortparallel }\widehat{\mathbf{\Gamma }}_{\ jk}^{a_{s}} &=&-%
\frac{i\kappa }{2}\overline{\mathcal{R}}_{\quad \quad \quad }^{n+i_{2}\
a_{s}~n+q_{2}}\ ^{\shortparallel }\mathbf{g}_{l_{2}q_{2}}\ ^{\shortparallel }%
\mathbf{e}_{i_{2}}\ ^{\shortparallel }\widehat{\mathbf{\Gamma }}_{\
j_{2}k_{2}}^{l_{2}}, \\
\ _{[10]}^{\shortparallel }\widehat{\mathbf{\Gamma }}_{\quad
k_{2}}^{i_{2}a_{s}} &=&\frac{i\kappa }{2}\overline{\mathcal{R}}_{\quad \quad
\quad }^{n+o_{2}\ a_{s}~n+l_{2}}\ \ ^{\shortparallel }\mathbf{g}%
^{q_{2}i_{2}}\ ^{\shortparallel }\mathbf{e}_{o_{2}}(\ ^{\shortparallel }%
\mathbf{g}_{q_{2}r_{2}}\ ^{\shortparallel }\widehat{\mathbf{\Gamma }}_{\
l_{2}k_{2}}^{r_{2}}), \\
\ _{[10]}^{\shortparallel }\widehat{\mathbf{\Gamma }}_{~k_{2}}^{i_{2}~a_{s}}
&=&\frac{i\kappa }{2}\overline{\mathcal{R}}_{\quad \quad \quad }^{n+o_{2}\
a_{s}~n+l_{s}}\ \ ^{\shortparallel }\mathbf{g}^{i_{2}q_{2}}\
^{\shortparallel }\mathbf{e}_{o_{2}}(\ ^{\shortparallel }\mathbf{g}%
_{q_{2}r_{2}}\ ^{\shortparallel }\widehat{\mathbf{\Gamma }}_{\
l_{2}k_{2}}^{r_{2}});
\end{eqnarray*}%
\begin{eqnarray*}
\ _{[11]}^{\shortparallel }\widehat{\mathbf{\Gamma }}_{\ j_{2}k_{2}}^{i_{2}}
&=&\frac{\hbar \kappa }{2}\overline{\mathcal{R}}_{\quad \quad \quad
}^{n+o_{2}\ n+q_{2}\ ~n+l_{2}}\ (\ ^{\shortparallel }\mathbf{e}_{o_{2}}\
^{\shortparallel }\mathbf{g}^{i_{2}s_{2}})(\ ^{\shortparallel }\mathbf{e}%
_{q_{2}}\ ^{\shortparallel }\mathbf{g}_{s_{2}r_{2}})(\ ^{\shortparallel }%
\mathbf{e}_{l_{2}}\ ^{\shortparallel }\widehat{\mathbf{\Gamma }}_{\
j_{2}k_{2}}^{r_{2}}), \\
\mbox{ where, for instance, }\ ^{\shortparallel }\mathbf{e}_{l_{2}} &=&\{%
\mathbf{e}_{l_{1}},e^{a_{2}}\},\ ^{\shortparallel }\mathbf{g}%
_{q_{2}r_{2}}=\{g_{i_{1}q_{1}},g^{a_{2}b_{2}}\},\ ^{\shortparallel }\mathbf{g%
}^{i_{2}q_{2}}=\{g^{i_{1}q_{1}},g_{a_{2}b_{2}}\},\
p_{a_{s}}=(p_{a_{3}},p_{a_{4}}).
\end{eqnarray*}%
We have to consider respective formulas (\ref{canhcs}) and (\ref%
{0canconnonas}) for the canonical s-connections, their star deformations and
parametric decompositions. Such s-coefficients result in nontrivial
distortions from the LC-connection. This mean that for general nonholonomic
distributions $\ _{[01]}^{\shortparallel }\widehat{\mathbf{\Gamma }}_{\
j_{s}k_{s}}^{i_{s}}$ (\ref{aux51}) and $\ _{[01]}^{\shortparallel }\widehat{%
\mathbf{R}}ic_{\beta _{s}\gamma _{s}}^{\star }$ (\ref{ric51}) being
proportional to $\hbar $ may be not zero even respective values for the
LC-connection are zero in certain coordinate frames, see formulas
(5.82)-(5.85) in \cite{aschieri17}.

Let us consider how the nonholonomic dyadic splitting on h1-v2 components
determine possible nontrivial spacetime R-flux contributions. We have such
nontrivial imaginary $[10]$ terms of the canonical Ricci s-tensor see in
explicit computations for formulas (\ref{ric52}), which for a corresponding
s-adapting structure with "small" canonical distortions is computed an
imaginary term%
\begin{eqnarray*}
\ _{[10]}^{\shortparallel }\widehat{\mathbf{R}}ic_{j_{2}k_{2}}^{\star }
&=&i\kappa \overline{\mathcal{R}}_{\quad \quad \quad }^{n+o\ a_{s}~n+l}\
p_{a_{s}}\{-\ ^{\shortparallel }\mathbf{e}_{i_{2}}[~^{\shortparallel }%
\mathbf{g}^{i_{2}q_{2}}(\ ^{\shortparallel }\mathbf{e}_{l_{2}}\
^{\shortparallel }\mathbf{g}_{q_{2}r_{2}})(\ ^{\shortparallel }\mathbf{e}%
_{o_{2}}\ ^{\shortparallel }\widehat{\mathbf{\Gamma }}_{\
j_{2}k_{2}}^{r_{2}})] \\
&&+\ ^{\shortparallel }\mathbf{e}_{k_{2}}[~^{\shortparallel }\mathbf{g}%
^{i_{2}q_{2}}(\ ^{\shortparallel }\mathbf{e}_{l_{2}}\ ^{\shortparallel }%
\mathbf{g}_{q_{2}r_{2}})(\ ^{\shortparallel }\mathbf{e}_{o_{2}}\
^{\shortparallel }\widehat{\mathbf{\Gamma }}_{\ j_{2}i_{2}}^{r_{2}})] \\
&&-\ ^{\shortparallel }\widehat{\mathbf{\Gamma }}_{\
j_{2}k_{2}}^{m_{2}}~^{\shortparallel }\mathbf{g}^{i_{2}q_{2}}(\
^{\shortparallel }\mathbf{e}_{l_{2}}\ ^{\shortparallel }\mathbf{g}%
_{q_{2}r_{2}})(\ ^{\shortparallel }\mathbf{e}_{o_{2}}\ ^{\shortparallel }%
\widehat{\mathbf{\Gamma }}_{m_{2}i_{2}}^{r_{2}})-\ ^{\shortparallel }%
\widehat{\mathbf{\Gamma }}_{\ i_{2}m_{2}}^{i_{2}}~^{\shortparallel }\mathbf{g%
}^{m_{2}q_{2}}(\ ^{\shortparallel }\mathbf{e}_{l_{2}}\ ^{\shortparallel }%
\mathbf{g}_{q_{2}r_{2}})(\ ^{\shortparallel }\mathbf{e}_{o_{2}}\
^{\shortparallel }\widehat{\mathbf{\Gamma }}_{j_{2}k_{2}}^{r_{2}}) \\
&&+\ ^{\shortparallel }\widehat{\mathbf{\Gamma }}_{\
j_{2}i_{2}}^{m_{2}}~^{\shortparallel }\mathbf{g}^{i_{2}q_{2}}(\
^{\shortparallel }\mathbf{e}_{l_{2}}\ ^{\shortparallel }\mathbf{g}%
_{q_{2}r_{2}})(\ ^{\shortparallel }\mathbf{e}_{o_{2}}\ ^{\shortparallel }%
\widehat{\mathbf{\Gamma }}_{mk}^{r_{2}})+\ ^{\shortparallel }\widehat{%
\mathbf{\Gamma }}_{\ m_{2}k_{2}}^{i_{2}}~^{\shortparallel }\mathbf{g}%
^{m_{2}q_{2}}(\ ^{\shortparallel }\mathbf{e}_{l_{2}}\ ^{\shortparallel }%
\mathbf{g}_{q_{2}r_{2}})(\ ^{\shortparallel }\mathbf{e}_{o_{2}}\
^{\shortparallel }\widehat{\mathbf{\Gamma }}_{j_{2}i_{2}}^{r_{2}}) \\
&&+(~^{\shortparallel }\mathbf{e}_{l_{2}}\ ^{\shortparallel }\widehat{%
\mathbf{\Gamma }}_{\ q_{2}i_{2}}^{i_{2}})(\ ^{\shortparallel }\mathbf{g}%
_{q_{2}r_{2}}\ ^{\shortparallel }\widehat{\mathbf{\Gamma }}_{\
l_{2}k_{2}}^{r_{2}})-(~^{\shortparallel }\mathbf{e}_{l_{2}}\
^{\shortparallel }\widehat{\mathbf{\Gamma }}_{\ q_{2}k_{2}}^{i_{2}})(\
^{\shortparallel }\mathbf{g}_{q_{2}r_{2}}\ ^{\shortparallel }\widehat{%
\mathbf{\Gamma }}_{\ l_{2}i_{2}}^{r_{2}})\}.
\end{eqnarray*}%
Nevertheless, there are also real nontrivial R-flux contributions:
\begin{eqnarray}
\ _{[11]}^{\shortparallel }\widehat{\mathbf{R}}ic_{j_{2}k_{2}}^{\star } &=&
\frac{\hbar \kappa }{2}\overline{\mathcal{R}}_{\quad \quad \quad }^{n+o_{2}\
n+q_{2}\ ~n+l_{2}}\{\ ^{\shortparallel }\mathbf{e}_{i_{2}}[(\
^{\shortparallel }\mathbf{e}_{o_{2}}~^{\shortparallel }\mathbf{g}%
^{i_{2}m_{2}})(\ ^{\shortparallel }\mathbf{e}_{q_{2}}\ ^{\shortparallel }%
\mathbf{g}_{m_{2}r_{2}})(\ ^{\shortparallel }\mathbf{e}_{l_{2}}\
^{\shortparallel }\widehat{\mathbf{\Gamma }}_{\ j_{2}k_{2}}^{r_{2}})]  \notag
\\
&&-\ ^{\shortparallel }\mathbf{e}_{k_{2}}[(\ ^{\shortparallel }\mathbf{e}%
_{o_{2}}~^{\shortparallel }\mathbf{g}^{i_{2}m_{2}})(\ ^{\shortparallel }%
\mathbf{e}_{q_{2}}\ ^{\shortparallel }\mathbf{g}_{m_{2}r_{2}})(\
^{\shortparallel }\mathbf{e}_{l_{2}}\ ^{\shortparallel }\widehat{\mathbf{%
\Gamma }}_{\ j_{2}i_{2}}^{r_{2}})]  \label{realrflux} \\
&&+(\ ^{\shortparallel }\mathbf{e}_{l_{2}}~^{\shortparallel }\mathbf{g}%
_{q_{2}r_{2}})[\ ^{\shortparallel }\mathbf{e}_{o_{2}}(\ ~^{\shortparallel }%
\mathbf{g}^{i_{2}q_{2}}\ ^{\shortparallel }\widehat{\mathbf{\Gamma }}_{\
i_{2}k_{2}}^{m_{2}})(\ ^{\shortparallel }\mathbf{e}_{q_{2}}\
^{\shortparallel }\widehat{\mathbf{\Gamma }}_{j_{2}m_{2}}^{r_{2}})-\
^{\shortparallel }\mathbf{e}_{o_{2}}(\ ~^{\shortparallel }\mathbf{g}%
^{i_{2}q_{2}}\ ^{\shortparallel }\widehat{\mathbf{\Gamma }}_{\
i_{2}m_{2}}^{m_{2}})(\ ^{\shortparallel }\mathbf{e}_{q_{2}}\
^{\shortparallel }\widehat{\mathbf{\Gamma }}_{j_{2}k_{2}}^{r_{2}})  \notag \\
&&+\left( \ ^{\shortparallel }\widehat{\mathbf{\Gamma }}_{\
j_{2}i_{2}}^{m_{2}}\ ^{\shortparallel }\mathbf{e}_{o_{2}}(\
~^{\shortparallel }\mathbf{g}^{i_{2}r_{2}})-\ ^{\shortparallel }\mathbf{e}%
_{o_{2}}(\ ^{\shortparallel }\widehat{\mathbf{\Gamma }}_{\
j_{2}i_{2}}^{m_{2}})\ ~^{\shortparallel }\mathbf{g}^{i_{2}r_{2}}\right) (\
^{\shortparallel }\mathbf{e}_{q_{2}}\ ^{\shortparallel }\widehat{\mathbf{%
\Gamma }}_{\ m_{2}k_{2}}^{r_{2}})  \notag \\
&&-\left( \ ^{\shortparallel }\widehat{\mathbf{\Gamma }}_{\
j_{2}k_{2}}^{m_{2}}\ ^{\shortparallel }\mathbf{e}_{o_{2}}(\
~^{\shortparallel }\mathbf{g}^{i_{2}r_{2}})-\ ^{\shortparallel }\mathbf{e}%
_{o_{2}}(\ ^{\shortparallel }\widehat{\mathbf{\Gamma }}_{\
j_{2}k_{2}}^{m_{2}})\ ~^{\shortparallel }\mathbf{g}^{i_{2}r_{2}}\right) (\
^{\shortparallel }\mathbf{e}_{q_{2}}\ ^{\shortparallel }\widehat{\mathbf{%
\Gamma }}_{\ m_{2}i_{2}}^{r_{2}})]\}.  \notag
\end{eqnarray}

Using the h1-v2 coefficients, we can write the respective s-coefficients of
nonassociative vacuum Einstein equations (\ref{purestarvacuum}) can be
written in a form distinguishing real contributions from string gravity with
R-fluxes,
\begin{equation}
\ \ ^{\shortparallel }\widehat{R}_{\ \star j_{2}k_{2}}(~_{s}x,\
_{s}^{\shortparallel }p)=\ \ ^{\shortparallel }\widehat{R}_{\
j_{2}k_{2}}(~_{s}x,\ _{s}^{\shortparallel }p)+\mathbf{\mathbf{\mathbf{%
\mathbf{\ \ }}}}_{[11]}^{\shortparallel }\widehat{\mathbf{\mathbf{\mathbf{%
\mathbf{R}}}}}ic_{j_{2}k_{2}}^{\star }(~_{s}x,\ _{s}^{\shortparallel }p)=0.
\label{aux61a}
\end{equation}%
Imposing additional nonholonomic constraints $\ _{s}^{\shortparallel }%
\widehat{\mathbf{D}}_{\mid \ _{s}^{\shortparallel }\widehat{\mathbf{T}}%
=0}^{\star }=\ ^{\shortparallel }\nabla ^{\star }$ (\ref{lccondnonass}) for $%
s=1,2,$ we extract from (\ref{aux61a}) horizontal LC-configurations of type (%
\ref{veinstlc}),%
\begin{equation*}
\ \ _{\nabla }^{\shortparallel }\mathcal{\Re }ic_{h_{2}j_{2}}^{\star
}(~_{s}x,\ _{s}^{\shortparallel }p)=\ \ _{\nabla }^{\shortparallel }R_{\
h_{2}j_{2}}(~_{s}x,\ _{s}^{\shortparallel }p)+\mathbf{\mathbf{\mathbf{%
\mathbf{\ \ }}}}_{[11]}^{\shortparallel \nabla }R_{h_{2}j_{2}}^{\star
}(~_{s}x,\ _{s}^{\shortparallel }p)=0.
\end{equation*}%
If we work only with holonomic structures when $\mathbf{g}_{j_{2}k_{2}}(\
_{s}x,\ _{s}^{\shortparallel }p)=\mathbf{g}_{j_{2}k_{2}}(\ _{s}x),$ we
generate on spacetime holonomic configurations of type (\ref{veinsthlc}),
\begin{equation*}
Ric_{h_{2}j_{2}}^{\circ }(\ _{s}x)=\ \ _{\nabla }Ric_{h_{2}j_{2}}(~_{s}x)+\
_{[11]}^{\shortparallel \nabla }R_{h_{2}j_{2}}^{\star }(~_{s}x)=0.
\end{equation*}%
In coordinate frames, these equations are equivalent to (5.90) from \cite%
{aschieri17}, when R-flux corrections to Ricci tensors determined on
spacetime by $\ ^{\shortparallel }\nabla $ are independent on $\hbar $. This
property holds true for some special classes of metrics $\mathbf{g}%
_{j_{2}k_{2}}(~_{s}x).$ Such a system of nonlinear PDEs can not decoupled
and integrated in a general form when the coefficients of the s-metric are
generic off-diagonal and depend on all spacetime coordinates with possible
one Killing symmetry.

Applying the AFDM, we can decouple and integrate such equations in arbitrary
form for various classes of s-metrics $\mathbf{g}_{j_{2}k_{2}}(~_{s}x,\
_{s}^{\shortparallel }p)\rightarrow \mathbf{g}_{j_{2}k_{2}}(~_{s}x)$ as
solutions of (\ref{aux61a}). considered as a system of (\ref%
{cannonsymparamc2}) with a nonholonomically constrained effective s-source $%
\ ^{\shortparallel }\mathbf{K}_{\ \beta _{s}}^{\alpha _{s}}$ (\ref%
{ansatzsourc}). We can search for nontrivial quasi-stationary configurations
with
\begin{eqnarray}
\ ^{\shortparallel }\mathbf{K}_{\ \beta _{s}}^{\alpha _{s}} &=&\{\
^{\shortparallel }\mathcal{K}_{\ j_{1}}^{i_{1}}=[\ \ _{1}^{\shortparallel
}\Upsilon (x^{k_{1}})+\ _{1}^{\shortparallel }\mathbf{K}(\kappa
,x^{k_{1}})]\delta _{j_{1}}^{i_{1}},\{\ ^{\shortparallel }\mathcal{K}_{\
j_{2}}^{i_{2}}=[\ \ _{2}^{\shortparallel }\Upsilon (x^{k_{1}},x^{3})+\
_{2}^{\shortparallel }\mathbf{K}(\kappa ,x^{k_{1}},x^{3})]\delta
_{b_{2}}^{a_{2}},  \label{ansatzsourchv} \\
&&\ ^{\shortparallel }\mathcal{K}_{\ a_{3}}^{b_{3}}=[\ \
_{3}^{\shortparallel }\Upsilon (x^{k_{2}},\ ^{\shortparallel }p_{6})+\
_{3}^{\shortparallel }\mathbf{K}(x^{k_{2}},\ ^{\shortparallel }p_{6})]\
\delta _{a_{3}}^{b_{3}},\ ^{\shortparallel }\mathcal{K}_{\ a_{4}}^{b_{4}}=[\
\ _{4}^{\shortparallel }\Upsilon (x^{k_{3}},\ ^{\shortparallel }p_{8})+\
_{4}^{\shortparallel }\mathbf{K}(x^{k_{3}},\ ^{\shortparallel }p_{8})]\delta
_{a_{4}}^{b_{4}}\},  \notag \\
&& \mbox{ where } \ ^{\shortparallel }\mathbf{K}_{\ j_{2}k_{2}}= - \
_{[11]}^{\shortparallel} \widehat{\mathbf{R}}ic_{j_{2}k_{2}}^{\star
}(x^{k_{1}},x^{3})\mbox{ as in }(\ref{realrflux}),  \notag \\
&& \mbox{ for }\mathbf{g}_{j_{2}k_{2}} =
\{g_{1}(x^{k_{1}}),g_{2}(x^{k_{1}}),g_{3}(x^{k_{1}},x^{3}),g_{4}(x^{k_{1}},x^{3})\}.
\notag
\end{eqnarray}%
For such h1-v2 configurations, we can prescribe the effective sources (\ref%
{ansatzsourchv}) satisfying conditions of type (\ref{nonhconstr1})
\begin{equation}
\mathbf{\mathbf{\mathbf{\mathbf{\ ^{\shortparallel }}}}}\lambda +\frac{1}{2}%
\mathbf{\mathbf{\mathbf{\mathbf{\ ^{\shortparallel }}}}}\widehat{\mathbf{%
\mathbf{\mathbf{\mathbf{R}}}}}sc^{\star }=\mathbf{\mathbf{\mathbf{\mathbf{\
^{\shortparallel }}}}}\lambda +\ ^{\shortparallel }\mathbf{K}_{\ \alpha
_{s}}^{\alpha _{s}}[\hbar \kappa \overline{\mathcal{R}}]~=0,
\label{nonhconstr1a}
\end{equation}%
where the contributions from shells $s=3,4$ are encoded into an effective
cosmological constant $\ ^{\shortparallel} \lambda ,$ but there are also
nontrivial R-flux effective sources on shells $s=1,2.$ We state such
dependencies as functionals [...].

The system of real spacetime s-adapted projections of parametric
nonassociative gravitational equations (\ref{cannonsymparamc2}) with sources
(\ref{ansatzsourchv}) transforms into%
\begin{eqnarray}
\ ^{\shortparallel }\widehat{\mathbf{R}}_{\ i_{2}j_{2}}~ &=&\
^{\shortparallel }\mathbf{K}_{_{i_{2}j_{2}}}[\hbar \kappa \overline{\mathcal{%
R}}]~,\mbox{ where }  \label{cannonsymparamc2hv} \\
&& \ ^{\shortparallel }\mathbf{K}_{~i_{2}}^{j_{2}} = [~_{1}^{\shortparallel }%
\mathcal{K}(\kappa ,x^{k_{1}})\delta _{i_{1}}^{j_{1}},~_{2}^{\shortparallel }%
\mathcal{K}(\kappa ,x^{k_{1}},x^{3})\delta _{b_{2}}^{a_{2}}]  \notag
\end{eqnarray}%
The AFCDM method for constructing exact and parametric solutions of
dimensions 2+2 and with a number of examples of (off-) diagonal black hole
and cosmological solutions is elaborated in \cite{bubuianu17}, see
references therein. Prescribing any generating sources $\
_{s}^{\shortparallel }\mathcal{K}=\left[\ _{1}^{\shortparallel }\mathcal{K}%
,\ _{2}^{\shortparallel }\mathcal{K}\right],$ for $s=1,2,$ in (\ref%
{cannonsymparamc2hv}), we can decouple and solve in certain general forms
such systems of nonlinear PDEs.

So, we conclude that for certain classes on nonholonomic constraints on
dyadic structure, the nonassociative vacuum gravitational equations on phase
spaces can be transformed into self-consistent h1-v2 Einstein type equations
for the canonical s-connection and with effective source from R-fluxes and
nontrivial cosmological constant on certain shells. Any s-metric $\mathbf{g}%
_{j_{2}k_{2}}(~_{s}x)$ solving (\ref{cannonsymparamc2hv}) on a spacetime
with R-flux effective sources defines on the total phase space respective
nonsymmetric (\ref{nsm1}) and symmetric (\ref{ssm1}) star deformed s-metric
structures. The s-adapted geometric objects on shells $s=1,2$ can be
subjected to general s-frame spacetime transforms when, for instance, a
parametric nonassociative and general non-diagonal source $\
^{\shortparallel }\widehat{\Upsilon }_{i_{2}^{\prime }j_{2}^{\prime }}= e_{\
i_{2}^{\prime }}^{i_{2}}e_{\ j_{2}^{\prime }}^{j_{2}}\ \ ^{\shortparallel }%
\mathbf{K}_{i_{2}j_{2}}$ is defined.

\section{Decoupling \& integrability of nonassociative vacuum gravitational
equations}

\label{sec5} In this section we extend and apply the anholonomic frame and
connection deformation method, AFCDM, for constructing exact and parametric
solutions in nonassociative gravity theories. In explicit form, our goal is
to prove a general decoupling and integrability property of star deformed
vacuum gravitational equations with parametric decomposition to systems of
nonlinear PDEs on 8-d phase spaces, see (\ref{cannonsymparamc2a}), and/or
4-d spacetime systems (\ref{cannonsymparamc2hv}), with R-flux corrections.
In a rigorous nonholonomic dyadic geometric form, the AFCDM is elaborated
for ral 8-d phase spaces modeled (co) tangent Lorentz bundles in Ref. \cite%
{bubuianu20} (see section 4 and respective appendices). It allows a trivial
extension of such geometric methods and generated solutions for momentum
like coordinates multiplied to the complex unity $i.$ We shall omit proofs
and cumbersome s-adapted computations because they are similar to those
presented in a series of theorems of the just cited work but for a different
type of effective sources, which in the nonassociative case contain
contributions from R-fluxes and star nonholonomic products.

\subsection{Off-diagonal ansatz for phase space parametric s-metrics}

We consider dyadic decompositions when the total phase space s-metric
structure may describe extensions of the BH stationary solutions in
associative and commutative gravity to s-metrics with Killing symmetry on $%
\partial _{4}=\partial /\partial y^{4}=\partial _{t}=\partial /\partial t$
on the shell $s=2;$ Killing symmetry $\ ^{\shortparallel}\partial
^{5}=\partial /\partial \ ^{\shortparallel }p_{5}$ on $s=3,$ and posses a
general and/or shell Killing symmetry on $\ ^{\shortparallel }\partial
^{7}=\partial /\partial \ ^{\shortparallel }p_{7}$ on $\ _{s}\mathbf{T}
_{\shortparallel }^{\ast }\mathbf{V.}$ The existence of such symmetries if
very important for performing "simplified" proofs of decoupling property of
modified Einstein equations in a number of (non) associative / commutative
modified gravity theories. Such solutions allow us, for instance, to
construct various extensions of the Schwarzschild and Kerr metrics in GR
(with a prime diagonal metric $%
g_{ij}(r)=diag[...,...,(1-r/r_{g})^{-1},(1-r/r_{g})].$ In this work, we
shall use standard spacetime spherical coordinates with $\
^{\shortmid}u^{1}= x^{1}=r,\ ^{\shortmid }u^{2}=x^{2}=\theta ,\
^{\shortmid}u^{3}=y^{3}= \varphi ,\ ^{\shortmid }u^{4}=y^{4}=t)$ when the
phase space solutions depend in explicit form on the energy type variable $\
^{\shortparallel }p_{8}=\ ^{\shortparallel }E$ and $\
^{\shortparallel}p_{6}, $ (also on $\ ^{\shortparallel }p_{5}$ via
N--connection coefficients) but not on $\ ^{\shortparallel }p_{7}.$ It is
supposed that the coefficients of s-metrics are such way parameterized that
possible terms with complex unity before coordinates are such way multiplied
to some coefficients with complex unity when the resulting s-adapted
coefficients are real. In certain cases, we shall use real phase coordinates
like $\ ^{\shortmid }p_{8}=E$ and $\ ^{\shortmid }p_{6},$ etc. We shall
construct and study explicit examples of nonassociative of BH hole
configurations gravity in our partner works. Such solutions are similar to
those provided in \cite{bubuianu19}.

\subsubsection{Quasi-stationary ansatz for s-metrics and R-flux sources}

For an associative and commutative s-metric $\ _{s}^{\shortparallel }\mathbf{%
g}= \ ^{\shortparallel }\mathbf{g}_{\alpha _{s}\beta _{s}}(\ _{s}x,\
_{s}^{\shortparallel }p) \ ^{\shortparallel }\mathbf{e}^{\alpha _{s}}
\otimes \ ^{\shortparallel } \mathbf{e}^{\beta _{s}}$ (\ref{commetr}), we
consider a quasi-stationary ansatz for a linear quadratic element with
additional parametric dependence on $\hbar $ and $\kappa $ of coefficients
(which will be computed in next sections),
\begin{eqnarray}
&&d\ ^{\shortparallel }\widehat{s}^{2}=\widehat{g}_{1}(r,\theta )dr^{2}+%
\widehat{g}_{2}(r,\theta )d\theta ^{2}+\widehat{g}_{3}(r,\theta ,\varphi
)\delta \varphi ^{2}+\widehat{g}_{4}(r,\theta ,\varphi )\delta t^{2}+\
^{\shortparallel }\widehat{g}^{5}(r,\theta ,\varphi ,t,\ ^{\shortparallel
}p_{6})(\ ^{\shortparallel }\widehat{\mathbf{e}}_{5})^{2}+  \notag \\
&&\ ^{\shortparallel }\widehat{g}^{6}(r,\theta ,\varphi ,t,\
^{\shortparallel }p_{6})(\ ^{\shortparallel }\widehat{\mathbf{e}}_{6})+\
^{\shortparallel }\widehat{g}^{7}(r,\theta ,\varphi ,t,\ ^{\shortparallel
}p_{6},\ ^{\shortparallel }E)(\ ^{\shortparallel }\widehat{\mathbf{e}}%
_{7})^{2}+\ ^{\shortparallel }\widehat{g}^{8}(r,\theta ,\varphi ,t,\
^{\shortparallel }p_{6},\ ^{\shortparallel }E)(\ ^{\shortparallel }\widehat{%
\mathbf{e}}_{8})^{2},  \label{ansatz1na}
\end{eqnarray}%
where the s-adapted coefficients for the phase metric are chosen in the form
\begin{eqnarray*}
\widehat{g}_{i_{1}j_{i}} &=&diag[\widehat{g}_{i_{1}}(x^{k_{1}})],\mbox{ for }%
i_{1},j_{1}=1,2\mbox{ and }x^{k_{1}}=(x^{1}=r,x^{2}=\theta ); \\
\widehat{g}_{a_{2}b_{2}} &=&diag[\widehat{g}_{a_{2}}(x^{k_{1}},y^{3})],%
\mbox{ for }a_{2},b_{2}=3,4\mbox{ and }y^{3}=x^{3}=\varphi ,y^{4}=x^{4}=t; \\
\ \ ^{\shortparallel }\widehat{g}^{a_{3}b_{3}} &=&diag[\ \ ^{\shortparallel }%
\widehat{g}^{a_{3}}(x^{k_{1}},y^{a_{2}},\ ^{\shortparallel }p_{6})],%
\mbox{
for }a_{3},b_{3}=5,6\mbox{ and
}\ \ ^{\shortparallel }u^{5}=\ ^{\shortparallel }p_{5},\ \ ^{\shortparallel
}u^{6}=\ ^{\shortparallel }p_{6}; \\
\ \ ^{\shortparallel }\widehat{g}^{a_{4}b_{4}} &=&diag[\ \ ^{\shortparallel }%
\widehat{g}^{a_{4}}(x^{k_{1}},y^{a_{2}},\ ^{\shortparallel }p_{a_{3}},\
^{\shortparallel }E)],\mbox{ for }a_{4},b_{4}=7,8\mbox{ and }\ \
^{\shortparallel }u^{7}=\ ^{\shortparallel }p_{7},\ \ ^{\shortparallel
}u^{8}=\ ^{\shortparallel }p_{8}=\ ^{\shortparallel }E.
\end{eqnarray*}

The nonholonomic s-adapted co-bases $\ ^{\shortparallel }\widehat{\mathbf{e}}%
^{\alpha _{s}}= (\ ^{\shortparallel }\mathbf{e}^{i_{s}}=dx^{i_{s}},\
^{\shortparallel}\widehat{\mathbf{e}}_{a_{s}} = d\
^{\shortparallel}p_{a_{s}}+\ ^{\shortparallel }\widehat{N}_{\
i_{s}a_{s}}dx^{i_{s}})$ (\ref{nadapbdss}), with
\begin{eqnarray*}
e^{1} &=&dr,e^{2}=d\theta ,\widehat{\mathbf{e}}^{3}=\delta \varphi =d\varphi
+w_{i_{1}}dx^{i_{1}},\widehat{\mathbf{e}}^{4}=\delta
t=dt+n_{i_{1}}dx^{i_{1}}, \\
\ ^{\shortparallel }\widehat{\mathbf{e}}_{5} &=&d\ ^{\shortparallel
}p_{5}+~^{\shortparallel }n_{j_{2}}dx^{j_{2}},\ ^{\shortparallel }\widehat{%
\mathbf{e}}_{5}=d\ ^{\shortparallel
}p_{5}+~^{\shortparallel}n_{j_{2}}dx^{j_{2}},
\end{eqnarray*}
are determined by N-connection s-adapted coefficients
\begin{eqnarray*}
\ \ ^{\shortparallel }\widehat{N}_{j_{1}}^{3} &=&\ _{2}w_{j_{1}}=\
w_{j_{1}}(r,\theta ,\varphi ),\ \ ^{\shortparallel }N_{j_{1}}^{4}=\
_{2}n_{j_{1}}=\ n_{j_{1}}(r,\theta ,\varphi ); \\
\ \ ^{\shortparallel }\widehat{N}_{j_{2}\ 5} &=&\ _{3}^{\shortparallel
}n_{j_{2}}(x^{k_{2}},\ ^{\shortparallel }p_{6})=~^{\shortparallel
}n_{j_{2}}(r,\theta ,\varphi ,t,\ ^{\shortparallel }p_{6}), \\
\ ^{\shortparallel }\widehat{N}_{j_{2}\ 6} &=&\ _{3}^{\shortparallel
}w_{j_{2}}(x^{k_{2}},\ ^{\shortparallel }p_{6})=\ ^{\shortparallel
}w_{j_{2}}(r,\theta ,\varphi ,t,~^{\shortparallel }p_{6})\mbox{ for }%
j_{2}=1,2,3,4; \\
\ \ ^{\shortparallel }\widehat{N}_{j_{3}7} &=&\ _{4}^{\shortparallel
}n_{j_{3}}(x^{k_{3}},\ ^{\shortparallel }p_{8})=\ ^{\shortparallel
}n_{j_{3}}(r,\theta ,\varphi ,t,\ ^{\shortparallel }p_{6},\
^{\shortparallel}E), \\
\ \ ^{\shortparallel }\widehat{N}_{j_{3}8} &=&\ _{4}^{\shortparallel
}w_{j_{3}}(x^{k_{3}},\ ^{\shortparallel }p_{6},\ ^{\shortparallel }p_{8})=\
\ ^{\shortparallel }w_{j_{3}}(r,\theta ,\varphi ,t,\ ^{\shortparallel
}p_{6},\ ^{\shortparallel }E)\mbox{ for }j_{3}=1,2,3,4,5,6.
\end{eqnarray*}%
In above formulas, we put "hat" labels on s-metric and N-connection
coefficients for a stationary ansatz which will be used for generating exact
and parametric solutions of some modified (non) associative / commutative
Einstein equations for a respective canonical s-connection $\
_{s}^{\shortparallel }\widehat{\mathbf{D}}$ (\ref{twocon}) with s-adapted
coefficients (\ref{canhcs}).

In this paper, the term "stationary" is used for metrics on phase space and
base spacetime manifolds, which in certain special coordinates do not depend
on respective time like coordinates but contain some off-diagonal terms. In
GR, such terms describe, for instance, rotating Kerr BH, see \cite%
{hawking73,misner,wald82,kramer03} but in our approach we work with general
nonholonomic configurations (not only with coordinate rotating frames). The
h1-v2 part (i.e. the first 4 terms for a Lorentz manifold base) in (\ref%
{ansatz1na}) is of stationary type. Nevertheless, the cofiber part (next 5-8
terms) of that ansatz may depend on a time like coordinate "t". This is
because various modified dispersion relations and R-flux sources induce, in
general, a nonholonomic cofiber dynamics with dependence on "t", and on "E"
(in literature, it is used the term "rainbow" metrics for a dependence on a
E-parameter). We use the term "quasi-stationary" for s-metric ansatz because
the associated N-connection s-structure is nonholonomic and depends in local
anisotropic form on phase space coordinates. Such ansatz allows us to extend
the AFCDM to R-flux sources when a corresponding nonholonomic shell dyadic
decompositions allows us to construct exact/parameteric quasi-stationary
solutions as we shall prove in section \ref{ssnonassocvsol}. Here we note
that an ansatz (\ref{ansatz1na}) is stationary if the s-metric and
N-connection coefficients are prescribed in such an adapted form that the
geometric s-objects do not depend explicitly on the time like variable $t.$
In dual form, such ansatz and corresponding geometric methods of
constructing exact solutions can be re-defined when the $t$-dependence is
important for constructing locally anisotropic cosmological solutions, see
details in \cite{bubuianu17}.

We shall be able to generate in explicit form exact off-diagonal solutions
of nonassociative modified vacuum Einstein equations on phase space for
effective sources $\ ^{\shortparallel }\mathbf{K}_{\ \beta _{s}}^{\alpha
_{s}}= \delta _{\ \beta _{s}}^{\alpha _{s}}\ _{s}^{\shortparallel }\mathcal{K%
}$, for $s=1,2,3,4$ in (\ref{cannonsymparamc2a}), or for respective real
R-flux spacetime contributions (\ref{realrflux}) in $\ ^{\shortparallel}
\mathbf{K}_{~j_{2}}^{i_{2}}=\ \delta _{~j_{2}}^{i_{2}}\ _{s}^{\shortparallel}%
\mathcal{K}$ for (\ref{cannonsymparamc2hv}), for $s=1,2.$ Those
parametrization of (effective) nonassociative and associative source were
stated in a quasi-stationary form. In explicit form, $\ _{s}^{\shortparallel
}\mathcal{K}$ should be considered as values determined by certain
phenomenological/ experimental/observational data or computed nonassociative
sectors of string theory.

\subsubsection{Nonassociative and parametric canonical Ricci s-tensors}

In this work, there are used also short notations for partial derivatives
when, for instance, $\partial _{1}q=q^{\bullet },\partial
_{2}q=q^{\prime},\partial _{3}q=\partial _{\varphi }q=q^{\diamond },\
^{\shortparallel}\partial ^{6}q=\partial q/\partial \ ^{\shortparallel
}p^{6},$ and $\ ^{\shortparallel }\partial ^{8}q=\partial q/\partial \
^{\shortparallel}p^{8}= \partial q/\partial \ ^{\shortparallel }E=\
^{\shortparallel}\partial _{E}q=q^{\shortparallel \ast }$ (we write $\
^{\shortparallel}A^{\ast }$ if a letter/formula has already a left label, $\
^{\shortparallel}A).$ If in some formulas, the momentum like coordinates
will be considered without complex unity, we shall write, respectively, $\
^{\shortmid}\partial ^{6}q= \partial q/\partial \ \ ^{\shortmid }p^{6},$ and
$\ ^{\shortmid }\partial ^{8}q=\partial q/\partial \
^{\shortmid}p^{8}=\partial q/\partial E=\ ^{\shortmid }\partial
_{E}q=q^{\shortmid \ast}.$

For quasi-stationary configurations, we can always define s-adapted frame
and coordinate transforms in ansatz (\ref{ansatz1na}) with $g_{4}^{\diamond
}\neq 0,\ ^{\shortparallel }\partial ^{6}\ ^{\shortparallel}g^{5}\neq 0$ and
$(\ ^{\shortparallel }g^{7})^{\shortparallel \ast }\neq 0.$ If such
conditions are note imposed, we can construct more special classes of exact
and parametric solutions but the formulas are more cumbersome and may not
allow an explicit integration of motion equations. A tedious computation of
the coefficients of the canonical s--connection and respective canonical
Ricci s-tensor (\ref{candricci}) of $\ _{s}^{\shortparallel }\widehat{%
\mathbf{D}}$ (\ref{canhcs}) for a s-metric (\ref{ansatz1na}), see details on
proof of Lemma 4.1 in \cite{bubuianu20} allows to write the system of vacuum
s-adapted gravitational equations (\ref{cannonsymparamc2a}) with R-flux
effective sources $\ _{s}^{\shortparallel}\mathcal{K}$ in the form
\begin{eqnarray}
\ \widehat{R}_{1}^{1} &=&\ \widehat{R}_{2}^{2}=\frac{1}{2g_{1}g_{2}}[\frac{%
g_{1}^{\bullet }g_{2}^{\bullet }}{2g_{1}}+\frac{(g_{2}^{\bullet })^{2}}{%
2g_{2}}-g_{2}^{\bullet \bullet }+\frac{g_{1}^{\prime }g_{2}^{\prime }}{2g_{2}%
}+\frac{\left( g_{1}^{\prime }\right) ^{2}}{2g_{1}}-g_{1}^{\prime \prime
}]=-\ \ _{1}^{\shortparallel }\mathcal{K}(\kappa ,r,\theta ),  \notag \\
\ \widehat{R}_{3}^{3} &=&\ \widehat{R}_{4}^{4}=\frac{1}{2g_{3}g_{4}}[\frac{%
\left( g_{4}^{\diamond }\right) ^{2}}{2g_{4}}+\frac{g_{3}^{\diamond
}g_{4}^{\diamond }}{2g_{3}}-g_{4}^{\diamond \diamond }]= -\
_{2}^{\shortparallel }\mathcal{K}(\kappa ,r,\theta ,\varphi ),
\label{riccist2} \\
\ ^{\shortmid }\widehat{R}_{3k_{1}} &=&\frac{\ w_{k_{1}}}{2g_{4}}%
[g_{4}^{\diamond \diamond }-\frac{\left( g_{4}^{\diamond }\right) ^{2}}{%
2g_{4}}-\frac{(g_{3}^{\diamond })(g_{4}^{\diamond })}{2g_{3}}]+\frac{%
g_{4}^{\diamond }}{4g_{4}}(\frac{\partial _{k_{1}}g_{3}}{g_{3}}+\frac{%
\partial _{k_{1}}g_{4}}{g_{4}})-\frac{\partial _{k_{1}}(g_{3}^{\diamond })}{%
2g_{3}}=0;  \notag \\
\ ^{\shortmid }\widehat{R}_{4k_{1}} &=&\frac{g_{4}}{2g_{3}}%
n_{k_{1}}^{\diamond \diamond }+\left( \frac{3}{2}g_{4}^{\diamond }-\frac{%
g_{4}}{g_{3}}g_{3}^{\diamond }\right) \frac{\ n_{k_{1}}^{\diamond }}{2g_{3}}%
=0,  \notag
\end{eqnarray}%
on shells $s=1$ and $s=2,$with $i_{1},k_{1}...=1,2;$
\begin{eqnarray}
\ ^{\shortparallel }\widehat{R}_{5}^{5}&=&\ ^{\shortparallel }\widehat{R}%
_{6}^{6}=-\frac{1}{2\ ^{\shortparallel }g^{5}\ ^{\shortparallel }g^{6}}[%
\frac{(\ ^{\shortparallel }\partial ^{6}(\ ^{\shortparallel }g^{5}))^{2}}{2\
^{\shortparallel }g^{5}}+\frac{(\ ^{\shortparallel }\partial ^{6}(\
^{\shortparallel }g^{5}))(\ ^{\shortparallel }\partial ^{6}(\
^{\shortparallel }g^{6}))}{2\ ^{\shortparallel }g^{6}}-\ ^{\shortparallel
}\partial ^{6}\ ^{\shortparallel }\partial ^{6}(\ ^{\shortparallel }g^{5})]=
\notag \\
&& -\ _{3}^{\shortparallel }\mathcal{K}(\kappa ,r,\theta ,\varphi ,t,\
^{\shortparallel }p_{6}),  \notag \\
\ ^{\shortparallel }\widehat{R}_{5k_{2}} &=&\frac{\ ^{\shortparallel }g^{5}}{%
2\ ^{\shortparallel }g^{6}}\ ^{\shortparallel }\partial ^{6}\
^{\shortparallel }\partial ^{6}\ ^{\shortparallel }n_{k_{2}}+\left( \frac{3}{%
2}\ ^{\shortparallel }\partial ^{6}\ ^{\shortparallel }g^{5}-\frac{\
^{\shortparallel }g^{5}}{\ ^{\shortparallel }g^{6}}\ ^{\shortparallel
}\partial ^{6}\ ^{\shortparallel }g^{6}\right) \frac{\ ^{\shortparallel
}\partial ^{6}\ ^{\shortparallel }n_{k_{2}}}{2\ ^{\shortparallel }g^{6}}=0,
\label{riccist3} \\
\ ^{\shortparallel }\widehat{R}_{6k_{2}} &=&\frac{\ ^{\shortparallel}
w_{k_{2}}}{2\ ^{\shortparallel }g^{5}}[\ ^{\shortparallel }\partial ^{6}\
^{\shortparallel }\partial ^{6}(\ ^{\shortparallel}g^{5})-\frac{\left( \
^{\shortparallel }\partial ^{6}\ ^{\shortparallel}g^{5}\right) ^{2}}{2\
^{\shortparallel }g^{5}}-\frac{(\ ^{\shortparallel}\partial ^{6}(\
^{\shortparallel }g^{5}))(\ ^{\shortparallel }\partial ^{6}(\
^{\shortparallel }g^{6}))}{2\ ^{\shortparallel }g^{6}}]  \notag \\
&& +\frac{\ ^{\shortparallel }\partial ^{6}(\ ^{\shortparallel }g^{5})}{4\
^{\shortparallel }g^{6}}(\frac{\ ^{\shortparallel }\partial _{k_{2}}\
^{\shortparallel }g^{5}}{\ ^{\shortparallel }g^{5}}+\frac{\partial _{k_{2}}\
^{\shortparallel }g^{6}}{\ ^{\shortparallel }g^{6}})-\frac{\partial
_{k_{2}}\ ^{\shortparallel }\partial ^{6}(\ ^{\shortparallel }g^{6})}{2\
^{\shortparallel }g^{6}}=0;  \notag
\end{eqnarray}%
on shell $s=3$ with $i_{2},k_{2}...=1,2,3,4;$
\begin{eqnarray}
\ ^{\shortparallel }\widehat{R}_{7}^{7} &=&\ ^{\shortparallel }\widehat{R}%
_{8}^{8}=-\frac{1}{2 \ ^{\shortparallel}g^{7} \ ^{\shortparallel}g^{8}}[%
\frac{(( \ ^{\shortparallel }g^{7})^{\ast })^{2}}{2\ ^{\shortparallel}g^{7}}+%
\frac{(\ ^{\shortparallel }g^{7})^{\ast }\ (\ ^{\shortparallel }g^{8})^{\ast
}}{2\ ^{\shortparallel }g^{8}}- (\ ^{\shortparallel }g^{7})^{\ast \ast }]=
\notag \\
&& -\ _{4}^{\shortparallel }\mathcal{K}(\kappa ,r,\theta ,\varphi ,t,\
^{\shortparallel }p_{5},\ ^{\shortparallel }p_{6},\ ^{\shortparallel }E),
\notag \\
\ ^{\shortparallel }\widehat{R}_{7k_{3}} &=&\frac{\ \ ^{\shortparallel }g^{7}%
}{2 \ ^{\shortparallel }g^{8}} \ ^{\shortparallel }n_{k_{3}}^{\ast \ast
}+\left( \frac{3}{2}(\ ^{\shortparallel }g^{7})^{\ast }-\frac{\
^{\shortparallel }g^{7}}{\ ^{\shortparallel }g^{8}}(\ ^{\shortparallel
}g^{8})^{\ast }\right) \frac{\ ^{\shortparallel }n_{k_{3}}^{\ast }}{2\ \
^{\shortparallel }g^{8}}=0,  \label{riccist4a} \\
\ ^{\shortparallel }\widehat{R}_{8k_{2}} &=&\frac{\ ^{\shortparallel
}w_{k_{3}}}{2\ ^{\shortparallel }g^{7}}[(\ ^{\shortparallel }g^{7})^{\ast
\ast }-\frac{\left( (\ ^{\shortparallel }g^{7})^{\ast }\right) ^{2}}{2 \
^{\shortparallel}g^{7}}-\frac{(\ ^{\shortparallel }g^{7})^{\ast }\ ( \
^{\shortparallel }g^{8})^{\ast }}{2 \ ^{\shortparallel }g^{8}}]+\frac{(\
^{\shortparallel }g^{7})^{\ast }} {4\ ^{\shortparallel }g^{8}}(\frac{%
\partial _{k_{3}}\ \ ^{\shortparallel }g^{7}}{ \ ^{\shortparallel}g^{7}}+%
\frac{\partial _{k_{3}} \ ^{\shortparallel }g^{8}}{ \ ^{\shortparallel}g^{8}}%
)-\frac{\partial _{k_{3}} (\ ^{\shortparallel }g^{8})^{\ast }}{2\ \
^{\shortparallel }g^{8}}=0  \notag
\end{eqnarray}%
on shell $s=4$ with $i_{3},k_{3}...=1,2,3,4,5,6.$

Using the above formulas, we can compute the Ricci scalar (\ref{canricscal1}%
) for $\ _{s}^{\shortparallel }\widehat{\mathbf{D}}$ (\ref{canhcs}), and
find such relations for shell components of quasi-stationary phase spaces,
\begin{equation*}
\ \ \ _{1}\widehat{R}sc=2(\widehat{R}_{1}^{1}),\ _{2}\widehat{R}sc=2(%
\widehat{R}_{1}^{1}+\widehat{R}_{3}^{3}),\ _{3}^{\shortparallel }\widehat{R}%
sc=2(\ ^{\shortparallel }\widehat{R}_{1}^{1}+\ ^{\shortparallel }\widehat{R}%
_{3}^{3}+\ ^{\shortparallel }\widehat{R}_{5}^{5}),\ \ _{4}^{\shortparallel }%
\widehat{R}sc=2(\ ^{\shortparallel }\widehat{R}_{1}^{1}+\ ^{\shortparallel }%
\widehat{R}_{3}^{3}+\ ^{\shortparallel }\widehat{R}_{5}^{5}+\
^{\shortparallel }\widehat{R}_{7}^{7}).
\end{equation*}%
Such formulas define additional s-adapted symmetries the Einstein canonical
s--tensor (\ref{cnsveinst1}) of the quasi-stationary phase spaces and their
parametric R-flux deformations. This allows to write the nonassociative
vacuum gravitational equations (\ref{cannonsymparamc2a}) in the form
\begin{eqnarray}
\ \widehat{R}_{1}^{1} &=&\widehat{R}_{2}^{2}=-\ \ \ _{1}^{\shortparallel }%
\mathcal{K};\ \widehat{R}_{3}^{3}=\widehat{R}_{4}^{4}=-\ \
_{2}^{\shortparallel }\mathcal{K};  \label{sourc1hv} \\
\ \ ^{\shortparallel }\widehat{R}_{5}^{5} &=&\ ^{\shortparallel }\widehat{R}%
_{6}^{6}=-\ \ _{3}^{\shortparallel }\mathcal{K};\ \ ^{\shortparallel }%
\widehat{R}_{7}^{7}=\ ^{\shortparallel }\widehat{R}_{8}^{8}= - \
_{4}^{\shortparallel }\mathcal{K}.  \label{sourc1cc}
\end{eqnarray}%
If the (effective) sources on shells $s=3,4$ are prescribed using formulas (%
\ref{assocsourc2}) and (\ref{nassocsourc2}) for respective parameterizations
that $\ _{3}^{\shortparallel }\mathcal{K}=\ _{4}^{\shortparallel}\mathcal{K}%
=0,$ we can consider trivial solutions for (\ref{sourc1cc}) on the cofibers
of phase spaces. In result the nonlinear system (\ref{cannonsymparamc2a})
reduces to (\ref{sourc1hv}) in a form equivalent to (\ref{cannonsymparamc2hv}%
) and (\ref{riccist2}). In result, we obtain a decoupling of respective
system of nonlinear PDEs for a quasi-stationary nonholonomic 4-d spacetime
involving parametric R-flux contributions (\ref{realrflux}) into the
effective source $\ ^{\shortparallel }\mathbf{K}_{~j_{2}}^{i_{2}}= \delta _{
j_{2}}^{i_{2}}\ _{s}^{\shortparallel }\mathcal{K}$, for $s=1,2.$ We study
the decoupling property of such phase and spacetime systems in details in
next subsection.

\subsection{Decoupling of nonassociative nonholonomic stationary vacuum
Einstein equations}

We can apply the Theorem 4.1 \ from \cite{bubuianu20} if we redefine those
geometric constructions for phase spaces with complex momenta variables and
effective sources involving parametric R-flux contributions. The priority of
the AFDCM is that for various generalizations and modifications we can take
certain known geometric results for nonlinear systems of PDEs (with a
rigorous proof in previous works) and apply a symbolic geometric calculus
when by analogy left labels change $\ _{s}^{\shortmid }\rightarrow \
_{3}^{\shortparallel }$ and associative sources change into nonassociative
parametric ones, $\ _{s}^{\shortmid }\widehat{\Upsilon }\rightarrow \
_{s}^{\shortparallel }\mathcal{K}$.

The nonassociative vacuum gravitational field equations (\ref%
{cannonsymparamc2a}) (represented in the form (\ref{sourc1hv}) and (\ref%
{sourc1cc}) with canonical Ricci s-tensors (\ref{riccist2}), (\ref{riccist3}%
) \ and (\ref{riccist4a}), computed for a quasi-stationary s-metric ansatz (%
\ref{ansatz1na})) decouple on respective shells of phase space:
\begin{eqnarray}
s &=&1\mbox{ with }g_{i_{1}}=e^{\psi (\hbar ,\kappa ;r,\theta )},i_{1}=1,2
\notag \\
&&\psi ^{\bullet \bullet }+\psi ^{\prime \prime }=2\ \ _{1}^{\shortparallel }%
\mathcal{K};  \label{eq1} \\
s &=&2\mbox{ with }\left\{
\begin{array}{c}
\alpha _{i_{1}}=g_{4}^{\diamond }\partial _{i_{1}}(\ _{2}\varpi ),\
_{2}\beta =g_{4}^{\diamond }(\ _{2}\varpi )^{\diamond },\ _{2}\gamma =(\ln
\frac{|g_{4}|^{3/2}}{|g_{3}|})^{\diamond } \\
\mbox{ for }\ _{2}\Psi =\exp (\ _{2}\varpi ),\ _{2}\varpi =\ln
|g_{4}^{\diamond }/\sqrt{|g_{3}g_{4}}|,%
\end{array}%
\right.  \notag \\
&&(\ _{2}\varpi )^{\diamond }g_{4}^{\diamond }=2g_{3}g_{4}\ \
_{2}^{\shortparallel }\mathcal{K},  \notag \\
&&\ _{2}\beta \ w_{j_{1}}-\alpha _{j_{1}}=0,  \notag \\
&&\ n_{k_{1}}^{\diamond \diamond }+\ _{2}\gamma \ n_{k_{1}}^{\diamond }=0;
\notag
\end{eqnarray}%
\begin{eqnarray}
s &=&3\mbox{ with }\left\{
\begin{array}{c}
\ ^{\shortparallel }\alpha _{i_{2}}=(\ ^{\shortparallel }\partial ^{6}\
^{\shortparallel }g^{5})\partial _{i_{2}}(\ _{3}^{\shortparallel }\varpi ),\
_{3}^{\shortparallel }\beta =(\ ^{\shortparallel }\partial ^{6}\
^{\shortparallel }g^{5})\ ^{\shortparallel }\partial ^{6}(\
_{3}^{\shortparallel }\varpi ),\ _{3}^{\shortparallel }\gamma =\
^{\shortparallel }\partial ^{6}(\ln \frac{|\ ^{\shortparallel }g^{5}|^{3/2}}{%
|\ ^{\shortparallel }g^{6}|}) \\
\mbox{ for }\ _{3}^{\shortparallel }\Psi =\exp (\ _{3}^{\shortparallel
}\varpi ),\ _{3}^{\shortparallel }\varpi =\ln |\ ^{\shortparallel }\partial
^{6}\ ^{\shortparallel }g^{5}/\sqrt{|\ ^{\shortparallel }g^{5}\
^{\shortparallel }g^{6}}|,(i_{2},j_{2},k_{2}=1,2,3,4)%
\end{array}%
\right.  \notag \\
&&\ ^{\shortparallel }\partial ^{6}(\ _{3}^{\shortparallel }\varpi )\
^{\shortparallel }\partial ^{6}\ ^{\shortparallel }g^{5}=2\ ^{\shortparallel
}g^{5}\ ^{\shortparallel }g^{6}\ \ _{3}^{\shortparallel }\mathcal{K},  \notag
\\
&&\ ^{\shortparallel }\partial ^{66}(\ \ ^{\shortparallel }n_{k_{2}})+\
_{3}^{\shortparallel }\gamma \ ^{\shortparallel }\partial ^{6}(\
^{\shortparallel }n_{k_{2}})=0,  \notag \\
&&\ _{3}^{\shortparallel }\beta \ \ ^{\shortparallel }w_{j_{2}}-\
^{\shortparallel }\alpha _{j_{2}}=0;  \notag \\
s &=&4\mbox{ with }\left\{
\begin{array}{c}
\ ^{\shortparallel }\alpha _{i_{3}}=(\ ^{\shortparallel }g^{7})^{\ast }\
^{\shortparallel }\partial _{i_{3}}(\ _{3}^{\shortparallel }\varpi ),\
_{4}^{\shortparallel }\beta =(\ ^{\shortparallel }g^{7})^{\ast }(\
_{4}^{\shortparallel }\varpi )^{\ast },\ _{4}^{\shortparallel }\gamma =(\ln
\frac{|\ ^{\shortparallel }g^{7}|^{3/2}}{|\ ^{\shortparallel }g^{8}|})^{\ast
} \\
\mbox{ for }\ _{4}^{\shortparallel }\Psi =\exp (\ _{4}^{\shortparallel
}\varpi ),\ _{4}^{\shortparallel }\varpi =\ln |(\ ^{\shortparallel
}g^{7})^{\ast }/\sqrt{|\ ^{\shortparallel }g^{7}\ ^{\shortparallel }g^{8}}%
|,(i_{3},j_{3},k_{3}=1,2,...,6)%
\end{array}%
\right.  \notag \\
&&\ ^{\shortparallel }\partial ^{6}(\ _{4}^{\shortparallel }\varpi )(\
^{\shortparallel }g^{7})^{\ast }=2\ ^{\shortparallel }g^{7}\
^{\shortparallel }g^{8}\ _{4}^{\shortmid }\widehat{\Upsilon },  \notag \\
&&(\ ^{\shortparallel }n_{k_{3}})^{\ast \ast }+\ _{4}^{\shortparallel
}\gamma (\ ^{\shortparallel }n_{k_{3}})^{\ast }=0,  \notag \\
&&\ _{4}^{\shortparallel }\beta \ \ ^{\shortparallel }w_{j_{3}}-\
^{\shortparallel }\alpha _{j_{3}}=0.  \notag
\end{eqnarray}%
Such a decoupling can be performed for real momentum variables $p_{a_{s}}$
using labels $\ _{s}^{\shortmid }$ geometric s-objects but in nonassociative
gravity/geometry certain coefficients contain the complex unity.

We can study independently a phase space dynamics and analyze it geometric
evolution in various nonholonomic vacuum gravitational models for a
quasi-stationary 4-d spacetime when parametric R-flux contributions (\ref%
{realrflux}) are encoded into the effective sources $\ ^{\shortparallel}%
\mathbf{K}_{~j_{2}}^{i_{2}}=\ \delta _{~j_{2}}^{i_{2}}\ _{s}^{\shortparallel
}\mathcal{K}$, for $s=1,2,$ and considering decoupling only of the system (%
\ref{sourc1hv}). In such cases, we have to solve recurrently the system of
h1-v2 equations (\ref{eq1}).

Above system of nonlinear PDEs possess an explicit decoupling property on
every shell. For instance (when $s=1$), the first equation in (\ref{eq1}) is
a 2-d Poisson equation which can be solved in certain general forms for any
prescribed source which, in this work, includes parametric R-flux
contributions. We can integrate recurrently on every shell $s=2,3,4$
respective equations containing only partial derivatives on $\
^{\shortparallel }y^{3},\ ^{\shortparallel }p_{6}, \ ^{\shortparallel }p_{8}$
for certain decoupled functions for the s-metric and N-connection
coefficients. This can be performed in explicit form if there are prescribed
any generating sources $\ _{s}^{\shortparallel }\mathcal{K}$ which for
nonassociative gravity theories involve in parametric forms various R-flux
and star deformation contributions.

\subsection{Generating stationary solutions for nonassociative vacuum phase
spaces}

\label{ssnonassocvsol} The nonassociative vacuum Einstein equations
formulated in canonical dyadic variables on phase spaces can be integrated
in some general forms following the results of \cite{bubuianu20}, see
Theorem 4.2 in section 4.3.1 and proofs from appendix A.5. Similar examples
and applications in 10-d string and Finsler-Lagrange-Hamilton geometry
modifications of gravity (with details on AFCDM, locally anisotropic
cosmological and BH solutions) are provided in Refs. \cite%
{bubuianu17b,bubuianu17,bubuianu19}. In this work, we omit proofs because
they can be obtained from respective constructions of cited work by
re-defining dyadic variables on $\ _{s}\mathbf{T}_{\shortparallel }^{\ast}%
\mathbf{V}$ and using parametric star deformed and R-flux induced sources $\
_{s}^{\shortparallel }\mathcal{K}$.

\subsubsection{Off-diagonal solutions with R-flux induced sources and
noholonomic s-torsion}

By tedious straightforward computations, we can check that quasi-stationary
solutions with Killing symmetry on $\partial /\ ^{\shortparallel }p_{7},$
for all shells $s=1,2,3,4$ and Killing symmetry on $\partial _{4}=\partial
_{t}$ on spacetime shells $s=1,2,$ for the nonassociative vacuum Einstein
equations (\ref{cannonsymparamc2a}), with canonical Ricci s-tensors and
effective sources decoupled in the form (\ref{eq1}), are defined by such
quadratic line elements on phase space:{\small
\begin{eqnarray}
d\widehat{s}^{2} &=&e^{\psi (\hbar ,\kappa
;x^{k_{1}})}[(dx^{1})^{2}+(dx^{2})^{2}]+\frac{[(\ _{2}\Psi )^{\diamond }]^{2}%
}{4(~_{2}^{\shortparallel }\mathcal{K})^{2}\{g_{4}^{[0]}-\int d\varphi
\lbrack (\ _{2}\Psi )^{2}]^{\diamond }/4(\ ~_{2}^{\shortparallel }\mathcal{K}%
)\}}\{d\varphi +\frac{\partial _{i_{1}}(\ _{2}\Psi )}{(\ _{2}\Psi
)^{\diamond }}dx^{i_{1}}\}^{2}+  \label{qeltors} \\
&&\{g_{4}^{[0]}-\int d\varphi \frac{\lbrack (\ _{2}\Psi )^{2}]^{\diamond }}{%
4(~_{2}^{\shortparallel }\mathcal{K})}\}\{dt+[\ _{1}n_{k_{1}}+\
_{2}n_{k_{1}}\int d\varphi \frac{\lbrack (\ _{2}\Psi )^{2}]^{\diamond }}{4(\
~_{2}^{\shortparallel }\mathcal{K})^{2}|g_{4}^{[0]}-\int d\varphi \lbrack (\
_{2}\Psi )^{2}]^{\diamond }/4(~_{2}^{\shortparallel }\mathcal{K})|^{5/2}}%
]dx^{k_{1}}\}+  \notag
\end{eqnarray}%
\begin{eqnarray*}
&&\{\ ^{\shortparallel }g_{5}^{[0]}-\int d\ ^{\shortparallel }p_{6} \frac{\
^{\shortparallel }\partial ^{6}[(\ _{3}^{\shortparallel}\Psi )^{2}]}{4(\
_{3}^{\shortparallel}\mathcal{K})}\}\{\ ^{\shortparallel}dp_{5}+ [\
_{1}^{\shortparallel}n_{k_{2}}+\ _{2}^{\shortparallel }n_{k_{2}}\int d\
^{\shortparallel}p_{6}\frac{\ ^{\shortparallel }\partial ^{6}[(\
_{3}^{\shortparallel}\Psi )^{2}]}{4(\ _{3}^{\shortparallel}\mathcal{K}%
)^{2}|\ ^{\shortparallel}g_{5}^{[0]}-\int d\ ^{\shortparallel}p_{6}\
^{\shortparallel}\partial ^{6}[(\ _{3}^{\shortparallel}\Psi)^{2}]/4 (\
_{3}^{\shortparallel}\mathcal{K})|^{5/2}}]dx^{k_{2}}\} \\
&&+ \frac{[\ ^{\shortparallel }\partial ^{6}(\ _{3}^{\shortparallel }\Psi
)]^{2}}{4(~_{3}^{\shortparallel }\mathcal{K})^{2}\{\ ^{\shortparallel
}g_{5}^{[0]}-\int d\ ^{\shortparallel }p_{6}\ ^{\shortparallel }\partial
^{6}[(\ _{3}^{\shortmid }\Psi )^{2}]/4(~_{3}^{\shortparallel }\mathcal{K})\}}%
\{d\ ^{\shortparallel }p_{6}+\frac{\partial _{i_{2}}(\ _{3}^{\shortparallel
}\Psi )}{\ ^{\shortparallel }\partial ^{6}(\ _{3}^{\shortmid }\Psi )}%
dx^{i_{2}}\}^{2}+ \\
&&\{\ ^{\shortparallel }g_{7}^{[0]}-\int d\ ^{\shortparallel }E\frac{[(\
_{4}^{\shortparallel }\Psi )^{2}]^{\ast }}{4(~_{4}^{\shortparallel }\mathcal{%
K})}\}\{d\ ^{\shortparallel }p_{7}+[\ _{1}^{\shortparallel }n_{k_{3}}+\
_{2}^{\shortparallel }n_{k_{3}}\int d\ ^{\shortparallel }E\frac{[(\
_{4}^{\shortparallel }\Psi )^{2}]^{\ast }}{4(~_{4}^{\shortparallel }\mathcal{%
K})^{2}|\ ^{\shortparallel }g_{7}^{[0]}-\int d\ ^{\shortparallel }E[(\
_{4}^{\shortparallel }\Psi )^{2}]^{\ast }/4(~_{4}^{\shortparallel }\mathcal{K%
})|^{5/2}}]d\ ^{\shortparallel }x^{k_{3}}\}- \\
&&\frac{[(\ _{4}^{\shortparallel }\Psi )^{\ast }]^{2}}{4(~_{4}^{%
\shortparallel }\mathcal{K})^{2}\{\ ^{\shortparallel }g_{7}^{[0]}-\int d\
^{\shortparallel }E[(\ _{4}^{\shortparallel }\Psi )^{2}]^{\ast
}/4(~_{4}^{\shortparallel }\mathcal{K})\}}\{d\ ^{\shortparallel }E+\frac{\
^{\shortparallel }\partial _{i_{3}}(\ _{4}^{\shortparallel }\Psi )}{(\
_{4}^{\shortparallel }\Psi )^{\ast }}d\ ^{\shortparallel }x^{i_{3}}\}^{2},
\end{eqnarray*}%
for indices $%
i_{1},j_{1},k_{1},...=1,2;i_{2},j_{2},k_{2},...=1,2,3,4;i_{3},j_{3},k_{3},...=1,2,...6; y^{3}=\varphi ,y^{4}=t,\ ^{\shortparallel }p_{8}=\ ^{\shortparallel }E;
$ and
\begin{eqnarray*}
&&\mbox{generating functions: }\psi (\hbar ,\kappa ;x^{k_{1}});\ _{2}\Psi
(\hbar ,\kappa ;x^{k_{1}},y^{3});\ \ _{3}^{\shortparallel }\Psi (\hbar
,\kappa ;x^{k_{2}},\ ^{\shortparallel }p_{6});\ \ _{4}^{\shortparallel }\Psi
(\hbar ,\kappa ;\ ^{\shortparallel }x^{k_{3}},\ ^{\shortparallel }E); \\
&&\mbox{generating sources:}\ ~_{1}^{\shortparallel }\mathcal{K}(\hbar
,\kappa ;x^{k_{1}});\ ~_{2}^{\shortparallel }\mathcal{K}(\hbar ,\kappa
;x^{k_{1}},y^{3});\ ~_{3}^{\shortparallel }\mathcal{K}(\hbar ,\kappa
;x^{k_{2}},\ ^{\shortparallel }p_{6});\ ~_{4}^{\shortparallel }\mathcal{K}%
(\hbar ,\kappa ;\ ^{\shortparallel }x^{k_{3}},\ ^{\shortparallel }E); \\
&&\mbox{integr. functions: }g_{4}^{[0]}(\hbar ,\kappa ;x^{k_{1}}),\
_{1}n_{k_{1}}(\hbar ,\kappa ;x^{j_{1}}),\ _{2}n_{k_{1}}(\hbar ,\kappa
;x^{j_{1}});\ ^{\shortparallel }g_{5}^{[0]}(\hbar ,\kappa ;x^{k_{2}}),\
_{1}n_{k_{2}}(\hbar ,\kappa ;x^{j_{2}}),\ _{2}n_{k_{2}}(\hbar ,\kappa
;x^{j_{2}}); \\
&&\ ^{\shortparallel }g_{7}^{[0]}(\hbar ,\kappa ;x^{j_{3}}), \
_{1}^{\shortparallel }n_{k_{3}}(\hbar ,\kappa ;x^{j_{3}}), \
_{2}^{\shortparallel }n_{k_{3}}(\hbar ,\kappa ;x^{j_{3}}).
\end{eqnarray*}%
}

Any generic off-diagonal solution $\ _{s}^{\shortparallel }\widehat{\mathbf{g%
}}=\ ^{\shortparallel }\widehat{\mathbf{g}}_{\alpha _{s}\beta _{s}}(~_{s}x,\
_{s}^{\shortparallel }p)\ ^{\shortparallel }\widehat{\mathbf{e}}^{\alpha
_{s}}\mathbf{\otimes \ ^{\shortparallel }}\widehat{\mathbf{e}}^{\beta _{s}}$
(\ref{qeltors}) is exact but depends in parametric form on $\hbar ,\kappa $
for any R-flux nonassociative data encoded in $~_{s}^{\shortparallel }%
\mathcal{K} $.

Let us discuss the difference between certain classes of solutions of
nonlinear systems of ODE (ordinary differential equations) and PDEs.
Usually, physically important solutions in gravity theories (for instance,
for BH solutions in GR) are found for certain diagonalize ansatz for metrics
depending on one space (radial) variable. For such configurations, there are
involved into considerations only four diagonal coefficients of a metric
(from six independent ones; in 4-d spacetimes, a symmetric metric tensor
contains 10 components but 4 of them can be transformed to zero values via
coordinate transforms, which is consequence of Bianchi identities), see
details in standard monographs on mathematical and general relativity and
exact solutions \cite{hawking73,misner,wald82,kramer03}. Under certain
symmetry assumptions, the corresponding system of nonlinear PDEs (for
instance, the vacuum Einstein equations) transforms into a second order
nonlinear system of ODEs with radial derivatives. This can be integrated in
exact form when the solutions are determined by two integration constants.
If we consider a cosmological constant, we also introduce an (effective)
source when the coefficients of diagonal static (or stationary, for rotating
BHs) also depend on such a constant. The values of the integration constants
are determined, for instance, from certain boundary/ asymptotic conditions,
i.e. from some physical considerations, that the time like coefficient of
the metric results in the Newton potential of a point mass.

When we consider systems of nonlinear PDEs of type (\ref{eq1}) with ansatz (%
\ref{qeltors}), such nonassociative vacuum Einstein equations are not
transformed into certain diagonal systems of nonlinear ODE. Such metrics
depend, in general, on all spacetime and phase space coordinates even
certain Killing symmetry conditions are imposed. As we have shown above, we
can decouple such equations in very general forms when there are involved
certain partial derivatives of the coefficients of metrics and if we work in
some well defined nonholonomic dyadic bases and for corresponding canonical
deformations of the nonlinear and linear connection structures. At the next
step, we can integrate in explicit form the corresponding s-adapted systems
of nonlinear PDEs. Performing such integrations we get dependencies not only
on integration constants but also on integration functions depending on
corresponding shell variables. More than that, such a geometric method of
constructing exact solutions involves also generating functions and
generating sources. Similar ideas are used in constructing solutions for 2-d
and 3-d nonlinear solitonic equations but using the AFCDM there are involved
different types of PDEs. Following such an approach, we construct certain
general classes of solutions when generic off-diagonal (non) symmetric
metrics and (generalized) connections depend, in principle, on all spacetime
and phase coordinates but via respective integration functions and
generating functions and sources. We have to consider additional physical
motivations (generalized symmetries, causality principles, Cauchy type
problems for certain evolution scenarios, or boundary/ asymptotic condition)
for prescribing certain classes of generating functions and sources and how
to determine the integration functions. One should be imposed certain
additional conditions of compatibility with experimental/ observational
data, to compute possible quantum corrections, various (non) associative/
commutative sources, star deformations etc. in order to elaborate on viable
physical models.

In references \cite%
{vacaru18,vacaru03,vacaru09a,bubuianu18a,bubuianu17b,bubuianu17,
bubuianu19,bubuianu20}, we study typical geometric properties and various
applications with constructing general integrals of systems of nonlinear
PDEs in (modified) gravity if such configurations are constructed in some
general forms and not only for special ansatz transforming PDEs into ODEs.
It should be noted that the AFCDM is different from the well known Cartan's
moving frame method and Newman-Penrose dyadic formalism considered in \cite%
{hawking73,misner,wald82,kramer03}. Our approach is more general (for
certain special conditions the AFCDM can be reduced to standard methods with
ODEs in GR and MGTs) because involves additional distortions of the (non)
linear connections structure. When certain classes of (off-) diagonal
solutions have been constructed in a general form, we can impose additional
nonholonomic constraints in order to extract, for instance,
LC-configurations, subclasses of diagonal solutions, study parametric
implications of some R-flux contributions, and to analyze how such
configurations can be extended for complex variables, nonassociative and
noncommutative symmetries, etc.

\subsubsection{Computing stationary coefficients for (non)symmetric star
deformed s-metrics}

A complete parametric solution of nonassociative nonholonomic vacuum
gravitational equations is determined also by nonsymmetric and symmetric
components of a star deformed s-metric. Such values decouple and can be
computed as induced by a solution $\left( \ ^{\shortparallel }\widehat{%
\mathbf{g}}_{\alpha _{s}\beta _{s}},\ ^{\shortparallel }\widehat{\mathbf{e}}
^{\alpha _{s}}\right) $ (\ref{qeltors}) if the effective sources $\
_{s}^{\shortparallel }\mathcal{K}$ are subjected to nonholonomic constraints
(\ref{nonhconstr1a}). Introducing such s-coefficients in formulas (\ref{nsm1}%
), we can compute the quasi-stationary canonical components of the
nonsymmetric part of respective star deformed s-metric $\ _{\star
}^{\shortparallel} \widehat{\mathfrak{g}}_{\alpha _{s}\beta _{s}},$
\begin{eqnarray}
\ _{\star }^{\shortparallel }\widehat{\mathfrak{a}}_{\alpha _{s}\beta
_{s}}:= &&[\ _{\star }^{\shortparallel }\widehat{\mathfrak{a}}%
_{i_{1}j_{1}}=0,\ _{\star }^{\shortparallel }\widehat{\mathfrak{a}}%
_{a_{2}b_{2}}=0,  \label{nsm1qs} \\
\ _{\star }^{\shortparallel }\widehat{\mathfrak{a}}^{a_{3}b_{3}} &=&\frac{%
i\kappa }{2}(\overline{\mathcal{R}}_{c_{3}\quad }^{\ n+k_{s}a_{3}}\ \mathbf{%
^{\shortparallel }}\widehat{\mathbf{e}}_{k_{s}}\ ^{\shortparallel }\widehat{%
\mathbf{g}}^{c_{3}b_{3}}-\overline{\mathcal{R}}_{c_{3}\quad }^{\
n+k_{s}b_{3}}\ \mathbf{^{\shortparallel }}\widehat{\mathbf{e}}_{k_{s}}\
^{\shortparallel }\widehat{\mathbf{g}}^{c_{3}a_{3}})=\ _{\star
}^{\shortparallel }\widehat{\mathfrak{a}}_{[1]}^{a_{3}b_{3}}(\kappa )=\frac{1%
}{2}(\ _{\star }^{\shortparallel }\widehat{\mathfrak{g}}_{[1]}^{a_{3}b_{3}}-%
\ _{\star }^{\shortparallel }\widehat{\mathfrak{g}}_{[1]}^{b_{3}a_{3}})(%
\kappa ),  \notag \\
\ _{\star }^{\shortparallel }\widehat{\mathfrak{a}}^{a_{4}b_{4}} &=&\frac{%
i\kappa }{2}(\overline{\mathcal{R}}_{c_{4}\quad }^{\ n+k_{s}a_{4}}\ \mathbf{%
^{\shortparallel }}\widehat{\mathbf{e}}_{k_{s}}\ ^{\shortparallel }\widehat{%
\mathbf{g}}^{c_{4}b_{4}}-\overline{\mathcal{R}}_{c_{4}\quad }^{\
n+k_{s}b_{4}}\ \mathbf{^{\shortparallel }}\widehat{\mathbf{e}}_{k_{s}}\
^{\shortparallel }\widehat{\mathbf{g}}^{c_{4}a_{4}})=\ _{\star
}^{\shortparallel }\widehat{\mathfrak{a}}_{[1]}^{a_{4}b_{4}}(\kappa )=\frac{1%
}{2}(\ _{\star }^{\shortparallel }\widehat{\mathfrak{g}}_{[1]}^{a_{4}b_{4}}-%
\ _{\star }^{\shortparallel }\widehat{\mathfrak{g}}_{[1]}^{b_{4}a_{4}})(%
\kappa )].  \notag
\end{eqnarray}%
We omit cumbersome explicit formulas involving generating functions and
sources, and integration functions from (\ref{qeltors}) which are encoded
into the coefficients of s-metrics and s-frames with hat labels.

In a similar form, using any data $\left( \ ^{\shortparallel }\widehat{%
\mathbf{g}}_{\alpha _{s}\beta _{s}},~^{\shortparallel }\widehat{\mathbf{e}}%
^{\alpha _{s}}\right) $ (\ref{qeltors}) for (\ref{ssm1}), we can compute the
symmetric part of a s-metric $\ _{\star }^{\shortparallel }\widehat{%
\mathfrak{g}}_{\alpha _{s}\beta _{s}},$
\begin{eqnarray}
\ ^{\shortparallel }\widehat{\mathfrak{\check{g}}}_{\alpha _{s}\beta _{s}}
&=&\ ^{\shortparallel }\widehat{\mathbf{g}}_{\alpha _{s}\beta _{s}}+\
_{\star }^{\shortparallel }\widehat{\mathfrak{\check{g}}}_{\alpha _{s}\beta
_{s}}^{[1]}(\kappa ),\mbox{ for }  \label{ssm1qs} \\
\ _{\star }^{\shortparallel }\widehat{\mathfrak{\check{g}}}_{\alpha
_{s}\beta _{s}}^{[1]}(\kappa ) &=&-\frac{i\kappa }{2}\left( \overline{%
\mathcal{R}}_{\quad \beta _{s}}^{\tau _{s}\xi _{s}}\ \mathbf{%
^{\shortparallel }}\widehat{\mathbf{e}}_{\xi _{s}}\ ^{\shortparallel }%
\widehat{\mathbf{g}}_{\tau _{s}\alpha _{s}}+\overline{\mathcal{R}}_{\quad
\alpha _{s}}^{\tau _{s}\xi _{s}}\ \mathbf{^{\shortparallel }}\widehat{%
\mathbf{e}}_{\xi _{s}}\ ^{\shortparallel }\widehat{\mathbf{g}}_{\beta
_{s}\tau _{s}}\right) .  \notag
\end{eqnarray}

Another important value which can be computed for quasi-stationary solutions
(\ref{qeltors}) is the real R-flux distortion on the spacetime which was
considered as an effective source $\ ^{\shortparallel }\widehat{\mathbf{K}}%
_{\ j_{2}k_{2}}= -\ _{[11]}^{\shortparallel }\widehat{\mathbf{R}}%
ic_{j_{2}k_{2}}^{\star }(x^{k_{1}},x^{3})$ where all terms in formulas (\ref%
{realrflux}) are explicitly computed for some data $\left(\
^{\shortparallel} \widehat{\mathbf{g}}_{\alpha _{s}\beta
_{s}},~^{\shortparallel }\widehat{\mathbf{e}}^{\alpha _{s}}\right).$ Such
way, we can define as determined by a solution (\ref{qeltors}) all s-adapted
coefficients of canonical Ricci s-tensor $\widehat{\mathbf{R}}ic_{\alpha
_{s}\beta _{s}}$ (\ref{candricci}), respective star deformations to $%
\widehat{\mathbf{R}}ic_{\alpha _{s}\beta _{s}}^{\star }$ (\ref%
{driccicanonstar1}), and corresponding parametric decompositions (\ref{ric50}%
)-(\ref{ric53}), when all s-operators and coefficients are used for "hat"
values in respective formulas. Certain values contain as multiples the
complex unity. It is not clear what is the physical importance of such
solutions which arise naturally in string gravity being induced by R-fluxes.
In principle, we can consider generalizations of GR for complex spacetimes
like in \cite{penrose1,penrose2} and then to involve complex models of
nonholonomic metric-affine and Finsler-Lagrange-Hamilton like geometries
\cite{vacaru05a} which were studied also in the framework of string theory
\cite{vacaru96a,vacaru96b,bubuianu17b}. The AFCDM can be extended for
constructing exact and parametric solutions in such complex gravity models.

We can consider a subclass of h1-v2 solutions (\ref{qeltors}) generated by
data%
\begin{equation*}
\ ^{\shortparallel }\widehat{\mathbf{g}}_{\alpha _{s}\beta _{s}}=[\
^{\shortparallel }\widehat{\mathbf{g}}_{i_{2}j_{2}}(\hbar ,\kappa
;x^{k_{1}},y^{3}),\ ^{\shortparallel }\eta ^{a_{3}b_{3}},\ ^{\shortparallel
}\eta ^{a_{4}b_{4}}],~^{\shortparallel }\widehat{\mathbf{e}}^{\alpha
_{s}}(\hbar ,\kappa ;x^{k_{1}},y^{3})=[~^{\shortparallel }\widehat{\mathbf{e}%
}_{2},~^{\shortparallel }\widehat{\mathbf{e}}^{\alpha _{3}}\rightarrow
d~^{\shortparallel }p_{a_{3}},~^{\shortparallel }\widehat{\mathbf{e}}%
^{\alpha _{4}}\rightarrow d~^{\shortparallel }p_{a_{4}}],  \label{qeltorshv}
\end{equation*}%
for $\ ^{\shortparallel }\eta ^{a_{3}b_{3}}=(1,1)$ and $\ ^{\shortparallel}
\eta ^{a_{3}b_{3}}=(1,-1),$ when the integration and generation functions on
shells $s=3,4$ are chosen for trivial N-connection structure, for instance
with $\ ^{\shortparallel}\widehat{N}_{j_{2}\ a_{3}}=0,\ ^{\shortparallel}
\widehat{N}_{j_{3}\ a_{4}}=0$ and prescribed c3-c4 sources $%
_{3}^{\shortparallel }\mathcal{K}= _{4}^{\shortparallel}\mathcal{K}=0.$ In
result, we obtain quasi-stationary spacetime quadratic elements {\small
\begin{eqnarray}
d\widehat{s}^{2} &=&e^{\psi (\hbar ,\kappa
;x^{k_{1}})}[(dx^{1})^{2}+(dx^{2})^{2}]+\frac{[(\ _{2}\Psi (\hbar ,\kappa
;x^{k_{1}},y^{3}))^{\diamond }]^{2}}{4(~_{2}^{\shortparallel }\mathcal{K}%
(\hbar ,\kappa ;x^{k_{1}},y^{3}))^{2}\{g_{4}^{[0]}-\int d\varphi \lbrack (\
_{2}\Psi (\hbar ,\kappa ;x^{k_{1}},y^{3}))^{2}]^{\diamond }/4(\
~_{2}^{\shortparallel }\mathcal{K}(\hbar ,\kappa ;x^{k_{1}},y^{3}))\}}
\notag \\
&&\{d\varphi +\frac{\partial _{i_{1}}(\ _{2}\Psi (\hbar ,\kappa
;x^{k_{1}},y^{3}))}{(\ _{2}\Psi (\hbar ,\kappa ;x^{k_{1}},y^{3}))^{\diamond }%
}dx^{i_{1}}\}^{2}+  \label{qeltorshv1} \\
&&\{g_{4}^{[0]}(\hbar ,\kappa ;x^{k_{1}})-\int d\varphi \frac{\lbrack (\
_{2}\Psi (\hbar ,\kappa ;x^{k_{1}},y^{3}))^{2}]^{\diamond }}{%
4(~_{2}^{\shortparallel }\mathcal{K}(\hbar ,\kappa ;x^{k_{1}},y^{3}))}%
\}\{dt+[\ _{1}n_{k_{1}}(\hbar ,\kappa ;x^{k_{1}})+  \notag \\
&&\ _{2}n_{k_{1}}(\hbar ,\kappa ;x^{k_{1}})\int d\varphi \frac{\lbrack (\
_{2}\Psi (\hbar ,\kappa ;x^{k_{1}},y^{3}))^{2}]^{\diamond }}{4(\
~_{2}^{\shortparallel }\mathcal{K}(\hbar ,\kappa
;x^{k_{1}},y^{3}))^{2}|g_{4}^{[0]}-\int d\varphi \lbrack (\ _{2}\Psi (\hbar
,\kappa ;x^{k_{1}},y^{3}))^{2}]^{\diamond }/4(~_{2}^{\shortparallel }%
\mathcal{K}(\hbar ,\kappa ;x^{k_{1}},y^{3}))|^{5/2}}]dx^{k_{1}}\}  \notag
\end{eqnarray}%
}defining exact solutions of the systems of nonassociative canonical
nonholonomic vacuum Einstein equations (\ref{cannonsymparamc2hv}) reduced to
(\ref{sourc1hv}) with trivial (\ref{sourc1cc}). The quadratic elements (\ref%
{qeltorshv1}) encode via effective generating sources $\
_{1}^{\shortparallel }\mathcal{K}$ and $\ _{2}^{\shortparallel }\mathcal{K}$
nontrivial R-flux contributions. Similar generic off-diagonal metrics were
studied in GR and various (non) commutative modifications \cite%
{vacaru03,vacaru09a,bubuianu17b} but for different types of effective
sources. Star deformations of such solutions are symmetric up to 1st order
on $\kappa $ as follow from formulas (\ref{nsm1qs}) and (\ref{ssm1qs}).

So, prescribing any generating quasi-stationary sources $\ ^{\shortparallel}%
\widehat{\mathbf{K}}_{\ \beta _{s}}^{\alpha _{s}}$ (\ref{ansatzsourc}) with
nonassociative parametric deformations and using block $[(2\times
2)+(2\times 2)]+[(2\times 2)+(2\times 2)]$ parametrization of symmetric and
nonsymmetric parts of star deformed s-metrics, $\ _{\star }^{\shortparallel }%
\widehat{\mathfrak{g}}_{\mu _{s}\nu _{s}}=(\ _{\star }^{\shortparallel }%
\widehat{\mathfrak{g}}_{i_{1}j_{1}},\ _{\star }^{\shortparallel} \widehat{%
\mathfrak{g}}_{a_{2}b_{2}},\ _{\star}^{\shortparallel}\widehat{\mathfrak{g}}%
^{a_{3}b_{3}}, \ _{\star }^{\shortparallel }\widehat{\mathfrak{g}}
^{a_{4}b_{4}})$ and $\ _{\star }^{\shortparallel }\widehat{\mathfrak{a}}%
_{\mu _{s}\nu _{s}}=(0,0,\ _{\star }^{\shortparallel }\widehat{\mathfrak{a}}%
_{c_{3}b_{3}},\ _{\star }^{\shortparallel }\widehat{\mathfrak{a}}%
_{c_{4}b_{4}}),$ we can compute R-flux deformations for any stated
associative (off-) diagonal quasi-stationary solution $\ ^{\shortparallel }%
\widehat{\mathbf{g}}_{\alpha _{s}\beta _{s}}.$ The corresponding quadratic
elements are considered on phase space to be of type (\ref{qeltors}) and
reduces on spacetime in the form (\ref{qeltorshv1}).

\subsubsection{Extracting nonassociative parametric LC-configurations}

Quasi-stationary solutions constructed in previous subsection are with
nontrivial canonical s-torsion structure $\ _{s}^{\shortmid }\widehat{%
\mathbf{T}}_{\ \alpha _{s}\beta _{s}}^{\gamma _{s}}$ of respective $\
_{s}^{\shortparallel }\widehat{\mathbf{D}}=\ ^{\shortparallel }\nabla +\
_{s}^{\shortparallel }\widehat{\mathbf{Z}}.$ We have to impose additional
nonholonomic constraints in the type of generating functions and effective
sources in order to satisfy the conditions $\ _{s}^{\shortparallel }\widehat{%
\mathbf{Z}}$ and extract zero torsion LC-configurations for $\
^{\shortparallel }\nabla $ defined on phase space and/ or for respective
constraints on spacetimes with nontrivial positive R-flux structure. The
solutions (\ref{qeltors}) and (\ref{qeltorshv1}) can be constrained in such
a form that up to orders $\hbar ,\kappa $ and $\hbar \kappa ,$ for star
deformations of respective connections, there are satisfied the conditions $%
\ _{\star }^{\shortparallel s}\widehat{\mathbf{Z}}=0,$ i. e. $\
_{s}^{\shortparallel }\widehat{\mathbf{D}}_{\mid \ _{s}^{\shortparallel}%
\widehat{\mathbf{T}}=0}^{\star }= \ ^{\shortparallel }\nabla ^{\star },$ see
formulas (\ref{lccondnonass}) and (\ref{candistrnas}) (we have to consider
additional nonholonomic shell by shell constraints for higher order
parametric terms).

By straightforward computations on cotangent bundles (see section 4.3.2 in
\cite{bubuianu20}, for similar details and reference on previous works), we
can verify that all canonical s-torsion coefficients vanish if the
coefficients of a quasi-stationary s-metric, and respective coefficients of
N-connections, see formulas for ansatz (\ref{ansatz1na}), are subjected to
satisfy such equations,
\begin{eqnarray}
s=2: &&\ w_{i_{1}}^{\diamond }=\mathbf{e}_{i_{1}}\ln \sqrt{|\ g_{3}|},%
\mathbf{e}_{i_{1}}\ln \sqrt{|\ g_{4}|}=0,\partial _{i_{1}}w_{j_{1}}=\partial
_{j_{1}}w_{i_{1}}\mbox{ and }n_{i_{1}}^{\diamond }=0;  \notag \\
s=3: &&\ \ ^{\shortparallel }\partial ^{6}\ ^{\shortparallel }n_{i_{2}}=\
\mathbf{e}_{i_{2}}\ln \sqrt{|\ ^{\shortparallel }g^{6}|},\ \mathbf{e}%
_{i_{2}}\ln \sqrt{|\ ^{\shortparallel }g^{5}|}=0,\partial
_{i_{2}}n_{j_{2}}=\partial _{j_{2}}n_{i_{2}}\mbox{ and }\ \ ^{\shortparallel
}\partial ^{6}w_{i_{2}}=0;  \label{zerot} \\
s=4: &&\ \ \ ^{\shortparallel }n_{i_{3}}^{\ast }=\ \ ^{\shortparallel }%
\mathbf{e}_{i_{3}}\ln \sqrt{|\ ^{\shortparallel }g^{8}|},\ \ \
^{\shortparallel }\mathbf{e}_{i_{3}}\ln \sqrt{|\ ^{\shortparallel }g^{7}|}%
=0,\ \ ^{\shortparallel }\partial _{i_{3}}\ \ ^{\shortparallel }n_{j_{3}}=\
\ ^{\shortparallel }\partial _{j_{3}}\ \ ^{\shortparallel }n_{i_{3}}%
\mbox{
and } \ ^{\shortparallel }n_{i_{3}}^{\ast }=0;  \notag
\end{eqnarray}%
\begin{eqnarray*}
\mbox{and }\ s=2: &&\ n_{k_{1}}(x^{i_{1}})=0\mbox{ and }\partial
_{i_{1}}n_{j_{1}}(x^{k_{1}})=\partial _{j_{1}}n_{i_{1}}(x^{k_{1}});
\label{expcondn} \\
s=3: &&\ w_{k_{2}}(x^{i_{2}})=0\mbox{ and }\partial _{i_{2}}\
w_{j_{2}}(x^{k_{2}})=\partial _{j_{2}}\ w_{i_{2}}(x^{k_{2}});  \notag \\
s=4: &&\ \ \ ^{\shortparallel }w_{k_{3}}(\ ^{\shortparallel }x^{i_{3}})=0%
\mbox{ and }\ \ ^{\shortparallel }\partial _{i_{3}}\ ^{\shortparallel
}w_{j_{3}}(\ ^{\shortparallel }x^{k_{3}})=\ ^{\shortparallel }\partial
_{j_{3}}\ ^{\shortparallel }w_{i_{3}}(\ ^{\shortparallel }x^{k_{3}}).  \notag
\end{eqnarray*}%
The solutions for $w$- and $n$-functions which can be obtained for (\ref%
{zerot}) depend in parametric form on $\hbar ,\kappa $ on the class of
vacuum or non--vacuum s-metrics encoding possible nonassociative effective
sources. So, they relate and constrain additionally such s-metric
coefficients to possible R-flux contributions. We state some important
conditions on respective classes of generating functions and generating
sources which result in zero s-torsions on every shell:

First, it should be noted that if we prescribe a shell generating function $%
\ _{2}\Psi =\ _{2}\check{\Psi}(\hbar ,\kappa ,x^{i_{1}},y^{3}),$ for which $%
[\partial _{i_{1}}(\ _{2}\check{\Psi})]^{\diamond }= \partial _{i_{1}}(\ _{2}%
\check{\Psi})^{\diamond },$ we solve the conditions for $w_{j_{1}}$ in (\ref%
{zerot}) in explicit form when the effective source $\ ~_{2}^{\shortparallel}%
\mathcal{K}=const,$ or if such an effective source can be expressed as a
functional $\ _{2}^{\shortparallel }\mathcal{K}(\hbar ,\kappa ,x^{i},y^{3})=
\ _{2}^{\shortparallel }\mathcal{K}[\ _{2}\check{\Psi}].$ Similar conditions
should be formulated for $s=3,4.$ For instance, we prescribe $\
_{3}^{\shortparallel }\Psi =\ _{3}^{\shortparallel }\check{\Psi}(\hbar
,\kappa ,x^{i_{2}},p^{6}),$ for which $\ ^{\shortparallel }\partial
^{6}[\partial _{i_{2}}(\ \ _{3}^{\shortparallel }\check{\Psi})]=\partial
_{i_{2}}~^{\shortparallel }\partial ^{6}(\ \ _{3}^{\shortparallel }\check{%
\Psi}).$

We note that the third conditions in (\ref{zerot}), for $s=2$, i.e. $%
\partial _{i_{1}}w_{j_{1}}=\partial _{j_{1}}w_{i_{1}},$ can be solved in
parametric form for any generating function $\ _{2}\check{A}=\ _{2}\check{A}%
(\hbar ,\kappa ,x^{k},y^{3})$ for which
\begin{equation*}
w_{i_{1}}=\check{w}_{i_{1}}=\partial _{i_{1}}\ _{2}\Psi /(\ _{2}\Psi
)^{\diamond }=\partial _{i_{1}}\ _{2}\check{A}.
\end{equation*}%
On the second shell, we consider similar formulas but with duality in order
to involve configurations with nontrivial dependence on energy type
variable,
\begin{equation*}
n_{i_{2}}=\check{n}_{i_{2}}=\partial _{i_{2}}\ _{3}^{\shortparallel }\Psi
/~^{\shortparallel }\partial ^{6}(\ _{3}^{\shortparallel }\Psi )=\partial
_{i_{2}}\ _{3}^{\shortparallel }\check{A}.
\end{equation*}

Summarizing above formulas, we obtain such LC-conditions for generating
functions and generating sources: {\small
\begin{eqnarray}
s=2: &&\ _{2}\Psi =\ _{2}\check{\Psi}(\hbar ,\kappa
,x^{i_{1}},y^{3}),(\partial _{i_{1}}\ _{2}\check{\Psi})^{\diamond }=\partial
_{i_{1}}(\ _{2}\check{\Psi})^{\diamond },\check{w}_{i_{1}}=\partial
_{i_{1}}(\ _{2}\check{\Psi})/(\ _{2}^{\shortmid }\check{\Psi})^{\diamond
}=\partial _{i_{1}}(\ _{2}\check{A});n_{i_{1}}=\partial _{i_{1}}[\
^{2}n(\hbar ,\kappa ,x^{k_{1}})];  \notag \\
&&\ ~_{2}^{\shortparallel }\mathcal{K}(\hbar ,\kappa
,x^{i},y^{3})=~_{2}^{\shortparallel }\mathcal{K}[\ _{2}^{\shortmid }\check{%
\Psi}],\mbox{ or
}\ ~_{2}^{\shortparallel }\mathcal{K}=const;  \label{expconda} \\
s=3: &&\ \ _{3}^{\shortparallel }\Psi =\ _{3}^{\shortmid }\check{\Psi}(\hbar
,\kappa ,x^{i_{2}},~^{\shortparallel }p^{6}),~^{\shortparallel }\partial
^{6}[\partial _{i_{2}}(\ _{3}^{\shortparallel }\check{\Psi})]=\partial
_{i_{2}}~^{\shortparallel }\partial ^{6}(\ _{3}^{\shortparallel }\check{\Psi}%
);\check{w}_{i_{2}}=\partial _{i_{2}}(\ _{3}^{\shortparallel }\Psi
)/~^{\shortparallel }\partial ^{6}(\ _{3}^{\shortparallel }\Psi )=\partial
_{i_{2}}(\ _{3}^{\shortparallel }\check{A});  \notag \\
&&\ ~\ n_{i_{2}}=\partial _{i_{2}}[\ ^{3}n(\hbar ,\kappa
,x^{k_{2}})];~_{3}^{\shortparallel }\mathcal{K}(\hbar ,\kappa
,x^{i_{2}},~^{\shortparallel }p^{6})=\ ~_{3}^{\shortparallel }\mathcal{K}[\
_{3}^{\shortmid }\check{\Psi}],\mbox{ or }\ ~_{3}^{\shortparallel }\mathcal{K%
}=const;  \notag \\
s=4: &&\ \ _{4}^{\shortparallel }\Psi =\ _{4}^{\shortmid }\check{\Psi}(\hbar
,\kappa ,~^{\shortparallel }x^{i_{3}},~^{\shortparallel
}E),[~^{\shortparallel }\partial _{i_{3}}(\ _{4}^{\shortparallel }\check{\Psi%
})]^{\ast }=~^{\shortparallel }\partial _{i_{3}}(\ _{4}^{\shortparallel }%
\check{\Psi})^{\ast };~^{\shortparallel }\check{w}_{i_{3}}=~^{\shortparallel
}\partial _{i_{3}}(\ _{4}^{\shortparallel }\Psi )/(\ _{4}^{\shortparallel
}\Psi )^{\ast }=~^{\shortparallel }\partial _{i_{3}}(\ _{4}^{\shortparallel }%
\check{A});  \notag \\
&&\ ~\ ~^{\shortparallel }n_{i_{3}}=~^{\shortparallel }\partial _{i_{3}}[\
^{4}n(\hbar ,\kappa ,~^{\shortparallel }x^{k_{3}})];~_{4}^{\shortparallel }%
\mathcal{K}(\hbar ,\kappa ,x^{i_{3}},~^{\shortparallel }E)=\
~_{4}^{\shortparallel }\mathcal{K}[\ _{4}^{\shortparallel }\check{\Psi}],%
\mbox{ or }\ ~_{4}^{\shortparallel }\mathcal{K}=const.  \notag
\end{eqnarray}%
}

For such subclasses of s-coefficients, the quadratic line element (\ref%
{qeltors}) transforms into
\begin{eqnarray}
d\widehat{s}_{LCst}^{2} &=&~^{\shortparallel }\widehat{\check{g}}_{\alpha
\beta }(\hbar ,\kappa ,~^{\shortparallel }x^{k_{3}},~^{\shortparallel
}E)d~^{\shortparallel }u^{\alpha }d~^{\shortparallel }u^{\beta }=e^{\psi
(\hbar ,\kappa ,x^{k_{1}})}[(dx^{1})^{2}+(dx^{2})^{2}]+  \label{qellc} \\
&&\frac{[(\ _{2}\check{\Psi})^{\diamond }]^{2}}{4(~_{2}^{\shortparallel }%
\mathcal{K}[\ _{2}\check{\Psi}])^{2}\{g_{4}^{[0]}-\int d\varphi \lbrack (\
_{2}\check{\Psi})^{2}]^{\diamond }/4(\ ~_{2}^{\shortparallel }\mathcal{K})\}}%
\{d\varphi +[\partial _{i_{1}}(\ _{2}\check{A})]dx^{i_{1}}\}+  \notag \\
&&\{g_{4}^{[0]}-\int d\varphi \frac{\lbrack (\ _{2}\check{\Psi}%
)^{2}]^{\diamond }}{4(\ ~_{2}^{\shortparallel }\mathcal{K}[\ _{2}\check{\Psi}%
])}\}\{dt+\partial _{i_{1}}[\ ^{2}n(x^{k_{1}})]dx^{i_{1}}\}+  \notag
\end{eqnarray}%
\begin{eqnarray*}
&&\{~^{\shortparallel }g_{5}^{[0]}-\int d~^{\shortparallel }p_{6}\frac{%
~^{\shortparallel }\partial ^{6}[(\ _{3}^{\shortparallel }\Psi )^{2}]}{%
4(~_{3}^{\shortparallel }\mathcal{K}[\ _{3}^{\shortmid }\check{\Psi}])}%
\}\{d~^{\shortparallel }p_{5}+\partial _{i_{2}}[\
^{3}n(x^{k_{2}})]dx^{i_{2}}\}+ \\
&&\frac{[~^{\shortparallel }\partial ^{6}(\ _{3}^{\shortparallel }\Psi )]^{2}%
}{4(\ ~_{3}^{\shortparallel }\mathcal{K}[\ _{3}^{\shortparallel }\check{\Psi}%
])^{2}\{~^{\shortparallel }g_{5}^{[0]}-\int d~^{\shortparallel
}p_{6}~^{\shortparallel }\partial ^{6}[(\ _{3}^{\shortparallel }\Psi
)^{2}]/4(\ ~_{3}^{\shortparallel }\mathcal{K}[\ _{3}^{\shortparallel }\check{%
\Psi}])\}}\{d~^{\shortparallel }p_{6}+[\partial _{i_{2}}(\
_{3}^{\shortparallel }\check{A})]dx^{i_{2}}\}^{2}+ \\
&&\{~^{\shortparallel }g_{7}^{[0]}-\int d~^{\shortparallel }E\frac{[(\
_{4}^{\shortparallel }\Psi )^{2}]^{\ast }}{4(\ ~_{4}^{\shortparallel }%
\mathcal{K}[\ _{4}^{\shortparallel }\check{\Psi}])}\}\{d~^{\shortparallel
}p_{7}+~^{\shortparallel }\partial _{i_{3}}[\ ^{4}n(~^{\shortparallel
}x^{k_{3}})]d~^{\shortparallel }x^{i_{3}}\}- \\
&&\frac{[(\ _{4}^{\shortparallel }\Psi )^{\ast }]^{2}}{4(\
~_{4}^{\shortparallel }\mathcal{K}[\ _{4}^{\shortparallel }\check{\Psi}%
])^{2}\{~^{\shortparallel }g_{7}^{[0]}-\int d~^{\shortparallel }E[(\
_{4}^{\shortparallel }\Psi )^{2}]^{\ast }/4(~_{4}^{\shortparallel }\mathcal{K%
}[\ _{4}^{\shortparallel }\check{\Psi}])\}}\{d~^{\shortparallel
}E+[~^{\shortparallel }\partial _{i_{3}}(\ _{4}^{\shortparallel }\check{A}%
)]d~^{\shortparallel }x^{i_{3}}\}^{2}.
\end{eqnarray*}

The first four terms for shells $s=1,2$ in (\ref{qellc}) define a LC-variant
of (\ref{qeltorshv1}) for quasi-stationary spacetime solutions with R-flux
contributions encoded in $\ _{1}^{\shortparallel}\mathcal{K}$ and $\
_{2}^{\shortparallel }\mathcal{K}.$ This way, we generate exact solutions of
the systems of nonassociative vacuum Einstein equations (\ref%
{cannonsymparamc2hv}) for $\ ^{\shortparallel }\nabla ^{\star }$ and reduced
to (\ref{sourc1hv}) with trivial (\ref{sourc1cc}) and $\
^{\shortparallel}\nabla ^{\star} \rightarrow \nabla ^{\star }.$ Such
nonassociative vacuum equations with nontrivial real R-flux contributions
were postulated in \cite{aschieri17}. Redefining and generalizing those
systems of nonlinear PDEs in nonholonomic dyadic frames, we can apply the
AFCDM and generate exact and parametric solutions as we explained above.

Finally, we emphasize that quasi-stationary s-metrics (\ref{qellc}) encoding
R-flux effective sources are generic off-diagonal if at least on one of the
four shells there are nontrivial anholonomy relations of type (\ref{anhrel}%
). On phase space, such exact solutions possess a Killing symmetry on $\
^{\shortparallel }\partial ^{7},$ i.e. there are such N-adapted coordinate
systems when the s-metrics do not depend on coordinate $\
^{\shortparallel}p_{7}$ but may depend generically on an energy type
(complex) variable $\ ^{\shortparallel }E.$ We generate quasi-stationary
configurations if the generating and integration functions and generating
sources encoding star and R-flux deformations on respective shells are
chosen to not depend on the $t$-coordinate. In dual form, redefining the
AFCDM in a form to generate solutions with Killing symmetry, for instance,
on $\partial _{3},$ we can construct various classes of locally cosmological
solutions, see details and references in \cite%
{bubuianu17,vacaru18,bubuianu18a}. For models of nonassociative gravity, we
shall construct and study nonassociative cosmological models in our partner
works.

\subsection{Nonassociative nonlinear symmetries for stationary generating
functions and R-flux sources}

Applying the AFCDM for generating exact and parametric solutions in (non)
associative / commutative gravity theories with (non) symmetric metrics and
generalized connections, we discover a new type of nonlinear symmetries of
nonholonomic gravitational and matter field systems with one Killing
symmetry \cite%
{vacaru08aa,vacaru08bb,vacaru08cc,bubuianu17,vacaru18,bubuianu18a}. In this
subsection, we analyze how such nonlinear symmetries relate generating
functions and (effective) sources in the case of quasi-stationary solutions (%
\ref{qeltors}), (\ref{qeltorshv1}), and restrictions to LC-configurations (%
\ref{qellc}). Considering nonlinear symmetries, we can redefine the
generating functions and introduce on dyadic shells some nontrivial
effective cosmological constants $\ _{s}^{\shortparallel }\Lambda _{0}=const$
instead of effective sources $\ _{s}^{\shortparallel }\mathcal{K}(\
_{s}^{\shortparallel }u).$ We can work also with effective polarizations of
cosmological constants for some functions $\ ^{s}\Lambda (\
_{s}^{\shortparallel }u),$ as we explain with respect to formulas (\ref%
{cannonsymparamc1}), (\ref{cannonsymparamcosm}), and (\ref{nonhconstr1}).
The redefined generating functions encode, for instance, R-flux
contributions and parametric star deformations. Fixing certain values $%
_{s}^{\shortparallel }\Lambda _{0}$, allows us to express off-diagonal
solutions in a more simplified form and compute possible parametric
nonassociative deformations of physically important solutions (for instance,
BHs, wormholes and homogeneous and isotropic cosmological configurations
nonholonomically modified into locally anisotropic ones, which is used in
our partner works).

On the shell $s=2,$ we can change the generating data,
\begin{equation*}
(\ _{2}\Psi (\hbar ,\kappa ,x^{i_{1}},y^{3}),\ ~_{2}^{\shortparallel }%
\mathcal{K}(\hbar ,\kappa ,x^{i_{1}},y^{3}))\leftrightarrow (\ _{2}\Phi
(\hbar ,\kappa ,x^{i_{1}},y^{3}),\ _{2}\Lambda _{0}),
\end{equation*}
following formulas%
\begin{eqnarray}
\frac{\lbrack (\ _{2}\Psi )^{2}]^{\diamond }}{\ ~_{2}^{\shortparallel }%
\mathcal{K}} &=&\frac{[(\ _{2}\Phi )^{2}]^{\diamond }}{\ _{2}\Lambda _{0}},%
\mbox{ which can be
integrated as  }  \label{ntransf1} \\
(_{2}\Phi )^{2} &=&\ _{2}\Lambda _{0}\int dx^{3}(~_{2}^{\shortparallel }%
\mathcal{K})^{-1}[(\ _{2}\Psi )^{2}]^{\diamond }\mbox{ and/or }(\ _{2}\Psi
)^{2}=(\ _{2}\Lambda _{0})^{-1}\int dx^{3}(~_{2}^{\shortparallel }\mathcal{K}%
)[(\ _{2}\Phi )^{2}]^{\diamond }.  \label{ntransf2}
\end{eqnarray}%
In result, we can express (see similar computations in detail in section 4.4
of \cite{bubuianu20}),
\begin{eqnarray*}
g_{3}[\ _{2}\Psi ] &=&-\frac{[(\ _{2}\Psi )^{\diamond }]^{2}}{%
4(~_{2}^{\shortparallel }\mathcal{K})^{2}g_{4}[\ _{2}\Psi ]}=g_{3}[\
_{2}\Phi ]=-\frac{1}{g_{4}[\ _{2}\Phi ]}\frac{(\ _{2}\Phi )^{2}[(\ _{2}\Phi
)^{\diamond }]^{2}}{|\ _{2}\Lambda _{0}\int dy^{3}(~_{2}^{\shortparallel }%
\mathcal{K})[(\ _{2}\Phi )^{2}]^{\diamond }|},\mbox{ where } \\
g_{4}[\ _{2}\Psi ] &=&g_{4}^{[0]}-\int d\varphi \frac{\lbrack (\ _{2}\Psi
)^{2}]^{\diamond }}{4(~_{2}^{\shortparallel }\mathcal{K})}=g_{4}[\ _{2}\Phi
]=g_{4}^{[0]}-\frac{(\ _{2}\Phi )^{2}}{4\ _{2}\Lambda }.
\end{eqnarray*}%
\begin{eqnarray*}
w_{i_{1}}(x^{k_{1}},y^{3}) &=&\frac{\partial _{i_{1}}(\ _{2}\Psi )}{(\
_{2}^{\shortmid }\Psi )^{\diamond }}=\frac{\partial _{i_{1}}[(\ _{2}\Psi
)^{2}]}{[(~_{2}^{\shortparallel }\mathcal{K})^{2}]^{\diamond }}=\frac{%
\partial _{i_{1}}\ \int dy^{3}(~_{2}^{\shortparallel }\mathcal{K})\ [(\
_{2}\Phi )^{2}]^{\diamond }}{(~_{2}^{\shortparallel }\mathcal{K})\ [(\
_{2}\Phi )^{2}]^{\diamond }};\mbox{
and } \\
n_{k_{1}}(x^{k_{1}},y^{3}) &=&\ _{1}n_{k_{1}}+\ _{2}n_{k_{1}}\int dy^{3}\
\frac{g_{3}[\ _{2}\Phi ]}{|\ g_{4}[\ _{2}\Phi ]|^{3/2}} \\
&=&\ _{1}n_{k_{1}}+\ _{2}n_{k_{1}}\int dy^{3}\left( \frac{(\ _{2}\Psi
)^{\diamond }}{2~_{2}^{\shortparallel }\mathcal{K}}\right) ^{2}\left\vert
g_{4}^{[0]}(x^{k_{1}})-\int dy^{3}\frac{[(\ _{2}\Psi )^{2}]^{\diamond }}{%
4(~_{2}^{\shortparallel }\mathcal{K})}\right\vert ^{-5/2} \\
&=&\ _{1}n_{k_{1}}+\ _{2}n_{k_{1}}\int dy^{3}\frac{(\ _{2}\Phi )^{2}[(\
_{2}\Phi )^{\diamond }]^{2}}{|\ _{2}\Lambda _{0}\int
dy^{3}(~_{2}^{\shortparallel }\mathcal{K})[(\ _{2}\Phi )^{2}]^{\diamond }|}%
\left\vert g_{4}^{[0]}-\frac{(\ _{2}\Phi )^{2}}{4\ _{2}\Lambda _{0}}%
\right\vert ^{-5/2}.
\end{eqnarray*}

Similar re-definitions of generating functions and sources (\ref{ntransf1})
and (\ref{ntransf2}) can be considered for $s=3,4$ which allows us to prove
that any quasi-stationary solution (\ref{qeltors}) of the nonassociative
nonholonomic dyadic vacuum gravitational equations on phase spaces for the
canonical s-connection is characterized by such important nonlinear
symmetries:{\small
\begin{eqnarray}
s=2: &&\frac{[(\ _{2}\Psi )^{2}]^{\diamond }}{~_{2}^{\shortparallel }%
\mathcal{K}}=\frac{[(\ _{2}\Phi )^{2}]^{\diamond }}{\ _{2}\Lambda _{0}},
\label{nonltransf} \\
&&\mbox{ i.e. }\ (\ _{2}\Phi )^{2}=\ _{2}\Lambda _{0}\int
dx^{3}(~_{2}^{\shortparallel }\mathcal{K})^{-1}[(\ _{2}\Psi )^{2}]^{\diamond
}\mbox{ and/or }(\ _{2}\Psi )^{2}=(\ _{2}\Lambda _{0})^{-1}\int
dx^{3}(~_{2}^{\shortparallel }\mathcal{K})[(\ _{2}\Phi )^{2}]^{\diamond };
\notag
\end{eqnarray}%
\begin{eqnarray}
s=3: &&\frac{~^{\shortparallel }\partial ^{6}[(\ _{3}^{\shortparallel }\Psi
)^{2}]}{~_{3}^{\shortparallel }\mathcal{K}}=\frac{~^{\shortparallel
}\partial ^{6}[(\ _{3}^{\shortparallel }\Phi )^{2}]}{\ _{3}^{\shortparallel
}\Lambda _{0}},  \notag \\
&&\mbox{ i.e. }\ (\ _{3}^{\shortparallel }\Phi )^{2}=\ _{3}^{\shortparallel
}\Lambda _{0}\int d~^{\shortparallel }p_{6}(~_{3}^{\shortparallel }\mathcal{K%
})^{-1}[(\ _{3}^{\shortparallel }\Psi )^{2}]\mbox{ and/or }(\
_{3}^{\shortparallel }\Psi )^{2}=(\ _{3}^{\shortparallel }\Lambda
_{0})^{-1}\int d~^{\shortparallel }p_{6}(~_{3}^{\shortparallel }\mathcal{K}%
)[(\ _{3}^{\shortparallel }\Phi )^{2}];  \notag
\end{eqnarray}%
} {\small
\begin{eqnarray}
s=4: &&\frac{[(\ _{4}^{\shortparallel }\Psi )^{2}]^{\ast }}{%
~_{4}^{\shortparallel }\mathcal{K}}=\frac{[(\ _{4}^{\shortparallel }\Phi
)^{2}]^{\ast }}{\ _{4}^{\shortparallel }\Lambda },  \notag \\
&&\mbox{ i.e. }\ (\ _{4}^{\shortparallel }\Phi )^{2}=\ _{4}^{\shortparallel
}\Lambda \int d~^{\shortparallel }E(~_{4}^{\shortparallel }\mathcal{K}%
)^{-1}[(\ _{4}^{\shortparallel }\Psi )^{2}]^{\ast }\mbox{ and/or }(\
_{4}^{\shortparallel }\Psi )^{2}=(\ _{4}^{\shortparallel }\Lambda )^{-1}\int
d~^{\shortparallel }E(~_{4}^{\shortparallel }\mathcal{K})[(\
_{4}^{\shortparallel }\Phi )^{2}]^{\ast }.  \notag
\end{eqnarray}%
}

Using nonlinear symmetries (\ref{nonltransf}), we can write the quadratic
element (\ref{qeltors}) for off-diagonal quasi-stationary nonholonomic
vacuum solutions with Killing symmetry on $\ ^{\shortparallel }p_{7}$ can be
written in equivalent form encoding the nonlinear symmetries of generating
functions, generating sources of R-flux origin and respective effective
shell cosmological constants,{\small
\begin{eqnarray}
d\widehat{s}^{2} &=&~^{\shortparallel }g_{\alpha _{s}\beta _{s}}(\hbar
,\kappa ,x^{k},y^{3},~^{\shortparallel }p_{a_{3}},~^{\shortparallel }E,\ \
_{s}^{\shortparallel }\Phi ,\ _{s}^{\shortparallel }\Lambda
_{0})d~^{\shortparallel }u^{\alpha _{s}}d~^{\shortparallel }u^{\beta
_{s}}=e^{\psi (\hbar ,\kappa ,x^{k_{1}})}[(dx^{1})^{2}+(dx^{2})^{2}]
\label{offdiagcosmcsh} \\
&&-\frac{(\ _{2}\Phi )^{2}[(\ _{2}\Phi )^{\diamond }]^{2}}{|\ _{2}\Lambda
_{0}\int dx^{3}(~_{2}^{\shortparallel }\mathcal{K})[(\ _{2}\Phi
)^{2}]^{\diamond }|[g_{4}^{[0]}-(\ _{2}\Phi )^{2}/4\ _{2}\Lambda _{0}]}%
\{dx^{3}+\frac{\partial _{i_{1}}\ \int dx^{3}(~_{2}^{\shortparallel }%
\mathcal{K})\ [(\ _{2}\Phi )^{2}]^{\diamond }}{(~_{2}^{\shortparallel }%
\mathcal{K})\ [(\ _{2}\Phi )^{2}]^{\diamond }}dx^{i_{1}}\}^{2}-  \notag \\
&&\{g_{4}^{[0]}-\frac{(\ _{2}\Phi )^{2}}{4\ _{2}\Lambda _{0}}\}\{dt+[\
_{1}n_{k_{1}}+\ _{2}n_{k_{1}}\int dy^{3}\frac{(\ _{2}\Phi )^{2}[(\ _{2}\Phi
)^{\diamond }]^{2}}{|\ _{2}\Lambda _{0}\int dy^{3}(~_{2}^{\shortparallel }%
\mathcal{K})[(\ _{2}\Phi )^{2}]^{\diamond }|[g_{4}^{[0]}-(\ _{2}\Phi
)^{2}/4\ _{2}\Lambda _{0}]^{5/2}}]\}+  \notag
\end{eqnarray}%
\begin{eqnarray*}
&&(\ ^{\shortparallel }g_{5}^{[0]}-\frac{(\ _{3}^{\shortparallel }\Phi )^{2}%
}{4\ _{3}^{\shortparallel }\Lambda _{0}})\{d~^{\shortparallel }p_{5}+[\
_{1}n_{k_{2}}+\ _{2}n_{k_{2}}\int d~^{\shortparallel }p_{6}\frac{(\
_{3}^{\shortparallel }\Phi )^{2}[~^{\shortparallel }\partial ^{6}(\
_{3}^{\shortparallel }\Phi )]^{2}}{|\ _{3}^{\shortparallel }\Lambda _{0}\int
d~^{\shortparallel }p_{6}\ (~_{3}^{\shortparallel }\mathcal{K}%
)~^{\shortparallel }\partial ^{6}[(\ _{3}^{\shortparallel }\Phi
)^{2}]|[~^{\shortparallel }g_{5}^{[0]}-(\ _{3}^{\shortparallel }\Phi
)^{2}/4\ _{3}^{\shortparallel }\Lambda _{0}]^{5/2}}]dx^{k_{2}}\}- \\
&&\frac{(\ _{3}^{\shortparallel }\Phi )^{2}[~^{\shortparallel }\partial
^{6}(\ _{3}^{\shortparallel }\Phi )]^{2}}{|\ _{3}^{\shortparallel }\Lambda
_{0}\int d~^{\shortparallel }p_{6}(~_{3}^{\shortparallel }\mathcal{K})\
~^{\shortparallel }\partial ^{6}[(\ _{3}^{\shortparallel }\Phi )^{2}]|\
[~^{\shortparallel }g_{5}^{[0]}-(\ _{3}^{\shortparallel }\Phi )^{2}/4\
_{3}^{\shortparallel }\Lambda _{0}]}\{d~^{\shortparallel }p_{6}+\frac{%
\partial _{i_{2}}\ \int d~^{\shortparallel }p_{6}(~_{3}^{\shortparallel }%
\mathcal{K})\ ~^{\shortparallel }\partial ^{6}[(\ _{3}^{\shortparallel }\Phi
)^{2}]}{(~_{3}^{\shortparallel }\mathcal{K})\ ~^{\shortparallel }\partial
^{6}[(\ _{3}^{\shortparallel }\Phi )^{2}]}dx^{i_{2}}\}^{2}+
\end{eqnarray*}%
}%
\begin{eqnarray*}
&&(\ ^{\shortparallel }g_{7}^{[0]}-\frac{(\ _{4}^{\shortparallel }\Phi )^{2}%
}{4\ _{4}^{\shortparallel }\Lambda _{0}})\{d~^{\shortparallel }p_{7}+[\
_{1}^{\shortparallel }n_{k_{3}}+\ _{2}^{\shortparallel }n_{k_{3}}\int
d~^{\shortparallel }E\frac{(\ _{4}^{\shortparallel }\Phi )^{2}[(\
_{4}^{\shortparallel }\Phi )^{\ast }]^{2}}{|\ \ _{4}^{\shortparallel
}\Lambda _{0}\int d~^{\shortparallel }E(~_{4}^{\shortparallel }\mathcal{K}%
)[(\ _{4}^{\shortparallel }\Phi )^{2}]^{\ast }|\ [~^{\shortparallel
}g_{7}^{[0]}-(\ _{4}^{\shortparallel }\Phi )^{2}/4\ \ _{4}^{\shortparallel
}\Lambda _{0}]^{5/2}}] \\
&& d\ ^{\shortparallel }x^{k_{3}}\}- \frac{(\ _{4}^{\shortparallel }\Phi
)^{2}[(\ _{4}^{\shortparallel }\Phi )^{\ast }]^{2}}{|\ _{4}^{\shortparallel
}\Lambda _{0}\int d~^{\shortparallel }E(~_{4}^{\shortparallel }\mathcal{K}%
)[(\ _{4}^{\shortparallel }\Phi )^{2}]^{\diamond }|[~^{\shortparallel
}g_{7}^{[0]}-(\ _{4}^{\shortmid }\Phi )^{2}/4\ \ _{4}^{\shortparallel
}\Lambda _{0}]}\{d~^{\shortparallel }E+\frac{~^{\shortparallel }\partial
_{i_{3}}\ \int d~^{\shortparallel }E(~_{4}^{\shortparallel }\mathcal{K})\
[(\ _{4}^{\shortparallel }\Phi )^{2}]^{\ast }}{(~_{4}^{\shortparallel }%
\mathcal{K})[(\ _{4}^{\shortparallel }\Phi )^{2}]^{\ast }}d~^{\shortparallel
}x^{i_{3}}\}^{2},
\end{eqnarray*}%
{\small for indices: $%
i_{1},j_{1},k_{1},...=1,2;i_{2},j_{2},k_{2},...=1,2,3,4;i_{3},j_{3},k_{3},...=1,2,...6;x^{3}=\varphi ,y^{4}=t,~^{\shortparallel }p_{8}=~^{\shortparallel }E;
$ and
\begin{eqnarray*}
&&\mbox{generating functions: }\psi (\hbar ,\kappa ,x^{k_{1}});\ _{2}\Phi
(\hbar ,\kappa ,x^{k_{1}}y^{3});\ _{3}^{\shortparallel }\Phi (\hbar ,\kappa
,x^{k_{2}},~^{\shortparallel }p_{6});\ _{4}^{\shortparallel }\Phi (\hbar
,\kappa ,~^{\shortparallel }x^{k_{3}},~^{\shortparallel }E); \\
&&\mbox{generating sources:}~_{1}^{\shortparallel }\mathcal{K}(\hbar ,\kappa
,x^{k_{1}});\ ~_{2}^{\shortparallel }\mathcal{K}(\hbar ,\kappa
,x^{k_{1}},y^{3});\ ~_{3}^{\shortparallel }\mathcal{K}(\hbar ,\kappa
,x^{k_{2}},~^{\shortparallel }p_{6});\ ~_{4}^{\shortparallel }\mathcal{K}%
(\hbar ,\kappa ,~^{\shortparallel }x^{k_{3}},~^{\shortparallel }E); \\
&&\mbox{integr. functions:}g_{4}^{[0]}(\hbar ,\kappa ,x^{k_{1}}),\
_{1}n_{k_{1}}(\hbar ,\kappa ,x^{j_{1}}),\ _{2}n_{k_{1}}(\hbar ,\kappa
,x^{j_{1}});~^{\shortparallel }g_{5}^{[0]}(\hbar ,\kappa ,x^{k_{2}}),\
_{1}n_{k_{2}}(\hbar ,\kappa ,x^{j_{2}}),\ _{2}n_{k_{2}}(\hbar ,\kappa
,x^{j_{2}}); \\
&&~^{\shortparallel }g_{7}^{[0]}(\hbar ,\kappa ,~^{\shortparallel
}x^{j_{3}}),\ _{1}^{\shortparallel }n_{k_{3}}(\hbar ,\kappa
,~^{\shortparallel }x^{j_{3}}),\ _{2}^{\shortparallel }n_{k_{3}}(\hbar
,\kappa ,~^{\shortparallel }x^{j_{3}}).
\end{eqnarray*}%
}

In a similar form, we can re-define the generating functions and introduce
effective cosmological constants $\ _{s}^{\shortparallel }\Lambda _{0}$ for
LC-configurations and quasi-stationary s-metrics (\ref{qellc}).

For convenience, we present the Appendix \ref{appendixb} where the procedure
of integrating nonassociative nonholonomic vacuum gravitational equations is
studied in details for the case of effective source encoding R-flux and star
deformation parametric terms. We provide a nonassociative generalization of
section 5 in \cite{bubuianu20} following such goals:

\begin{enumerate}
\item We show how certain coefficients of quasi-stationary s-metrics can be
chosen as shell generating functions which is important for constructing new
classes of exact and parametric solutions with R-flux deformations from
certain nonholonomic but associative/ commutative physically important
solutions.

\item The AFCDM is re-formulated for conventional gravitational polarization
functions for nonholonomic deformations of arbitrary s-metrics (they can be
nonassociative or associative/commutative ones) into certain new classes of
exact solutions determining general quasi-stationary nonassociative vacuum
configurations.

\item A small parametric decomposition formalism is elaborated for
off-diagonal deformations and diagonalization procedures for generating new
classes of exact and parametric solutions encoding nonassociative R-flux
contributions on phase spaces and spacetime configurations.

\item There are studied three examples of generic off-diagonal metrics for
canonical and Levi-Civita "pure" vacuum configurations (\ref{purestarvacuum}%
) when possible real cosmological constants are compensated by effective
ones, for instance, by R-flux contributions.
\end{enumerate}

\section{Conclusions}

\label{sec6} In this work, we have developed the anholonomic frame and
connection deformation method, AFCDM \cite%
{bubuianu20,bubuianu19,bubuianu17,vacaru19,vacaru20}, in a nonholonomic
dyadic form which allows us to prove a general decoupling and integration
property of nonassociative vacuum gravitational equations introduced in \cite%
{blumenhagen16,aschieri17}. The star product deformation approach to
geometric and "non-geometric" models with R-flux contributions from string
theories was completed with a nonassociative nonholonomic geometric
formalism involving nonsymmetric and symmetric metric structures and
generalized (non) linear connections in the first our partner paper \cite%
{partner01}. A very important result proven in \cite{aschieri17} for
parametric decompositions on $\hbar $ and $\kappa $ (respectively, the
Planck and string constants) is that there are nontrivial real R-flux
contributions on background (associative and commutative) pseudo-Riemannian
spacetimes. In our approach, such nonassociative parametric deformations are
encoded in effective sources and, via nonlinear symmetries, related to
corresponding classes of generating function and prescribed values of shell
constants.

Nonassociative extensions of \ GR with parametric R-flux deformations are
defined on phase spaces which can be modelled as cotangent Lorents bundles
enabled with additional algebraic (for instance, quasi-Hopf / octonion /
quaternion structures), star products, complexified momentum like
coordinates, etc. On real, and commutative background, phase spaces, such
extensions of general relativity and string gravity theories are
characterized by modified dispersion relations, MDRs, of type $c^{2}%
\overrightarrow{\mathbf{p}}^{2}-E^{2}+c^{4}m^{2}=\varpi (E,\overrightarrow{%
\mathbf{p}},m;\hbar ,\kappa )$, see details and motivations for
nonassociative gravity in \cite{partner01}. Such modified gravity theories,
MGTs, with MDRs determined by indicators $\varpi (x^{i},E,\overrightarrow{%
\mathbf{p}},m;\ell _{P}),$ where $m$ is a mass parameter and $\ell _{P}$ is
the Planck lengths, were axiomatized and studied in \cite%
{vacaru18,bubuianu18a}, see also references therein. On commutative and
noncommutative generalized Einstein, Finsler-Lagrange-Hamilton and
Einsten-Eisenhart-Moffat models, we cite respectively \cite%
{einstein25,einstein45,eisenhart51,eisenhart52,moffat79,moffat95,
moffat95a,moffat00,vacaru08aa,vacaru08bb,vacaru08cc}, etc.). The metric
structure on such (non) associative/ commutative phase spaces depends not
only on spacetime coordinates but also on some conventional velocity /
momentum like variables, for instance, with effective metrics $g_{\alpha
\beta }(x^{k},p_{a}),$ which for R-flux contributions deform into symmetric
and nonsymmetric metric structures adapted to certain classes of nonlinear
connections determined by respective MDR indicators.

To construct exact and parametric solutions in nonassociative gravity and
geometric flow theories, which for various classes of above mentioned (non)
associative/ commutative MGTs are characterized by generic diagonal metrics,
auxiliary canonical d- and s-connections and respective nonholonomic
constraints to zero torsion LC-configurations, we can apply the results of
Theorems 4.1- 4.3 and 5.1, 5.2 from \cite{bubuianu20}. Considering phase
spaces with momentum variables multiplied to complex unity and effective
R-flux sources, we summarize the main results of this article:

\begin{enumerate}
\item Nonassociative vacuum Einstein equations can be decoupled in general
off-diagonal forms with symmetric and nonsymmetric star metrics determined
by star parametric R-flux deformations, when the solutions depend on all
phase and spacetime coordinates being determined by respective classes of
generating functions and (effective) sources and integration functions.

\item Quasi-stationary off-diagonal solutions with one phase space Killing
symmetry and time-like spacetime Killing symmetry on the first two
nonholonomic dyadic shell can be constructed in splicit forms. Using
additional nonholonomic constraints, we can extract LC-configurations,
define spacetime projections encoding R-flux contributions, perform
diagonalization procedures etc.

\item We defined new types of nonassociative nonlinear parametric symmetries
relating generating functions, generating effective R-flux sources, and
certain effective shell cosmological constants which are important for
constructing exact and parametric solutions with generic off-diagonal
symmetric and nonsymmetric background and star deformed metrics, generalized
connections and zero torsion configurations.

\item Small parametric decompositions on $\hbar $ and $\kappa $ in
nonassociative vacuum gravity have a natural R-flux deformation
interpretation and can be performed in a self-consistent form defining star
nonholonomic deformations of certain prime (associative) metrics into new
classes of target symmetric and nonsymetric metrics.
\end{enumerate}

We note that the AFCDM can be applied for constructing real (with all types
local signature) and complex solutions for vacuum and non-vacuum generalized
(non) associative/ commutative nonholonomic gravitational and matter field
equations as we reviewed in \cite{vacaru18,bubuianu18a,partner01}. In
general, it is not clear how to evaluate the physical importance of complex
R-flux contributions into parametric decompositions of Ricci tensors,
effective sources, and (non) symmetric metrics. Nevertheless, we can always
impose such nonholonomic constraints when complex terms vanish for certain
N- / s-adapted configurations and the star R-flux deformations result in
real terms and decoupling of respective physically important nonlinear
systems of PDEs. An alternative way, is to work in so called
Finsler-Lagrange-Hamilton variables with almost complex/ symplectic
structures which allow a real relativistic interpretation of nonassociative
nonholonomic gravity and geometric flow theories. Here we also note that
such constructions can be used for deformation, nonholonomic gauge like and
A-brane quantization of nonassociative gravity theories as we considered in
our previous works, see reviews of results in \cite%
{vacaru03,vacaru05a,vacaru09a,vacaru16,vacaru18}.

Finally, we note that the geometric formalism and analytic methods on
decoupling and integrating in general form nonassociative gravity theories
elaborated in this and first partner work \cite{partner01} will be applied
in a series of related partner papers (under elaboration) on nonassociative
black hole solutions, locally anisotroic cosmological solutions with
nonassociatvie quasi-periods and phase space/ spacetime quasi-crystal
structure and theory of nonassociative geometric and quantum information
flows. For associative models and geometric methods, we cite our previous
works \cite{bubuianu17b,bubuianu17,vacaru19,vacaru20}.

\vskip6pt

\textbf{Acknowledgments:} This work develops for nonassociative geometry and
gravity some research programs on geometry and physics, during 2006-2015,
supported by senior fellowships at the Perimeter Institute and Fields
Institute (Ontario, Canada), CERN (Geneva, Switzerland) and Max Planck
Institut f\"{u}r Physik / Werner Heisenberg Institut, M\"{u}nchen (Germany).
SV is grateful to professors V. G. Kupriyanov, D. L\"{u}st, N. Mavromatos,
J. Moffat, D. Singleton and P. Stavrinos for respective hosting of short/
long terms visits, seminars, and/or discussing important ideas and
preliminary results.

\appendix

\setcounter{equation}{0} \renewcommand{\theequation}
{A.\arabic{equation}} \setcounter{subsection}{0}
\renewcommand{\thesubsection}
{A.\arabic{subsection}}

\section{Nonassociative differential geometry with s-connections \&
non\-symmet\-r\-ic met\-rics}

\label{appendixa}

In this appendix, we re-formulate the nonassociative differential geometry
with N-connections and nonsymmetric metrics \cite{partner01} in dyadic
s-variables involving quasi-Hopf structure. For holonomic configurations
with zero cosmological constants and effective sources, we obtain geometric
models from \cite{aschieri17}. We omit constructions with s-transforms from
some subsections of section 3 in \cite{partner01}, which define nonholonomic
generalizations of nonassociative geometry elaborated in \cite{blumenhagen16}%
. Such formulas can be obtained if we re-define respective frame, tensor
product and star product transforms (i.e. of $\ ^{\shortparallel}\mathbf{e}%
_{\alpha _{s}}\rightarrow \ ^{\shortparallel}\partial _{\alpha _{s}},
\otimes _{\star s}\rightarrow \otimes _{\star },\star _{s}\rightarrow \star $
etc.). There are provided abstract (with s-labels) and s-adapted coefficient
formulas, with respect to nonholonomic dyadic frames, for star deformed
connections, s-connections, and sketched the proofs of formulas for their
torsion and curvature dyadic shell distinguished tensors, s-tensors.

\subsection{Dyadic s-adapted linear connections and quasi-Hopf s-structures}

\label{ssdconstar}The (non) associative d-connection formalism elaborated in
section 3 of partner work \cite{partner01} can be extended for s-tensors and
differential forms on the $\mathcal{A}_{s}^{\star }$--bimodule $\emph{Vec}%
_{\star _{s}}$ of s-vector fields and the $\Omega _{s\star }^{\natural }$%
--bimodule $\emph{Vec}_{\star _{s}}^{\natural }=\emph{Vec}_{\star _{s}}$ $%
\otimes _{\star _{s}}\Omega _{s\star }^{\natural }(\mathcal{M},\
_{s}^{\shortparallel }N)$ as we considered in subsection \ref{ssdtqh}.

For a dyadic with $s=1,2,3,4$, a star s-connection can be defined as a
s-adapted linear map $h$%
\begin{eqnarray}
\ _{s}^{\shortparallel }\mathbf{D}^{\star } &:&\ \emph{Vec}_{\star
_{s}}\rightarrow \emph{Vec}_{\star _{s}}\otimes _{\star _{s}}\mathbf{\ }%
\Omega _{\star }^{1},\mbox{  i. e. }\mathbf{v}\rightarrow \
_{s}^{\shortparallel }D^{\star }\mathbf{\mathbf{\mathbf{\mathbf{\
^{\shortparallel }}}}v}=\mathbf{\mathbf{\mathbf{\mathbf{\ ^{\shortparallel }}%
}}v}^{\alpha _{s}}\otimes _{\star _{s}}\mathbf{\mathbf{\mathbf{\mathbf{\
^{\shortparallel }}}}}\omega _{\alpha _{s}}=(v^{i_{s}}\otimes _{\star
_{s}}\omega _{i_{s}},\mathbf{\mathbf{\mathbf{\mathbf{\ ^{\shortparallel }}}}}%
v_{a_{s}}\otimes _{\star _{s}}\mathbf{\mathbf{\mathbf{\mathbf{\
^{\shortparallel }}}}}\omega ^{a_{s}});  \notag \\
\ _{s}^{\shortparallel }\mathbf{D}^{\star } &=&(\ _{h_{1}}^{\shortparallel }%
\mathbf{D}^{\star },\ _{v_{2}}^{\shortparallel }\mathbf{D}^{\star },\
_{c_{3}}^{\shortparallel }\mathbf{D}^{\star },\ _{c_{4}}^{\shortparallel }%
\mathbf{D}^{\star }):\ \emph{Vec}_{\star _{s}}=(h_{1}\emph{Vec}_{\star
_{s}},v_{2}\emph{Vec}_{\star _{s}},c_{3}\emph{Vec}_{\star _{s}},c_{4}\emph{%
Vec}_{\star _{s}})\rightarrow  \notag \\
&&(h_{1}\emph{Vec}_{\star _{s}},v_{2}\emph{Vec}_{\star _{s}},c_{3}\emph{Vec}%
_{\star _{s}},c_{4}\emph{Vec}_{\star _{s}})\otimes _{\star _{s}}(\
h_{1}\Omega _{\star }^{\natural },\ v_{2}\Omega _{\star }^{\natural },\
c_{3}\Omega _{\star }^{\natural },c_{4}\Omega _{\star }^{\natural }),
\label{sconhopf}
\end{eqnarray}%
with h- / v-, c-splitting and for $\mathbf{\mathbf{\mathbf{\mathbf{\
^{\shortparallel }}}}v}^{\alpha _{s}}\otimes _{\star _{s}}\mathbf{\mathbf{%
\mathbf{\mathbf{\ ^{\shortparallel }}}}}\omega _{\alpha _{s}}\in \emph{Vec}%
_{\star _{s}}\otimes _{\star _{s}}\mathbf{\ }\Omega _{s\star }^{1}.$

Here we note that the adjoint action of a s-vector $\
_{s}^{\shortparallel}\xi \in U\emph{Vec}^{\mathcal{F}}(\mathcal{M},\
_{s}^{\shortparallel }N)$ results in a s-adapted linear map $\
_{s\xi}^{\shortparallel }\mathbf{D}^{\star }\mathbf{:\ }\emph{Vec}_{\star
_{s}}\rightarrow \emph{Vec}_{\star _{s}}\otimes _{\star _{s}}\ \Omega
_{\star }^{1},$ which is also a s-connection. This property exists for all
types of s-connections in nonassociative nonholonomic geometry when an
infinite number of covariant star nonholonomic calculi can be elaborated. We
shall follow a "minimal" way for a s-connection $\ _{s}^{\shortparallel }%
\mathbf{D}^{\star }$ (\ref{sconhopf}) with corresponding extensions to a
s-adapted covariant derivative for s-vector filed values in the exterior
s-algebra $\emph{Vec}_{\star _{s}}^{\natural }=\emph{Vec}_{\star s}\otimes
_{\star s}\ \Omega _{s\star }^{\natural}$, see details for N-configurations
in section 3.1.2 of \cite{partner01}. We omit such a tedious geometric
dyadic calculus which in local coordinate (co) bases is similar to formula
(4.8) in \cite{aschieri17}. Usually, it is possible to follow a geometric
symbolic rule when the constructions with quasi-Hopf algebras and respective
nonassociative covariant calculus can be extended in N- and s-adapted forms
when (using notations from that and our works), for instance, $\partial
\rightarrow \ ^{\shortparallel }\mathbf{e\rightarrow }\ _{s}^{\shortparallel
}\mathbf{e}, \bigtriangledown _{z}^{\star }\rightarrow \ ^{\shortparallel }%
\mathbf{D}\rightarrow \ _{s}^{\shortparallel }\mathbf{D}$ and with actions
of absolute differential operators $d_{\bigtriangledown _{z}^{\star }}
\rightarrow \ _{\mathbf{D}^{\star }}^{\shortparallel } \mathbf{d\rightarrow
\ _{s\mathbf{D}^{\star }}^{\shortparallel }\mathbf{d}.}$

For a star deformed d-connection (\ref{sconhopf}), the s-adapted
coefficients $\ _{s}^{\shortparallel }\mathbf{D}^{\star }=\{\mathbf{\
^{\shortparallel }\Gamma }_{\star \beta _{s}\gamma _{s}}^{\alpha _{s}}\in
\mathcal{A}_{s}^{\star }\}$ are computed following such definitions and
formulas
\begin{eqnarray}
\ _{s}^{\shortparallel }\mathbf{D}^{\star }\ ^{\shortparallel }\mathbf{e}%
_{\alpha _{s}}:= &&\ ^{\shortparallel }\mathbf{e}_{\beta _{s}}\otimes
_{\star _{s}}\mathbf{\ ^{\shortparallel }\Gamma }_{\star \alpha _{s}}^{\beta
_{s}},\mbox{ where }  \label{aux300} \\
&&\mathbf{\ ^{\shortparallel }\Gamma }_{\star \beta _{s}}^{\alpha _{s}}=%
\mathbf{\ ^{\shortparallel }\Gamma }_{\star \beta _{s}\gamma _{s}}^{\alpha
_{s}}\star _{s}\ ^{\shortparallel }\mathbf{e}^{\gamma _{s}}%
\mbox{ is a star
s-connection 1-form; in brief,} \mathbf{\ _{s\star }^{\shortparallel }\Gamma
}=\mathbf{\ _{\star }^{\shortparallel }\Gamma }_{\gamma _{s}}\ \star _{s}\
^{\shortparallel }\mathbf{e}^{\gamma _{s}}.  \notag
\end{eqnarray}%
Using $\mathbf{\mathbf{\mathbf{\mathbf{^{\shortparallel }}}}D}_{\
^{\shortparallel }\mathbf{e}_{\beta _{s}}}^{\star }:=\mathbf{\
^{\shortparallel }D}_{\ \beta _{s}}^{\star },$ we obtain
\begin{equation}
\mathbf{\mathbf{\mathbf{\mathbf{^{\shortparallel }}}}D}_{\ \alpha
_{s}}^{\star }\ ^{\shortparallel }\mathbf{e}_{\beta _{s}}:=\left\langle
\mathbf{\mathbf{\mathbf{\mathbf{\ }}}}_{s}^{\shortparallel }\mathbf{D}%
^{\star }\ ^{\shortparallel }\mathbf{e}_{\beta _{s}},\mathbf{\mathbf{\mathbf{%
\mathbf{\ }}}}\ ^{\shortparallel }\mathbf{e}_{\alpha _{s}}\right\rangle
_{\star _{s}}=\left\langle \left( \ ^{\shortparallel }\mathbf{e}_{\gamma
_{s}}\otimes _{\star _{s}}(\mathbf{\ ^{\shortparallel }\Gamma }_{\star \beta
_{s}\tau _{s}}^{\gamma _{s}}\star _{s}\ ^{\shortparallel }\mathbf{e}^{\tau
_{s}})\right) ,\mathbf{\mathbf{\mathbf{\mathbf{\ }}}}\ ^{\shortparallel }%
\mathbf{e}_{\alpha _{s}}\right\rangle _{\star _{s}}=\ ^{\shortparallel }%
\mathbf{e}_{\gamma _{s}}\star _{s}\mathbf{\ ^{\shortparallel }\Gamma }%
_{\star \beta _{s}\alpha _{s}}^{\gamma _{s}}.  \notag
\end{equation}%
The s-connection coefficients $\mathbf{\ ^{\shortparallel }\Gamma }%
_{\star\beta _{s}\gamma _{s}}^{\alpha _{s}}$ and respective dyadic 1-form $\
_{s}^{\shortparallel }\mathbf{\Gamma }_{\star }$ are used for computing
s-adapted covariant derivatives:%
\begin{eqnarray}
&& \ _{s}^{\shortparallel }\mathbf{D}^{\star }\ _{s}^{\shortparallel }%
\mathbf{v} =\ ^{\shortparallel }\mathbf{e}_{\gamma _{s}}\otimes _{\star
_{s}}(\ _{s}^{\shortparallel }\mathbf{d}\ ^{\shortparallel }\mathbf{v}%
^{\gamma _{s}}+\mathbf{\ ^{\shortparallel }\Gamma }_{\star \alpha
_{s}}^{\gamma _{s}}\star _{s}\mathbf{\ ^{\shortparallel }v}^{\alpha _{s}}),%
\mbox{ for s-vector }\ _{s}^{\shortparallel }\mathbf{v=\ ^{\shortparallel }%
\mathbf{e}}_{\alpha _{s}}\star _{s}\mathbf{\ ^{\shortparallel }v}^{\alpha
_{s}},\mathbf{\ ^{\shortparallel }v}^{\alpha _{s}}\in \mathcal{A}_{s}^{\star
};  \notag \\
&& \ _{s\mathbf{D}^{\star }}^{\shortparallel }\mathbf{d}(\ ^{\shortparallel }%
\mathbf{e}_{\gamma _{s}}\otimes _{\star _{s}}\ \mathbf{\mathbf{\mathbf{%
\mathbf{^{\shortparallel }}}}}\omega ^{\gamma _{s}}) = \ ^{\shortparallel }%
\mathbf{e}_{\gamma _{s}}\otimes _{\star _{s}}(\ ^{\shortparallel }\mathbf{d%
\mathbf{\mathbf{\mathbf{\ ^{\shortparallel }}}}}\omega ^{\gamma _{s}}+\
^{\shortparallel }\mathbf{\Gamma }_{\star \alpha _{s}}^{\gamma _{s}}\star
_{s}\mathbf{\mathbf{\mathbf{\mathbf{\ ^{\shortparallel }}}}}\omega ^{\alpha
_{s}}),\mbox{
for }\mathbf{\mathbf{\mathbf{\mathbf{\ ^{\shortparallel }}}}}\omega ^{\alpha
_{s}}\in \mathbf{\ }\Omega _{s\star }^{\natural }.  \label{aux302}
\end{eqnarray}

We define s-components of a star deformed s-connection by introducing $\
^{\shortparallel }\mathbf{e}_{i_{s}}$ and $\ ^{\shortparallel }e^{b_{s}}$ in
formulas (\ref{aux300}) - (\ref{aux302}). For instance, we obtain a dyadic
splitting for $\mathbf{\mathbf{\mathbf{\mathbf{^{\shortparallel }}}}D}_{\
k_{s}}^{\star }\ ^{\shortparallel }e^{b_{s}}=\ ^{\shortparallel}
e^{a_{s}}\star _{s} \mathbf{\ ^{\shortparallel }}L_{\star a_{s}\ k_{s}}^{\
b_{s}}.$ This way, acting on respective adapted (co) bases, we obtain a
s-decomposition,
\begin{eqnarray}
\ \ _{s}^{\shortparallel }\mathbf{D}^{\star } &=&\{\ \ ^{\shortparallel }%
\mathbf{\Gamma }_{\star \alpha _{s}\beta _{s}}^{\gamma _{s}}=(\ \
^{\shortparallel }L_{\star j_{1}k_{1}}^{i_{1}},\ \ ^{\shortparallel
}L_{\star b_{2}\ k_{1}}^{a_{2}},\ \ ^{\shortparallel }C_{\star \
j_{1}c_{2}}^{i_{1}\ },\ \ ^{\shortparallel }C_{\star b_{2}c_{2}}^{a_{2}};\ \
\label{irevndecomdc} \\
&&\ \ ^{\shortparallel }L_{\star j_{2}k_{2}}^{i_{2}},\ \ ^{\shortparallel
}L_{\star a_{3}k_{2}}^{\ b_{3}},\ \ ^{\shortparallel }C_{\star \
j_{2}}^{i_{2}\ c_{3}},\ \ ^{\shortparallel }C_{\star a_{3}b_{3}}^{\quad
c_{3}};\ \ ^{\shortparallel }L_{\star j_{3}k_{3}}^{i_{3}},\ \
^{\shortparallel }L_{\star a_{4}\ k_{3}}^{\ b_{4}},\ \ ^{\shortparallel
}C_{\star \ j_{3}}^{i_{3}\ c_{4}},\ \ ^{\shortparallel }C_{\star
a_{4}b_{4}}^{\quad c_{4}})\},  \notag
\end{eqnarray}%
where $i_{1},j_{1},k_{1}=1,2$ (such indices are h-indices)$%
;a_{2},b_{2},c_{2}=3,4$ (such indices are $v$-indices)$%
;i_{2},j_{2},;k_{2}=1,2,3,4$ (such indices for 1,2 are h-indices and, for
3,4, are v-indices)$;a_{2},b_{2},c_{2}=5,6$ (such indices a c-indices); $%
i_{3},j_{3},k_{3}=1,2,3,4,5,6$ (such indices for 1,2 are h-indices; for 3,4,
are v-indices; for 5,6, are c-indices)$;a_{4},b_{4},c_{4}=7,8$ (such indices
are also c-indices).

Similar s-adapted coefficients can be computed for star s-deformations of $%
T_{s}\mathbf{TV}$ and $T_{s}\mathbf{T}^{\ast }\mathbf{V}$ but we omit such
formulas which are similar to (\ref{irevndecomdc}) with respective changing
of v- to c-indices, and labels $\ \ $"$^{\shortparallel }$"$\rightarrow \ $"$%
^{\shortmid }$", for respective shall, or not putting such duality symbols
if not necessary.

For quasi-Hopf s-structures, we can also consider left actions of covariant
s-derivatives, which can be formulated in terms of the \textbf{dual star
s-connection} $\ _{s\star}^{\shortparallel}\mathbf{D.}$ Such details were
considered in a subsection related to formula (66) in \cite{partner01} for
the case of dual star d-connections. We omit dubbing of those constructions
for s-structures because all formulas can be obtained by geometric
similarity. In abstract geometric form but using only holonomic bases, such
constructions are elaborated in sections 4.2 and 4.3 of \cite{aschieri17}.
Respective s-adapted coefficient formulas can be written by geometric
analogy and/or derived for explicit actions on s-elongated (co) bases as we
computed above.

\subsection{Star generalizations of the torsion, Riemann and Ricci s-tensors}

We show how such geometric constructions can be performed in abstract
s-adapted coefficient forms using N-adapted coefficient formulas from
section 3 in \cite{partner01} generalizing in noholonomic (if necessary,
dyadic) forms respective definitions and holonomic formulas from \cite%
{aschieri17}.

\subsubsection{Nonassociative star s-torsions for s-connections and
quasi-Hopf s-structures}

A star deformed s-torsion $\ _{s\star }^{\shortparallel }\mathcal{T}\in
\emph{Vec}_{\star_{N}} \otimes _{\star _{N}}\Omega _{\star }^{2}$ of
nonassociative s-connection $\ _{s}^{\shortparallel }\mathbf{D}^{\star }$ (%
\ref{sconhopf}) can be defined as a s-adapted map with identity $\langle \
^{\shortparallel}\mathbf{e}_{\alpha _{s}}\otimes _{\star_{s}}\
^{\shortparallel}\mathbf{e}^{\alpha _{s}},\ \mathbf{\rangle }_{\star _{s}}:
\emph{Vec}_{\star _{s}}\rightarrow \emph{Vec}_{\star _{s}}$, which can be
expanded to any s-vector field $\mathbf{\ ^{\shortparallel }v=\
^{\shortparallel }\mathbf{e}}_{\alpha _{s}}\star _{s}\mathbf{\
^{\shortparallel }v}^{\alpha _{s}},\mathbf{\ ^{\shortparallel }v}^{\alpha
_{s}}\in \mathcal{A}_{N}^{\star },$ when the associator acts trivially on
any basis (co) s-vector $\mathbf{\ }$ $^{\shortparallel }\mathbf{e}_{\alpha
_{s}}=(\ ^{\shortparallel }\mathbf{e}_{i_{s}},\ ^{\shortparallel }e^{a_{s}})$
(\ref{nadapdc}). Using the s-adapted absolute differential $\ _{s\mathbf{D}%
^{\star }}^{\shortparallel }\mathbf{d}$ $\ $(\ref{aux302}), we define and
compute%
\begin{eqnarray}
\ _{s}^{\shortparallel }\mathcal{T}^{\star }:= &&\ _{s\mathbf{D}^{\star
}}^{\shortparallel }\mathbf{d}\ (\mathbf{\ ^{\shortparallel }\mathbf{e}}%
_{\alpha _{s}}\star _{s}\mathbf{\ ^{\shortparallel }e}^{\alpha _{s}})=\
_{s}^{\shortparallel }\mathbf{D}^{\star }\mathbf{\ ^{\shortparallel }\mathbf{%
e}}_{\alpha _{s}}{\wedge _{\star s}\ ^{\shortparallel }\mathbf{e}^{\alpha
_{s}}=\ ^{\shortparallel }\mathbf{e}}_{\alpha _{s}}\otimes _{\star
_{s}}\left( \mathbf{\ ^{\shortparallel }\Gamma }_{\star \beta _{s}\gamma
_{s}}^{\alpha _{s}}\star _{s}(\ ^{\shortparallel }\mathbf{e}^{\gamma _{s}}{%
\wedge _{\star s}}\ ^{\shortparallel }\mathbf{e}^{\beta _{s}})\right)  \notag
\\
:= &&\ {^{\shortparallel }\mathbf{e}}_{\alpha _{s}}\otimes _{\star _{s}}%
\mathbf{\mathbf{\mathbf{\mathbf{\ _{\star }^{\shortparallel }}}}}\mathcal{T}{%
^{\alpha _{s}},\mbox{ where }}  \label{dtorshform} \\
&&\mathbf{\mathbf{\mathbf{\mathbf{\ _{\star }^{\shortparallel }}}}}\mathcal{T%
}{^{\alpha _{s}}}:=\mathbf{\ ^{\shortparallel }\Gamma }_{\star \beta
_{s}\gamma _{s}}^{\alpha _{s}}\star _{s}(\ ^{\shortparallel }\mathbf{e}%
^{\gamma _{s}}{\wedge _{\star s}}\ ^{\shortparallel }\mathbf{e}^{\beta
_{s}})=(\mathbf{\ ^{\shortparallel }\Gamma }_{\star \beta _{s}\gamma
_{s}}^{\alpha _{s}}-\mathbf{\ ^{\shortparallel }\Gamma }_{\star \gamma
_{s}\beta _{s}}^{\alpha _{s}}+\mathbf{\ ^{\shortparallel }w}_{\star \gamma
_{s}\beta _{s}}^{\alpha _{s}})\star _{s}(\ ^{\shortparallel }\mathbf{e}
^{\gamma _{s}}\star _{s}\ ^{\shortparallel }\mathbf{e}^{\beta _{s}}),%
\mbox{
for }  \notag \\
&&[\ ^{\shortparallel }\mathbf{e}_{\gamma _{s}},\ ^{\shortparallel }\mathbf{e%
}_{\beta _{s}}]_{\star _{s}}=\ ^{\shortparallel }\mathbf{e}_{\gamma _{s}}{%
\wedge _{\star s}}\ ^{\shortparallel }\mathbf{e}_{\beta _{s}}=\
^{\shortparallel }\mathbf{e}_{\gamma _{s}}\star _{s}\ ^{\shortparallel }%
\mathbf{e}_{\beta _{s}}-\ ^{\shortparallel }\mathbf{e}_{\beta _{s}}\star
_{s}\ ^{\shortparallel }\mathbf{e}_{\gamma _{s}}=\mathbf{\ ^{\shortparallel
}w}_{\star \gamma _{s}\beta _{s}}^{\alpha _{s}}\star _{s}\ {^{\shortparallel
}\mathbf{e}}_{\alpha _{s}}.  \notag
\end{eqnarray}

A star s-torsion is also a s-adapted map $\ _{s}^{\shortparallel }\mathcal{T}%
^{\star }:$ $\emph{Vec}_{\star _{s}}\otimes _{\star _{s}}\emph{Vec}_{\star
_{s}}\rightarrow \emph{Vec}_{\star _{s}},$ when
\begin{equation*}
\mathbf{\mathbf{\mathbf{\mathbf{\ }}}}_{s}^{\shortparallel }\mathcal{T}%
^{\star }(\mathbf{\ ^{\shortparallel }z},\mathbf{\ ^{\shortparallel }v})=%
\mathbf{\langle }\ _{s}^{\shortparallel }\mathbf{\mathbf{\mathcal{T}}}%
^{\star },\mathbf{\mathbf{\mathbf{\mathbf{\ }}\ ^{\shortparallel }z}}\otimes
_{\star _{s}}\mathbf{\mathbf{\ ^{\shortparallel }v}\rangle }_{\star _{s}}={\
^{\shortparallel }\mathbf{e}}_{\alpha _{s}}\star _{s}\mathbf{\langle \mathbf{%
\mathbf{\mathbf{\ }}}}\ _{\star }^{\shortparallel }\mathcal{T}^{\alpha _{s}},%
\mathbf{\mathbf{\mathbf{\mathbf{\ }}\ ^{\shortparallel }z}}\otimes _{\star
_{s}}\mathbf{\mathbf{\ ^{\shortparallel }v}\rangle }_{\star _{s}},
\end{equation*}%
for some s-vectors $\mathbf{\ ^{\shortparallel }z}$ and $\mathbf{\
^{\shortparallel }v.}$ The conditions of right $\mathcal{A}_{s}^{\star }$%
--linearity and star antisymmetry, $\ _{s\star }^{\shortparallel }\mathcal{T}%
(\mathbf{\ ^{\shortparallel }z},\mathbf{\ ^{\shortparallel }v})=-\mathbf{%
\mathbf{\mathbf{\mathbf{\ }}}}\ _{s\star }^{\shortparallel }\mathcal{T}(%
\mathbf{\ _{\intercal }^{\shortparallel }v},\mathbf{\ _{\intercal
}^{\shortparallel }z}),$ are also satisfied, where the \textbf{braiding}
d-operator "$\mathbf{\ _{\intercal }^{{}}}$"\ is used for actions of type
\begin{equation}
\tau _{\mathcal{R}}(\mathbf{\ ^{\shortparallel }z\otimes _{\star _{s}}\
^{\shortparallel }v})=(\mathbf{\ _{\intercal }^{\shortparallel }v\otimes
_{\star _{s}}\ _{\intercal }^{\shortparallel }z}).  \label{braidop}
\end{equation}%
This results in formulas,
\begin{eqnarray*}
\mathbf{\langle }\ _{\intercal }^{\shortparallel }\mathbf{e}^{\alpha
_{s}}\otimes _{\star _{s}}\mathbf{\ _{\intercal }^{\shortparallel }\mathbf{e}%
}^{\beta _{s}},\mathbf{\mathbf{\mathbf{\mathbf{\ }}\ ^{\shortparallel }z}}%
\otimes _{\star _{s}}\mathbf{\mathbf{\ ^{\shortparallel }v}\rangle }_{\star
_{s}} &=&\mathbf{\langle }\ ^{\shortparallel }\mathbf{e}^{\beta _{s}}\otimes
_{\star _{s}}\mathbf{\ ^{\shortparallel }\mathbf{e}}^{\alpha _{s}},\mathbf{%
\mathbf{\mathbf{\mathbf{\ }}\ _{\intercal }^{\shortparallel }v}}\otimes
_{\star _{s}}\mathbf{\mathbf{\ _{\intercal }^{\shortparallel }z}\rangle }%
_{\star _{s}},\mbox{ and } \\
\mathbf{\langle }\ ^{\shortparallel }\mathbf{e}^{\alpha _{s}}{\wedge }%
_{\star _{s}}\mathbf{\ ^{\shortparallel }\mathbf{e}}^{\beta _{s}},\mathbf{%
\mathbf{\mathbf{\mathbf{\ }}\ ^{\shortparallel }z}}\otimes _{\star _{s}}%
\mathbf{\mathbf{\ ^{\shortparallel }v}\rangle }_{\star _{s}} &=&\mathbf{%
\langle }\ ^{\shortparallel }\mathbf{e}^{\alpha _{s}}\otimes _{\star _{s}}%
\mathbf{\ ^{\shortparallel }\mathbf{e}}^{\beta _{s}},\mathbf{\mathbf{\mathbf{%
\mathbf{\ }}\ ^{\shortparallel }z}}{\wedge }_{\star _{s}}\mathbf{\mathbf{\
^{\shortparallel }v}\rangle }_{\star _{s}}.
\end{eqnarray*}

Using the s-operator $\mathbf{\mathbf{\mathbf{\mathbf{\ }}}}\
_{s}^{\shortparallel }\mathcal{T}^{\star }$ (\ref{dtorshform}) on $\
^{\shortparallel }\mathbf{e}_{\alpha _{s}}=(\ ^{\shortparallel }\mathbf{e}%
_{i_{s}},\ ^{\shortparallel }e^{a_{s}})$ (\ref{nadapdc}), we compute the
d-torsion coefficients for $\mathbf{\ ^{\shortparallel }z=}\
^{\shortparallel }\mathbf{e}_{\alpha _{s}}$ and$\mathbf{\ ^{\shortparallel
}v=}\ ^{\shortparallel }\mathbf{e}_{\beta _{s}},$%
\begin{eqnarray}
\mathbf{\mathbf{\mathbf{\mathbf{\ \ }}}}_{s}^{\shortparallel }\mathcal{T}%
^{\star }(\mathbf{\ ^{\shortparallel }e}_{\alpha _{s}},\mathbf{\
^{\shortparallel }e}_{\beta _{s}}) &=&\mathbf{\ ^{\shortparallel }e}_{\gamma
_{s}}\star _{s}\mathbf{\langle \mathbf{\mathbf{\mathbf{\mathbf{\mathbf{%
\mathbf{\ }}}}}}}\ _{\star }^{\shortparallel }T^{\alpha _{s}}\mathbf{,}\
^{\shortparallel }\mathbf{e}_{\alpha _{s}}{\wedge }_{\star _{s}}\mathbf{\
^{\shortparallel }\mathbf{e}}_{\beta _{s}}\mathbf{\rangle }_{\star _{s}}=%
\mathbf{\ ^{\shortparallel }e}_{\gamma _{s}}\star _{s}(\mathbf{\mathbf{%
\mathbf{\mathbf{\ ^{\shortparallel }}}}\Gamma }_{\star \beta _{s}\alpha
_{s}}^{\gamma _{s}}-\mathbf{\mathbf{\mathbf{\mathbf{\ ^{\shortparallel }}}}%
\Gamma }_{\star \alpha _{s}\beta _{s}}^{\gamma _{s}}+\ \mathbf{\mathbf{%
\mathbf{\mathbf{\ ^{\shortparallel }}}}}w_{\star \alpha _{s}\beta
_{s}}^{\gamma _{s}})  \label{aux303} \\
&=&\mathbf{\ ^{\shortparallel }e}_{\gamma _{s}}\star _{s}\mathbf{\mathbf{%
\mathbf{\mathbf{\ ^{\shortparallel }}}}T}_{\star \beta _{s}\alpha
_{s}}^{\gamma _{s}},  \notag
\end{eqnarray}%
see formulas (\ref{dtorshform}). The s-coefficient N-adapted decomposition
of s-torsion is parameterized in the form
\begin{eqnarray*}
\mathbf{\mathbf{\mathbf{\mathbf{\ \ \ }}}}_{s}^{\shortparallel }\mathcal{T}%
^{\star } &=&\{\mathbf{\mathbf{\mathbf{\mathbf{\ ^{\shortparallel }}}}T}%
_{\star \alpha _{s}\beta _{s}}^{\gamma _{s}}=(\mathbf{\mathbf{\mathbf{%
\mathbf{\ ^{\shortparallel }}}}}T_{\star \ j_{1}k_{1}}^{i_{1}},\mathbf{%
\mathbf{\mathbf{\mathbf{\ ^{\shortparallel }}}}}T_{\star a_{2}j_{1}}^{i_{1}\
},\mathbf{\mathbf{\mathbf{\mathbf{\ \ ^{\shortparallel }}}}}T_{\star
j_{1}c_{2}}^{i_{1}\ },\mathbf{\mathbf{\mathbf{\mathbf{\ ^{\shortparallel }}}}%
}T_{\star j_{1}i_{1}}^{c_{2}},\mathbf{\mathbf{\mathbf{\mathbf{\
^{\shortparallel }}}}}T_{\star a_{2}\ j_{1}}^{\ c_{2}},\mathbf{\mathbf{%
\mathbf{\mathbf{\ \ ^{\shortparallel }}}}}T_{\star b_{2}c_{2}\ }^{\ a_{2}},%
\mathbf{\mathbf{\mathbf{\mathbf{\ ^{\shortparallel }}}}}T_{\star \
j_{2}}^{i_{2}\ a_{3}},\ \mathbf{\mathbf{\mathbf{\mathbf{\ ^{\shortparallel }}%
}}}T_{\star a_{3}j_{2}i_{2}},\mathbf{\mathbf{\mathbf{\mathbf{\ \
^{\shortparallel }}}}}T_{\star c_{3}\ j_{2}}^{\ a_{3}} \\
&&\mathbf{\mathbf{\mathbf{\mathbf{\ \ ^{\shortparallel }}}}}T_{\star a_{3}\
}^{\ b_{3}c_{3}},\mathbf{\mathbf{\mathbf{\mathbf{\ ^{\shortparallel }}}}}%
T_{\star \ j_{3}}^{i_{3}\ a_{4}},\mathbf{\mathbf{\mathbf{\mathbf{\
^{\shortparallel }}}}}T_{\star a_{4}j_{3}i_{3}},\mathbf{\mathbf{\mathbf{%
\mathbf{\ \ ^{\shortparallel }}}}}T_{\star c_{4}\ j_{3}}^{\ a_{4}},\mathbf{%
\mathbf{\mathbf{\mathbf{\ \ ^{\shortparallel }}}}}T_{\star a_{4}\ }^{\
b_{4}c_{4}})\},
\end{eqnarray*}%
where the s-adapted coefficients are computed
\begin{eqnarray}
\mathbf{\mathbf{\mathbf{\mathbf{\ \ ^{\shortparallel }}}}}T_{\star \
j_{1}k_{1}}^{i_{1}} &=&\mathbf{\mathbf{\mathbf{\mathbf{\ \ ^{\shortparallel }%
}}}}L_{\star j_{1}k_{1}}^{i_{1}}-\mathbf{\mathbf{\mathbf{\mathbf{\ \
^{\shortparallel }}}}}L_{\star k_{1}j_{1}}^{i_{1}},\mathbf{\mathbf{\mathbf{%
\mathbf{\ ^{\shortparallel }}}}}T_{\star a_{2}j_{1}}^{i_{1}\ }=\mathbf{%
\mathbf{\mathbf{\mathbf{\ ^{\shortparallel }}}}}C_{\star
a_{2}j_{1}}^{i_{1}},\ \mathbf{\mathbf{\mathbf{\mathbf{\ \ ^{\shortparallel }}%
}}}T_{\star j_{1}c_{2}}^{i_{1}\ }=\mathbf{\mathbf{\mathbf{\mathbf{\
^{\shortparallel }}}}}C_{\star j_{1}c_{2}}^{i_{1}},\mathbf{\mathbf{\mathbf{%
\mathbf{\ ^{\shortparallel }}}}}T_{\star j_{1}i_{1}}^{c_{2}}=-\ \mathbf{%
\mathbf{\mathbf{\mathbf{\ \ ^{\shortparallel }}}}}\Omega _{\star
j_{1}i_{1}}^{c_{2}},  \notag \\
\mathbf{\mathbf{\mathbf{\mathbf{\ \ ^{\shortparallel }}}}}T_{\star a_{2}\
j_{1}}^{\ c_{2}} &=&\mathbf{\mathbf{\mathbf{\mathbf{\ \ ^{\shortparallel }}}}%
}L_{\star a_{2}\ j_{1}}^{c_{2}}-\mathbf{\mathbf{\mathbf{\mathbf{\
^{\shortparallel }}}}}e_{a_{2}}(\mathbf{\mathbf{\mathbf{\mathbf{\
^{\shortparallel }}}}}N_{\star j_{1}}^{c_{2}}),\mathbf{\mathbf{\mathbf{%
\mathbf{\ \ ^{\shortparallel }}}}}T_{\star b_{2}c_{2}\ }^{\ a_{2}}=\mathbf{%
\mathbf{\mathbf{\mathbf{\ ^{\shortparallel }}}}}C_{\star b_{2}c_{2}\
}^{a_{2}}-\mathbf{\mathbf{\mathbf{\mathbf{\ \ ^{\shortparallel }}}}}C_{\star
c_{2}b_{2}\ }^{a_{2}};  \label{storsnonassoc} \\
\mathbf{\mathbf{\mathbf{\mathbf{\ \ ^{\shortparallel }}}}}T_{\star \
j_{2}}^{i_{2}\ a_{3}} &=&\mathbf{\mathbf{\mathbf{\mathbf{\ ^{\shortparallel }%
}}}}C_{\star j_{2}}^{i_{2}a_{3}},\ \mathbf{\mathbf{\mathbf{\mathbf{\
^{\shortparallel }}}}}T_{\star a_{3}j_{2}i_{2}}=-\ \mathbf{\mathbf{\mathbf{%
\mathbf{\ \ ^{\shortparallel }}}}}\Omega _{\star a_{3}j_{2}i_{2}},\mathbf{%
\mathbf{\mathbf{\mathbf{\ \ ^{\shortparallel }}}}}T_{\star c_{3}\ j_{2}}^{\
a_{3}}=\mathbf{\mathbf{\mathbf{\mathbf{\ \ ^{\shortparallel }}}}}L_{\star
c_{3}\ j_{2}}^{a_{3}}-\mathbf{\mathbf{\mathbf{\mathbf{\ ^{\shortparallel }}}}%
}e^{a_{3}}(\mathbf{\mathbf{\mathbf{\mathbf{\ ^{\shortparallel }}}}}N_{\star
c_{3}j_{2}}),  \notag \\
\mathbf{\mathbf{\mathbf{\mathbf{\ \ ^{\shortparallel }}}}}T_{\star a_{3}\
}^{\ b_{3}c_{3}} &=&\mathbf{\mathbf{\mathbf{\mathbf{\ ^{\shortparallel }}}}}%
C_{\star a_{3}}^{\ b_{3}c_{3}}-\mathbf{\mathbf{\mathbf{\mathbf{\ \
^{\shortparallel }}}}}C_{\star a_{3}}^{\ c_{3}b_{3}},  \notag \\
\mathbf{\mathbf{\mathbf{\mathbf{\ ^{\shortparallel }}}}}T_{\star \
j_{3}}^{i_{3}\ a_{4}} &=&\mathbf{\mathbf{\mathbf{\mathbf{\ ^{\shortparallel }%
}}}}C_{\star j_{3}}^{i_{3}a_{4}},\ \mathbf{\mathbf{\mathbf{\mathbf{\
^{\shortparallel }}}}}T_{\star a_{4}j_{3}i_{3}}=-\ \mathbf{\mathbf{\mathbf{%
\mathbf{\ \ ^{\shortparallel }}}}}\Omega _{\star a_{4}j_{3}i_{3}},\mathbf{%
\mathbf{\mathbf{\mathbf{\ \ ^{\shortparallel }}}}}T_{\star c_{4}\ j_{3}}^{\
a_{4}}=\mathbf{\mathbf{\mathbf{\mathbf{\ \ ^{\shortparallel }}}}}L_{\star
c_{4}\ j_{3}}^{~a_{4}}-\mathbf{\mathbf{\mathbf{\mathbf{\ ^{\shortparallel }}}%
}}e^{a_{4}}(\mathbf{\mathbf{\mathbf{\mathbf{\ ^{\shortparallel }}}}}N_{\star
c_{4}j_{3}}),  \notag \\
\mathbf{\mathbf{\mathbf{\mathbf{\ \ ^{\shortparallel }}}}}T_{\star a_{4}\
}^{\ b_{4}c_{4}} &=&\mathbf{\mathbf{\mathbf{\mathbf{\ ^{\shortparallel }}}}}%
C_{\star a_{4}}^{\ b_{4}c_{4}}-\mathbf{\mathbf{\mathbf{\mathbf{\ \
^{\shortparallel }}}}}C_{\star a_{4}}^{\ c_{4}b_{4}},  \notag
\end{eqnarray}%
when $\ \ _{s}^{\shortparallel }\mathbf{D}^{\star }$ is determined by
s-components as in formulas (\ref{irevndecomdc}).

For $\mathbf{^{\shortparallel }z=}\ ^{\shortparallel }\mathbf{e}_{\alpha
_{s}},$ $\mathbf{\ ^{\shortparallel }v=}\ ^{\shortparallel }\mathbf{e}%
_{\beta _{s}}$ and $\mathbf{\mathbf{\mathbf{\mathbf{\ ^{\shortparallel }}}}D}%
_{\intercal \alpha _{s}}^{\star }=\mathbf{\mathbf{\mathbf{\mathbf{\
^{\shortparallel }}}}D}_{\mathbf{_{\intercal }^{\shortparallel }e}_{\alpha
_{s}}}^{\star },$ we compute
\begin{equation*}
\mathbf{\mathbf{\mathbf{\mathbf{\ _{s}^{\shortparallel }}}}}\mathcal{T}%
^{\star }(\mathbf{\ ^{\shortparallel }e}_{\alpha _{s}},\mathbf{\
^{\shortparallel }e}_{\beta _{s}})=\mathbf{\mathbf{\mathbf{\mathbf{\
^{\shortparallel }}}}D}_{\beta _{s}}^{\star }\mathbf{\ ^{\shortparallel }e}%
_{\alpha _{s}}-\mathbf{\mathbf{\mathbf{\mathbf{\ ^{\shortparallel }}}}D}%
_{\intercal \alpha _{s}}^{\star }\mathbf{\ _{\intercal }^{\shortparallel }e}%
_{\beta _{s}}+[\mathbf{\ ^{\shortparallel }e}_{\alpha _{s}},\mathbf{\
^{\shortparallel }e}_{\beta _{s}}]_{\star _{s}}=\mathbf{\ ^{\shortparallel }e%
}_{\gamma _{s}}\star _{s}(\mathbf{\ ^{\shortparallel }\Gamma }_{\star \alpha
_{s}\beta _{s}}^{\gamma _{s}}-\mathbf{\ ^{\shortparallel }\Gamma }_{\star
\beta _{s}\alpha _{s}}^{\gamma _{s}}+\mathbf{\ ^{\shortparallel }}w_{\star
\alpha _{s}\beta _{s}}^{\gamma _{s}}).
\end{equation*}%
This defines a $U\emph{Vec}^{\mathcal{F}}(\mathcal{M},\ _{s}^{\shortparallel
}N)$-equivariant s-operator and can be written in terms of s-adapted
associative composition $\bullet ,$ pairing and braiding $\tau _{\mathcal{R}%
} $ (\ref{braidop}),%
\begin{equation*}
\ _{s}^{\shortparallel }\mathcal{T}^{\star }(,)=\mathbf{\langle }\ ,\mathbf{%
\mathbf{\mathbf{\mathbf{\ }}}\rangle }_{\star s}\bullet \left( \mathbf{%
\mathbf{\mathbf{\mathbf{\ _{s}^{\shortparallel }}}}D}^{\star }\otimes
_{\star _{s}}id\right) -\mathbf{\langle }\ ,\mathbf{\mathbf{\mathbf{\mathbf{%
\ }}}\rangle }_{\star _{s}}\bullet \left( \ _{s}^{\shortparallel }\mathbf{D}%
^{\star }\otimes _{\star _{s}}id\right) \bullet \tau _{\mathcal{R}%
}+[,]_{\star _{s}},
\end{equation*}%
when $\ _{s}^{\shortparallel }\mathcal{T}^{\star }(\mathbf{\
^{\shortparallel }z},\mathbf{\ ^{\shortparallel }v})=\ _{s}^{\shortparallel }%
\mathcal{T}^{\star }(\mathbf{\ ^{\shortparallel }z\otimes _{\star _{s}}\
^{\shortparallel }v}).$

\subsubsection{Star deformed dyadic curvature of quasi-Hopf s-adapted
structures}

We extend for s-connections and s-adapted frames the definitions and
formulas the star deformed curvature s-tensors. For holonomic basic
structures such constructions are provided in section 4.5 of \cite%
{aschieri17} and for N-decompositions in section 3.2 of \cite{partner01}.

\paragraph{Abstract definition of the curvature s-tensor and s-adapted
differential forms: \newline
}

We can define the curvature of a N-adapted covariant derivative $\ _{s
\mathbf{D}^{\star }}^{\shortparallel }\mathbf{d}$ (\ref{aux302}) as the
s-operator%
\begin{equation}
\ _{s}^{\shortparallel }\mathcal{\Re }^{\star }:=\ _{s\mathbf{D}^{\star
}}^{\shortparallel }\mathbf{d}\ \mathbf{\bullet }\ \ _{s\mathbf{D}^{\star
}}^{\shortparallel }\mathbf{d}\ \mathbf{:\ }\emph{Vec}_{\star
_{s}}\rightarrow \emph{Vec}_{\star _{s}}\otimes _{\star _{s}}\Omega _{\star
}^{2},  \label{stardcurvh}
\end{equation}
which is $\mathcal{A}_{s}^{\star }$--linear and determined by a star
s-connection $\ _{s}^{\shortparallel }\mathbf{D}^{\star }$ (\ref{sconhopf}).
A $\ _{s}^{\shortparallel }\mathcal{\Re }^{\star }$ is also $U\emph{Vec}^{%
\mathcal{F}}(\mathcal{M},\ _{s}^{\shortparallel }N)$-equivariant and can be
written in terms of N-adapted associative composition $\bullet ,$ pairing
and braiding $\tau _{\mathcal{R}}$ (\ref{braidop}), and associator $\Phi ,$
\begin{eqnarray}
\mathbf{\mathbf{\mathbf{\mathbf{\ }}}}\ _{s}^{\shortparallel }\mathcal{\Re }%
^{\star }:= &&\mathbf{\langle }\ ,\mathbf{\ \rangle }_{\star _{a}}\bullet (\
_{s}^{\shortparallel }\mathbf{D}^{\star }\otimes _{\star s}id)\bullet (%
\mathbf{\langle }\ ,\mathbf{\mathbf{\mathbf{\mathbf{\ }}}\rangle }_{\star
_{s}}\otimes _{\star _{s}}id)\bullet \Phi _{Vec_{\star _{s}}\otimes _{\star
s}\Omega _{\star }^{1},Vec_{\star _{s}},Vec_{\star _{s}}}^{-1}\bullet (\
_{s}^{\shortparallel }\mathbf{D}^{\star }\otimes _{\star _{s}}id^{\otimes
_{\star _{s}}^{2}})  \notag \\
&&\bullet (id^{\otimes _{\star _{s}}^{3}}-id\otimes _{\mathcal{R}}\tau _{%
\mathcal{R}})+\mathbf{\langle }\ ,\mathbf{\mathbf{\mathbf{\mathbf{\ }}}%
\rangle }_{\star _{s}}\bullet \left( \mathbf{\mathbf{\mathbf{\mathbf{\
^{\shortparallel }}}}D}^{\star }\otimes _{\star _{s}}id\right) \bullet
(id\otimes _{\mathcal{R}}[,]_{\star _{s}}).  \label{stardcurvhopf}
\end{eqnarray}%
Using formulas (\ref{aux300}), the action of s-operator $\
_{s}^{\shortparallel }\mathcal{\Re }^{\star }$ (\ref{stardcurvh}) on $%
^{\shortparallel }\mathbf{e}_{\alpha _{s}}$ can be computed and expressed in
the form%
\begin{equation*}
\ \ _{s}^{\shortparallel }\mathcal{\Re }^{\star }(^{\shortparallel }\mathbf{e%
}_{\alpha _{s}})=\ _{s\mathbf{D}^{\star }}^{\shortparallel }\mathbf{d}(\
_{s}^{\shortparallel }\mathbf{D}^{\star }\ ^{\shortparallel }\mathbf{e}%
_{\alpha _{s}})=\ _{s\mathbf{D}^{\star }}^{\shortparallel }\mathbf{d}%
(^{\shortparallel }\mathbf{e}_{\beta _{s}}\otimes _{\star _{s}}\mathbf{\
^{\shortparallel }\Gamma }_{\star \alpha _{s}}^{\beta _{s}})=\
^{\shortparallel }\mathbf{e}_{\gamma _{s}}\otimes _{\star _{s}}\mathbf{%
\mathbf{\mathbf{\mathbf{\ ^{\shortparallel }}}}}\mathcal{R}_{\quad \alpha
_{s}}^{\star \gamma _{s}},
\end{equation*}%
where the the matrix valued describing a Hopf s-algebra star deformed
s-curvature 2-form is%
\begin{equation}
\mathbf{\mathbf{\mathbf{\mathbf{\ ^{\shortparallel }}}}}\mathcal{\Re }%
_{\quad \alpha _{s}}^{\star \gamma _{s}}:=\ _{s}^{\shortparallel }\mathbf{d\
^{\shortparallel }}\Gamma _{\star \alpha _{s}}^{\gamma
_{s}}+^{\shortparallel }\mathbf{\Gamma }_{\star \beta _{s}}^{\gamma _{s}}{%
\wedge }_{\star _{s}}\ ^{\shortparallel }\mathbf{\Gamma }_{\star \alpha
_{s}}^{\beta _{s}}.  \label{strdcurvhf}
\end{equation}

\paragraph{Coefficient formulas for the nonassociative Riemann s-tensor and
quasi-Hopf s-algebras: \newline
}

Such formulas are expressed for the star deformed curvature s-tensor of $\
_{s}^{\shortparallel }\mathbf{D}^{\star },$ i.e. for $\
_{s}^{\shortparallel} \mathcal{\Re }_{\quad \alpha _{s}}^{\star \gamma _{s}}$
(\ref{strdcurvhf}), with respect to $^{\shortparallel }\mathbf{e}_{\alpha
_{s}}$ (\ref{nadapdc}),%
\begin{eqnarray}
\ _{s}^{\shortparallel }\mathcal{\Re }^{\star } &&(\ ^{\shortparallel }%
\mathbf{e}_{\alpha _{s}},\ ^{\shortparallel }\mathbf{e}_{\beta _{s}},\
^{\shortparallel }\mathbf{e}_{\gamma _{s}})=\langle \ \mathbf{%
^{\shortparallel }e}_{\mu _{s}}\otimes _{\star _{s}}\mathbf{\mathbf{\mathbf{%
\mathbf{\ _{s}^{\shortparallel }}}}}\mathcal{\Re }_{\quad \alpha
_{s}}^{\star \mu _{s}},\mathbf{\ ^{\shortparallel }e}_{\beta _{s}}\mathbf{%
\mathbf{\mathbf{\mathbf{\ {\wedge }_{\star _{s}}}\ ^{\shortparallel }e}}}%
_{\gamma _{s}}\mathbf{\mathbf{\rangle }}_{\star _{s}}  \label{aux15} \\
&=&\ \mathbf{^{\shortparallel }e}_{\mu _{s}}\star _{s}\langle \ \mathbf{%
^{\shortparallel }e}_{\nu _{s}}\mathbf{\ ^{\shortparallel }}\Gamma _{\star
\alpha _{s}\varphi _{s}}^{\mu _{s}}\star _{s}(\ \mathbf{^{\shortparallel }e}%
^{\nu _{s}}\mathbf{\mathbf{\mathbf{\mathbf{{\wedge }_{\star _{s}}}}}}\
\mathbf{^{\shortparallel }e}^{\varphi _{s}})+  \notag \\
&&\ ^{\shortparallel }\mathbf{\Gamma }_{\star \nu _{s}\varphi _{s}}^{\mu
_{s}}\star _{s}(\delta _{\ \tau _{s}}^{\varphi _{s}}\ ^{\shortparallel }%
\mathbf{\Gamma }_{\star \alpha _{s}\lambda _{s}}^{\nu _{s}}+i\kappa \mathcal{%
R}_{\quad \tau _{s}}^{\varphi _{s}\xi _{s}}\mathbf{\ ^{\shortparallel }e}%
_{\xi _{s}}\mathbf{\Gamma }_{\star \alpha _{s}\lambda _{s}}^{\nu _{s}})\star
_{s}\ \mathbf{^{\shortparallel }e}^{\tau _{s}}\mathbf{\mathbf{\mathbf{%
\mathbf{{\wedge }_{\star _{s}}}}}}\ \mathbf{^{\shortparallel }e}^{\lambda
_{s}},\mathbf{\ ^{\shortparallel }e}_{\beta _{s}}\mathbf{\mathbf{\mathbf{%
\mathbf{\ {\wedge }_{\star _{s}}}\ ^{\shortparallel }e}}}_{\gamma _{s}}%
\mathbf{\mathbf{\rangle }}_{\star _{s}}  \notag \\
&=&\ \mathbf{^{\shortparallel }e}_{\mu _{s}}\star _{s}\mathbf{\mathbf{%
\mathbf{\mathbf{\ ^{\shortparallel }}}}}\mathcal{\Re }_{\quad \alpha
_{s}\beta _{s}\gamma _{s}}^{\star \mu _{s}},  \notag
\end{eqnarray}%
where $\delta _{\ \tau _{s}}^{\varphi _{s}}$ is the Kronecker delta symbol
and the nonassociative Riemann s-tensor for the quasi-Hopf s-algebra can be
written in the form%
\begin{eqnarray}
\mathbf{\mathbf{\mathbf{\mathbf{\ ^{\shortparallel }}}}}\mathcal{\Re }%
_{\quad \alpha _{s}\beta _{s}\gamma _{s}}^{\star \mu _{s}} &=&\mathbf{%
\mathbf{\mathbf{\mathbf{\ _{1}^{\shortparallel }}}}}\mathcal{\Re }_{\quad
\alpha _{s}\beta _{s}\gamma _{s}}^{\star \mu _{s}}+\mathbf{\mathbf{\mathbf{%
\mathbf{\ _{2}^{\shortparallel }}}}}\mathcal{\Re }_{\quad \alpha _{s}\beta
_{s}\gamma _{s}}^{\star \mu _{s}},\mbox{ where }  \label{nadriemhopf} \\
&&\mathbf{\mathbf{\mathbf{\mathbf{\ _{1}^{\shortparallel }}}}}\mathcal{\Re }%
_{\quad \alpha _{s}\beta _{s}\gamma _{s}}^{\star \mu _{s}}=\ \mathbf{%
^{\shortparallel }e}_{\gamma _{s}}\mathbf{\ ^{\shortparallel }}\Gamma
_{\star \alpha _{s}\beta _{s}}^{\mu _{s}}-\ \mathbf{^{\shortparallel }e}%
_{\beta _{s}}\mathbf{\ ^{\shortparallel }}\Gamma _{\star \alpha _{s}\gamma
_{s}}^{\mu _{s}}+\mathbf{\ ^{\shortparallel }}\Gamma _{\star \nu _{s}\tau
_{s}}^{\mu _{s}}\star _{s}(\delta _{\ \gamma _{s}}^{\tau _{s}}\mathbf{\
^{\shortparallel }}\Gamma _{\star \alpha _{s}\beta _{s}}^{\nu _{s}}-\delta
_{\ \beta _{s}}^{\tau _{s}}\mathbf{\ ^{\shortparallel }}\Gamma _{\star
\alpha _{s}\gamma _{s}}^{\nu _{s}})  \notag \\
&&+\mathbf{\ ^{\shortparallel }}w_{\beta _{s}\gamma _{s}}^{\tau _{s}}\star
_{s}\mathbf{\ ^{\shortparallel }}\Gamma _{\star \alpha _{s}\tau _{s}}^{\mu
_{s}}  \notag \\
&&\mathbf{\mathbf{\mathbf{\mathbf{\ _{2}^{\shortparallel }}}}}\mathcal{\Re }%
_{\quad \alpha _{s}\beta _{s}\gamma _{s}}^{\star \mu _{s}}=i\kappa \mathbf{\
^{\shortparallel }}\Gamma _{\star \nu _{s}\tau _{s}}^{\mu _{s}}\star _{s}(%
\mathcal{R}_{\quad \gamma _{s}}^{\tau _{s}\xi _{s}}\ \mathbf{%
^{\shortparallel }e}_{\xi _{s}}\mathbf{\ ^{\shortparallel }}\Gamma _{\star
\alpha _{s}\beta _{s}}^{\nu _{s}}-\mathcal{R}_{\quad \beta _{s}}^{\tau
_{s}\xi _{s}}\ \mathbf{^{\shortparallel }e}_{\xi _{s}}\mathbf{\
^{\shortparallel }}\Gamma _{\star \alpha _{s}\gamma _{s}}^{\nu _{s}}).
\notag
\end{eqnarray}

With respect to $^{\shortparallel }\mathbf{e}_{\alpha _{s}}$ (\ref{nadapbdss}%
) and for $\ ^{\shortparallel }\mathbf{\Gamma }_{\star \alpha _{s}\beta
_{s}}^{\gamma _{s}}$ (\ref{irevndecomdc}), we find such a s-decomposition:
\begin{eqnarray}
\ \mathbf{\mathbf{\mathbf{\mathbf{\ _{1}^{\shortparallel }}}}}\mathcal{\Re }%
^{\star } &=&\mathbf{\{}\ \mathbf{\mathbf{\mathbf{\mathbf{\
_{1}^{\shortparallel }}}}}\mathcal{\Re }_{\quad \alpha \beta \gamma }^{\star
\mu }=(\ \ _{1}^{\shortparallel }R_{\ \star h_{1}j_{1}k_{1}}^{i_{1}},\
_{1}^{\shortparallel }R_{\star b_{2}j_{1}k_{1}}^{a_{2}\ },\ \
_{1}^{\shortparallel }P_{\star \ h_{1}j_{1}a_{2}}^{i_{1}\ \ \ },\
_{1}^{\shortparallel }P_{\star b_{2}k_{1}a_{1}}^{c_{2}\ \ },\
_{1}^{\shortparallel }S_{\star \ j_{1}b_{2}c_{2}}^{i_{1}\quad },\
_{1}^{\shortparallel }S_{\star b_{2}c_{2}d_{2}}^{a_{2}\quad };
\label{nadriemhopf1} \\
&&\ _{1}^{\shortparallel }R_{\star a_{3}\ j_{2}k_{2}}^{\ b_{3}},\ \
_{1}^{\shortparallel }P_{\star \ k_{2}j_{2}}^{i_{2}\ \ \ a_{3}},\
_{1}^{\shortparallel }P_{\star c_{3}\ k_{2}}^{\quad b_{3}\ a_{3}},\
_{1}^{\shortparallel }S_{\star \ j_{2}}^{i_{2}\quad b_{3}c_{3}},\ \
_{1}^{\shortparallel }S_{\star a_{3}\ }^{\quad b_{3}c_{3}d_{3}};  \notag \\
&&\ _{1}^{\shortparallel }R_{\star a_{4}\ j_{3}k_{3}}^{\ b_{4}},\ \
_{1}^{\shortparallel }P_{\star \ k_{3}j_{3}}^{i_{3}\ \ \ a_{4}},\
_{1}^{\shortparallel }P_{\star c_{4}\ k_{3}}^{\quad b_{4}\ a_{4}},\
_{1}^{\shortparallel }S_{\star \ j_{3}}^{i_{3}\quad b_{4}c_{4}},\ \
_{1}^{\shortparallel }S_{\star a_{4}\ }^{\quad b_{4}c_{4}d_{4}})\},%
\mbox{
where }  \notag
\end{eqnarray}%
\begin{eqnarray*}
\ \ _{1}^{\shortparallel }R_{\ \star h_{1}j_{1}k_{1}}^{i_{1}} &=&\ \
^{\shortparallel }\mathbf{e}_{k_{1}}\ ^{\shortparallel }L_{\star
h_{1}j_{1}}^{i_{1}}-\ \ ^{\shortparallel }\mathbf{e}_{j_{1}}\
^{\shortparallel }L_{\star h_{1}k_{1}}^{i_{1}}+\ \ ^{\shortparallel
}L_{\star h_{1}j_{1}}^{m_{1}}\star _{s}\quad ^{\shortparallel }L_{\star
m_{1}k_{1}}^{i_{1}}-\ \ ^{\shortparallel }L_{\star h_{1}k_{1}}^{m_{1}}\star
_{s}\ ^{\shortparallel }L_{\star m_{1}j_{1}}^{i_{1}} \\
&&-\ ^{\shortparallel }C_{\star \ h_{1}a_{2}}^{i_{1}\ }\ \star _{s}\
^{\shortparallel }\Omega _{\star k_{1}j_{1}}^{a_{2}}, \\
\ _{1}^{\shortparallel }R_{\star b_{2}j_{1}k_{1}}^{a_{2}\ } &=&\
^{\shortparallel }\mathbf{e}_{k_{1}}\ ^{\shortparallel }\acute{L}_{\star
b_{2}j_{1}}^{a_{2}}-\ ^{\shortparallel }\mathbf{e}_{j_{1}}\ ^{\shortparallel
}\acute{L}_{\star b_{2}k_{1}}^{a_{2}}+\ ^{\shortparallel }\acute{L}_{\star
c_{2}j_{1}}^{a_{2}}\star _{s}\ ^{\shortparallel }\acute{L}_{\star
b_{2}k_{1}}^{c_{2}}-\ ^{\shortparallel }\acute{L}_{\star
c_{2}k_{1}}^{a_{2}}\star _{s}\ ^{\shortparallel }\acute{L}_{\star
b_{2}j_{1}}^{c_{2}} \\
&&-\ ^{\shortparallel }C_{\star b_{2}c_{2}\ }^{a_{2}}\star _{s}\
^{\shortparallel }\Omega _{\star k_{1}j_{1}}^{c_{2}}, \\
\ \ _{1}^{\shortparallel }P_{\star \ h_{1}j_{1}a_{2}}^{i_{1}\ \ \ } &=&\
^{\shortparallel }e_{a_{2}}\ ^{\shortparallel }L_{\star \
j_{1}k_{1}}^{i_{1}}-\ \ ^{\shortparallel }D_{k_{1}}^{\star }\ \star _{s}\
^{\shortparallel }\acute{C}_{\star j_{1}a_{2}}^{i_{1}}+\ ^{\shortparallel }%
\acute{C}_{\star j_{1}b_{2}}^{i_{1}\ }\star _{s}\ ^{\shortparallel }T_{\star
k_{1}a_{2}}^{b_{2}\ \ \ },\  \\
\ _{1}^{\shortparallel }P_{\star b_{2}k_{1}a_{1}}^{c_{2}\ \ } &=&\
^{\shortparallel }e_{a_{2}}\ ^{\shortparallel }\acute{L}_{\star
b_{2}k_{1}}^{c_{2}}-\ ^{\shortparallel }D_{k_{1}}^{\star }\ \star _{s}\
^{\shortparallel }C_{\star b_{2}a_{2}\ }^{c_{2}}+\ ^{\shortparallel
}C_{\star b_{2}d_{2}}^{c_{2}}\star _{s}\ ^{\shortparallel }T_{\star
k_{1}a_{2}}^{d_{2}}, \\
\ _{1}^{\shortparallel }S_{\star \ j_{1}b_{2}c_{2}}^{i_{1}\quad } &=&\
^{\shortparallel }e_{c_{2}}\ ^{\shortparallel }\acute{C}_{\star
j_{1}b_{2}}^{i_{1}\ }-\ \ ^{\shortparallel }e_{b_{2}}\ ^{\shortparallel }%
\acute{C}_{\star j_{1}c_{2}}^{i_{1}}+\ \ ^{\shortparallel }\acute{C}_{\star
h_{1}b_{2}}^{i_{1}}\ \star _{s}\ ^{\shortparallel }\acute{C}_{\star
j_{1}c_{2}}^{h_{1}}-\ \ ^{\shortparallel }\acute{C}_{\star
h_{1}c_{2}}^{i_{1}\quad }\star _{s}\ \ ^{\shortparallel }\acute{C}_{\star
j_{1}b_{2}}^{h_{1}\quad }, \\
\ \ _{1}^{\shortparallel }S_{\star b_{2}c_{2}d_{2}}^{a_{2}\quad } &=&\
^{\shortparallel }e_{d_{2}}\ ^{\shortparallel }C_{\star b_{2}c_{2}\
}^{a_{2}}-\ \ ^{\shortparallel }e_{c_{2}}\ ^{\shortparallel }C_{\star
b_{2}d_{2}\ }^{a_{2}\quad }+\ \ ^{\shortparallel }C_{\star b_{2}c_{2}\
}^{e_{2}}\ \star _{s}\ ^{\shortparallel }C_{\star e_{2}d_{2}\ }^{a_{2}\quad
}-\ ^{\shortparallel }C_{\star b_{2}d_{2}}^{e_{2}\quad }\ \star _{s}\
^{\shortparallel }C_{\star e_{2}c_{2}}^{a_{2}},
\end{eqnarray*}%
\begin{eqnarray*}
\ _{1}^{\shortparallel }R_{\star a_{3}\ j_{2}k_{2}}^{\ b_{3}} &=&\
^{\shortparallel }\mathbf{e}_{k_{2}}\ ^{\shortparallel }\acute{L}_{\star
a_{3}\ j_{2}}^{\ b_{3}}-\ ^{\shortparallel }\mathbf{e}_{j_{2}}\
^{\shortparallel }\acute{L}_{\star a_{3}\ k_{2}}^{\ b_{3}}+\
^{\shortparallel }\acute{L}_{\star c_{3}\ j_{2}}^{\ b_{3}}\star _{s}\
^{\shortparallel }\acute{L}_{\star a_{3}\ k_{2}}^{\ c_{3}}-\
^{\shortparallel }\acute{L}_{\star c_{3}\ k_{2}}^{\ b_{3}}\star _{s}\
^{\shortparallel }\acute{L}_{\star a_{3}\ j_{2}}^{\ c_{3}} \\
&&-\ ^{\shortparallel }C_{\star a_{3}\ }^{\ b_{3}c_{3}}\ \star _{s}\
^{\shortparallel }\Omega _{\star c_{3}k_{2}j_{2}}, \\
\ \ _{1}^{\shortparallel }P_{\star \ k_{2}j_{2}}^{i_{2}\ \ \ a_{3}} &=&\
^{\shortparallel }e^{a_{3}}\ \ ^{\shortparallel }L_{\star \
j_{2}k_{2}}^{i_{2}}-\ \ ^{\shortparallel }D_{k_{2}}^{\star }\ \star _{s}\
^{\shortparallel }\acute{C}_{\star \ j_{2}}^{i_{2}\ a_{3}}+\
^{\shortparallel }\acute{C}_{\star \ j_{2}}^{i_{2}\ b_{3}}\star _{s}\
^{\shortparallel }T_{\star b_{3}k_{2}}^{\ \ \ a_{3}},\  \\
\ _{1}^{\shortparallel }P_{\star c_{3}\ k_{2}}^{\quad b_{3}\ a_{3}} &=&\
^{\shortparallel }e^{a_{3}}\ ^{\shortparallel }\acute{L}_{\star c_{3}\
k_{2}}^{\ b_{3}}-\ ^{\shortparallel }D_{k_{2}}^{\star }\ \star _{s}\
^{\shortparallel }C_{\star c_{3}\ }^{\ \quad b_{3}a_{3}}+\ ^{\shortparallel
}C_{\star c_{3}}^{\quad \ b_{3}d_{3}}\ \star _{s}\ ^{\shortparallel
}T_{\star \ d_{3}k_{2}}^{\qquad a_{3}}, \\
\ _{1}^{\shortparallel }S_{\star \ j_{2}}^{i_{2}\quad b_{3}c_{3}} &=&\
^{\shortparallel }e^{c_{3}}\ ^{\shortparallel }\acute{C}_{\star \
j_{2}}^{i_{2}\ b_{3}}-\ \ ^{\shortparallel }e^{b_{3}}\ ^{\shortparallel }%
\acute{C}_{\star \ j_{2}}^{i_{2}\quad c_{3}}+\ \ ^{\shortparallel }\acute{C}%
_{\star \ j_{2}}^{h_{2}\ b_{3}}\ \star _{s}\ ^{\shortparallel }\acute{C}%
_{\star \ h_{2}}^{i_{2}\ c_{3}}-\ \ ^{\shortparallel }\acute{C}_{\star \
j_{2}}^{h_{2}\quad c_{3}}\star _{s}\ \ ^{\shortparallel }\acute{C}_{\star \
h_{2}}^{i_{2}\quad b_{3}}, \\
\ \ \ _{1}^{\shortparallel }S_{\star a_{3}\ }^{\quad b_{3}c_{3}d_{3}} &=&\
^{\shortparallel }e^{d_{3}}\ \ ^{\shortparallel }C_{\star a_{3}\ }^{\quad
b_{3}c_{3}}-\ \ ^{\shortparallel }e^{c_{3}}\ \ ^{\shortparallel }C_{\star
a_{3}\ }^{\quad b_{3}d_{3}}+\ \ ^{\shortparallel }C_{\star e_{3}\ }^{\quad
b_{3}c_{3}}\ \star _{s}\ ^{\shortparallel }C_{\star a_{3}\ }^{\quad
e_{3}d_{3}}-\ \ ^{\shortparallel }C_{\star e_{3}}^{\quad b_{3}d_{3}}\ \star
_{s}\ ^{\shortparallel }C_{\star a_{3}}^{\quad e_{3}c_{3}},
\end{eqnarray*}%
\begin{eqnarray}
\ \ _{1}^{\shortparallel }R_{\star a_{4}\ j_{3}k_{3}}^{\ b_{4}} &=&\
^{\shortparallel }\mathbf{e}_{k_{3}}\ ^{\shortparallel }\acute{L}_{\star
a_{4}\ j_{3}}^{\ b_{4}}-\ ^{\shortparallel }\mathbf{e}_{j_{3}}\
^{\shortparallel }\acute{L}_{\star a_{4}\ k_{3}}^{\ b_{4}}+\
^{\shortparallel }\acute{L}_{\star c_{4}\ j_{3}}^{\ b_{4}}\star _{s}\
^{\shortparallel }\acute{L}_{\star a_{4}\ k_{3}}^{\ c_{4}}-\
^{\shortparallel }\acute{L}_{\star c_{4}\ k_{3}}^{\ b_{4}}\star _{s}\
^{\shortparallel }\acute{L}_{\star a_{4}\ j_{3}}^{\ c_{4}}  \notag \\
&&-\ ^{\shortparallel }C_{\star a_{4}\ }^{\ b_{4}c_{4}}\ \star _{s}\
^{\shortparallel }\Omega _{\star c_{4}k_{3}j_{3}},  \notag \\
\ \ _{1}^{\shortparallel }P_{\star \ k_{3}j_{3}}^{i_{3}\ \ \ a_{4}} &=&\
^{\shortparallel }e^{a_{4}}\ \ ^{\shortparallel }L_{\star \
j_{3}k_{3}}^{i_{3}}-\ \ ^{\shortparallel }D_{k_{3}}^{\star }\ \star _{s}\
^{\shortparallel }\acute{C}_{\star \ j_{3}}^{i_{3}\ a_{4}}+\
^{\shortparallel }\acute{C}_{\star \ j_{4}}^{i_{3}\ b_{4}}\star _{s}\
^{\shortparallel }T_{\star b_{4}k_{3}}^{\ \ \ a_{4}},\
\label{nadriemhopf1c} \\
\ _{1}^{\shortparallel }P_{\star c_{4}\ k_{3}}^{\quad b_{4}\ a_{4}} &=&\
^{\shortparallel }e^{a_{4}}\ ^{\shortparallel }\acute{L}_{\star c_{4}\
k_{3}}^{\ b_{4}}-\ ^{\shortparallel }D_{k_{3}}^{\star }\ \star _{s}\
^{\shortparallel }C_{\star c_{4}\ }^{\ \quad b_{4}a_{4}}+\ ^{\shortparallel
}C_{\star c_{4}}^{\quad \ b_{4}d_{4}}\ \star _{s}\ ^{\shortparallel
}T_{\star \ d_{4}k_{3}}^{\qquad a_{4}},  \notag \\
\ _{1}^{\shortparallel }S_{\star \ j_{3}}^{i_{3}\quad b_{4}c_{4}} &=&\
^{\shortparallel }e^{c_{4}}\ ^{\shortparallel }\acute{C}_{\star \
j_{3}}^{i_{3}\ b_{4}}-\ \ ^{\shortparallel }e^{b_{4}}\ ^{\shortparallel }%
\acute{C}_{\star \ j_{3}}^{i_{3}\quad c_{4}}+\ \ ^{\shortparallel }\acute{C}%
_{\star \ j_{3}}^{h_{3}\ b_{4}}\ \star _{s}\ ^{\shortparallel }\acute{C}%
_{\star \ h_{3}}^{i_{3}\ c_{4}}-\ \ ^{\shortparallel }\acute{C}_{\star \
j_{3}}^{h_{3}\quad c_{4}}\star _{s}\ \ ^{\shortparallel }\acute{C}_{\star \
h_{3}}^{i_{3}\quad b_{4}},  \notag \\
\ \ \ _{1}^{\shortparallel }S_{\star a_{4}\ }^{\quad b_{4}c_{4}d_{4}} &=&\
^{\shortparallel }e^{d_{4}}\ \ ^{\shortparallel }C_{\star a_{4}\ }^{\quad
b_{4}c_{4}}-\ \ ^{\shortparallel }e^{c_{4}}\ \ ^{\shortparallel }C_{\star
a_{4}\ }^{\quad b_{4}d_{4}}+\ \ ^{\shortparallel }C_{\star e_{4}\ }^{\quad
b_{4}c_{4}}\ \star _{s}\ ^{\shortparallel }C_{\star a_{4}\ }^{\quad
e_{4}d_{4}}-\ \ ^{\shortparallel }C_{\star e_{4}}^{\quad b_{4}d_{4}}\ \star
_{s}\ ^{\shortparallel }C_{\star a_{4}}^{\quad e_{4}c_{4}}.  \notag
\end{eqnarray}%
Tedious computations when (\ref{nadapbdss}) and (\ref{irevndecomdc}) are
used for $\ _{2}^{\shortparallel } \mathcal{\Re }_{\quad \alpha \beta
\gamma}^{\star \mu }$ allow us to compute s-coefficients in explicit form
(we omit such formulas because they are not used in this paper).

We emphasize that the nonassociative Riemann s-tensor (\ref{nadriemhopf}) is
constructed for an arbitrary star deformed s-connection $\ ^{\shortparallel}%
\mathbf{D}^{\star }$ (\ref{sconhopf}), i.e. we have a dyadic model of
nonassociative nonholonomic geometry of affine (linear) connections adapted
to quasi-Hopf s-structures.

\paragraph{The star deformed Ricci s-tensor for quasi-Hopf s-adapted
structures: \newline
}

Contracting the first and forth indices in the Riemann s-tensor formulas $\
_{s}^{\shortparallel }\mathcal{\Re }^{\star }$ (\ref{stardcurvh}), (\ref%
{stardcurvhopf}), and/or (\ref{aux15}), we can define and compute a
respective Ricci s-tensor. Such Riemann and Ricci s-tensors consist
respective dyadic variants of $\ ^{\shortparallel }\mathcal{\Re }^{\star }$
and $\ ^{\shortparallel }\mathcal{\Re }ic^{\star }$ introduced in
subsections of 3.1 and 3.2 of \cite{partner01}. Here we nota that similar
N-and s-structures were studied for nonholonomic and noncommutative models
elaborated in our previous works \cite{vacaru03,vacaru09a,vacaru16}, see
also parts II and III of monograph \cite{vacaru05a}. Using star curvature
s-operator (\ref{stardcurvhopf}), we define
\begin{equation*}
\ _{s}^{\shortparallel }\mathcal{\Re }ic^{\star }(\mathbf{\ ^{\shortparallel
}z},\mathbf{\ ^{\shortparallel }v}):=\langle \ _{s}^{\shortparallel }%
\mathcal{\Re }^{\star }(\mathbf{\ ^{\shortparallel }z},\mathbf{\
^{\shortparallel }v,}\ ^{\shortparallel }\mathbf{e}_{\alpha _{s}})\ ,\mathbf{%
\mathbf{\ ^{\shortparallel }\mathbf{e}}}^{\alpha _{s}}\rangle _{\star _{s}},
\end{equation*}%
for any s-vectors $\mathbf{\ ^{\shortparallel }z\rightarrow }\
_{s}^{\shortparallel }\mathbf{z},\ ^{\shortparallel }v \rightarrow \
_{s}^{\shortparallel }v\in \emph{Vec}_{\star _{s}}$ and s-adapted (co)
frames $\ ^{\shortparallel }\mathbf{e}_{\alpha _{s}}$ and $\mathbf{\
^{\shortparallel }\mathbf{e}}^{\beta _{s}}.$

The star Ricci s-tensor is defined by following such formulas
\begin{eqnarray}
\ _{s}^{\shortparallel }\mathcal{\Re }ic^{\star } &=&\mathbf{\mathbf{\mathbf{%
\mathbf{\ ^{\shortparallel }R}}}}ic_{\alpha _{s}\beta _{s}}^{\star }\star
_{s}(\ \mathbf{^{\shortparallel }e}^{\alpha _{s}}\otimes _{\star _{s}}\
\mathbf{^{\shortparallel }e}^{\beta _{s}}),\mbox{ where }  \label{driccina}
\\
&&\mathbf{\mathbf{\mathbf{\mathbf{\ ^{\shortparallel }R}}}}ic_{\alpha
_{s}\beta _{s}}^{\star }:=\ _{s}^{\shortparallel }\mathcal{\Re }ic^{\star }(%
\mathbf{\ }\ ^{\shortparallel }\mathbf{e}_{\alpha _{s}},\ ^{\shortparallel }%
\mathbf{e}_{\beta _{s}})=\mathbf{\langle }\ \mathbf{\mathbf{\mathbf{\mathbf{%
\ ^{\shortparallel }R}}}}ic_{\mu _{s}\nu _{s}}^{\star }\star _{s}(\ \mathbf{%
^{\shortparallel }e}^{\mu _{s}}\otimes _{\star _{s}}\ \mathbf{%
^{\shortparallel }e}^{\nu _{s}}),\mathbf{\mathbf{\ }\ ^{\shortparallel }%
\mathbf{e}}_{\alpha _{s}}\mathbf{\otimes _{\star _{s}}\ ^{\shortparallel }%
\mathbf{e}}_{\beta _{s}}\mathbf{\rangle }_{\star _{s}}.  \notag
\end{eqnarray}%
Using s-coefficient formulas for $\ ^{\shortparallel }\mathcal{\Re }^{\star
}(\ ^{\shortparallel }\mathbf{e}_{\alpha },\ ^{\shortparallel }\mathbf{e}%
_{\beta },\ ^{\shortparallel }\mathbf{e}_{\gamma })$ (\ref{aux15}) and the
property that the associator acts trivially on basis s-vectors and s-adapted
forms (and contracting respective indices in (\ref{nadriemhopf}) and (\ref%
{nadriemhopf1})), we can calculate explicitly all s-adapted coefficients
\begin{eqnarray*}
\mathbf{\mathbf{\mathbf{\mathbf{\ ^{\shortparallel }R}}}}ic_{\alpha
_{s}\beta _{s}}^{\star } &=&\mathbf{\mathbf{\mathbf{\mathbf{\
^{\shortparallel }}}}}\mathcal{\Re }_{\quad \alpha _{s}\beta _{s}\mu
_{s}}^{\star \mu _{s}}=\mathbf{\mathbf{\mathbf{\mathbf{\
_{1}^{\shortparallel }R}}}}ic_{\alpha _{s}\beta _{s}}^{\star }+\mathbf{%
\mathbf{\mathbf{\mathbf{\ _{2}^{\shortparallel }R}}}}ic_{\alpha _{s}\beta
_{s}}^{\star },\mbox{ for } \\
&&\mathbf{\mathbf{\mathbf{\mathbf{\ _{1}^{\shortparallel }R}}}}ic_{\alpha
_{s}\beta _{s}}^{\star }=\mathbf{\mathbf{\mathbf{\mathbf{\
_{1}^{\shortparallel }}}}}\mathcal{\Re }_{\quad \alpha _{s}\beta _{s}\mu
_{s}}^{\star \mu _{s}}\mbox{ and }\ _{2}^{\shortparallel }\mathbf{\mathbf{%
\mathbf{\mathbf{R}}}}ic_{\alpha _{s}\beta _{s}}^{\star }=\
_{2}^{\shortparallel }\mathcal{\Re }_{\quad \alpha _{s}\beta _{s}\mu
_{s}}^{\star \mu _{s}}.
\end{eqnarray*}%
Such computations are presented in details via associative and commutative
formulas in (A1) of \cite{vacaru18,bubuianu18a}), see also references
therein. In dyadic form, we obtain such s-coefficients:%
\begin{eqnarray}
\ \mathbf{\mathbf{\mathbf{\mathbf{^{\shortparallel }R}}}}ic_{\alpha
_{s}\beta _{s}}^{\star } &=&\{\ ^{\shortparallel }R_{\ \star h_{1}j_{1}}=\ \
^{\shortparallel }\mathcal{\Re }_{\ \star h_{1}j_{1}i_{1}}^{i_{1}},\ \
^{\shortparallel }P_{\star j_{1}a_{2}}^{\ }=-\ ^{\shortparallel }\mathcal{%
\Re }_{\star \ j_{1}i_{1}a_{2}}^{i_{1}\quad }\ ,\ ^{\shortparallel }P_{\star
\ b_{2}k_{1}}=\ ^{\shortparallel }\mathcal{\Re }_{\star
b_{2}k_{1}c_{2}}^{c_{2}\ },\ ^{\shortparallel }S_{\star b_{2}c_{2}\ }=\
^{\shortparallel }\mathcal{\Re }_{\star b_{2}c_{2}a_{2}\ }^{a_{2}\ },  \notag
\\
\ ^{\shortparallel }P_{\star j_{2}}^{\ a_{3}} &=&-\ ^{\shortparallel }%
\mathcal{\Re }_{\star \ j_{2}i_{2}}^{i_{2}\quad a_{3}}\ ,\ ^{\shortparallel
}P_{\star \ k_{2}}^{b_{3}\quad }=\ ^{\shortparallel }\mathcal{\Re }_{\star
c_{3}\ k_{2}}^{\ b_{3}\ c_{3}},\ ^{\shortparallel }S_{\star \
}^{b_{3}c_{3}\quad }=\ ^{\shortparallel }\mathcal{\Re }_{\star a_{3}\
}^{\quad b_{3}c_{3}a_{3}}\ ,  \label{driccinahc} \\
\ ^{\shortparallel }P_{\star j_{3}}^{\ a_{4}} &=&-\ ^{\shortparallel }%
\mathcal{\Re }_{\star \ j_{3}i_{3}}^{i_{3}\quad a_{4}}\ ,\ ^{\shortparallel
}P_{\star \ k_{3}}^{b_{4}\quad }=\ ^{\shortparallel }\mathcal{\Re }_{\star
c_{4}\ k_{3}}^{\ b_{4}\ c_{4}},\ ^{\shortparallel }S_{\star \
}^{b_{4}c_{4}\quad }=\ ^{\shortparallel }\mathcal{\Re }_{\star a_{4}\
}^{\quad b_{4}c_{4}a_{4}}\ \}.  \notag
\end{eqnarray}%
We can decompose this s-tensor in powers of $i\kappa ,$
\begin{equation*}
\ \mathbf{\mathbf{\mathbf{\mathbf{^{\shortparallel }R}}}}ic_{\alpha
_{s}\beta _{s}}^{\star }=(i\kappa )^{0}\ _{[0]}^{\shortparallel }\mathbf{%
\mathbf{\mathbf{\mathbf{R}}}}ic_{\alpha _{s}\beta _{s}}^{\star }+(i\kappa
)^{1}\ _{[1]}^{\shortparallel }\mathbf{\mathbf{\mathbf{\mathbf{R}}}}%
ic_{\alpha _{s}\beta _{s}}^{\star }-\kappa ^{2}\ _{[2]}^{\shortparallel }%
\mathbf{\mathbf{\mathbf{\mathbf{R}}}}ic_{\alpha _{s}\beta _{s}}^{\star },
\end{equation*}%
when $\ _{[0]}^{\shortparallel }\mathbf{R}ic_{\alpha \beta }^{\star }$
include nonassociative and noncommutative contributions from star products
of type (\ref{starpn}) and their generalizations for quasi-Hopf
s-structures. These decompositions are similar to formulas (4.83) in \cite%
{aschieri17} providing a holonomic version of such a decomposition on powers
of $\kappa .$ In this work, we generalize in nonhollonomic dyadic form those
formulas considering that $\partial \rightarrow \ _{s}^{\shortparallel }%
\mathbf{e}$ and nontrivial s-connection structure result in non-zero
anholonomy coefficients $\ ^{\shortparallel }w_{\ \alpha _{s}\gamma
_{s}}^{\nu _{s}}.$ Such nonholonomic s-structures allow us to apply the
AFCDM.

\subsection{Nonassociative s-metrics and canonical s-connections with small
parameters}

\label{assparamet}Decompositions of star nonholonomic deformations of
d-metrics and canonical d-connections on small parameters $\hbar $ and $%
\kappa $ were studied in section 3.3.2 of \cite{partner01}. Here, we
extended in N-adapted form the coordinate frame approach to nonassociative
geometry with quasi-Hopf structures, (non) symmetric metrics and
LC-connections with parametric decompositions provided in section 5.3 of
\cite{aschieri17}.

We can transform all (non) commutative and/ or nonassociative geometric
d-objects into s-objects (we shall omit dubbing coordinate formulas from
just cited works) using nonholonomic dyadic decompositions with two linear
star deformed connection structures (\ref{twoconsstar}), s-frames and (non)
symmetric metrics and respective canonical s-distortions (\ref{candistrnas})
determined by fundamental geometric data:
\begin{eqnarray}
(\ \mathbf{^{\shortparallel }\partial }_{\alpha };\ _{\star
}^{\shortparallel }\mathsf{G}_{\alpha \beta } &=&\ _{\star }^{\shortparallel
}g_{\alpha \beta }-i\kappa \mathcal{R}_{\quad \alpha }^{\tau \xi }\ \
\mathbf{^{\shortparallel }}\partial _{\xi }\ _{\star }^{\shortparallel
}g_{\beta \tau }:=\ _{\star }^{\shortparallel }\mathsf{G}_{\alpha \beta
}^{[0]}+\ _{\star }^{\shortparallel }\mathsf{G}_{\alpha \beta }^{[1]}(\kappa
),\mbox{see (\ref{offdns1}) and (\ref{offdns1inv})},  \label{conv2a} \\
\ _{\star }^{\shortparallel }\mathsf{G}^{\alpha \beta } &=&\
^{\shortparallel }g^{\alpha \beta }-i\kappa \ ^{\shortparallel }g^{\alpha
\tau }\mathcal{R}_{\quad \tau }^{\mu \nu }(\ \mathbf{^{\shortparallel }}%
\partial _{\mu }\ ^{\shortparallel }g_{\nu \varepsilon })\ ^{\shortparallel
}g^{\varepsilon \beta }+O(\kappa ^{2}):=\ _{\star }^{\shortparallel }\mathsf{%
G}_{[0]}^{\alpha \beta }+\ _{\star }^{\shortparallel }\mathsf{G}%
_{[1]}^{\alpha \beta }(\kappa )+O(\kappa ^{2});  \notag \\
&&\ ^{\shortparallel }\nabla ^{\star },\mbox{ see (\ref{twoconsstar}}))
\notag \\
&&\Updownarrow  \notag \\
(\ \mathbf{^{\shortparallel }e}_{\alpha },\ ^{\shortparallel }w_{\alpha
\beta }^{\gamma },&& \ _{\star }^{\shortparallel }\mathfrak{g}_{\alpha \beta
},\ ^{\shortparallel }\widehat{\mathbf{D}}^{\star } =\ ^{\shortparallel
}\nabla ^{\star }+\ _{\star }^{\shortparallel }\widehat{\mathbf{Z}}%
)\longleftrightarrow (\ \mathbf{^{\shortparallel }e}_{\alpha _{s}},\
^{\shortparallel }w_{\alpha _{s}\beta _{s}}^{\gamma _{s}},\ _{\star
}^{\shortparallel }\mathfrak{g}_{\alpha \beta },\ _{s}^{\shortparallel }%
\widehat{\mathbf{D}}^{\star }=\ ^{\shortparallel }\nabla ^{\star }+\ _{\star
s}^{\shortparallel }\widehat{\mathbf{Z}}).  \notag
\end{eqnarray}%
Such transforms allow to compute additionally to holonomic geometric objects
with $\ \mathbf{^{\shortparallel }\partial }_{\alpha }$ certain new star
nonholonomic terms of the d-/ s-curvature and d-/s-torsion, and Ricci d-/
s-tensor. The nonholonomic star distortions appear, for instance, with
s-adapted frames $\ ^{\shortparallel }\mathbf{e}_{\alpha _{s}}$ and their
nontrivial anholonomy coefficients $\ ^{\shortparallel }w_{\alpha _{s}\beta
_{s}}^{\gamma _{s}}.$

We provide in this subsection certain important s-formulas which show how
the nonassociative geometric dyadic formalism differs from the cases with
(non) holonomic and N-connection structures. There will be introduced
certain parametric on $\hbar $ and $\kappa $ formulas written in generic
off-diagonal and/or s-adapted forms.

Let us introduce in a coordinate local basis
\begin{equation}
\ _{\star }^{\shortparallel }\mathsf{W}_{\gamma \alpha \beta }=\ _{\star
}^{\shortparallel }g_{\gamma \mu }\mathbf{\mathbf{\ ^{\shortparallel }}}%
\Gamma _{\star \alpha \beta }^{\nu }+i\kappa \mathcal{R}_{\quad \alpha
}^{\tau \xi }\ \ \mathbf{^{\shortparallel }}\partial _{\xi }\ _{\star
}^{\shortparallel }g_{\beta \tau }:=\ _{\star }^{\shortparallel }\mathsf{W}%
_{\gamma \alpha \beta }^{[0]}+\ _{\star }^{\shortparallel }\mathsf{W}%
_{\gamma \alpha \beta }^{[1]}(\kappa ),  \label{aux36}
\end{equation}%
where, for respective zero parameters, $\ ^{\shortparallel }\Gamma _{\star
\alpha \beta \mid \hbar ,\kappa =0}^{\nu }=\ ^{\shortparallel }\Gamma
_{\alpha \beta }^{\nu }$ are taken as the coefficients of the commutative
LC-connection $\ ^{\shortparallel }\nabla $ in (\ref{twocon}). For
nontrivial $\hbar $ and $\kappa ,$ we work with a nonassociative $\
^{\shortparallel }\nabla ^{\star }.$ The geometric techniques of
decomposition on small parameters and (\ref{aux36}) can be extended for
nonholonomic dyadic canonical data $(\ _{\star}^{\shortparallel }\mathfrak{g}%
_{\alpha _{s}\beta _{s}},\ _{s}^{\shortparallel }\widehat{\mathbf{D}}^{\star
})$ following Convention 2 (\ref{conv2s}) and rules (\ref{conv2a}), working
in s-adapted (co) bases. We express
\begin{eqnarray}
\ _{\star }^{\shortparallel }\mathfrak{g}_{\alpha _{s}\beta _{s}} &=&\
_{\star }^{\shortparallel }\mathbf{g}_{\alpha _{s}\beta _{s}}-i\kappa
\overline{\mathcal{R}}_{\quad \alpha _{s}}^{\tau _{s}\xi _{s}}\ \mathbf{%
^{\shortparallel }e}_{\xi _{s}}\ _{\star }^{\shortparallel }\mathbf{g}%
_{\beta _{s}\tau _{s}}:=\ _{\star }^{\shortparallel }\mathfrak{g}_{\alpha
_{s}\beta _{s}}^{[0]}+\ _{\star }^{\shortparallel }\mathfrak{g}_{\alpha
_{s}\beta _{s}}^{[1]}(\kappa )\mbox{ and }  \notag \\
\ _{\star }^{\shortparallel }\mathfrak{g}^{\alpha _{s}\beta _{s}} &=&\ \
_{\star }^{\shortparallel }\mathbf{g}^{\alpha _{s}\beta _{s}}-i\kappa \ \
_{\star }^{\shortparallel }\mathbf{g}^{\alpha _{s}\tau _{s}}\overline{%
\mathcal{R}}_{\quad \tau _{s}}^{\mu _{s}\nu _{s}}(\ \mathbf{^{\shortparallel
}e}_{\mu _{s}}\ \ _{\star }^{\shortparallel }\mathbf{g}_{\nu _{s}\varepsilon
_{s}})\ \ _{\star }^{\shortparallel }\mathbf{g}^{\varepsilon _{s}\beta
_{s}}+O(\kappa ^{2})\ :=\ _{\star }^{\shortparallel }\mathfrak{g}%
_{[0]}^{\alpha _{s}\beta _{s}}+\ _{\star }^{\shortparallel }\mathfrak{g}%
_{[1]}^{\alpha _{s}\beta _{s}}(\kappa )+O(\kappa ^{2}),  \notag \\
\ _{\ast }^{\shortparallel }\mathfrak{g}^{\alpha _{s}\beta _{s}} &=&2\
_{\star }^{\shortparallel }\mathfrak{g}^{\alpha _{s}\beta _{s}}-\ _{\star
}^{\shortparallel }\mathfrak{g}^{\alpha _{s}\gamma _{s}}\ast \ _{\star
}^{\shortparallel }\mathfrak{g}_{\gamma _{s}\tau _{s}}\ast \ _{\star
}^{\shortparallel }\mathfrak{g}^{\tau _{s}\beta _{s}}+O(\hbar ^{2}),
\label{aux37}
\end{eqnarray}%
with possible explicit s-splitting of the star s-metric $\
_{\star}^{\shortparallel }\mathfrak{g}_{\gamma _{s}\tau _{s}}$ (\ref{dmss1})
into h1-v2-c3-c4 components for respective shells $s=1,2,3,4.$ For a star
canonical s-connection $\ ^{\shortparallel}\widehat{\Gamma }_{\star \alpha
_{s}\beta _{s}}^{\nu _{s}}$ (\ref{eqnasdmdc}), we consider similarly to (\ref%
{aux36}) that
\begin{equation}
\ _{\star }^{\shortparallel }\widehat{\mathsf{W}}_{\gamma _{s}\alpha
_{s}\beta _{s}}=\ _{\star }^{\shortparallel }\mathbf{g}_{\gamma _{s}\mu _{s}}%
\mathbf{\mathbf{\ ^{\shortparallel }}}\widehat{\Gamma }_{\star \alpha
_{s}\beta _{s}}^{\nu _{s}}+i\kappa \overline{\mathcal{R}}_{\quad \alpha
_{s}}^{\tau _{s}\xi _{s}}\ \mathbf{^{\shortparallel }e}_{\xi _{s}}\ _{\star
}^{\shortparallel }\mathbf{g}_{\beta _{s}\tau _{s}}:=\ _{\star
}^{\shortparallel }\widehat{\mathsf{W}}_{\gamma _{s}\alpha _{s}\beta
_{s}}^{[0]}+\ _{\star }^{\shortparallel }\widehat{\mathsf{W}}_{\gamma
_{s}\alpha _{s}\beta _{s}}^{[1]}(\kappa ).  \label{aux38}
\end{equation}

We can re-define the N-adapted formulas to s-adapted ones\footnote{%
for holonomic configurations, see in details the Section 5.2, formulas
(5.55)-(5.65), of \cite{aschieri17} on decomposition of the coefficients of
metrics and linear connections} with coefficients up to parameters $\hbar ,$
$\kappa $ and $\hbar $ $\kappa $ for the star s-metrics (symmetric and
nonsymmetric) and star canonical s-connections. The coordinate formulas for
the LC-connections can be distorted following (\ref{candistrnas}).

\subsubsection{Parametric decomposition of inverse nonsymmetric star
s-metrics}

\label{assinvsm}Considering in explicit form the s-coefficients proportional
to both parameters, we obtain such formulas for the nonsymmetric d-metric
from (\ref{aux37}):
\begin{equation}
\ _{\ast }^{\shortparallel }\mathfrak{g}_{[0]}^{\alpha _{s}\beta _{s}}=\
_{\star }^{\shortparallel }\mathfrak{g}_{[00]}^{\alpha _{s}\beta _{s}}+\
_{\star }^{\shortparallel }\mathfrak{g}_{[01]}^{\alpha _{s}\beta _{s}}(\hbar
)+O(\hbar ^{2},\kappa ^{2})\mbox{ and }\ _{\ast }^{\shortparallel }\mathfrak{%
g}_{[1]}^{\alpha \beta }=\ _{\star }^{\shortparallel }\mathfrak{g}%
_{[10]}^{\alpha \beta }(\kappa )+\ _{\star }^{\shortparallel }\mathfrak{g}%
_{[11]}^{\alpha \beta }(\hbar \kappa )+O(\hbar ^{2},\kappa ^{2}),
\label{aux38a}
\end{equation}%
\begin{eqnarray*}
\mbox{ where }\ _{\star }^{\shortparallel }\mathfrak{g}_{[00]}^{\alpha
_{s}\beta _{s}} &=&\ ^{\shortparallel }\mathbf{g}^{\alpha _{s}\beta _{s}}, \\
\ _{\star }^{\shortparallel }\mathfrak{g}_{[01]}^{\alpha _{s}\beta _{s}} &=&%
\frac{i\hbar }{2}(\ ^{\shortparallel }\mathbf{e}^{n+i_{1}}\ ^{\shortparallel
}\mathbf{g}^{\alpha _{s}\gamma _{s}}\ ^{\shortparallel }\mathbf{e}_{i_{1}}\
^{\shortparallel }\mathbf{g}_{\gamma _{s}\tau _{s}}-\ ^{\shortparallel }%
\mathbf{e}_{i_{1}}\ ^{\shortparallel }\mathbf{g}^{\alpha _{s}\gamma _{s}}\
^{\shortparallel }\mathbf{e}^{n+i_{1}}\ ^{\shortparallel }\mathbf{g}_{\gamma
_{s}\tau _{s}}+ \\
&&~^{\shortparallel }\partial ^{n+i_{2}}\ ^{\shortparallel }\mathbf{g}%
^{\alpha _{s}\gamma _{s}}\ \ ^{\shortparallel }\mathbf{e}_{i_{2}}\
^{\shortparallel }\mathbf{g}_{\gamma _{s}\tau _{s}}-\ ^{\shortparallel }%
\mathbf{e}_{i_{2}}\ ^{\shortparallel }\mathbf{g}^{\alpha _{s}\gamma _{s}}\
^{\shortparallel }\partial ^{n+i_{2}}\ ^{\shortparallel }\mathbf{g}_{\gamma
_{s}\tau _{s}})\ ^{\shortparallel }\mathbf{g}^{\tau _{s}\beta _{s}}, \\
\ _{\star }^{\shortparallel }\mathfrak{g}_{[10]}^{\alpha _{s}\beta _{s}}
&=&i\kappa \overline{\mathcal{R}}_{\quad \gamma _{s}}^{\tau _{s}\xi _{s}}\
^{\shortparallel }\mathbf{g}^{\alpha _{s}\gamma _{s}}(\ \mathbf{%
^{\shortparallel }e}_{\xi _{s}}\ ^{\shortparallel }\mathbf{g}_{\mu _{s}\tau
_{s}})\ ^{\shortparallel }\mathbf{g}^{\mu _{s}\beta _{s}},
\end{eqnarray*}%
\begin{eqnarray*}
\ _{\star }^{\shortparallel }\mathfrak{g}_{[11]}^{\alpha _{1}\beta _{1}} &=&%
\frac{\hbar \kappa }{2}\overline{\mathcal{R}}_{\quad \gamma _{1}}^{\tau
_{1}\xi _{1}}\{(\ \mathbf{^{\shortparallel }e}_{\xi _{1}}\ ^{\shortparallel }%
\mathbf{g}_{\mu _{1}\tau _{1}})[\ ^{\shortparallel }\mathbf{g}^{\nu
_{s}\gamma _{s}}\ ^{\shortparallel }\mathbf{g}^{\mu _{s}\beta _{1}}(\
\mathbf{^{\shortparallel }e}_{i_{1}}\ ^{\shortparallel }\mathbf{g}^{\alpha
_{s}\lambda _{s}}\ ^{\shortparallel }\mathbf{e}^{n+i_{1}}\ ^{\shortparallel }%
\mathbf{g}_{\lambda _{s}\nu _{s}}- \\
&&\ ^{\shortparallel }\mathbf{e}^{n+i_{1}}\ ^{\shortparallel }\mathbf{g}%
^{\alpha _{s}\lambda _{s}}\ \ \mathbf{^{\shortparallel }e}_{i_{1}}\
^{\shortparallel }\mathbf{g}_{\lambda _{s}\nu _{s}}+\ \mathbf{%
^{\shortparallel }e}_{i_{2}}\ ^{\shortparallel }\mathbf{g}^{\alpha
_{s}\lambda _{s}}\ ^{\shortparallel }\partial ^{n+i_{2}}\ ^{\shortparallel }%
\mathbf{g}_{\lambda _{s}\nu _{s}}-\ ^{\shortparallel }\partial ^{n+i_{2}}\
^{\shortparallel }\mathbf{g}^{\alpha _{s}\lambda _{s}}\ \mathbf{%
^{\shortparallel }e}_{i_{2}}\ ^{\shortparallel }\mathbf{g}_{\lambda _{s}\nu
_{s}})+ \\
&&\ ^{\shortparallel }\mathbf{g}^{\alpha _{s}\gamma _{s}}\ ^{\shortparallel }%
\mathbf{g}^{\nu _{s}\beta _{s}}(\ \mathbf{^{\shortparallel }e}_{i_{1}}\
^{\shortparallel }\mathbf{g}^{\mu _{s}\lambda _{s}}\ ^{\shortparallel }%
\mathbf{e}^{n+i_{1}}\ ^{\shortparallel }\mathbf{g}_{\lambda _{s}\nu _{s}}-\
^{\shortparallel }\mathbf{e}^{n+i_{1}}\ ^{\shortparallel }\mathbf{g}^{\mu
_{s}\lambda _{s}}\ \ \mathbf{^{\shortparallel }e}_{i_{1}}\ ^{\shortparallel }%
\mathbf{g}_{\lambda _{s}\nu _{s}}+ \\
&&\ \mathbf{^{\shortparallel }e}_{i_{2}}\ ^{\shortparallel }\mathbf{g}^{\mu
_{s}\lambda _{s}}\ ^{\shortparallel }\partial ^{n+i_{2}}\ ^{\shortparallel }%
\mathbf{g}_{\lambda _{2}\nu _{2}}-\ ^{\shortparallel }\partial ^{n+i_{2}}\
^{\shortparallel }\mathbf{g}^{\mu _{s}\lambda _{s}}\ \ \mathbf{%
^{\shortparallel }e}_{i_{2}}\ ^{\shortparallel }\mathbf{g}_{\lambda _{s}\nu
_{s}})+(\ \mathbf{^{\shortparallel }e}_{i_{1}}\ ^{\shortparallel }\mathbf{g}%
^{\alpha _{s}\gamma _{s}}\ ^{\shortparallel }\mathbf{e}^{n+i_{1}}\
^{\shortparallel }\mathbf{g}^{\mu _{s}\beta _{s}}- \\
&&\ ^{\shortparallel }\mathbf{e}^{n+i_{1}}\ ^{\shortparallel }\mathbf{g}%
^{\alpha _{s}\gamma _{s}}\ \ \mathbf{^{\shortparallel }e}_{i_{1}}\
^{\shortparallel }\mathbf{g}^{\mu _{s}\beta _{s}}+\ \mathbf{^{\shortparallel
}e}_{i_{2}}\ ^{\shortparallel }\mathbf{g}^{\alpha _{s}\gamma _{s}}\
^{\shortparallel }\partial ^{n+i_{2}}\ ^{\shortparallel }\mathbf{g}^{\mu
_{s}\beta _{s}}-\ ^{\shortparallel }\partial ^{n+i_{2}}\ ^{\shortparallel }%
\mathbf{g}^{\alpha _{s}\gamma _{s}}\ \ \mathbf{^{\shortparallel }e}_{i_{2}}\
^{\shortparallel }\mathbf{g}^{\mu _{s}\beta _{s}})]- \\
&&(\ \mathbf{^{\shortparallel }e}_{i_{1}}\ \mathbf{^{\shortparallel }e}_{\xi
_{s}}\ ^{\shortparallel }\mathbf{g}_{\nu _{s}\tau _{s}})(\ ^{\shortparallel }%
\mathbf{g}^{\alpha _{s}\gamma _{s}}\ ^{\shortparallel }\mathbf{e}^{n+i_{1}}\
^{\shortparallel }\mathbf{g}^{\nu _{s}\beta _{s}}-\ ^{\shortparallel }%
\mathbf{g}^{\nu _{s}\beta _{s}}\ ^{\shortparallel }\mathbf{e}^{n+i_{1}}\
^{\shortparallel }\mathbf{g}^{\alpha _{s}\gamma _{s}})- \\
&&(\ \ \mathbf{^{\shortparallel }e}_{i_{2}}\ \mathbf{^{\shortparallel }e}%
_{\xi _{s}}\ ^{\shortparallel }\mathbf{g}_{\nu _{s}\tau _{s}})(\
^{\shortparallel }\mathbf{g}^{\alpha _{s}\gamma _{s}}\ ^{\shortparallel
}\partial ^{n+i_{2}}\ ^{\shortparallel }\mathbf{g}^{\nu _{s}\beta _{s}}-\
^{\shortparallel }\mathbf{g}^{\nu _{s}\beta _{s}}\ ^{\shortparallel
}\partial ^{n+i_{2}}\ ^{\shortparallel }\mathbf{g}^{\alpha _{s}\gamma _{s}})+
\\
&&(\ ^{\shortparallel }\mathbf{e}^{n+i_{1}}\ \mathbf{^{\shortparallel }e}%
_{\xi _{s}}\ ^{\shortparallel }\mathbf{g}_{\nu _{s}\tau _{s}})(\
^{\shortparallel }\mathbf{g}^{\alpha _{s}\gamma _{s}}\ \mathbf{%
^{\shortparallel }e}_{i_{1}}\ ^{\shortparallel }\mathbf{g}^{\nu _{s}\beta
_{s}}-\ ^{\shortparallel }\mathbf{g}^{\nu _{s}\beta _{s}}\ \ \mathbf{%
^{\shortparallel }e}_{i_{1}}\ ^{\shortparallel }\mathbf{g}^{\alpha
_{s}\gamma _{s}})+ \\
&&(\ \ ^{\shortparallel }\partial ^{n+i_{2}}\ \mathbf{^{\shortparallel }e}%
_{\xi _{s}}\ ^{\shortparallel }\mathbf{g}_{\nu _{s}\tau _{s}})(\
^{\shortparallel }\mathbf{g}^{\alpha _{s}\gamma _{s}}\ \mathbf{%
^{\shortparallel }e}_{i_{2}}\ ^{\shortparallel }\mathbf{g}^{\nu _{s}\beta
_{s}}-\ ^{\shortparallel }\mathbf{g}^{\nu _{s}\beta _{s}}\ \mathbf{%
^{\shortparallel }e}_{i_{2}}\ ^{\shortparallel }\mathbf{g}^{\alpha
_{s}\gamma _{s}})\}+O(\hbar ^{2},\kappa ^{2}),
\end{eqnarray*}%
\begin{equation*}
\mbox{ for }\ ^{\shortparallel }\mathbf{e}^{n+i_{1}}=\frac{\partial }{%
\partial \ ^{\shortparallel }p_{n+i_{1}}}+\ ^{\shortparallel
}N_{n+i_{2}}^{n+i_{1}}\frac{\partial }{\partial \ ^{\shortparallel
}p_{n+i_{2}}},\ ^{\shortparallel }\partial ^{n+i_{2}}=\frac{\partial }{%
\partial \ ^{\shortparallel }p_{n+i_{2}}},\mbox{ when }i_{1}=1,2\mbox{ and }%
i_{2}=3,4.
\end{equation*}%
We can prescribe the nonholonomic s-frame structure when for $\
^{\shortparallel }\mathbf{e}_{\alpha _{s}}\rightarrow \ ^{\shortparallel} {%
\partial }_{\alpha}$ and, inversely, following the Convention 2 (\ref{conv2s}%
), all above formulas are equivalent correspondingly to coordinate base
formulas (5.59) - (5.61) computed in \cite{aschieri17}, see also N-adapted
formulas (118) in \cite{partner01}.

\subsubsection{Nonholonomic dyadic parametric decomposition of star
canonical s-connections}

Computing s-adapted parametric decompositions for $\ _{\star
}^{\shortparallel }\mathfrak{g}_{\alpha _{s}\beta _{s}}=\ _{\star
}^{\shortparallel }\mathfrak{g}_{\alpha _{s}\beta _{s}}^{[0]}+\ _{\star
}^{\shortparallel }\mathfrak{g}_{\alpha _{s}\beta _{s}}^{[1]},$ $\ _{\ast
}^{\shortparallel }\mathfrak{g}^{\alpha _{s}\beta _{s}}=\ _{\ast
}^{\shortparallel }\mathfrak{g}_{[0]}^{\alpha _{s}\beta _{s}}+\ _{\ast
}^{\shortparallel }\mathfrak{g}_{[1]}^{\alpha _{s}\beta _{s}}$ (\ref{aux38})
and $\ ^{\shortparallel }\widehat{\mathbf{\Gamma }}_{\star \alpha _{s}\beta
_{s}}^{\gamma _{s}}$ (\ref{eqnasdmdc}), we obtain%
\begin{eqnarray}
\ _{[0]}^{\shortparallel }\widehat{\Gamma }_{\ast \alpha _{s}\beta
_{s}}^{\nu _{s}} &=&\ _{\ast }^{\shortparallel }\mathfrak{g}_{[0]}^{\nu
_{s}\gamma _{s}}\ast \ _{\ast }^{\shortparallel }\widehat{\mathsf{W}}%
_{\gamma _{s}\alpha \beta }^{[0]}:=\ _{[00]}^{\shortparallel }\widehat{%
\Gamma }_{\star \alpha \beta }^{\nu _{s}}+\ _{[01]}^{\shortparallel }%
\widehat{\Gamma }_{\star \alpha \beta }^{\nu _{s}}(\hbar )+O(\hbar ^{2}),
\label{aux39} \\
\ _{[1]}^{\shortparallel }\widehat{\Gamma }_{\ast \alpha _{s}\beta
_{s}}^{\nu _{s}} &=&\ _{\ast }^{\shortparallel }\mathfrak{g}_{[0]}^{\nu
_{s}\gamma _{s}}\ast \ _{\ast }^{\shortparallel }\widehat{\mathsf{W}}%
_{\gamma _{s}\alpha _{s}\beta _{s}}^{[1]}+\ _{\ast }^{\shortparallel }%
\mathfrak{g}_{[1]}^{\nu _{s}\gamma _{s}}\ast \ _{\ast }^{\shortparallel }%
\widehat{\mathsf{W}}_{\gamma _{s}\alpha _{s}\beta _{s}}^{[0]}:=\
_{[10]}^{\shortparallel }\widehat{\Gamma }_{\star \alpha _{s}\beta
_{s}}^{\nu _{s}}(\kappa )+\ _{[11]}^{\shortparallel }\widehat{\Gamma }%
_{\star \alpha _{s}\beta _{s}}^{\nu _{s}}(\hbar \kappa )+O(\hbar ^{2}),
\notag
\end{eqnarray}%
\begin{eqnarray}
\mbox{ where }\ _{[00]}^{\shortparallel }\widehat{\Gamma }_{\star \alpha
_{s}\beta _{s}}^{\nu _{s}} &=&\ ^{\shortparallel }\widehat{\Gamma }_{\
\alpha _{s}\beta _{s}}^{\nu _{s}},\ _{[01]}^{\shortparallel }\widehat{\Gamma
}_{\star \alpha _{s}\beta _{s}}^{\nu _{s}}=-\frac{i\hbar }{2}\
^{\shortparallel }\mathbf{g}^{\nu _{s}\gamma _{s}}(\ \mathbf{%
^{\shortparallel }e}_{i_{1}}\ ^{\shortparallel }\mathbf{g}_{\gamma _{s}\tau
_{s}}\ ^{\shortparallel }\partial ^{n+i_{1}}\ ^{\shortparallel }\widehat{%
\Gamma }_{\ \alpha _{s}\beta _{s}}^{\tau _{s}}\ -  \label{aux51} \\
&&\ ^{\shortparallel }\mathbf{e}^{n+i_{1}}\ \ ^{\shortparallel }\mathbf{g}%
_{\gamma _{s}\tau _{s}}\ \ \mathbf{^{\shortparallel }e}_{i_{1}}\
^{\shortparallel }\widehat{\Gamma }_{\ \alpha _{s}\beta _{s}}^{\tau _{s}}+\
\mathbf{^{\shortparallel }e}_{i}\ ^{\shortparallel }\mathbf{g}_{\gamma
_{s}\tau _{s}}\ ^{\shortparallel }\partial ^{n+i_{2}}\ ^{\shortparallel }%
\widehat{\Gamma }_{\ \alpha _{s}\beta _{s}}^{\tau _{s}}\ -\ ^{\shortparallel
}\partial ^{n+i_{2}}\ \ ^{\shortparallel }\mathbf{g}_{\gamma _{s}\tau _{s}}\
\ \mathbf{^{\shortparallel }e}_{i}\ ^{\shortparallel }\widehat{\Gamma }_{\
\alpha _{s}\beta _{s}}^{\tau _{s}}).  \notag
\end{eqnarray}%
To simplify s-formulas with contracting of spacetime indices form $s=1,2$
with respective $s=3,4,$ we shall use also such a convention:
\begin{equation}
^{\shortparallel }\mathbf{e}^{n+i_{s}}=(^{\shortparallel }\mathbf{e}%
^{n+i_{1}}=\frac{\partial }{\partial \ ^{\shortparallel }p_{n+i_{1}}}+\
^{\shortparallel }N_{n+k_{2}}^{n+i_{1}}\frac{\partial }{\partial \
^{\shortparallel }p_{n+k_{2}}},\ ^{\shortparallel }\partial ^{n+i_{2}}=\frac{%
\partial }{\partial \ ^{\shortparallel }p_{n+i_{2}}}),\mbox{ when }i_{1}=1,2%
\mbox{ and }i_{2}=3,4.  \label{aux52}
\end{equation}%
So, we compute and write%
\begin{equation*}
\ _{[10]}^{\shortparallel }\widehat{\Gamma }_{\star \alpha _{s}\beta
_{s}}^{\nu _{s}}=i\kappa \overline{\mathcal{R}}_{\quad \quad \quad
}^{n+i_{s}\ n+j_{s}\ n+k_{s}}\ (\ \ ^{\shortparallel }\mathbf{g}%
_{~n+k_{s}}^{\nu _{s}}\quad ^{\shortparallel }\mathbf{g}_{j_{s}m_{s}}(\
\mathbf{^{\shortparallel }e}_{i_{s}}\ ^{\shortparallel }\widehat{\Gamma }_{\
\alpha _{s}\beta _{s}}^{m_{s}})-\ ^{\shortparallel }\mathbf{g}_{\ }^{\nu
_{s}\mu _{s}}\ \ \mathbf{^{\shortparallel }}p_{n+j_{s}}(\ \ \mathbf{%
^{\shortparallel }e}_{k_{s}}\ ^{\shortparallel }\mathbf{g}_{\mu _{s}\tau
_{s}})(\ \mathbf{^{\shortparallel }e}_{i_{s}}\ ^{\shortparallel }\widehat{%
\Gamma }_{\ \alpha _{s}\beta _{s}}^{\tau _{s}})),
\end{equation*}%
\begin{eqnarray}
\ _{[11]}^{\shortparallel }\widehat{\Gamma }_{\star \alpha _{s}\beta
_{s}}^{\nu _{s}} &=&i\frac{\kappa \hbar }{2}\overline{\mathcal{R}}_{\quad
\quad \quad }^{n+i_{s}\ n+j_{s}\ n+k_{s}}\{-(\ \mathbf{^{\shortparallel }e}%
_{l_{s}}\ ^{\shortparallel }\mathbf{g}_{~n+k_{s}}^{\nu _{s}})\
^{\shortparallel }\mathbf{e}^{n+l_{s}}[\ \mathbf{^{\shortparallel }e}%
_{i_{s}}(~^{\shortparallel }\mathbf{g}_{j_{s}m_{s}}\ ^{\shortparallel }%
\widehat{\Gamma }_{\ \alpha _{s}\beta _{s}}^{m_{s}})]+  \label{aux39a} \\
&&(\ ^{\shortparallel }\mathbf{e}^{n+l_{s}}\ ^{\shortparallel }\mathbf{g}%
_{~~n+k_{s}}^{\nu _{s}})[~\mathbf{^{\shortparallel }e}_{l_{s}}(\ \ \mathbf{%
^{\shortparallel }e}_{i_{s}}~^{\shortparallel }\mathbf{g}_{j_{s}m_{s}}\
^{\shortparallel }\widehat{\Gamma }_{\ \alpha _{s}\beta _{s}}^{m_{s}})]+
\notag
\end{eqnarray}%
\begin{eqnarray*}
&&[(\ ^{\shortparallel }\mathbf{e}_{l_{s}}\ ^{\shortparallel }\mathbf{g}%
_{~}^{\nu _{s}\gamma _{s}})(\ ^{\shortparallel }\mathbf{e}^{n+l_{s}}\
^{\shortparallel }\mathbf{g}_{\gamma _{s}\tau _{s}})-(\ ^{\shortparallel }%
\mathbf{e}^{n+l_{s}}~^{\shortparallel }\mathbf{g}_{~}^{\nu _{s}\gamma
_{s}})(\ \mathbf{^{\shortparallel }e}_{l_{s}}\ ^{\shortparallel }\mathbf{g}%
_{\gamma _{s}\tau _{s}})]\ ^{\shortparallel }\mathbf{g}_{~n+k_{s}}^{\tau
_{s}}(\ \ \mathbf{^{\shortparallel }e}_{i_{s}}~^{\shortparallel }\mathbf{g}%
_{j_{s}m_{s}}\ ^{\shortparallel }\widehat{\Gamma }_{\ \alpha _{s}\beta
_{s}}^{m_{s}})+ \\
&&\ \mathbf{^{\shortparallel }e}_{l_{s}}[\ ^{\shortparallel }\mathbf{g}%
_{~n+k_{s}}^{\nu _{s}}~^{\shortparallel }\mathbf{g}_{~}^{q_{s}m_{s}}(\
\mathbf{^{\shortparallel }e}_{i_{s}}\ ^{\shortparallel }\mathbf{g}%
_{q_{s}j_{s}})]\ ^{\shortparallel }\mathbf{e}^{n+l_{s}}(\ ~^{\shortparallel }%
\mathbf{g}_{m_{s}o_{s}}\ ^{\shortparallel }\widehat{\Gamma }_{\ \alpha
_{s}\beta _{s}}^{o_{s}})- \\
&&\ ^{\shortparallel }\mathbf{e}^{n+l_{s}}[(\ ^{\shortparallel }\mathbf{g}%
_{~n+k_{s}}^{\nu _{s}}~^{\shortparallel }\mathbf{g}_{~}^{q_{s}m_{s}}(\
\mathbf{^{\shortparallel }e}_{i_{s}}\ ^{\shortparallel }\mathbf{g}%
_{q_{s}j_{s}})]\ \mathbf{^{\shortparallel }e}_{l_{s}}(\ ~^{\shortparallel }%
\mathbf{g}_{m_{s}o_{s}}\ ^{\shortparallel }\widehat{\Gamma }_{\ \alpha
_{s}\beta _{s}}^{o_{s}})-
\end{eqnarray*}%
\begin{eqnarray*}
&&(\ \mathbf{^{\shortparallel }e}_{i_{s}}\ ^{\shortparallel }\mathbf{g}%
_{q_{s}j_{s}})[\ ^{\shortparallel }\mathbf{g}_{~n+k_{s}}^{\tau _{s}}\left(
(\ \mathbf{^{\shortparallel }e}_{l_{s}}~^{\shortparallel }\mathbf{g}%
_{~}^{\nu _{s}\gamma _{s}})(\ ^{\shortparallel }\mathbf{e}^{n+l_{s}}\
^{\shortparallel }\mathbf{g}_{\gamma _{s}\tau _{s}})-(\ ^{\shortparallel }%
\mathbf{e}^{n+l_{s}}~^{\shortparallel }\mathbf{g}_{~}^{\nu _{s}\gamma
_{s}})(\ \mathbf{^{\shortparallel }e}_{l_{s}}\ ^{\shortparallel }\mathbf{g}%
_{\gamma _{s}\tau _{s}})\right) \ ^{\shortparallel }\widehat{\Gamma }_{\
\alpha _{s}\beta _{s}}^{q_{s}}+ \\
&&\ ^{\shortparallel }\mathbf{g}_{~n+k_{s}}^{\nu _{s}}\left( (\ \mathbf{%
^{\shortparallel }e}_{l_{s}}~^{\shortparallel }\mathbf{g}_{~}^{\nu
_{s}\gamma _{s}})(\ ^{\shortparallel }\mathbf{e}^{n+k_{s}}\ ^{\shortparallel
}\mathbf{g}_{\gamma _{s}\tau _{s}})-(\ ^{\shortparallel }\mathbf{e}%
^{n+k_{s}}~^{\shortparallel }\mathbf{g}_{~}^{\nu _{s}\gamma _{s}})(\ \mathbf{%
^{\shortparallel }e}_{l_{s}}\ ^{\shortparallel }\mathbf{g}_{\gamma _{s}\tau
_{s}})\right) \ ^{\shortparallel }\widehat{\Gamma }_{\ \alpha _{s}\beta
_{s}}^{\tau _{s}}+ \\
&&\left( \ (~\mathbf{^{\shortparallel }e}_{l_{s}}\ ^{\shortparallel }\mathbf{%
g}_{~n+k_{s}}^{\nu _{s}})(\ ^{\shortparallel }\mathbf{e}^{n+l_{s}}~^{%
\shortparallel }\mathbf{g}_{~}^{\mu _{s}\gamma _{s}})-(\ ^{\shortparallel }%
\mathbf{e}^{n+k_{s}}~\ ^{\shortparallel }\mathbf{g}_{~n+k_{s}}^{\nu _{s}})(~%
\mathbf{^{\shortparallel }e}_{l_{s}}\ ^{\shortparallel }\mathbf{g}_{~}^{\mu
_{s}\gamma _{s}})\ \right) ~^{\shortparallel }\mathbf{g}_{\gamma _{s}\tau
_{s}}\ ^{\shortparallel }\widehat{\Gamma }_{\ \alpha _{s}\beta _{s}}^{\tau
_{s}}]+
\end{eqnarray*}%
\begin{eqnarray*}
&&(~\mathbf{^{\shortparallel }e}_{l_{s}}~\mathbf{^{\shortparallel }e}%
_{i_{s}}\ ^{\shortparallel }\mathbf{g}_{q_{s}j_{s}})[(\ ^{\shortparallel }%
\mathbf{g}_{~n+k_{s}}^{\nu _{s}})(\ ^{\shortparallel }\mathbf{e}%
^{n+l_{s}}~^{\shortparallel }\mathbf{g}_{~}^{q_{s}s_{s}})\ ^{\shortparallel }%
\mathbf{g}_{s_{s}p_{s}}\ ^{\shortparallel }\widehat{\Gamma }_{\ \alpha
_{s}\beta _{s}}^{p_{s}}-(\ ^{\shortparallel }\mathbf{e}^{n+l_{s}}~\
^{\shortparallel }\mathbf{g}_{~n+k_{s}}^{\nu _{s}})\ ^{\shortparallel }%
\widehat{\Gamma }_{\ \alpha _{s}\beta _{s}}^{q_{s}}]- \\
&&(\ ^{\shortparallel }\mathbf{e}^{n+l_{s}}~\mathbf{^{\shortparallel }e}%
_{i_{s}}\ ^{\shortparallel }\mathbf{g}_{q_{s}j_{s}})[(\ ^{\shortparallel }%
\mathbf{g}_{~n+k_{s}}^{\nu _{s}})(~\mathbf{^{\shortparallel }e}_{l_{s}}\
~^{\shortparallel }\mathbf{g}_{~}^{q_{s}s_{s}})\ ^{\shortparallel }\mathbf{g}%
_{s_{s}p_{s}}\ ^{\shortparallel }\widehat{\Gamma }_{\ \alpha _{s}\beta
_{s}}^{p_{s}}-(~\mathbf{^{\shortparallel }e}_{l_{s}}~\ ^{\shortparallel }%
\mathbf{g}_{~n+k_{s}}^{\nu _{s}})\ ^{\shortparallel }\widehat{\Gamma }_{\
\alpha _{s}\beta _{s}}^{q_{s}}]+ \\
&&p_{n+j_{s}}[~\mathbf{^{\shortparallel }e}_{l_{s}}\left( ~^{\shortparallel }%
\mathbf{g}_{~}^{\nu _{s}\tau _{s}}(~\mathbf{^{\shortparallel }e}%
_{k_{s}}~^{\shortparallel }\mathbf{g}_{\tau _{s}\mu _{s}})\right) \
^{\shortparallel }\mathbf{e}^{n+l_{s}}(~\mathbf{^{\shortparallel }e}%
_{i_{s}}\ \ ^{\shortparallel }\widehat{\Gamma }_{\ \alpha _{s}\beta
_{s}}^{\mu _{s}})- \\
&&\ ^{\shortparallel }\mathbf{e}^{n+l_{s}}\left( ~^{\shortparallel }\mathbf{g%
}_{~}^{\nu _{s}\tau _{s}}(~\mathbf{^{\shortparallel }e}_{k_{s}}~^{%
\shortparallel }\mathbf{g}_{\tau _{s}\mu _{s}})\right) ~\mathbf{%
^{\shortparallel }e}_{l_{s}}(~\mathbf{^{\shortparallel }e}_{i_{s}}\
^{\shortparallel }\widehat{\Gamma }_{\ \alpha _{s}\beta _{s}}^{\mu _{s}})]+
\\
&&p_{n+j_{s}}~^{\shortparallel }\mathbf{g}_{~}^{\nu _{s}\tau _{s}}(~\mathbf{%
^{\shortparallel }e}_{k_{s}}~^{\shortparallel }\mathbf{g}_{\tau _{s}\mu
_{s}})~\mathbf{^{\shortparallel }e}_{i_{s}}\ [((~\mathbf{^{\shortparallel }e}%
_{l_{s}}~^{\shortparallel }\mathbf{g}_{~}^{\mu _{s}\gamma _{s}})\
^{\shortparallel }\mathbf{e}^{n+l_{s}}(~^{\shortparallel }\mathbf{g}_{\gamma
_{s}\tau _{s}}\ ^{\shortparallel }\widehat{\Gamma }_{\ \alpha _{s}\beta
_{s}}^{\tau _{s}})- \\
&&(\ ^{\shortparallel }\mathbf{e}^{n+l_{s}}~^{\shortparallel }\mathbf{g}%
_{~}^{\mu _{s}\gamma _{s}})~\mathbf{^{\shortparallel }e}_{l_{s}}(~^{%
\shortparallel }\mathbf{g}_{\gamma _{s}\tau _{s}}\ ^{\shortparallel }%
\widehat{\Gamma }_{\ \alpha _{s}\beta _{s}}^{\tau _{s}}))-
\end{eqnarray*}%
\begin{eqnarray*}
&&\left( (~\mathbf{^{\shortparallel }e}_{l_{s}}~^{\shortparallel }\mathbf{g}%
_{~}^{\mu _{s}\gamma _{s}})(\ ^{\shortparallel }\mathbf{e}%
^{n+l_{s}}~^{\shortparallel }\mathbf{g}_{\gamma _{s}\tau _{s}})-(\
^{\shortparallel }\mathbf{e}^{n+l_{s}}~^{\shortparallel }\mathbf{g}_{~}^{\mu
_{s}\gamma _{s}})(~\mathbf{^{\shortparallel }e}_{l_{s}}~^{\shortparallel }%
\mathbf{g}_{\gamma _{s}\tau _{s}})\right) \ ^{\shortparallel }\widehat{%
\Gamma }_{\ \alpha _{s}\beta _{s}}^{\tau _{s}}]- \\
&&p_{n+j_{s}}\left( (~\mathbf{^{\shortparallel }e}_{l_{s}}~^{\shortparallel }%
\mathbf{g}_{~}^{\nu _{s}\tau _{s}})(\ ^{\shortparallel }\mathbf{e}%
^{n+l_{s}}~~\mathbf{^{\shortparallel }e}_{k_{s}}~^{\shortparallel }\mathbf{g}%
_{\tau _{s}\mu _{s}})-(\ ^{\shortparallel }\mathbf{e}^{n+l_{s}}~^{%
\shortparallel }\mathbf{g}_{~}^{\nu _{s}\tau _{s}})(~~\mathbf{%
^{\shortparallel }e}_{l_{s}}~\mathbf{^{\shortparallel }e}_{k_{s}}~^{%
\shortparallel }\mathbf{g}_{\tau _{s}\mu _{s}})\right) (~~\mathbf{%
^{\shortparallel }e}_{i_{s}}\ ^{\shortparallel }\widehat{\Gamma }_{\ \alpha
_{s}\beta _{s}}^{\mu _{s}})- \\
&&p_{n+j_{s}}((~\mathbf{^{\shortparallel }e}_{l_{s}}~^{\shortparallel }%
\mathbf{g}_{~}^{\nu _{s}\tau _{s}})(\ ^{\shortparallel }\mathbf{e}%
^{n+l_{s}}~~\mathbf{^{\shortparallel }e}_{k_{s}}~^{\shortparallel }\mathbf{g}%
_{\tau _{s}\varepsilon _{s}})- \\
&&(\ ^{\shortparallel }\mathbf{e}^{n+l_{s}}~^{\shortparallel }\mathbf{g}%
_{~}^{\nu _{s}\tau _{s}})(~~\mathbf{^{\shortparallel }e}_{l_{s}}~^{%
\shortparallel }\mathbf{g}_{\tau _{s}\varepsilon _{s}}))~^{\shortparallel }%
\mathbf{g}_{~}^{\varepsilon _{s}\lambda _{s}}(~\mathbf{^{\shortparallel }e}%
_{k_{s}}~^{\shortparallel }\mathbf{g}_{\lambda _{s}\mu _{s}})(~\mathbf{%
^{\shortparallel }e}_{i_{s}}\ ^{\shortparallel }\widehat{\Gamma }_{\ \alpha
_{s}\beta _{s}}^{\mu _{s}})+ \\
&&(~\mathbf{^{\shortparallel }e}_{l_{s}}\ ^{\shortparallel }\mathbf{g}%
_{~}^{\nu _{s}\tau _{s}})(~\mathbf{^{\shortparallel }e}_{j_{s}}~^{%
\shortparallel }\mathbf{g}_{\tau _{s}\mu _{s}})(~\mathbf{^{\shortparallel }e}%
_{k_{s}}\ ^{\shortparallel }\widehat{\Gamma }_{\ \alpha _{s}\beta _{a}}^{\mu
_{s}})\}.
\end{eqnarray*}%
These formulas are for the canonical s-connection $\ ^{\shortparallel }%
\widehat{\mathbf{D}}^{\star }$ (\ref{candistrnas}). We can extract formulas
for the nonassociative LC-connection for zero distortion s-tensors when $\
_{s}^{\shortparallel }\widehat{\mathbf{D}}_{\mid \ ^{\shortparallel }%
\widehat{\mathbf{T}}=0}^{\star }=\ ^{\shortparallel }\nabla ^{\star },$ see (%
\ref{lccondnonass}). We can prescribe the nonholonomic s-frame structure in
such a form when for $\ \mathbf{^{\shortparallel }e}_{\alpha
_{s}}\rightarrow $ $\ \mathbf{^{\shortparallel }\partial }_{\alpha }$ and,
inversely, following the Convention 2 (\ref{conv2s}). This way we obtain
both coordinate basis formulas (5.62) - (5.65) from \cite{aschieri17} and
formulas (119)-(120) from \cite{partner01} if we redefine the constructions
in N-adapted form.

Above formulas can be subjected also to further to
h1-v2-c3-c4-decompositions with respect to s-frames $\ ^{\shortparallel }%
\mathbf{e}^{\alpha _{s}}$ and $\ ^{\shortparallel }\mathbf{e}_{\alpha _{s}}$%
, when $\ _{[00]}^{\shortparallel }\widehat{\Gamma }_{\ast \alpha _{s}\beta
_{s}}^{\nu _{s}}=\ ^{\shortparallel }\widehat{\Gamma }_{\alpha _{s}\beta
_{s}}^{\nu _{s}}$ are used for taking the canonical s-connection from (\ref%
{twocon}) and then for considering R-flux star deformations of nonholonomic
dyadic structure into nonassociative geometric models determined by
nonassociative canonical s-connection from (\ref{twoconsstar}).

\subsubsection{Parametric coefficients for the nonassocative canonical Ricci
s-tensor}

\label{assnaricci}For the [00] term in (\ref{driccicanonstar1}), we compute
\begin{eqnarray}
\ _{[00]}^{\shortparallel }\widehat{\mathbf{R}}ic_{\beta _{s}\gamma
_{s}}^{\star } &=&\ ^{\shortparallel }\widehat{\mathbf{R}}ic_{\beta
_{s}\gamma _{s}}=  \label{ric50} \\
\ ^{\shortparallel }\widehat{\mathbf{R}}_{\beta _{s}\gamma _{s}} &=&\mathbf{%
\mathbf{\mathbf{\mathbf{\ ^{\shortparallel }}}}e}_{\alpha _{s}}\mathbf{%
\mathbf{\mathbf{\mathbf{\ ^{\shortparallel }}}}}\widehat{\mathbf{\Gamma }}%
_{\ \beta _{s}\gamma _{s}}^{\alpha _{s}}-\mathbf{\mathbf{\mathbf{\mathbf{\
^{\shortparallel }}}}e}_{\beta _{s}}\mathbf{\mathbf{\mathbf{\mathbf{\
^{\shortparallel }}}}}\widehat{\mathbf{\Gamma }}_{\ \gamma _{s}\alpha
_{s}}^{\alpha _{s}}+\mathbf{\mathbf{\mathbf{\mathbf{\ ^{\shortparallel }}}}}%
\widehat{\mathbf{\Gamma }}_{\ \beta _{s}\gamma _{s}}^{\mu _{s}}\mathbf{%
\mathbf{\mathbf{\mathbf{\ ^{\shortparallel }}}}}\widehat{\mathbf{\Gamma }}%
_{\ \mu _{s}\alpha _{s}}^{\alpha _{s}}-\mathbf{\mathbf{\mathbf{\mathbf{\
^{\shortparallel }}}}}\widehat{\mathbf{\Gamma }}_{\ \beta _{s}\alpha
_{s}}^{\mu _{s}}\mathbf{\mathbf{\mathbf{\mathbf{\ ^{\shortparallel }}}}}%
\widehat{\mathbf{\Gamma }}_{\ \mu _{s}\gamma _{s}}^{\alpha _{s}}+\mathbf{%
\mathbf{\mathbf{\mathbf{\ ^{\shortparallel }}}}}w_{\ \gamma _{s}\alpha
_{s}}^{\mu _{s}}\mathbf{\mathbf{\mathbf{\mathbf{\ ^{\shortparallel }}}}}%
\widehat{\mathbf{\Gamma }}_{\ \beta _{s}\mu _{s}}^{\alpha _{s}},  \notag
\end{eqnarray}%
when further s-decompositions are stated by associative and commutative
values (\ref{candricci}).

The terms with labels [0,1] determine the order $\hbar $ contribution to
star deformations of the Ricci s-tensor for respective parametric
decompositions of the canonical s-connection,
\begin{eqnarray}
\mathbf{\mathbf{\mathbf{\mathbf{\ \ }}}}_{[01]}^{\shortparallel }\widehat{%
\mathbf{\mathbf{\mathbf{\mathbf{R}}}}}ic_{\beta _{s}\gamma _{s}}^{\star } &=&%
\mathbf{\mathbf{\mathbf{\mathbf{\ ^{\shortparallel }}}}e}_{\alpha _{s}}%
\mathbf{\mathbf{\mathbf{\mathbf{\ }}}}_{[01]}^{\shortparallel }\widehat{%
\mathbf{\Gamma }}_{\ \beta _{s}\gamma _{s}}^{\alpha _{s}}-\mathbf{\mathbf{%
\mathbf{\mathbf{\ ^{\shortparallel }}}}e}_{\beta _{s}}\
_{[01]}^{\shortparallel }\widehat{\mathbf{\Gamma }}_{\ \gamma _{s}\alpha
_{s}}^{\alpha _{s}}+\mathbf{\mathbf{\mathbf{\mathbf{\ }}}}%
_{[01]}^{\shortparallel }\widehat{\mathbf{\Gamma }}_{\ \beta _{s}\gamma
_{s}}^{\mu _{s}}\mathbf{\mathbf{\mathbf{\mathbf{\ ^{\shortparallel }}}}}%
\widehat{\mathbf{\Gamma }}_{\ \mu _{s}\alpha _{s}}^{\alpha _{s}}+\mathbf{%
\mathbf{\mathbf{\mathbf{\ ^{\shortparallel }}}}}\widehat{\mathbf{\Gamma }}%
_{\ \beta _{s}\gamma _{s}}^{\mu _{s}}\mathbf{\mathbf{\mathbf{\mathbf{\ \ }}}}%
_{[01]}^{\shortparallel }\widehat{\mathbf{\Gamma }}_{\ \mu _{s}\alpha
_{s}}^{\alpha _{s}}  \notag \\
&&-\mathbf{\mathbf{\mathbf{\mathbf{\ \ }}}}_{[01]}^{\shortparallel }\widehat{%
\mathbf{\Gamma }}_{\ \beta _{s}\alpha _{s}}^{\mu _{s}}\mathbf{\mathbf{%
\mathbf{\mathbf{\ ^{\shortparallel }}}}}\widehat{\mathbf{\Gamma }}_{\ \mu
_{s}\gamma _{s}}^{\alpha _{s}}-\mathbf{\mathbf{\mathbf{\mathbf{\
^{\shortparallel }}}}}\widehat{\mathbf{\Gamma }}_{\ \beta _{s}\alpha
_{s}}^{\mu _{s}}\mathbf{\mathbf{\mathbf{\mathbf{\ }}}}_{[01]}^{%
\shortparallel }\widehat{\mathbf{\Gamma }}_{\ \mu _{s}\gamma _{s}}^{\alpha
_{s}}+\mathbf{\mathbf{\mathbf{\mathbf{\ ^{\shortparallel }}}}}w_{\ \gamma
_{s}\alpha _{s}}^{\mu _{s}}\mathbf{\mathbf{\mathbf{\mathbf{\ }}}}%
_{[01]}^{\shortparallel }\widehat{\mathbf{\Gamma }}_{\ \beta _{s}\mu
_{s}}^{\alpha _{s}}  \label{ric51} \\
&&+\frac{i\hbar }{2}[(\ \mathbf{^{\shortparallel }e}_{l_{s}}~\mathbf{\mathbf{%
\mathbf{\mathbf{^{\shortparallel }}}}}\widehat{\mathbf{\Gamma }}_{\ \mu
_{s}\alpha _{s}}^{\alpha _{s}})(\ ^{\shortparallel }\mathbf{e}^{n+l_{s}}\
\mathbf{\mathbf{\mathbf{\mathbf{\ ^{\shortparallel }}}}}\widehat{\mathbf{%
\Gamma }}_{\ \beta _{s}\gamma _{s}}^{\mu _{s}})-(\ ^{\shortparallel }\mathbf{%
e}^{n+l_{s}}\ ~\mathbf{\mathbf{\mathbf{\mathbf{^{\shortparallel }}}}}%
\widehat{\mathbf{\Gamma }}_{\ \mu _{s}\alpha _{s}}^{\alpha _{s}})(~\mathbf{%
^{\shortparallel }e}_{l_{s}}\mathbf{\mathbf{\mathbf{\mathbf{\
^{\shortparallel }}}}}\widehat{\mathbf{\Gamma }}_{\ \beta _{s}\gamma
_{s}}^{\mu _{s}})]  \notag \\
&&-\frac{i\hbar }{2}[(\ \mathbf{^{\shortparallel }e}_{l_{s}}\mathbf{\mathbf{%
\mathbf{\mathbf{\ ^{\shortparallel }}}}}\widehat{\mathbf{\Gamma }}_{\ \mu
_{s}\gamma _{s}}^{\alpha _{s}})(\ ^{\shortparallel }\mathbf{e}^{n+l_{s}}\
\mathbf{\mathbf{\mathbf{\mathbf{\ ^{\shortparallel }}}}}\widehat{\mathbf{%
\Gamma }}_{\ \beta _{s}\alpha _{s}}^{\mu _{s}})-(\ ^{\shortparallel }\mathbf{%
e}^{n+l_{s}}\ \mathbf{\mathbf{\mathbf{\mathbf{\ ^{\shortparallel }}}}}%
\widehat{\mathbf{\Gamma }}_{\ \mu _{s}\gamma _{s}}^{\alpha _{s}})(~\mathbf{%
^{\shortparallel }e}_{l_{s}}\mathbf{\mathbf{\mathbf{\mathbf{\ \
^{\shortparallel }}}}}\widehat{\mathbf{\Gamma }}_{\ \beta _{s}\alpha
_{s}}^{\mu _{s}})].  \notag
\end{eqnarray}%
If we introduce in (\ref{ric51}) the formulas for s-adapted coefficients of
type $\ _{[01]}^{\shortparallel }\widehat{\Gamma }_{\star \alpha _{s}\beta
_{s}}^{\nu _{s}}$ (\ref{aux51}), we can complete the procedure of
computation of such s-terms. We do not provide such cumbersome formulas, and
possible explicit h1-v2-c3-c4 decompositions and similar ones with labels
[1,0], [1,1] etc., because they will be not used in this work.

Now, we show how to compute the [1,0] terms of order $\kappa .$ For $\
_{[00]}^{\shortparallel }\widehat{\Gamma }_{\star \alpha _{s}\beta
_{s}}^{\nu _{s}}=$ $\ ^{\shortparallel }\widehat{\Gamma }_{~\alpha _{s}\beta
_{s}}^{\nu _{s}}$ and $\ _{[10]}^{\shortparallel }\widehat{\Gamma }_{\star
\alpha _{s}\beta _{s}}^{\nu _{s}}$ from (\ref{aux39}) and (\ref{aux51}).
Using, Kronecker s-adapted symbols of type $\delta _{\gamma _{s}k_{s}}$ and $%
\delta _{a_{s}}^{\gamma _{s}}$, we obtain such s-adapted components of the
nonassociative Ricci s-tensor:
\begin{eqnarray}
\ _{[10]}^{\shortparallel }\widehat{\mathbf{\mathbf{\mathbf{\mathbf{R}}}}}%
ic_{\beta _{s}\gamma _{s}}^{\star } &=& \mathbf{\mathbf{\mathbf{\mathbf{\
^{\shortparallel }}}}e}_{\alpha _{s}}\mathbf{\mathbf{\mathbf{\mathbf{\ }}}}%
_{[10]}^{\shortparallel }\widehat{\mathbf{\Gamma }}_{\ \beta _{s}\gamma
_{s}}^{\alpha _{s}}-\mathbf{\mathbf{\mathbf{\mathbf{\ ^{\shortparallel }}}}e}%
_{\beta _{s}}\ _{[10]}^{\shortparallel }\widehat{\mathbf{\Gamma }}_{\ \gamma
_{s}\alpha _{s}}^{\alpha _{s}}+\mathbf{\mathbf{\mathbf{\mathbf{\ }}}}%
_{[10]}^{\shortparallel }\widehat{\mathbf{\Gamma }}_{\ \beta _{s}\gamma
_{s}}^{\mu _{s}}\mathbf{\mathbf{\mathbf{\mathbf{\ ^{\shortparallel }}}}}%
\widehat{\mathbf{\Gamma }}_{\ \mu _{s}\alpha _{s}}^{\alpha _{s}}+\mathbf{%
\mathbf{\mathbf{\mathbf{\ ^{\shortparallel }}}}}\widehat{\mathbf{\Gamma }}%
_{\ \beta _{s}\gamma _{s}}^{\mu _{s}}\mathbf{\mathbf{\mathbf{\mathbf{\ \ }}}}%
_{[10]}^{\shortparallel }\widehat{\mathbf{\Gamma }}_{\ \mu _{s}\alpha
_{s}}^{\alpha _{s}}-  \notag \\
&&\mathbf{\mathbf{\mathbf{\mathbf{\ }}}}_{[10]}^{\shortparallel }\widehat{%
\mathbf{\Gamma }}_{\ \beta _{s}\alpha _{s}}^{\mu _{s}}\mathbf{\mathbf{%
\mathbf{\mathbf{\ ^{\shortparallel }}}}}\widehat{\mathbf{\Gamma }}_{\ \mu
_{s}\gamma _{s}}^{\alpha _{s}}-\mathbf{\mathbf{\mathbf{\mathbf{\
^{\shortparallel }}}}}\widehat{\mathbf{\Gamma }}_{\ \beta _{s}\alpha
_{s}}^{\mu _{s}}\mathbf{\mathbf{\mathbf{\mathbf{\ }}}}_{[10]}^{%
\shortparallel }\widehat{\mathbf{\Gamma }}_{\ \mu _{s}\gamma _{s}}^{\alpha
_{s}}+\mathbf{\mathbf{\mathbf{\mathbf{\ ^{\shortparallel }}}}}w_{\ \gamma
_{s}\alpha _{s}}^{\mu _{s}}\mathbf{\mathbf{\mathbf{\mathbf{\ }}}}%
_{[10]}^{\shortparallel }\widehat{\mathbf{\Gamma }}_{\ \beta _{s}\mu
_{s}}^{\alpha _{s}}+  \label{ric52} \\
&&i\kappa \overline{\mathcal{R}}_{\quad \quad \quad }^{n+i_{s}\ b_{s}\
n+k_{s}}\mathbf{\mathbf{\mathbf{\mathbf{\ }}}}\ ^{\shortparallel }p_{b_{s}}[(%
\mathbf{\mathbf{\mathbf{\mathbf{\ ^{\shortparallel }}}}e}_{k_{s}}\mathbf{%
\mathbf{\mathbf{\mathbf{\ ^{\shortparallel }}}}}\widehat{\mathbf{\Gamma }}%
_{\ \mu _{s}\alpha _{s}}^{\alpha _{s}})(\mathbf{\mathbf{\mathbf{\mathbf{\
^{\shortparallel }}}}e}_{i_{s}}\mathbf{\mathbf{\mathbf{\mathbf{\
^{\shortparallel }}}}}\widehat{\mathbf{\Gamma }}_{\ \beta _{s}\gamma
_{s}}^{\mu _{s}})-(\mathbf{\mathbf{\mathbf{\mathbf{\ ^{\shortparallel }}}}e}%
_{k_{s}}\mathbf{\mathbf{\mathbf{\mathbf{\ ^{\shortparallel }}}}}\widehat{%
\mathbf{\Gamma }}_{\ \mu _{s}\gamma _{s}}^{\alpha _{s}})(\mathbf{\mathbf{%
\mathbf{\mathbf{\ ^{\shortparallel }}}}e}_{i_{s}}\mathbf{\mathbf{\mathbf{%
\mathbf{\ ^{\shortparallel }}}}}\widehat{\mathbf{\Gamma }}_{\ \beta
_{s}\alpha _{s}}^{\mu _{s}})]-  \notag \\
&&i\kappa \overline{\mathcal{R}}_{\quad \quad \quad }^{n+i_{s}\ n+j_{s}\
n+k_{s}}\delta _{\gamma _{s}k_{s}}\mathbf{\mathbf{\mathbf{\mathbf{\
^{\shortparallel }}}}}\widehat{\mathbf{\Gamma }}_{\ \mu _{s}i_{s}}^{\alpha
_{s}}(\mathbf{\mathbf{\mathbf{\mathbf{\ ^{\shortparallel }}}}e}_{j_{s}}%
\mathbf{\mathbf{\mathbf{\mathbf{\ ^{\shortparallel }}}}}\widehat{\mathbf{%
\Gamma }}_{\ \beta _{s}\alpha _{s}}^{\mu _{s}})+i\kappa \overline{\mathcal{R}%
}_{\quad \quad \quad }^{n+i_{s}\ n+j_{s}\ a_{s}}\delta _{a_{s}}^{\gamma
_{s}}[\mathbf{\mathbf{\mathbf{\mathbf{\ ^{\shortparallel }}}}e}_{j_{s}}%
\mathbf{\mathbf{\mathbf{\mathbf{\ ^{\shortparallel }}}}e}_{\gamma _{s}}%
\widehat{\mathbf{\Gamma }}_{\ \beta _{s}i_{s}}^{\alpha _{s}}-  \notag \\
&&(\mathbf{\mathbf{\mathbf{\mathbf{\ ^{\shortparallel }}}}e}_{j_{s}}\mathbf{%
\mathbf{\mathbf{\mathbf{\ ^{\shortparallel }}}}}\widehat{\mathbf{\Gamma }}%
_{\ \mu _{s}i_{s}}^{\alpha _{s}})\mathbf{\mathbf{\mathbf{\mathbf{\
^{\shortparallel }}}}}\widehat{\mathbf{\Gamma }}_{\ \beta _{s}\gamma
_{s}}^{\mu _{s}}+(\mathbf{\mathbf{\mathbf{\mathbf{\ ^{\shortparallel }}}}e}%
_{j_{s}}\mathbf{\mathbf{\mathbf{\mathbf{\ ^{\shortparallel }}}}}\widehat{%
\mathbf{\Gamma }}_{\ \mu _{s}\gamma _{s}}^{\alpha _{s}})\mathbf{\mathbf{%
\mathbf{\mathbf{\ ^{\shortparallel }}}}}\widehat{\mathbf{\Gamma }}_{\ \beta
_{s}i_{s}}^{\mu _{s}}+~\mathbf{\mathbf{\mathbf{\mathbf{^{\shortparallel }}}}}%
\widehat{\mathbf{\Gamma }}_{\ \mu _{s}\gamma _{s}}^{\alpha _{s}}(\mathbf{%
\mathbf{\mathbf{\mathbf{\ ^{\shortparallel }}}}e}_{j_{s}}\mathbf{\mathbf{%
\mathbf{\mathbf{\ ^{\shortparallel }}}}}\widehat{\mathbf{\Gamma }}_{\ \beta
_{s}i_{s}}^{\mu _{s}})].  \notag
\end{eqnarray}%
We note that the terms (\ref{ric51}) and (\ref{ric52}) for the star deformed
Ricci s-tensor are imaginary as typical N-adapted configurations with
commutative real N-connection structure studied in \cite{partner01}. Such
properties were found (in our case, for zero torsion distortions to
LC-configurations) in \cite{aschieri17}. We can consider also inverse
transforms from holonomic configurations to N- and/ s-adapted ones. For
instance, the coordinate formulas (5.78) - (5.79) from \cite{aschieri17},
for the LC-connection, are modified into respective s-formulas (\ref{ric51})
and (\ref{ric52}) for the canonical s-connection if we follow the Convention
2 (\ref{conv2s}) with such rules:
\begin{eqnarray}
1) \ ^{\shortparallel }\partial _{\alpha } &\rightarrow &\ \ \mathbf{%
^{\shortparallel }e}_{\alpha _{s}}\mbox{ and }\lbrack \ \mathbf{%
^{\shortparallel }\partial }_{\alpha },\ \ \mathbf{^{\shortparallel
}\partial }_{\beta }]\rightarrow \lbrack \ \mathbf{^{\shortparallel }e}%
_{\alpha _{s}},\ \mathbf{^{\shortparallel }e}_{\beta _{s}}]+w_{\alpha
_{s}\beta _{s}}^{\gamma _{s}}\ \mathbf{^{\shortparallel }e}_{\gamma _{s}};
\label{rules12} \\
2)\ ^{\shortparallel }\Gamma _{\ \beta \gamma }^{\mu } &\rightarrow &\
^{\shortparallel}\widehat{\mathbf{\Gamma }}_{\ \beta _{s}\gamma _{s}}^{\mu
_{s}},\mbox{ i.e. }\mathbf{\mathbf{\mathbf{\mathbf{\ ^{\shortparallel }}}}}%
\nabla \rightarrow \mathbf{\mathbf{\mathbf{\mathbf{\ _{s}^{\shortparallel }}}%
}}\widehat{\mathbf{D}};\mbox{ and }\mathbf{\mathbf{\mathbf{\mathbf{\
^{\shortparallel }}}}}g_{\alpha \beta }\rightarrow \mathbf{\mathbf{\mathbf{%
\mathbf{\ ^{\shortparallel }}}}g}_{\alpha _{s}\beta _{s}},\mbox{etc.}  \notag
\end{eqnarray}%
In result, we can deform nonholonomically the formula (5.80) in \cite%
{aschieri17} and compute the [11] $\hbar \kappa =\ell _{s}^{3}/6$
contributions in dyadic form for the nonassociative canonical Ricci s-tensor
(\ref{driccicanonstar}),
\begin{eqnarray}
\mathbf{\mathbf{\mathbf{\mathbf{\ \ }}}}_{[11]}^{\shortparallel }\widehat{%
\mathbf{\mathbf{\mathbf{\mathbf{R}}}}}ic_{\beta _{s}\gamma _{s}}^{\star } &=&%
\mathbf{\mathbf{\mathbf{\mathbf{\ ^{\shortparallel }}}}e}_{\alpha _{s}}%
\mathbf{\mathbf{\mathbf{\mathbf{\ }}}}_{[11]}^{\shortparallel }\widehat{%
\mathbf{\Gamma }}_{\ \beta _{s}\gamma _{s}}^{\alpha _{s}}-\mathbf{\mathbf{%
\mathbf{\mathbf{\ ^{\shortparallel }}}}e}_{\beta _{s}}\
_{[11]}^{\shortparallel }\widehat{\mathbf{\Gamma }}_{\ \gamma _{s}\alpha
_{s}}^{\alpha _{s}}+\mathbf{\mathbf{\mathbf{\mathbf{\ }}}}%
_{[11]}^{\shortparallel }\widehat{\mathbf{\Gamma }}_{\ \beta _{s}\gamma
_{s}}^{\mu _{s}}\mathbf{\mathbf{\mathbf{\mathbf{\ ^{\shortparallel }}}}}%
\widehat{\mathbf{\Gamma }}_{\ \mu _{s}\alpha _{s}}^{\alpha _{s}}+\mathbf{%
\mathbf{\mathbf{\mathbf{\ }}}}_{[10]}^{\shortparallel }\widehat{\mathbf{%
\Gamma }}_{\ \beta _{s}\gamma _{s}}^{\mu _{s}}\ _{[01]}^{\shortparallel }%
\widehat{\mathbf{\Gamma }}_{\ \mu _{s}\alpha _{s}}^{\alpha _{s}}  \notag \\
&&+\mathbf{\mathbf{\mathbf{\mathbf{\ }}}}_{[01]}^{\shortparallel }\widehat{%
\mathbf{\Gamma }}_{\ \beta _{s}\gamma _{s}}^{\mu _{s}}\mathbf{\mathbf{%
\mathbf{\mathbf{\ }}}}_{[10]}^{\shortparallel }\widehat{\mathbf{\Gamma }}_{\
\mu _{s}\alpha _{s}}^{\alpha _{s}}+\mathbf{\mathbf{\mathbf{\mathbf{\
^{\shortparallel }}}}}\widehat{\mathbf{\Gamma }}_{\ \beta _{s}\gamma
_{s}}^{\mu _{s}}\mathbf{\mathbf{\mathbf{\mathbf{\ \ }}}}_{[11]}^{%
\shortparallel }\widehat{\mathbf{\Gamma }}_{\ \mu _{s}\alpha _{s}}^{\alpha
_{s}}-\mathbf{\mathbf{\mathbf{\mathbf{\ \ }}}}_{[11]}^{\shortparallel }%
\widehat{\mathbf{\Gamma }}_{\ \beta _{s}\alpha _{s}}^{\mu _{s}}\mathbf{%
\mathbf{\mathbf{\mathbf{\ ^{\shortparallel }}}}}\widehat{\mathbf{\Gamma }}%
_{\ \mu _{s}\gamma _{s}}^{\alpha _{s}}-\mathbf{\mathbf{\mathbf{\mathbf{\ \ }}%
}}_{[10]}^{\shortparallel }\widehat{\mathbf{\Gamma }}_{\ \beta _{s}\alpha
_{s}}^{\mu _{s}}\mathbf{\mathbf{\mathbf{\mathbf{\ }}}}_{[01]}^{%
\shortparallel }\widehat{\mathbf{\Gamma }}_{\ \mu _{s}\gamma _{s}}^{\alpha
_{s}}  \notag \\
&&-\mathbf{\mathbf{\mathbf{\mathbf{\ \ }}}}_{[01]}^{\shortparallel }\widehat{%
\mathbf{\Gamma }}_{\ \beta _{s}\alpha _{s}}^{\mu _{s}}\mathbf{\mathbf{%
\mathbf{\mathbf{\ }}}}_{[10]}^{\shortparallel }\widehat{\mathbf{\Gamma }}_{\
\mu _{s}\gamma _{s}}^{\alpha _{s}}-\mathbf{\mathbf{\mathbf{\mathbf{\
^{\shortparallel }}}}}\widehat{\mathbf{\Gamma }}_{\ \beta _{s}\alpha
_{s}}^{\mu _{s}}\mathbf{\mathbf{\mathbf{\mathbf{\ }}}}_{[11]}^{%
\shortparallel }\widehat{\mathbf{\Gamma }}_{\ \mu _{s}\gamma _{s}}^{\alpha
_{s}}+\mathbf{\mathbf{\mathbf{\mathbf{\ ^{\shortparallel }}}}}w_{\ \gamma
_{s}\alpha _{s}}^{\mu _{s}}\mathbf{\mathbf{\mathbf{\mathbf{\ }}}}%
_{[11]}^{\shortparallel }\widehat{\mathbf{\Gamma }}_{\ \beta _{s}\mu
_{s}}^{\alpha _{s}}  \label{ric53}
\end{eqnarray}%
\begin{eqnarray*}
&&+\frac{i\hbar }{2}[(\ \mathbf{^{\shortparallel }e}_{l_{s}}\
_{[10]}^{\shortparallel }\widehat{\mathbf{\Gamma }}_{\ \mu _{s}\alpha
_{s}}^{\alpha _{s}})(\ ^{\shortparallel }\mathbf{e}^{n+l_{s}}\mathbf{\mathbf{%
\mathbf{\mathbf{\ }}}}^{\shortparallel }\widehat{\mathbf{\Gamma }}_{\ \beta
_{s}\gamma _{s}}^{\mu _{s}})+(\ \mathbf{^{\shortparallel }e}_{l_{s}}\
^{\shortparallel }\widehat{\mathbf{\Gamma }}_{\ \mu _{s}\alpha _{s}}^{\alpha
_{s}})(\ ^{\shortparallel }\mathbf{e}^{n+l_{s}}\mathbf{\mathbf{\mathbf{%
\mathbf{\ }}}}_{[10]}^{\shortparallel }\widehat{\mathbf{\Gamma }}_{\ \beta
_{s}\gamma _{s}}^{\mu _{s}}) \\
&&-(\ ^{\shortparallel }\mathbf{e}^{n+l_{s}}\mathbf{\mathbf{\mathbf{\mathbf{%
\ }}}}\ _{[10]}^{\shortparallel }\widehat{\mathbf{\Gamma }}_{\ \mu
_{s}\alpha _{s}}^{\alpha _{s}})(\ \mathbf{^{\shortparallel }e}_{l_{s}}%
\mathbf{~}^{\shortparallel }\widehat{\mathbf{\Gamma }}_{\ \beta _{s}\gamma
_{s}}^{\mu _{s}})-(\ ^{\shortparallel }\mathbf{e}^{n+l_{s}}\
^{\shortparallel }\widehat{\mathbf{\Gamma }}_{\ \mu _{s}\alpha _{s}}^{\alpha
_{s}})(\ \mathbf{^{\shortparallel }e}_{l_{s}}\mathbf{\mathbf{\mathbf{\mathbf{%
\ }}}}_{[10]}^{\shortparallel }\widehat{\mathbf{\Gamma }}_{\ \beta
_{s}\gamma _{s}}^{\mu _{s}})] \\
&&-\frac{i\hbar }{2}[(\ \mathbf{^{\shortparallel }e}_{l_{s}}\mathbf{\mathbf{%
\mathbf{\mathbf{\ }}}}_{[10]}^{\shortparallel }\widehat{\mathbf{\Gamma }}_{\
\mu _{s}\gamma _{s}}^{\alpha _{s}})(\ ^{\shortparallel }\mathbf{e}^{n+l_{s}}%
\mathbf{\mathbf{\mathbf{\mathbf{\ ^{\shortparallel }}}}}\widehat{\mathbf{%
\Gamma }}_{\ \beta _{s}\alpha _{s}}^{\mu _{s}})+\ \mathbf{^{\shortparallel }e%
}_{l_{s}}\mathbf{\mathbf{\mathbf{\mathbf{\ }}}}^{\shortparallel }\widehat{%
\mathbf{\Gamma }}_{\ \mu _{s}\gamma _{s}}^{\alpha _{s}})(\ ^{\shortparallel }%
\mathbf{e}^{n+l_{s}}\ _{[10]}^{\shortparallel }\widehat{\mathbf{\Gamma }}_{\
\beta _{s}\alpha _{s}}^{\mu _{s}}) \\
&&-(\ ^{\shortparallel }\mathbf{e}^{n+l_{s}}\mathbf{\mathbf{\mathbf{\mathbf{%
\ }}}}_{[10]}^{\shortparallel }\widehat{\mathbf{\Gamma }}_{\ \mu _{s}\gamma
_{s}}^{\alpha _{s}})(\ \mathbf{^{\shortparallel }e}_{l_{s}}\mathbf{\mathbf{%
\mathbf{\mathbf{\ ^{\shortparallel }}}}}\widehat{\mathbf{\Gamma }}_{\ \beta
_{s}\alpha _{s}}^{\mu _{s}})-(\ ^{\shortparallel }\mathbf{e}^{n+l_{s}}%
\mathbf{\mathbf{\mathbf{\mathbf{\ }}}}^{\shortparallel }\widehat{\mathbf{%
\Gamma }}_{\ \mu _{s}\gamma _{s}}^{\alpha _{s}})(\ \mathbf{^{\shortparallel
}e}_{l_{s}}\ _{[10]}^{\shortparallel }\widehat{\mathbf{\Gamma }}_{\ \beta
_{s}\alpha _{s}}^{\mu _{s}})]
\end{eqnarray*}%
\begin{eqnarray*}
&&+i\kappa \overline{\mathcal{R}}_{\quad \quad \quad }^{n+i_{s}\ b_{s}\
n+k_{s}}\mathbf{\mathbf{\mathbf{\mathbf{\ }}}}\ ^{\shortparallel
}p_{b_{s}}[(\ \mathbf{^{\shortparallel }e}_{k_{s}}\mathbf{\mathbf{\mathbf{%
\mathbf{\ }}}}\ _{[01]}^{\shortparallel }\widehat{\mathbf{\Gamma }}_{\ \mu
_{s}\alpha _{s}}^{\alpha _{s}})(\ \mathbf{^{\shortparallel }e}_{i_{s}}%
\mathbf{~}^{\shortparallel }\widehat{\mathbf{\Gamma }}_{\ \beta _{s}\gamma
_{s}}^{\mu _{s}})+(\ \mathbf{^{\shortparallel }e}_{k_{s}}\ ^{\shortparallel }%
\widehat{\mathbf{\Gamma }}_{\ \mu _{s}\alpha _{s}}^{\alpha _{s}})(\ \mathbf{%
^{\shortparallel }e}_{i_{s}}\mathbf{~}_{[01]}^{\shortparallel }\widehat{%
\mathbf{\Gamma }}_{\ \beta _{s}\gamma _{s}}^{\mu _{s}}) \\
&&-(\ \mathbf{^{\shortparallel }e}_{k_{s}}\mathbf{\mathbf{\mathbf{\mathbf{\ }%
}}}_{[01]}^{\shortparallel }\widehat{\mathbf{\Gamma }}_{\ \mu _{s}\gamma
_{s}}^{\alpha _{s}})(\ \ \mathbf{^{\shortparallel }e}_{i_{s}}\mathbf{\mathbf{%
\mathbf{\mathbf{\ ^{\shortparallel }}}}}\widehat{\mathbf{\Gamma }}_{\ \beta
_{s}\alpha _{s}}^{\mu _{s}})-(\ \mathbf{^{\shortparallel }e}_{k_{s}}\mathbf{%
\mathbf{\mathbf{\mathbf{\ }}}}^{\shortparallel }\widehat{\mathbf{\Gamma }}%
_{\ \mu _{s}\gamma _{s}}^{\alpha _{s}})(\ \ \mathbf{^{\shortparallel }e}%
_{i_{s}}\mathbf{\mathbf{\mathbf{\mathbf{\ }}}}_{[01]}^{\shortparallel }%
\widehat{\mathbf{\Gamma }}_{\ \beta _{s}\alpha _{s}}^{\mu _{s}})]
\end{eqnarray*}%
\begin{eqnarray*}
&&-\frac{\hbar \kappa }{2}\overline{\mathcal{R}}_{\quad \quad \quad
}^{n+i_{s}\ b_{s}\ n+k_{s}}\mathbf{\mathbf{\mathbf{\mathbf{\ }}}}\
^{\shortparallel }p_{b_{s}}[(\ \mathbf{^{\shortparallel }e}_{l_{s}}\mathbf{~}%
\ \mathbf{^{\shortparallel }e}_{k_{s}}\mathbf{\mathbf{\mathbf{\mathbf{\ }}}}%
\ ^{\shortparallel }\widehat{\mathbf{\Gamma }}_{\ \mu _{s}\alpha
_{s}}^{\alpha _{s}})(\ ^{\shortparallel }\mathbf{e}^{n+l_{s}}\ \mathbf{%
^{\shortparallel }e}_{i_{s}}\mathbf{~}^{\shortparallel }\widehat{\mathbf{%
\Gamma }}_{\ \beta _{s}\gamma _{s}}^{\mu _{s}}) \\
&&-(\ ^{\shortparallel }\mathbf{e}^{n+l_{s}}\ \mathbf{^{\shortparallel }e}%
_{k_{s}}\ ^{\shortparallel }\widehat{\mathbf{\Gamma }}_{\ \mu _{s}\alpha
_{s}}^{\alpha _{s}})(\ \mathbf{^{\shortparallel }e}_{l_{s}}\mathbf{~}\
\mathbf{^{\shortparallel }e}_{i_{s}}\mathbf{~}^{\shortparallel }\widehat{%
\mathbf{\Gamma }}_{\ \beta _{s}\gamma _{s}}^{\mu _{s}}) \\
&&-(\ \mathbf{^{\shortparallel }e}_{l_{s}}\mathbf{~}\ \mathbf{%
^{\shortparallel }e}_{k_{s}}\mathbf{\mathbf{\mathbf{\mathbf{\ }}}}%
^{\shortparallel }\widehat{\mathbf{\Gamma }}_{\ \mu _{s}\gamma _{s}}^{\alpha
_{s}})(\ ^{\shortparallel }\mathbf{e}^{n+l_{s}}\ \ \mathbf{^{\shortparallel
}e}_{i_{s}}\mathbf{\mathbf{\mathbf{\mathbf{\ ^{\shortparallel }}}}}\widehat{%
\mathbf{\Gamma }}_{\ \beta _{s}\alpha _{s}}^{\mu _{s}})+(\ ^{\shortparallel }%
\mathbf{e}^{n+l_{s}}\ \mathbf{^{\shortparallel }e}_{k_{s}}\mathbf{\mathbf{%
\mathbf{\mathbf{\ }}}}^{\shortparallel }\widehat{\mathbf{\Gamma }}_{\ \mu
_{s}\gamma _{s}}^{\alpha _{s}})(\ \mathbf{^{\shortparallel }e}_{l_{s}}\ \
\mathbf{^{\shortparallel }e}_{i_{s}}\mathbf{\mathbf{\mathbf{\mathbf{\
^{\shortparallel }}}}}\widehat{\mathbf{\Gamma }}_{\ \beta _{s}\alpha
_{s}}^{\mu _{s}})]
\end{eqnarray*}%
\begin{eqnarray*}
&&-i\kappa \overline{\mathcal{R}}_{\quad \quad \quad }^{n+i_{s}\ n+j_{s}\
n+k_{s}}[\delta _{\alpha _{s}k_{s}}\left( (\ \mathbf{^{\shortparallel }e}%
_{j_{s}}\mathbf{\mathbf{\mathbf{\mathbf{\ }}}}^{\shortparallel }\widehat{%
\mathbf{\Gamma }}_{\ \mu _{s}i_{s}}^{\alpha _{s}})\ _{[01]}^{\shortparallel }%
\widehat{\mathbf{\Gamma }}_{\ \beta _{s}\gamma _{s}}^{\mu _{s}}+(\ \mathbf{%
^{\shortparallel }e}_{j_{s}}\mathbf{\mathbf{\mathbf{\mathbf{\ }}}}%
_{[01]}^{\shortparallel }\widehat{\mathbf{\Gamma }}_{\ \mu
_{s}i_{s}}^{\alpha _{s}})\mathbf{\mathbf{\mathbf{\mathbf{\ ^{\shortparallel }%
}}}}\widehat{\mathbf{\Gamma }}_{\ \beta _{s}\gamma _{s}}^{\mu _{s}}\right) \\
&&+\delta _{\gamma _{s}k_{s}}\left( \mathbf{\mathbf{\mathbf{\mathbf{\ }}}}%
^{\shortparallel }\widehat{\mathbf{\Gamma }}_{\ \mu _{s}i_{s}}^{\alpha
_{s}}(\ \mathbf{^{\shortparallel }e}_{j}\ _{[01]}^{\shortparallel }\widehat{%
\mathbf{\Gamma }}_{\ \beta _{s}\alpha _{s}}^{\mu _{s}})+\mathbf{\mathbf{%
\mathbf{\mathbf{\ }}}}_{[01]}^{\shortparallel }\widehat{\mathbf{\Gamma }}_{\
\mu _{s}i_{s}}^{\alpha _{s}}(\ \mathbf{^{\shortparallel }e}_{j_{s}}\mathbf{%
\mathbf{\mathbf{\mathbf{\ ^{\shortparallel }}}}}\widehat{\mathbf{\Gamma }}%
_{\ \beta _{s}\alpha _{s}}^{\mu _{s}})\right) ]
\end{eqnarray*}%
\begin{eqnarray*}
&&+\frac{\hbar \kappa }{2}\overline{\mathcal{R}}_{\quad \quad \quad
}^{n+i_{s}\ n+j_{s}\ n+k_{s}}[\delta _{\alpha _{s}k_{s}}\left( (\ \mathbf{%
^{\shortparallel }e}_{l_{s}}\ \mathbf{^{\shortparallel }e}_{j_{s}}\mathbf{%
\mathbf{\mathbf{\mathbf{\ }}}}^{\shortparallel }\widehat{\mathbf{\Gamma }}%
_{\ \mu _{s}i_{s}}^{\alpha _{s}})(\ ^{\shortparallel }\mathbf{e}^{n+l_{s}}\
^{\shortparallel }\widehat{\mathbf{\Gamma }}_{\ \beta _{s}\gamma _{s}}^{\mu
_{s}})-(\ ^{\shortparallel }\mathbf{e}^{n+l_{s}}\ \mathbf{^{\shortparallel }e%
}_{j_{s}}\mathbf{\mathbf{\mathbf{\mathbf{\ }}}}^{\shortparallel }\widehat{%
\mathbf{\Gamma }}_{\ \mu _{s}i_{s}}^{\alpha _{s}})(\ \mathbf{%
^{\shortparallel }e}_{l_{s}}\mathbf{\mathbf{\mathbf{\mathbf{\
^{\shortparallel }}}}}\widehat{\mathbf{\Gamma }}_{\ \beta _{s}\gamma
_{s}}^{\mu _{s}})\right) \\
&&+\delta _{\gamma _{s}k_{s}}\left( (\ \mathbf{^{\shortparallel }e}_{l_{s}}%
\mathbf{~\mathbf{\mathbf{\mathbf{\ }}}}^{\shortparallel }\widehat{\mathbf{%
\Gamma }}_{\ \mu _{s}i_{s}}^{\alpha _{s}})(\ ^{\shortparallel }\mathbf{e}%
^{n+l_{s}}\ \mathbf{^{\shortparallel }e}_{j_{s}}\ ^{\shortparallel }\widehat{%
\mathbf{\Gamma }}_{\ \beta _{s}\alpha _{s}}^{\mu _{s}})-(\ ^{\shortparallel }%
\mathbf{e}^{n+l_{s}}\mathbf{\mathbf{\mathbf{\mathbf{\ }}}}^{\shortparallel }%
\widehat{\mathbf{\Gamma }}_{\ \mu _{s}i_{s}}^{\alpha _{s}})(\ \mathbf{%
^{\shortparallel }e}_{l_{s}}\ \mathbf{^{\shortparallel }e}_{j_{s}}\mathbf{%
\mathbf{\mathbf{\mathbf{\ ^{\shortparallel }}}}}\widehat{\mathbf{\Gamma }}%
_{\ \beta _{s}\alpha _{s}}^{\mu _{s}})\right) ]
\end{eqnarray*}%
\begin{eqnarray*}
&&+i\kappa \overline{\mathcal{R}}_{\quad \quad \quad }^{n+i_{s}\ n+j_{s}\
n+k_{s}}\delta _{\alpha _{s}k_{s}}[\left( (\ \mathbf{^{\shortparallel }e}%
_{j_{s}}\ \mathbf{^{\shortparallel }e}_{\gamma _{s}}\mathbf{\mathbf{\mathbf{%
\mathbf{\ }}}}_{[01]}^{\shortparallel }\widehat{\mathbf{\Gamma }}_{\ \mu
_{s}i_{s}}^{\alpha _{s}})+(\ \mathbf{^{\shortparallel }e}_{j_{s}}\mathbf{%
\mathbf{\mathbf{\mathbf{\ }}}}^{\shortparallel }\widehat{\mathbf{\Gamma }}%
_{\ \mu _{s}\gamma _{s}}^{\alpha _{s}})\ _{[01]}^{\shortparallel }\widehat{%
\mathbf{\Gamma }}_{\ \beta _{s}i_{s}}^{\mu _{s}}+(\ \mathbf{^{\shortparallel
}e}_{j_{s}}\mathbf{\mathbf{\mathbf{\mathbf{\ }}}}_{[01]}^{\shortparallel }%
\widehat{\mathbf{\Gamma }}_{\ \mu _{s}\gamma _{s}}^{\alpha _{s}})\
^{\shortparallel }\widehat{\mathbf{\Gamma }}_{\ \beta _{s}i_{s}}^{\mu
_{s}}\right) \\
&&+\mathbf{\mathbf{\mathbf{\mathbf{\ }}}}^{\shortparallel }\widehat{\mathbf{%
\Gamma }}_{\ \mu _{s}\gamma _{s}}^{\alpha _{s}}(\ \mathbf{^{\shortparallel }e%
}_{j_{s}}\ _{[01]}^{\shortparallel }\widehat{\mathbf{\Gamma }}_{\ \beta
_{s}i_{s}}^{\mu _{s}})+\mathbf{\mathbf{\mathbf{\mathbf{\ }}}}%
_{[01]}^{\shortparallel }\widehat{\mathbf{\Gamma }}_{\ \mu _{s}\gamma
_{s}}^{\alpha _{s}}(\ \mathbf{^{\shortparallel }e}_{j_{s}}\ ^{\shortparallel
}\widehat{\mathbf{\Gamma }}_{\ \beta _{s}i_{s}}^{\mu _{s}})]
\end{eqnarray*}
\begin{eqnarray*}
&&-\frac{\hbar \kappa }{2}\overline{\mathcal{R}}_{\quad \quad \quad
}^{n+i_{s}\ n+j_{s}\ n+k_{s}}\delta _{\alpha _{s}k_{s}}[(\ \mathbf{%
^{\shortparallel }e}_{l_{s}}\ \mathbf{^{\shortparallel }e}_{j_{s}}\mathbf{%
\mathbf{\mathbf{\mathbf{\ }}}}^{\shortparallel }\widehat{\mathbf{\Gamma }}%
_{\ \mu _{s}\gamma _{s}}^{\alpha _{s}})(\ ^{\shortparallel }\mathbf{e}%
^{n+l_{s}}\ ^{\shortparallel }\widehat{\mathbf{\Gamma }}_{\ \beta
_{s}i_{s}}^{\mu _{s}})-(\ ^{\shortparallel }\mathbf{e}^{n+l_{s}}\ \mathbf{%
^{\shortparallel }e}_{j_{s}}\mathbf{\mathbf{\mathbf{\mathbf{\ }}}}%
^{\shortparallel }\widehat{\mathbf{\Gamma }}_{\ \mu _{s}\gamma _{s}}^{\alpha
_{s}})(\ \mathbf{^{\shortparallel }e}_{l_{s}}\mathbf{\mathbf{\mathbf{\mathbf{%
\ ^{\shortparallel }}}}}\widehat{\mathbf{\Gamma }}_{\ \beta _{s}i_{s}}^{\mu
_{s}}) \\
&&+(\ \mathbf{^{\shortparallel }e}_{l_{s}}\mathbf{~\mathbf{\mathbf{\mathbf{\
}}}}^{\shortparallel }\widehat{\mathbf{\Gamma }}_{\ \mu _{s}\gamma
_{s}}^{\alpha _{s}})(\ ^{\shortparallel }\mathbf{e}^{n+l_{s}}\ \mathbf{%
^{\shortparallel }e}_{j_{s}}\ ^{\shortparallel }\widehat{\mathbf{\Gamma }}%
_{\ \beta _{s}i_{s}}^{\mu _{s}})-(\ ^{\shortparallel }\mathbf{e}^{n+l_{s}}%
\mathbf{\mathbf{\mathbf{\mathbf{\ }}}}^{\shortparallel }\widehat{\mathbf{%
\Gamma }}_{\ \mu _{s}\gamma _{s}}^{\alpha _{s}})(\ \mathbf{^{\shortparallel
}e}_{l_{s}}\ \mathbf{^{\shortparallel }e}_{j_{s}}\mathbf{\mathbf{\mathbf{%
\mathbf{\ ^{\shortparallel }}}}}\widehat{\mathbf{\Gamma }}_{\ \beta
_{s}i_{s}}^{\mu _{s}})].
\end{eqnarray*}%
To complete the computation of such real nonassociative and noncommutative
contributions we have to insert formulas for coefficients of type $\
_{[01]}^{\shortparallel }\widehat{\Gamma }_{\star \alpha \beta }^{\nu },\
_{[10]}^{\shortparallel }\widehat{\Gamma }_{\star \alpha \beta }^{\nu }$ and
$\ _{[11]}^{\shortparallel }\widehat{\Gamma }_{\star \alpha \beta }^{\nu },$
see s-coefficients (\ref{aux39}), (\ref{aux51}) and (\ref{aux39a}). We omit
cumbersome explicit formulas for [01],[10], and [11] because such
nonassociative R-flux contributions can be encoded into an effective source
for nonholonomic canonically deformed Einstein equations. Using effective
sources with respective parameters and associative and commutative [00]
canonical Ricci s-tensors, the solutions of physically important systems of
nonlinear PDEs can be computed using the AFCDM outlined in \cite%
{bubuianu17,bubuianu19}.

Finally, we note that nonassociative LC-connections can be extracted for
nonholonomic constraints resulting in $\ _{s}^{\shortparallel }\widehat{%
\mathbf{D}}_{\mid \ _{s}^{\shortparallel }\widehat{\mathbf{T}}=0}^{\star }=\
^{\shortparallel }\nabla ^{\star },$ see (\ref{lccondnonass}). In coordinate
bases, we obtain the formulas for parametric Ricci tensor (5.80) from \cite%
{aschieri17}. It is important to have a prescription with Rules 1-2 (\ref%
{rules12}) in order to move geometric s-objects from local coordinate bases
to s-adapted bases, and inversely, determined by canonical data $(\
^{\shortparallel}\mathbf{e}_{\alpha _{s}}, \ ^{\shortparallel}\mathbf{g}
_{\alpha _{s}\beta _{s}},\ ^{\shortparallel }\widehat{\mathbf{\Gamma }}_{\
\beta _{s}\gamma _{s}}^{\mu _{s}}).$

\setcounter{equation}{0} \renewcommand{\theequation}
{B.\arabic{equation}} \setcounter{subsection}{0}
\renewcommand{\thesubsection}
{B.\arabic{subsection}}

\section{Nonholonomic dyadic deformations into parametric solutions}

\label{appendixb} In this appendix, we reformulate for nonassociative
gravity models the results on section 5 from \cite{bubuianu20}. Such results
are important for constructing exact and parametric locally anisotropic BH
and cosmological solutions generated by nonassociative star nonholonomic
deformations and related effective R-flux sources.

\subsection{Using s-metric coefficients as generating functions}

Using nonlinear symmetries (\ref{nonltransf}), we obtain formulas:
\begin{equation*}
\begin{array}{ccccc}
\mbox{ shell }s=2: &  & [(\ _{2}\Psi )^{2}]^{\diamond }=-\int
dy^{3}(~_{2}^{\shortparallel }\mathcal{K})g_{4}^{\diamond }, &  & (\
_{2}\Phi )^{2}=-4\ _{2}\Lambda _{0}g_{4}; \\
&  &  &  &  \\
\mbox{ shell }s=3: &  & ~^{\shortparallel }\partial ^{6}[(\
~_{3}^{\shortparallel }\Psi )^{2}]=-\int d~^{\shortparallel
}p_{6}(~_{3}^{\shortparallel }\mathcal{K})\ ~^{\shortparallel }\partial
^{6}~^{\shortparallel }g^{5}, &  & (\ _{3}^{\shortparallel }\Phi )^{2}=-4\
_{3}^{\shortparallel }\Lambda _{0}~^{\shortparallel }g^{6}; \\
&  &  &  &  \\
\mbox{ shell }s=4: &  & [(~_{4}^{\shortparallel }\Psi )^{2}]^{\ast }=-\int
d~^{\shortparallel }E(~_{4}^{\shortparallel }\mathcal{K})\
(~^{\shortparallel }g^{7})^{\ast }, &  & (\ _{4}^{\shortparallel }\Phi
)^{2}=-4\ \ _{4}^{\shortparallel }\Lambda ~^{\shortparallel }g^{8}.%
\end{array}%
\end{equation*}%
We conclude that the quadratic elements for quasi-periodic solutions (\ref%
{qeltors}) and/or (\ref{offdiagcosmcsh}) can be rewritten equivalently in
terms of generating data $(g_{4},~^{\shortparallel}g^{5},~^{%
\shortparallel}g^{7};\ _{s}^{\shortparallel }\Lambda )$ and respective
nonlinear symmetries involving generating sources $\ _{s}^{\shortparallel }%
\mathcal{K},$
\begin{eqnarray}
&&d\ ~^{\shortparallel }\widehat{s}^{2}=\ ~^{\shortparallel }g_{\alpha
_{s}\beta _{s}}(\hbar ,\kappa ,x^{k},y^{3},\ ~^{\shortparallel }p_{a_{3}},\
~^{\shortparallel }E;g_{4},\ ~^{\shortparallel }g^{5},\ ~^{\shortparallel
}g^{7},\ ~_{s}^{\shortparallel }\Lambda _{0};\ ~_{s}^{\shortparallel }%
\mathcal{K})d\ ~^{\shortparallel }u^{\alpha _{s}}d\ ~^{\shortparallel
}u^{\beta _{s}}=  \label{offdsolgenfgcosmc} \\
&&e^{\psi (\hbar ,\kappa ,x^{k_{1}})}[(dx^{1})^{2}+(dx^{2})^{2}]-\frac{%
(g_{4}^{\diamond })^{2}}{|\int dy^{3}[(~_{2}^{\shortparallel }\mathcal{K}%
)g_{4}]^{\diamond }|\ g_{4}}\{dy^{3}+\frac{\partial _{i_{1}}[\int
dy^{3}(~_{2}^{\shortparallel }\mathcal{K})\ g_{4}^{\diamond }]}{%
(~_{2}^{\shortparallel }\mathcal{K})\ g_{4}^{\diamond }}dx^{i_{1}}\}^{2}+
\notag \\
&&g_{4}\{dt+[\ _{1}n_{k_{1}}+\ _{2}n_{k_{1}}\int dy^{3}\frac{%
(g_{4}^{\diamond })^{2}}{|\int dy^{3}[(~_{2}^{\shortparallel }\mathcal{K}%
)g_{4}]^{\diamond }|\ [g_{4}]^{5/2}}]dx^{\acute{k}_{1}}\}+  \notag
\end{eqnarray}%
\begin{eqnarray*}
&&~^{\shortparallel }g^{5}\{d~^{\shortparallel }p_{5}+[\
_{1}^{\shortparallel }n_{k_{2}}+\ _{2}^{\shortparallel }n_{k_{2}}\int
d~^{\shortparallel }p_{6}\frac{[~^{\shortparallel }\partial
^{6}(~^{\shortparallel }g^{5})]^{2}}{|\int d~^{\shortparallel }p_{6}\
~^{\shortparallel }\partial ^{6}[(~_{3}^{\shortparallel }\mathcal{K}%
)~^{\shortparallel }g^{5}]|\ [~^{\shortparallel }g^{5}]^{5/2}}]dx^{k_{2}}\}-
\\
&&\frac{[~^{\shortparallel }\partial ^{6}(~^{\shortparallel }g^{5})]^{2}}{%
|\int d~^{\shortparallel }p_{6}\ ~^{\shortparallel }\partial
^{6}[(~_{3}^{\shortparallel }\mathcal{K})~^{\shortparallel }g^{5}]\ |\
~^{\shortparallel }g^{5}}\{d~^{\shortparallel }p_{6}+\frac{\partial
_{i_{2}}[\int d~^{\shortparallel }p_{6}(~_{3}^{\shortparallel }\mathcal{K})\
~^{\shortparallel }\partial ^{6}(~^{\shortparallel }g^{5})]}{%
(~_{3}^{\shortparallel }\mathcal{K})~^{\shortparallel }\partial
^{6}(~^{\shortparallel }g^{5})}dx^{i_{2}}\}^{2}+
\end{eqnarray*}%
\begin{eqnarray*}
&&~^{\shortparallel }g^{7}\{d~^{\shortparallel }p_{7}+[\
_{1}^{\shortparallel }n_{k_{3}}+\ _{2}^{\shortparallel }n_{k_{3}}\int
d~^{\shortparallel }E\frac{[(~^{\shortparallel }g^{7})^{\ast }]^{2}}{|\int
d~^{\shortparallel }E\ [(~_{4}^{\shortparallel }\mathcal{K}%
)~^{\shortparallel }g^{7}]^{\ast }|\ [~^{\shortparallel }g^{7}]^{5/2}}%
]d~^{\shortparallel }x^{k_{3}}\}- \\
&&\frac{\lbrack (~^{\shortparallel }g^{7})^{\ast }]^{2}}{|\int
d~^{\shortparallel }E\ [(~_{4}^{\shortparallel }\mathcal{K}%
)~^{\shortparallel }g^{7}]^{\ast }\ |\ ~^{\shortparallel }g^{7}}%
\{d~^{\shortparallel }E+\frac{~^{\shortparallel }\partial _{i_{3}}[\int
d~^{\shortparallel }E(\ ~_{4}^{\shortparallel }\mathcal{K})\
(~^{\shortparallel }g^{7})^{\ast }]}{(~_{4}^{\shortparallel }\mathcal{K}%
)(~^{\shortparallel }g^{7})^{\ast }}d~^{\shortparallel }x^{i_{3}}\}^{2}.
\end{eqnarray*}%
The parametric dependence, signs and integration functions/constants in
above formulas should to be chosen some forms which are compatible with
experimental/ observational data or corrections from R-fluxes in string
theories.

In a similar form we can generate LC-configurations using data
\begin{equation*}
\check{g}_{4}(\hbar ,\kappa ,x^{i_{1}},y^{3}),~^{\shortparallel }\check{g}%
^{5}(\hbar ,\kappa ,x^{i_{1}},x^{a_{2}},~^{\shortparallel
}p_{6}),~^{\shortparallel }\check{g}^{7}(\hbar ,\kappa
,x^{i_{1}},x^{a_{2}},~^{\shortparallel }p_{a_{2}},~^{\shortparallel }E)
\end{equation*}
and nonlinear symmetries (\ref{expconda}) for quasi-stationary solutions (%
\ref{qellc}). We omit such formulas (for associative and commutative
configurations, similar results are stated pe Remark 5.1 and formula (64) in
\cite{bubuianu20}). The R-flux contributions of effective sources $%
_{s}^{\shortparallel }\mathcal{K}$ are encoded correspondingly in
N-connection coefficients $\ _{s}^{\shortparallel }\check{A}$ from (\ref%
{qellc}).

\subsection{Polarization parametric functions for nonassociative prime and
target s-metrics}

On a phase space (for simplicity, we can consider an associative/commutative
variant modelled by a dual Lorentz bundle enabled with nonholonomic dyadic
structure, $\ ^{\shortparallel }\mathbf{e}^{\alpha _{s}}\in T_{s}^{\ast }%
\mathbf{T}_{\shortparallel }^{\ast }\mathbf{V}$ (\ref{nadapbdsc}), we
consider a \textbf{prime } s-metric $\ _{s}^{\shortmid }\mathbf{\mathring{g}}
$ structure (\ref{sdm}), with possible star deformations to respective
symmetric and nonsymmetric s-metrics of type (\ref{ssdm}) and (\ref{nssdm}).
In s-adapted form, we write
\begin{eqnarray}
\ \ ~^{\shortparallel }\mathbf{\mathring{g}} &=&\ _{s}^{\shortmid }\mathbf{%
\mathring{g}}=\ ^{\shortmid }\mathring{g}_{\alpha _{s}\beta
_{s}}(x^{i_{s}},p_{a_{s}})d\ ~^{\shortparallel }u^{\alpha _{s}}\otimes d\
~^{\shortparallel }u^{\beta _{s}}=\ ~^{\shortparallel }\mathbf{\mathring{g}}%
_{\alpha _{s}\beta _{s}}(\ _{s}^{\shortmid }u)\ ~^{\shortparallel }\mathbf{%
\mathbf{\mathring{e}}}^{\alpha _{s}}\mathbf{\otimes \ ~^{\shortparallel }%
\mathbf{\mathring{e}}}^{\beta _{s}}  \label{primedm} \\
&=&~^{\shortparallel }\mathring{g}_{i_{s}j_{s}}(x^{k_{s}})~^{\shortparallel
}e^{i_{s}}\otimes ~^{\shortparallel }e^{j_{s}}+\ ~^{\shortparallel }\mathbf{%
\mathring{g}}^{a_{s}b_{s}}(x^{i_{s}},p_{a_{s}})\ ~^{\shortparallel }\mathbf{%
\mathring{e}}_{a_{s}}\otimes \ ~^{\shortparallel }\mathbf{\mathring{e}}%
_{b_{s}},\mbox{ for }  \notag \\
\ ~^{\shortparallel }\mathbf{\mathring{e}}_{\alpha _{s}} &=&(\
~^{\shortparallel }\mathbf{\mathring{e}}_{i_{s}}=~^{\shortparallel }\partial
_{i_{s}}-\ ~^{\shortparallel }\mathring{N}_{i_{s}}^{b_{s}}(~^{\shortparallel
}u)~^{\shortparallel }\partial _{b_{s}},\ ~^{\shortparallel }{e}%
_{a_{s}}=~^{\shortparallel }\partial _{a_{s}})\mbox{ and }  \notag \\
\ ~^{\shortparallel }\mathbf{\mathring{e}}^{\alpha _{s}}
&=&(d~^{\shortparallel }x^{i_{s}},~^{\shortparallel }\mathbf{\mathring{e}}%
^{a_{s}}=d~^{\shortparallel }y^{a_{s}}+\ ~^{\shortparallel }\mathring{N}%
_{i_{s}}^{a_{s}}(~^{\shortparallel }u)d~^{\shortparallel }x^{i_{s}}).  \notag
\end{eqnarray}%
We label prime s-metrics and related geometric s-objects with a small circle
on the left/right/up of corresponding symbols. Such a prime $\
_{s}^{\shortparallel }\mathbf{\mathring{g}}$ (\ref{primedm}) my be, or not,
a solution of certain gravitational field equations in a MGT or GR with
phase space extension, which may result in parametric deformations to
nonassociative configurations following formulas $\ _{\star}^{\shortparallel}%
\mathfrak{\check{g}}_{\mu _{s}\nu _{s}}^{\circ }=(\ _{\star}^{\shortparallel
}\mathfrak{\check{g}}_{i_{1}j_{1}}^{\circ },\ _{\star }^{\shortparallel }%
\mathfrak{\check{g}}_{a_{2}b_{2}}^{\circ },\ _{\star}^{\shortparallel }%
\mathfrak{\check{g}}_{\circ }^{a_{3}b_{3}},\ _{\star}^{\shortparallel }%
\mathfrak{\check{g}}_{\circ }^{a_{4}b_{4}})$ (\ref{ssm1}) and $\ _{\star
}^{\shortparallel }\mathfrak{a}_{\mu _{s}\nu _{s}}^{\circ }=(0,0,\ _{\star
}^{\shortparallel }\mathfrak{a}_{c_{3}b_{3}}^{\circ }, \ _{\star
}^{\shortparallel }\mathfrak{a} _{c_{4}b_{4}}^{\circ })$ (\ref{nsm1}).

For nontrivial R-flux and star deformations, we can study nonassociative
parametric nonholonomic deformations to nonlinear quadratic elements
determined by \textbf{target} quasi-stationary s-metrics of type $\
_{s}^{\shortparallel }\mathbf{g}$ (\ref{qeltors}) as solutions of
nonassociative vacuum Einstein equations (\ref{cannonsymparamc2a}). Such
nonholonomic star s-deformations can be described in terms of so-called
gravitational polarization ($\eta $-polarization) functions,
\begin{equation*}
\ ~_{s}^{\shortparallel }\mathbf{\mathring{g}}\rightarrow \
_{s}^{\shortparallel }\mathbf{g}=[~^{\shortparallel }g_{\alpha _{s}}=\
^{\shortparallel }\eta _{\alpha _{s}}\ ~^{\shortparallel }\mathring{g}
_{\alpha _{s}},\ ^{\shortparallel }N_{i_{s-1}}^{a_{s}}=\
~^{\shortparallel}\eta _{i_{s-1}}^{a_{s}}\ ~^{\shortparallel }\mathring{N}%
_{i_{s-1}}^{a_{s}}],
\end{equation*}%
when the target s-metrics are written
\begin{eqnarray}
\ _{s}^{\shortparallel }\mathbf{g} &=&~^{\shortparallel }g_{i_{s}}(\hbar
,\kappa ,~^{\shortparallel }x^{k_{s}})d~^{\shortparallel }x^{i_{s}}\otimes
d~^{\shortparallel }x^{i_{s}}+~^{\shortparallel }g_{a_{s}}(\hbar ,\kappa
,~^{\shortparallel }x^{i_{s}},~^{\shortparallel }p_{b_{s}})\
~^{\shortparallel }\mathbf{e}^{a_{s}}\otimes ~^{\shortparallel }\mathbf{e}%
^{a_{s}}  \label{dmpolariz} \\
&=&~^{\shortparallel }\eta _{i_{k}}(\hbar ,\kappa
,x^{i_{1}},y^{a_{2}},~^{\shortparallel }p_{a_{3}},~^{\shortparallel
}p_{a_{4}})\ ~^{\shortparallel }\mathring{g}_{i_{s}}(\hbar ,\kappa
,x^{i_{1}},y^{a_{2}},~^{\shortparallel }p_{a_{3}},~^{\shortparallel
}p_{a_{4}})d~^{\shortparallel }x^{i_{s}}\otimes d~^{\shortparallel }x^{i_{s}}
\notag \\
&&+\ ~^{\shortparallel }\eta _{b_{s}}(\hbar ,\kappa
,x^{i_{1}},y^{a_{2}},~^{\shortparallel }p_{a_{3}},~^{\shortparallel
}p_{a_{4}})\ ~^{\shortparallel }\mathring{g}_{b_{s}}(\hbar ,\kappa
,x^{i_{1}},y^{a_{2}},~^{\shortparallel }p_{a_{3}},~^{\shortparallel
}p_{a_{4}})\ ~^{\shortparallel }\mathbf{e}^{b_{s}}[\eta ]\otimes \
~^{\shortparallel }\mathbf{e}^{b_{s}}[\eta ],  \notag \\
~^{\shortparallel }\mathbf{e}^{\alpha _{s}}[\eta ] &=&(d~^{\shortparallel
}x^{i_{s}},\ ~^{\shortparallel }\mathbf{e}^{a_{s}}=d~^{\shortparallel
}y^{a_{s}}+\ ~^{\shortparallel }\eta _{i_{s}}^{a_{s}}(\hbar ,\kappa
,x^{i_{1}},y^{a_{2}},~^{\shortparallel }p_{a_{3}},~^{\shortparallel
}p_{a_{4}})\ ~^{\shortparallel }\mathring{N}_{i_{s}}^{a_{s}}(\hbar ,\kappa
,x^{i_{1}},y^{a_{2}},~^{\shortparallel }p_{a_{3}},~^{\shortparallel
}p_{a_{4}})d~^{\shortparallel }x^{i_{s}}).  \notag
\end{eqnarray}

We emphasize that any of multiples in a $\ ^{\shortparallel}\eta \
^{\shortparallel }\mathring{g}$ may depend, in principle, on mixed shell
coordinates and various parameters, sources, etc. To apply the AFCDM we have
to use products subjected to the condition that the target s-metrics (with
the coefficients in the left sides) are adapted to ordered shell
coordinates. We use the term "gravitational polarizations" because for $\eta$%
-deformations with a small parameter, we can generate solutions, for
instance, of black hole/ ellipsoid type but with effective polarization of
fundamental physical constants (for (non) commutative MGTs, see solutions
from \cite{vacaru03,vacaru05a,vacaru09a,vacaru16,vacaru18,
bubuianu17,bubuianu18a,bubuianu19,bubuianu20} and reference therein).

For any prescribed prime s-metric $\ _{s}^{\shortparallel }\mathbf{\mathring{%
g},}$ we can consider as generating functions some data given by $\eta $%
-polarizations, for instance, by $\ ^{\shortparallel }\eta _{4}(\hbar,
\kappa,x^{i_{1}},y^{3}),$ \newline
$~^{\shortparallel }\eta ^{5}(\hbar
,\kappa,x^{i_{1}},y^{a_{1}},~^{\shortparallel }p_{6}),\ ~^{\shortparallel
}\eta ^{7}(\hbar ,\kappa ,x^{i_{1}},y^{a_{1}},~^{\shortparallel
}p_{a_{2}},~^{\shortparallel }E)$ which should be defined from the condition
that the target s-metric is a quasi-stationary solution equivalent to $\
_{s}^{\shortparallel }\mathbf{g}$ (\ref{qeltors}). By straightforward
computations, we can check such solutions parameterized as (\ref{dmpolariz})
ad described by nonlinear quadratic elements with explicit dependence on $%
\eta $-polarizations,
\begin{eqnarray}
&&d\ ^{\shortparallel }\widehat{s}^{2}=\ ^{\shortparallel }g_{\alpha
_{s}\beta _{s}}(\hbar ,\kappa ,x^{k},y^{3},\ ^{\shortparallel }p_{a_{3}},\
^{\shortparallel }p_{a_{4}};\ g_{4},\ ^{\shortparallel
}g^{5},~^{\shortparallel }g^{7},\ _{s}^{\shortparallel }\Lambda _{0};\
_{s}^{\shortparallel }\mathcal{K})d~^{\shortparallel }u^{\alpha
_{s}}d~^{\shortparallel }u^{\beta _{s}}  \label{offdiagpolf} \\
&=&e^{\psi (\hbar ,\kappa ,x^{k_{1}})}[(dx^{1})^{2}+(dx^{2})^{2}]-\frac{[(\
^{\shortparallel }\eta _{4}\ ^{\shortparallel }\mathring{g}_{4})^{\diamond
}]^{2}}{|\int dy^{3}(~_{2}^{\shortparallel }\mathcal{K})(\ ^{\shortparallel
}\eta _{4}\ ^{\shortparallel }\mathring{g}_{4})^{\diamond }|\ (\
^{\shortparallel }\eta _{4}\ ^{\shortparallel }\mathring{g}_{4})}\{dy^{3}+%
\frac{\partial _{i_{1}}[\int dy^{3}(\ _{2}^{\shortparallel }\mathcal{K})\
(~^{\shortparallel }\eta _{4}\ ^{\shortparallel }\mathring{g}_{4})^{\diamond
}]}{(~_{2}^{\shortparallel }\mathcal{K})(\ ^{\shortparallel }\eta _{4}\
^{\shortparallel }\mathring{g}_{4})^{\diamond }}dx^{i_{1}}\}^{2}  \notag \\
&&+~^{\shortparallel }\eta _{4}\ ^{\shortparallel }\mathring{g}_{4})\{dt+[\
_{1}n_{k_{1}}+\ _{2}n_{k_{1}}\int dy^{3}\frac{[(\ \ ^{\shortparallel }\eta
_{4}\ ~^{\shortparallel }\mathring{g}_{4})^{\diamond }]^{2}}{|\int dy^{3}(\
_{2}^{\shortparallel }\mathcal{K})(\ ^{\shortparallel }\eta _{4}\
^{\shortparallel }\mathring{g}_{4})^{\diamond }|\ (\ ~^{\shortparallel }\eta
_{4}\ ~^{\shortparallel }\mathring{g}_{4})^{5/2}}]dx^{\acute{k}_{1}}\}
\notag
\end{eqnarray}%
\begin{eqnarray*}
&&+(~^{\shortparallel }\eta ^{5}\ \ ^{\shortparallel }\mathring{g}^{5})\{d\
~^{\shortparallel }p_{5}+[\ _{1}^{\shortparallel }n_{k_{2}}+\
_{2}^{\shortparallel }n_{k_{2}}\int d\ ~^{\shortparallel }p_{6}\frac{[\
~^{\shortparallel }\partial ^{6}(~^{\shortparallel }\eta ^{5}\
~^{\shortparallel }\mathring{g}^{5})]^{2}}{|\int d\ ~^{\shortparallel
}p_{6}(~_{3}^{\shortparallel }\mathcal{K})\ \partial ^{6}(\
~^{\shortparallel }\eta ^{5}\ \ ~^{\shortparallel }\mathring{g}^{5})|\
(~^{\shortparallel }\eta ^{5}\ ~^{\shortparallel }\mathring{g}^{5})^{5/2}}%
]dx^{k_{2}}\} \\
&&-\frac{[\ ~^{\shortparallel }\partial ^{6}(~^{\shortparallel }\eta ^{5}\ \
~^{\shortparallel }\mathring{g}^{5})]^{2}}{|\int d\ ~^{\shortparallel
}p_{6}\ (~_{3}^{\shortparallel }\mathcal{K})\ ~^{\shortparallel }\partial
^{6}(~^{\shortparallel }\eta ^{5}\ ~^{\shortparallel }\mathring{g}^{5})\ |\
(\ ~^{\shortparallel }\eta ^{5}\ \ ~^{\shortparallel }\mathring{g}^{5})}\{d\
~^{\shortparallel }p_{6}+\frac{\ ~^{\shortparallel }\partial _{i_{2}}[\int
d\ ~^{\shortparallel }p_{6}(\ _{3}^{\shortmid }\widehat{\Upsilon })\ \
~^{\shortparallel }\partial ^{6}(\ ~^{\shortparallel }\eta ^{5}\ \
~^{\shortparallel }\mathring{g}^{5})]}{(~_{3}^{\shortparallel }\mathcal{K})\
~^{\shortparallel }\partial ^{6}(\ ~^{\shortparallel }\eta ^{5}\ \
~^{\shortparallel }\mathring{g}^{5})}dx^{i_{2}}\}^{2}
\end{eqnarray*}%
\begin{eqnarray*}
+ &&(\ ^{\shortparallel }\eta ^{7}\ ^{\shortparallel }\mathring{g}%
^{7})\{d~^{\shortparallel }p_{7}+[\ _{1}n_{k_{3}}+\ _{2}n_{k_{3}}\int d\
^{\shortparallel }E\frac{[(\ ^{\shortparallel }\eta ^{7}\ ^{\shortparallel}%
\mathring{g}^{7})^{\ast }]^{2}}{|\int d\ ^{\shortparallel }E\ (\
_{4}^{\shortparallel }\mathcal{K})[(\ ^{\shortparallel} \eta ^{7}\
^{\shortparallel }\mathring{g}^{7})]^{\ast }|\ [(\ ^{\shortparallel }\eta
^{7}\ ~^{\shortparallel }\mathring{g}^{7})]^{5/2}}]d\
^{\shortparallel}x^{k_{3}}\} \\
&&-\frac{[(~^{\shortparallel }\eta ^{7}\ ~^{\shortparallel }\mathring{g}%
^{7})^{\ast }]^{2}}{|\int d~^{\shortparallel }E\ (\ ~_{4}^{\shortparallel }%
\mathcal{K})(~^{\shortparallel }\eta ^{7}\ ~^{\shortparallel }\mathring{g}%
^{7})^{\ast }\ |\ (~^{\shortparallel }\eta ^{7}\ ~^{\shortparallel }%
\mathring{g}^{7})}\{d~^{\shortparallel }E+\frac{~^{\shortparallel }\partial
_{i_{3}}[\int d~^{\shortparallel }E(\ ~_{4}^{\shortparallel }\mathcal{K})\
(~^{\shortparallel }\eta ^{7}\ ~^{\shortparallel }\mathring{g}^{7})^{\ast }]%
}{(~_{4}^{\shortparallel }\mathcal{K})(~^{\shortparallel }\eta ^{7}\
~^{\shortparallel }\mathring{g}^{7})^{\ast }}d\
^{\shortparallel}x^{i_{3}}\}^{2},
\end{eqnarray*}%
where the polarization functions are determined by generating data $[\psi
(\hbar ,\kappa ,x^{i_{1}}),\ ^{\shortparallel }\eta _{4}(\hbar ,\kappa
,x^{i_{1}},y^{3}),\ ^{\shortparallel }$ \newline
$\eta ^{5}(\hbar ,\kappa ,x^{i_{1}},y^{a_{a}},\ ^{\shortparallel }p_{6}),$ $%
\ ^{\shortmid }\eta ^{7}(\hbar ,\kappa
,x^{i_{1}},y^{a_{2}},~^{\shortparallel }p_{a_{3}},\ ^{\shortparallel }E)].$

We can express $\eta $-polarizations in terms of generating functions $\
_{s}^{\shortparallel }\Psi $ and $\ _{s}^{\shortparallel }\check{\Psi}$
using nonlinear quadratic elements (\ref{qeltors}) and/or (\ref%
{offdiagcosmcsh}) (for LC-configurations, see (\ref{qellc})). Such formulas
and related ones for s-metric and N-connection coefficients with $\eta $%
-dependence are given in \cite{bubuianu20}), see there details related to
formula (69). In this work, we have to use $\ "_{s}^{\shortparallel}"$%
-variables and respective nonassociative parametric sources $\
_{s}^{\shortparallel}\mathcal{K}$.

\subsection{Parametric nonassociative transforms to quasi-stationary
s-metrics}

The nonassociative vacuum gravitational field equations (\ref%
{cannonsymparamc2a}), which can be represented in the form (\ref{sourc1hv})
and (\ref{sourc1cc}), in both cases involving in parametrical form an
effective R-flux source (\ref{realrflux}), are geometrically similar to
those considered in our previous works \cite%
{vacaru05a,vacaru09a,vacaru18,bubuianu17, bubuianu18a,bubuianu19,bubuianu20}
with parametric deformations on a small parameter $\varepsilon ,0\leq
\varepsilon <1.$ The same AFCDM can be applied for a quasi-stationary
s-metric ansatz (\ref{ansatz1na}) when as a small parameter it is considered
the sting constant, $\varepsilon \rightarrow \kappa .$ We have to use also a
second small parameter, $\hbar $ when our purpose is to distinguish between
possible noncommutative and nonassociative effects, to extract real
configurations etc. For simplicity, in this subsection, we shall study only
line approximations on $\kappa $ considering that $\hbar $ is encoded into
generation functions and sources and (if necessary) in integration
functions. We respectively re-define for nonassociative phase spaces and
effective sources $\ _{s}^{\shortparallel }\mathcal{K}$ the results of
section 5.3 of \cite{bubuianu20}.

We can use such parametric $\kappa $--decompositions of the $\eta $%
-polarization functions in a s-metric (\ref{offdiagpolf}) resulting in
quasi-stationary solutions of type (\ref{qeltors}) and/or (\ref%
{offdiagcosmcsh}):
\begin{eqnarray*}
\ ^{\shortparallel }g_{i_{1}}(\kappa ,x^{k_{1}}) &=&~^{\shortparallel }\eta
_{i_{i}}\ ~^{\shortparallel }\mathring{g}_{i_{1}}=~^{\shortparallel }\zeta
_{i_{1}}(1+\kappa ~^{\shortparallel }\chi _{i_{1}})\ ~^{\shortparallel }%
\mathring{g}_{i_{1}}= \\
&=&\{~^{\shortparallel }\zeta _{i_{1}}(x^{i_{1}},y^{a_{2}},~^{\shortparallel
}p_{a_{3}},~^{\shortparallel }p_{a_{4}})[1+\kappa ~^{\shortparallel }\chi
_{i_{1}}(x^{i_{1}},y^{a_{2}},~^{\shortparallel }p_{a_{3}},~^{\shortparallel
}p_{a_{4}})]\}\ ~^{\shortparallel }\mathring{g}%
_{i_{1}}(x^{i_{1}},y^{a_{2}},~^{\shortparallel }p_{a_{3}},~^{\shortparallel
}p_{a_{4}}), \\
\ ~^{\shortparallel }g_{b_{2}}(\kappa ,x^{i_{1}},y^{3}) &=&\
~^{\shortparallel }\eta _{b_{2}}\ ~^{\shortparallel }\mathring{g}%
_{b_{1}}=~^{\shortparallel }\zeta _{b_{2}}(1+\kappa \ ~^{\shortparallel
}\chi _{b_{2}})\ ~^{\shortparallel }\mathring{g}_{b_{1}}= \\
&=&\{~^{\shortparallel }\zeta _{b_{2}}(x^{i_{1}},y^{a_{2}},~^{\shortparallel
}p_{a_{3}},~^{\shortparallel }p_{a_{4}})[1+\kappa \ ~^{\shortparallel }\chi
_{b_{2}}(x^{i_{1}},y^{a_{2}},~^{\shortparallel }p_{a_{3}},~^{\shortparallel
}p_{a_{4}})]\}\ ~^{\shortparallel }\mathring{g}%
_{b_{1}}(x^{i_{1}},y^{a_{2}},~^{\shortparallel }p_{a_{3}},~^{\shortparallel
}p_{a_{4}}), \\
\ ~^{\shortparallel }g^{a_{3}}(\kappa ,x^{i_{2}},~^{\shortparallel }p_{6})
&=&\ ~^{\shortparallel }\eta ^{a_{3}}\ ~^{\shortparallel }\mathring{g}%
^{a_{3}}=~^{\shortparallel }\zeta ^{a_{3}}(1+\kappa \ ~^{\shortparallel
}\chi ^{a_{3}})\ ~^{\shortparallel }\mathring{g}^{a_{3}}= \\
&=&\{~^{\shortparallel }\zeta ^{a_{3}}(x^{i_{1}},y^{b_{2}},~^{\shortparallel
}p_{b_{3}},~^{\shortparallel }p_{b_{4}})\ [1+\kappa \ ~^{\shortparallel
}\chi ^{a_{3}}(x^{i_{1}},y^{b_{2}},~^{\shortparallel
}p_{b_{3}},~^{\shortparallel }p_{b_{4}})]\}\ ^{\shortmid }\mathring{g}%
^{a_{3}}(x^{i_{1}},y^{b_{2}},~^{\shortparallel }p_{b_{3}},~^{\shortparallel
}p_{b_{4}}), \\
~^{\shortparallel }g^{a_{4}}(\kappa ,~^{\shortparallel
}x^{i_{3}},~^{\shortparallel }E) &=&\ ~^{\shortparallel }\eta ^{a_{4}}\
~^{\shortparallel }\mathring{g}^{a_{4}}=~^{\shortparallel }\zeta
^{a_{4}}(1+\kappa \ ~^{\shortparallel }\chi ^{a_{4}})\ ~^{\shortparallel }%
\mathring{g}^{a_{4}}= \\
&=&\{~^{\shortparallel }\zeta ^{a_{4}}(x^{i_{1}},y^{b_{2}},~^{\shortparallel
}p_{b_{3}},~^{\shortparallel }p_{b_{4}})[1+\kappa \ ~^{\shortparallel }\chi
^{a_{4}}(x^{i_{1}},y^{b_{2}},~^{\shortparallel }p_{b_{3}},~^{\shortparallel
}p_{b_{4}})]\}\ ~^{\shortparallel }\mathring{g}%
^{a_{4}}(x^{i_{1}},y^{a_{2}},~^{\shortparallel }p_{a_{3}},~^{\shortparallel
}p_{a_{4}}).
\end{eqnarray*}%
Similar $\kappa $--decompositions can be used for N-connection
s-coefficients
\begin{eqnarray*}
&&\ ^{\shortparallel }N_{i_{1}}^{a_{2}}(\kappa
,x^{k_{1}},y^{3})=~^{\shortparallel }\eta _{i_{1}}^{a_{2}}\
~^{\shortparallel }\mathring{N}_{i_{1}}^{a_{2}}=~^{\shortparallel }\zeta
_{i_{1}}^{a_{2}}(1+\kappa \ ^{\shortparallel }\chi _{i_{1}}^{a_{2}})\
~^{\shortparallel }\mathring{N}_{i_{1}}^{a_{2}}= \\
&=&\{~^{\shortparallel }\zeta
_{i_{1}}^{a_{2}}(x^{i_{1}},y^{b_{2}},~^{\shortparallel
}p_{b_{3}},~^{\shortparallel }p_{b_{4}})[1+\kappa \ ~^{\shortparallel }\chi
_{i_{1}}^{a_{2}}(x^{i_{1}},y^{b_{2}},~^{\shortparallel
}p_{b_{3}},~^{\shortparallel }p_{b_{4}})]\}\ ~^{\shortparallel }\mathring{N}%
_{i_{1}}^{a_{2}}(x^{i_{1}},y^{b_{2}},~^{\shortparallel
}p_{b_{3}},~^{\shortparallel }p_{b_{4}}), \\
&&\ ^{\shortparallel }N_{i_{2}a_{3}}(\kappa
,x^{k_{1}},y^{b_{2}},~^{\shortparallel }p_{6})=\ ~^{\shortparallel }\eta
_{i_{2}a_{3}}\ ~^{\shortparallel }\mathring{N}_{i_{2}a_{3}}=~^{%
\shortparallel }\zeta _{i_{2}a_{3}}(1+\kappa ~^{\shortparallel }\chi
_{i_{2}a_{3}})\ ~^{\shortparallel }\mathring{N}_{i_{2}a_{3}}= \\
&=&\{\ ~^{\shortparallel }\zeta
_{i_{2}a_{3}}(x^{i_{1}},y^{b_{2}},~^{\shortparallel
}p_{b_{3}},~^{\shortparallel }p_{b_{4}})[1+\kappa ~^{\shortparallel }\chi
_{i_{2}a_{3}}(x^{i_{1}},y^{b_{2}},~^{\shortparallel
}p_{b_{3}},~^{\shortparallel }p_{b_{4}})]\}\ ^{\shortmid }\mathring{N}%
_{i_{2}a_{3}}(x^{i_{1}},y^{b_{2}},~^{\shortparallel
}p_{b_{3}},~^{\shortparallel }p_{b_{4}}), \\
&&\ ~^{\shortparallel }N_{i_{3}a_{4}}(\kappa
,x^{k_{1}},y^{b_{2}},~^{\shortparallel }p_{a_{3}},~^{\shortparallel }E)=\
^{\shortparallel }\eta _{i_{3}a_{4}}\ ~^{\shortparallel }\mathring{N}%
_{i_{3}a_{4}}=~^{\shortparallel }\zeta _{i_{3}a_{4}}(1+\kappa
~^{\shortparallel }\chi _{i_{3}a_{4}})\ ~^{\shortparallel }\mathring{N}%
_{i_{3}a_{4}}= \\
&=&\{~^{\shortparallel }\zeta
_{i_{3}a_{4}}(x^{i_{1}},y^{b_{2}},~^{\shortparallel
}p_{b_{3}},~^{\shortparallel }p_{b_{4}})[1+\kappa \chi
_{i_{3}a_{4}}(x^{i_{1}},y^{b_{2}},~^{\shortparallel
}p_{b_{3}},~^{\shortparallel }p_{b_{4}})]\}\ ~^{\shortparallel }\mathring{N}%
_{i_{3}a_{4}}(x^{i_{1}},y^{b_{2}},~^{\shortparallel
}p_{b_{3}},~^{\shortparallel }p_{b_{4}}).
\end{eqnarray*}%
In brief, we write above formulas on star parametric deformations of a
s-metric and N-connection structure on $_{s}\mathbf{T}_{\shortparallel}^{%
\ast }\mathbf{V}$ in the form
\begin{equation}
\ _{s}^{\shortparallel }\mathbf{\mathring{g}}\rightarrow \
_{s}^{\shortparallel \varepsilon }\mathbf{g}=[\ ^{\shortparallel }g_{\alpha
_{s}}=\ ^{\shortparallel }\zeta _{\alpha _{s}}(1+\kappa \ ^{\shortparallel
}\chi _{\alpha _{s}})\ \ ^{\shortparallel }\mathring{g}_{\alpha _{s}},\ \
^{\shortparallel }N_{i_{s}}^{a_{s}}=\ ^{\shortparallel }\zeta
_{i_{s-1}}^{a_{s}}(1+\kappa \ \ ^{\shortparallel }\chi _{i_{s-1}}^{a_{s}})\
\ ^{\shortparallel }\mathring{N}_{i_{s-1}}^{a_{s}}].  \label{epstargsm}
\end{equation}

The $\zeta $- and $\chi $-coefficients for deformations (\ref{epstargsm})
are generated by shell data
\begin{equation}
\ \ \ ^{\shortparallel }\eta _{2}=\ ^{\shortparallel }\zeta _{2}(1+\kappa \
^{\shortparallel }\chi _{2}),\ \ \ ^{\shortparallel }\eta _{4}=\
^{\shortparallel }\zeta _{4}(1+\kappa \ ^{\shortparallel }\chi _{4}),\ \
^{\shortparallel }\eta ^{5}=\ ^{\shortparallel }\zeta ^{5}(1+\kappa \ \
^{\shortparallel }\chi ^{5}),\ \ ^{\shortparallel }\eta ^{7}=\
^{\shortparallel }\zeta ^{7}(1+\kappa \ \ ^{\shortparallel }\chi ^{7}).
\label{epsilongenfdecomp}
\end{equation}%
Let us explain below how such parametric deformations are performed on
respective shells:

For $s=1,$ we can consider $\ \ ^{\shortparallel }\zeta _{i_{1}}=(\
^{\shortparallel }\mathring{g}_{i_{1}})^{-1}e^{\psi _{0}(x^{k_{1}})}$ and $\
^{\shortparallel }\chi _{i_{1}}=(\ ^{\shortparallel }\mathring{g}%
_{i_{1}})^{-1}\ ^{\psi }\ ^{\shortparallel }\chi (x^{k_{1}})$, \ where%
\begin{equation*}
\ ^{\shortparallel }\zeta _{i_{1}}(1+\kappa \ ^{\shortparallel }\chi
_{i_{1}})\ \ ^{\shortparallel }\mathring{g}_{i_{1}}=e^{\psi
(x^{k_{1}})}\approx e^{\psi _{0}(x^{k_{1}})(1+\kappa \ ^{\psi }\chi
(x^{k_{1}}))}\approx e^{\psi _{0}(x^{k_{1}})}(1+\kappa \ ^{\psi }\
^{\shortparallel }\chi )
\end{equation*}%
for $\psi _{0}(x^{k_{1}})$ and $\ ^{\shortparallel }\chi (x^{k_{1}})$
defined by a solution of a 2-d Poisson equation (\ref{eq1}).

Then, for $s=2$ (with generating functions $\ ^{\shortparallel }\zeta _{4}$
and $\ ^{\shortparallel }\chi _{4};$ generating source and cosmological
constant,$\ ~_{2}^{\shortparallel }\mathcal{K},$ $\
_{2}^{\shortparallel}\Lambda _{0};$ integration functions$,\
_{1}^{\shortparallel }n_{k_{1}},\ _{2}^{\shortparallel }n_{k_{1}};$
prescribed data for a prime s-metric, $(\ ^{\shortparallel }\mathring{g}%
_{3},\ ^{\shortparallel }\mathring{g}_{4}; \ ^{\shortparallel }\mathring{N}%
_{i_{1}}^{3}, \ ^{\shortparallel }\mathring{N}_{k_{1}}^{4})),$ after tedious
computations, we express
\begin{eqnarray*}
\ ^{\shortparallel }\zeta _{3} &=&-\frac{4}{\ \ ^{\shortparallel }\mathring{g%
}_{3}}\frac{[(|\ \ ^{\shortparallel }\zeta _{4}\ \ ^{\shortparallel }%
\mathring{g}_{4}|^{1/2})^{\diamond }]^{2}}{|\int
dy^{3}\{(~_{2}^{\shortparallel }\mathcal{K})(\ ^{\shortparallel }\zeta _{4}\
^{\shortparallel }\mathring{g}_{4})^{\diamond }\}|}\mbox{ and }\
^{\shortparallel }\chi _{3}=\frac{(\ ^{\shortparallel }\chi _{4}|\ \
^{\shortparallel }\zeta _{4}\ \ ^{\shortparallel }\mathring{g}%
_{4}|^{1/2})^{\diamond }}{4(|\ \ ^{\shortparallel }\zeta _{4}\ \
^{\shortparallel }\mathring{g}_{4}|^{1/2})^{\diamond }}-\frac{\int
dy^{3}\{[(~_{2}^{\shortparallel }\mathcal{K})\ (\ ^{\shortparallel }\zeta
_{4}\ ^{\shortparallel }\mathring{g}_{4})\ ^{\shortparallel }\chi
_{4}]^{\diamond }\}}{\int dy^{3}\{(~_{2}^{\shortparallel }\mathcal{K})(\
^{\shortparallel }\zeta _{4}\ \ ^{\shortparallel }\mathring{g}%
_{4})^{\diamond }\}}, \\
\ ^{\shortparallel }\zeta _{i_{1}}^{3} &=&\frac{\partial _{i_{1}}\ \int
dy^{3}(~_{2}^{\shortparallel }\mathcal{K})\ (\ ^{\shortparallel }\zeta
_{4})^{\diamond }}{(\ ^{\shortparallel }\mathring{N}_{i_{1}}^{3})(~_{2}^{%
\shortparallel }\mathcal{K})(\ ^{\shortparallel }\zeta _{4})^{\diamond }}%
\mbox{ and }\ ^{\shortparallel }\chi _{i_{1}}^{3}=\frac{\partial
_{i_{1}}[\int dy^{3}(~_{2}^{\shortparallel }\mathcal{K})(\ ^{\shortparallel
}\zeta _{4}\ ^{\shortparallel }\chi _{4})^{\diamond }]}{\partial _{i_{1}}\
[\int dy^{3}(~_{2}^{\shortparallel }\mathcal{K})(\ ^{\shortparallel }\zeta
_{4})^{\diamond }]}-\frac{(\ ^{\shortparallel }\zeta _{4}\ ^{\shortparallel
}\chi _{4})^{\diamond }}{(\ ^{\shortparallel }\zeta _{4})^{\diamond }}, \\
\ ^{\shortparallel }\zeta _{k_{1}}^{4} &=&\ (\ \ ^{\shortparallel }\mathring{%
N}_{k_{1}}^{4})^{-1}[\ _{1}^{\shortparallel }n_{k_{1}}+16\
_{2}^{\shortparallel }n_{k_{1}}[\int dy^{3}\{\frac{\left( [(\
^{\shortparallel }\zeta _{4}\ \ ^{\shortparallel }\mathring{g}%
_{4})^{-1/4}]^{\diamond }\right) ^{2}}{|\int dy^{3}(~_{2}^{\shortparallel }%
\mathcal{K})(\ ^{\shortparallel }\zeta _{4}\ \ ^{\shortparallel }\mathring{g}%
_{4})^{\diamond }|}]\mbox{ and } \\
\ ^{\shortparallel }\chi _{k_{1}}^{4} &=&\ -\frac{16\ _{2}^{\shortparallel
}n_{k_{1}}\int dy^{3}\frac{\left( [(\ \ ^{\shortparallel }\zeta _{4}\ \
^{\shortparallel }\mathring{g}_{4})^{-1/4}]^{\diamond }\right) ^{2}}{|\int
dy^{3}(~_{2}^{\shortparallel }\mathcal{K})[(\zeta _{4}\ ^{\shortmid }%
\mathring{g}_{4})]^{\diamond }|}(\frac{[(\ \ ^{\shortparallel }\zeta _{4}\ \
^{\shortparallel }\mathring{g}_{4})^{-1/4}\chi _{4})]^{\diamond }}{2[(\ \
^{\shortparallel }\zeta _{4}\ \ ^{\shortparallel }\mathring{g}%
_{4})^{-1/4}]^{\diamond }}+\frac{\int dy^{3}[(~_{2}^{\shortparallel }%
\mathcal{K})(\ ^{\shortparallel }\zeta _{4}\ ^{\shortparallel }\chi _{4}\ \
^{\shortparallel }\mathring{g}_{4})]^{\diamond }}{\int
dy^{3}(~_{2}^{\shortparallel }\mathcal{K})(\zeta _{4}\ \ ^{\shortparallel }%
\mathring{g}_{4})^{\diamond }})}{\ _{1}^{\shortparallel }n_{k_{1}}+16\
_{2}^{\shortparallel }n_{k_{1}}[\int dy^{3}\frac{\left( [(\ \
^{\shortparallel }\zeta _{4}\ \ ^{\shortparallel }\mathring{g}%
_{4})^{-1/4}]^{\diamond }\right) ^{2}}{|\int dy^{3}(~_{2}^{\shortparallel }%
\mathcal{K})[(\ ^{\shortparallel }\zeta _{4}\ \ ^{\shortparallel }\mathring{g%
}_{4})]^{\diamond }|}].}.
\end{eqnarray*}%
Such formulas can be obtained for $s=3$ (with generating functions, $\
^{\shortparallel }\zeta ^{5},\ ^{\shortparallel }\chi ^{5};$ generating
source and cosmological constant $,\ _{3}^{\shortparallel}\mathcal{K},\ \
_{3}^{\shortparallel }\Lambda _{0};$ integration functions, $\
_{1}^{\shortparallel }n_{k_{3}},\ _{2}^{\shortparallel }n_{k_{3}};$
prescribed data for a prime s-metric, $(\ ^{\shortparallel }\mathring{g}%
^{5},\ ^{\shortparallel }\mathring{g}^{6};\ ^{\shortparallel }\mathring{N}%
_{k_{2}5},\ ^{\shortparallel }\mathring{N}_{i_{2}6}):$%
\begin{eqnarray*}
\ ^{\shortparallel }\zeta ^{6} &=&-\frac{4}{\ ^{\shortparallel }\mathring{g}%
^{6}} \frac{[\ ^{\shortparallel }\partial ^{6}(|\ ^{\shortparallel }\zeta
^{5}\ \ ^{\shortparallel }\mathring{g}^{5}|^{1/2})]^{2}}{|\int d\
^{\shortparallel }p_{6}\{(~_{3}^{\shortparallel }\mathcal{K})\
^{\shortparallel }\partial ^{6}(\ ^{\shortparallel }\zeta ^{5}\ \
^{\shortparallel }\mathring{g}^{5})\}|}\mbox{ and } \\
\ ^{\shortparallel }\chi ^{6} &=&\frac{\ ^{\shortparallel }\partial ^{6}(\
^{\shortparallel }\chi ^{5}|\ \ ^{\shortparallel }\zeta ^{5}\ \
^{\shortparallel }\mathring{g}^{5}|^{1/2})}{4\ ^{\shortparallel }\partial
^{6}(|\ \ ^{\shortparallel }\zeta ^{5}\ \ ^{\shortparallel }\mathring{g}%
^{5}|^{1/2})}-\frac{\int d\ ^{\shortparallel }p_{6}\{\ ^{\shortparallel
}\partial ^{6}[(\ ~_{3}^{\shortparallel }\mathcal{K})\ (\ ^{\shortparallel
}\zeta ^{5}\ \ ^{\shortparallel }\mathring{g}^{5})\ ^{\shortparallel }\chi
^{5}]\}}{\int d\ ^{\shortparallel }p_{6}\{(\ \ ~_{3}^{\shortparallel }%
\mathcal{K})\ ^{\shortparallel }\partial ^{6}(\zeta ^{5} \ ^{\shortparallel }%
\mathring{g}^{5})\}}, \\
\ ^{\shortparallel }\zeta _{i_{2}5} &=&\ (\ ^{\shortparallel }\mathring{N}%
_{i_{2}5})^{-1}[\ _{1}^{\shortparallel }n_{i_{2}}+16\ _{2}^{\shortparallel
}n_{i_{2}}[\int d\ ^{\shortparallel }p_{6}\{\frac{\left( \ ^{\shortparallel
}\partial ^{6}[(\ ^{\shortparallel }\zeta ^{5}\ \ ^{\shortparallel }%
\mathring{g}^{5})^{-1/4}]\right) ^{2}}{|\int d\ ^{\shortparallel }p_{6}\
(~_{3}^{\shortparallel }\mathcal{K})\ ^{\shortparallel }\partial ^{6}(\
^{\shortparallel }\zeta ^{5}\ \ ^{\shortparallel }\mathring{g}^{5})|}]%
\mbox{
and } \\
\ ^{\shortparallel }\chi _{i_{2}5} &=&\ -\frac{16\ _{2}^{\shortparallel
}n_{i_{2}}\int d\ ^{\shortparallel }p_{6}\frac{\left( \ ^{\shortparallel
}\partial ^{6}[(\ ^{\shortparallel }\zeta ^{5}\ ^{\shortparallel }\mathring{g%
}^{5})^{-1/4}]\right) ^{2}}{|\int d\ ^{\shortparallel }p_{6}\ (\
~_{3}^{\shortparallel }\mathcal{K})\ ^{\shortparallel }\partial ^{6}(\
^{\shortparallel }\zeta ^{5}\ \ ^{\shortparallel }\mathring{g}^{5})|}(\frac{%
\ ^{\shortparallel }\partial ^{6}[(\ ^{\shortparallel }\zeta ^{5}\
^{\shortparallel }\mathring{g}^{5})^{-1/4}\ ^{\shortparallel }\chi ^{5})]}{%
2\ \ ^{\shortparallel }\partial ^{6}[(\ ^{\shortparallel }\zeta ^{5}\
^{\shortparallel }\mathring{g}^{5})^{-1/4}]}+\frac{\int d\ ^{\shortparallel
}p_{6}\ ^{\shortparallel }\partial ^{6}[(~_{3}^{\shortparallel }\mathcal{K}%
)(~^{\shortparallel }\zeta ^{5}\ ^{\shortparallel }\mathring{g}^{5})\
^{\shortparallel }\chi ^{5}]}{\int d\ ^{\shortparallel }p_{6}\
(~_{3}^{\shortparallel }\mathcal{K})\ ^{\shortparallel }\partial ^{6}(\
^{\shortparallel }\zeta ^{5}\ ^{\shortparallel }\mathring{g}^{5})})}{\
_{1}^{\shortparallel }n_{i_{2}}+16\ _{2}^{\shortparallel }n_{i_{2}}[\int d\
^{\shortparallel }p_{6}\frac{\left( \ ^{\shortparallel }\partial ^{6}[(\
^{\shortparallel }\zeta ^{5}\ \ ^{\shortparallel }\mathring{g}%
^{5})^{-1/4}]\right) ^{2}}{|\int d\ ^{\shortparallel }p_{6}\
(~_{3}^{\shortparallel }\mathcal{K})\ ^{\shortparallel }\partial ^{6}(\
^{\shortparallel }\zeta ^{5}\ \ ^{\shortparallel }\mathring{g}^{5})|}]};
\end{eqnarray*}%
\begin{equation*}
\ ^{\shortparallel }\zeta _{i_{2}6}=\frac{\ ^{\shortparallel }\partial
_{i_{2}}\ \int d\ ^{\shortparallel }p_{6}(~_{3}^{\shortparallel }\mathcal{K}%
)\ \ ^{\shortparallel }\partial ^{6}(\ ^{\shortparallel }\zeta ^{5})}{(\ \
^{\shortparallel }\mathring{N}_{i_{2}6})(\ ~_{3}^{\shortparallel }\mathcal{K}%
)\ ^{\shortparallel }\partial ^{6}(\ ^{\shortparallel }\zeta ^{5})}%
\mbox{
and }\ ^{\shortparallel }\chi _{i_{2}6}=\frac{\ ^{\shortparallel }\partial
_{i_{2}}[\int d\ ^{\shortparallel }p_{6}(\ \ ~_{3}^{\shortparallel }\mathcal{%
K})\ ^{\shortparallel }\partial ^{6}(\ ^{\shortparallel }\zeta ^{5}\ \
^{\shortparallel }\mathring{g}^{5})]}{\ ^{\shortparallel }\partial _{i_{2}}\
[\int d\ ^{\shortparallel }p_{6}(\ \ ~_{3}^{\shortparallel }\mathcal{K})\
^{\shortparallel }\partial ^{6}(\ ^{\shortparallel }\zeta ^{5})]}-\frac{\
^{\shortparallel }\partial ^{6}(\ ^{\shortparallel }\zeta ^{5}\ \
^{\shortparallel }\mathring{g}^{5})}{\ ^{\shortparallel }\partial ^{6}(\
^{\shortparallel }\zeta ^{5})}.
\end{equation*}

Finally, we provide the formulas for the shell $s=4$ (with generating
functions, $\ ^{\shortparallel }\zeta ^{7},\ ^{\shortparallel }\chi ^{7};$
generating source and cosmological constant$,\ ~_{4}^{\shortparallel }
\mathcal{K},\ \ _{4}^{\shortparallel }\Lambda _{0};$ integration functions $%
,\ _{1}^{\shortparallel }n_{k_{4}},\ _{2}^{\shortparallel }n_{k_{4}};$
prescribed data for a prime s-metric, $(\ ^{\shortparallel }\mathring{g}%
^{7},\ ^{\shortparallel }\mathring{g}^{8};\ ^{\shortparallel }\mathring{N}%
_{k_{3}7},\ ^{\shortparallel }\mathring{N}_{i_{3}8}):$%
\begin{eqnarray*}
\ ^{\shortparallel }\zeta ^{8} &=&-\frac{4}{\ \ ^{\shortparallel }\mathring{g%
}^{8}}\frac{[(|\ ^{\shortparallel }\zeta ^{7}\ ^{\shortparallel }\mathring{g}%
^{7}|^{1/2})^{\ast }]^{2}}{|\int d\ ^{\shortparallel
}E\{(~_{4}^{\shortparallel }\mathcal{K})[(\ ^{\shortparallel }\zeta ^{7}\
^{\shortparallel }\mathring{g}^{7})]^{\ast }\}|}\mbox{ and } \\
\ ^{\shortparallel }\chi ^{8} &=&\frac{(~^{\shortparallel }\chi ^{7}|\
^{\shortparallel }\zeta ^{7}\ ^{\shortparallel }\mathring{g}%
^{7}|^{1/2})^{\ast }}{4(|\ ^{\shortparallel }\zeta ^{7}\ ^{\shortparallel }%
\mathring{g}^{7}|^{1/2})^{\ast }}-\frac{\int d\ ^{\shortparallel
}E\{[(~_{4}^{\shortparallel }\mathcal{K})\ (~^{\shortparallel }\zeta ^{7}\
^{\shortparallel }\mathring{g}^{7})\ ^{\shortparallel }\chi ^{7}]^{\ast }\}}{%
\int d\ ^{\shortparallel }E\{(~_{4}^{\shortparallel }\mathcal{K})(\
^{\shortparallel }\zeta ^{4}\ ^{\shortparallel }\mathring{g}^{4})^{\ast }\}},
\\
\ ^{\shortparallel }\zeta _{i_{3}7} &=&\ (\ ^{\shortparallel }\mathring{N}%
_{i_{3}7})^{-1}[\ _{1}^{\shortparallel }n_{i_{3}}+16\ _{2}^{\shortparallel
}n_{i_{3}}[\int d\ ^{\shortparallel }E\{\frac{\left( [(~^{\shortparallel
}\zeta ^{7}\ ^{\shortparallel }\mathring{g}^{7})^{-1/4}]^{\ast }\right) ^{2}%
}{|\int d\ ^{\shortparallel }E\ (~_{4}^{\shortparallel }\mathcal{K})(\
^{\shortparallel }\zeta ^{7}\ ^{\shortparallel }\mathring{g}^{7})^{\ast }|}]%
\mbox{ and } \\
\ ^{\shortparallel }\chi _{i_{3}7} &=&\ -\frac{16\ _{2}^{\shortparallel
}n_{i_{3}}\int d\ ^{\shortparallel }E\frac{\left( [(\ ^{\shortparallel }\
\zeta ^{7}\ \ ^{\shortparallel }\mathring{g}^{7})^{-1/4}]^{\ast }\right) ^{2}%
}{|\int d\ ^{\shortparallel }E\ (\ ~_{4}^{\shortparallel }\mathcal{K})(\ \
^{\shortparallel }\zeta ^{7}\ \ ^{\shortparallel }\mathring{g}^{7})^{\ast }|}%
(\frac{[(\ ^{\shortparallel }\zeta ^{7}\ \ ^{\shortparallel }\mathring{g}%
^{7})^{-1/4}\ ^{\shortparallel }\chi ^{7})]^{\ast }}{2\ [(\ \
^{\shortparallel }\zeta ^{7}\ \ ^{\shortparallel }\mathring{g}%
^{7})^{-1/4}]^{\ast }}+\frac{\int d\ ^{\shortparallel }E\ (\
~_{4}^{\shortparallel }\mathcal{K})[(\ ^{\shortparallel }\zeta ^{7}\ \
^{\shortparallel }\mathring{g}^{7})\ ^{\shortparallel }\chi ^{7}]^{\ast }}{%
\int d\ ^{\shortparallel }E\ (\ ~_{4}^{\shortparallel }\mathcal{K})(\
^{\shortparallel }\zeta ^{7}\ \ ^{\shortparallel }\mathring{g}^{7})^{\ast }})%
}{\ _{1}^{\shortparallel }n_{i_{3}}+16\ _{2}^{\shortparallel }n_{i_{3}}[\int
d\ ^{\shortparallel }E\frac{\left( \ [(\ ^{\shortparallel }\zeta ^{7}\
^{\shortparallel }\mathring{g}^{7})^{-1/4}]^{\ast }\right) ^{2}}{|\int d\
^{\shortparallel }E\ (\ ~_{4}^{\shortparallel }\mathcal{K})(\
^{\shortparallel }\zeta ^{7}\ \ ^{\shortparallel }\mathring{g}^{7})^{\ast }|}%
]}, \\
\ ^{\shortparallel }\zeta _{i_{3}8} &=&\frac{\ ^{\shortparallel }\partial
_{i_{3}}\int d\ ^{\shortparallel }E(~_{4}^{\shortparallel }\mathcal{K})\
(~^{\shortparallel }\zeta ^{7})^{\ast }}{(\ ^{\shortparallel }\mathring{N}%
_{i_{3}8})(~_{4}^{\shortparallel }\mathcal{K})(\ ^{\shortparallel }\zeta
^{7})^{\ast }}\mbox{ and }\ ^{\shortparallel }\chi _{i_{3}7}=\frac{\
^{\shortparallel }\partial _{i_{3}}[\int d\ ^{\shortparallel
}E(~_{4}^{\shortparallel }\mathcal{K})(\ ^{\shortparallel }\zeta ^{7}\
^{\shortparallel }\mathring{g}^{7})^{\ast }]}{\ ^{\shortparallel }\partial
_{i_{3}}\ [\int d\ ^{\shortparallel }E(~_{4}^{\shortparallel }\mathcal{K})(\
^{\shortparallel }\zeta ^{7})^{\ast }]}-\frac{(\ ^{\shortparallel }\zeta
^{7}\ ^{\shortparallel }\mathring{g}^{7})^{\ast }}{(\ ^{\shortparallel
}\zeta ^{7})^{\ast }}.
\end{eqnarray*}

Similar formulas for associative and commutative phase spaces are stated by
Theorem 5.1 and proven in appendix A.6 of \cite{bubuianu20} in terms of
different type dyadic variables and effective sources.

Introducing above $\kappa $-coefficients instead of $\eta $-coefficients of (%
\ref{offdiagpolf}), we obtain such nonlinear quadratic elements for
quasi-stationary solutions,
\begin{equation*}
d\ ^{\shortparallel }\widehat{s}^{2}=\ ^{\shortparallel }\widehat{g}_{\alpha
_{s}\beta _{s}}(x^{k},y^{3},\ ^{\shortparallel }p_{a_{3}},\ ^{\shortparallel
}p_{a_{4}};g_{4},\ ^{\shortparallel }g^{5},\ ^{\shortparallel }g^{7},\
~_{s}^{\shortparallel }\mathcal{K},\ \ _{s}^{\shortparallel }\Lambda
_{0},\kappa )d\ ^{\shortparallel }u^{\alpha _{s}}d\ ^{\shortparallel
}u^{\beta _{s}}=e^{\psi _{0}(x^{k_{1}})}(1+\kappa \ ^{\psi }\
^{\shortparallel }\chi )[(dx^{1})^{2}+(dx^{2})^{2}]
\end{equation*}%
\begin{eqnarray*}
&&-\{\frac{4[(|\ ^{\shortparallel }\zeta _{4}\ \ ^{\shortparallel }\mathring{%
g}_{4}|^{1/2})^{\diamond }]^{2}}{\ ^{\shortparallel }\mathring{g}_{3}|\int
dy^{3}\{(~_{2}^{\shortparallel }\mathcal{K})(\ ^{\shortparallel }\zeta _{4}\
^{\shortparallel }\mathring{g}_{4})^{\diamond }\}|}-\kappa \lbrack \frac{(\
^{\shortparallel }\chi _{4}|\ ^{\shortparallel }\zeta _{4}\ ^{\shortparallel
}\mathring{g}_{4}|^{1/2})^{\diamond }}{4(|\ ^{\shortparallel }\zeta _{4}\
^{\shortparallel }\mathring{g}_{4}|^{1/2})^{\diamond }}-\frac{\int
dy^{3}\{(~_{2}^{\shortparallel }\mathcal{K})[(~^{\shortparallel }\zeta _{4}\
^{\shortparallel }\mathring{g}_{4})\ ^{\shortparallel }\chi _{4}]^{\diamond
}\}}{\int dy^{3}\{(~_{2}^{\shortparallel }\mathcal{K})(\ ^{\shortparallel
}\zeta _{4}\ ^{\shortparallel }\mathring{g}_{4})^{\diamond }\}}]\}\
^{\shortparallel }\mathring{g}_{3} \\
&&+\{dy^{3}+[\frac{\partial _{i_{1}}\ \int dy^{3}(~_{2}^{\shortparallel }%
\mathcal{K})\ (\ ^{\shortparallel }\zeta _{4})^{\diamond }}{(\
^{\shortparallel }\mathring{N}_{i_{1}}^{3})(~_{2}^{\shortparallel }\mathcal{K%
})(\ ^{\shortparallel }\zeta _{4})^{\diamond }}+\kappa (\frac{\partial
_{i_{1}}[\int dy^{3}(~_{2}^{\shortparallel }\mathcal{K})(\ ^{\shortparallel
}\zeta _{4}\ ^{\shortparallel }\chi _{4})^{\diamond }]}{\partial _{i_{1}}\
[\int dy^{3}(~_{2}^{\shortparallel }\mathcal{K})(\ ^{\shortparallel }\zeta
_{4})^{\diamond }]}-\frac{(\ ^{\shortparallel }\zeta _{4}\ ^{\shortparallel
}\chi _{4})^{\diamond }}{(\ ^{\shortparallel }\zeta _{4})^{\diamond }})](\
^{\shortparallel }\mathring{N}_{i_{1}}^{3})dx^{i_{1}}\}^{2}
\end{eqnarray*}%
\begin{eqnarray}
&&+\ ^{\shortparallel }\zeta _{4}(1+\kappa \ ^{\shortparallel }\chi _{4})\
^{\shortparallel }\mathring{g}_{4}\{dt+[(\ ^{\shortparallel }\mathring{N}%
_{k_{1}}^{4})^{-1}[\ _{1}^{\shortparallel }n_{k_{1}}+16\
_{2}^{\shortparallel }n_{k_{1}}[\int dy^{3}\{\frac{\left( [(\
^{\shortparallel }\zeta _{4}\ ^{\shortparallel }\mathring{g}%
_{4})^{-1/4}]^{\diamond }\right) ^{2}}{|\int dy^{3}[(~_{2}^{\shortparallel }%
\mathcal{K})(\ ^{\shortparallel }\zeta _{4}\ \ ^{\shortparallel }\mathring{g}%
_{4})]^{\diamond }|}]  \label{offdncelepsilon} \\
&&-\kappa \frac{16\ _{2}^{\shortparallel }n_{k_{1}}\int dy^{3}\frac{\left(
[(\ ^{\shortparallel }\ \zeta _{4}\ \ ^{\shortparallel }\mathring{g}%
_{4})^{-1/4}]^{\diamond }\right) ^{2}}{|\int dy^{3}[(~_{2}^{\shortparallel }%
\mathcal{K})(\ ^{\shortparallel }\zeta _{4}\ ^{\shortparallel }\mathring{g}%
_{4})]^{\diamond }|}(\frac{[(\ ^{\shortparallel }\ \zeta _{4}\ \
^{\shortparallel }\mathring{g}_{4})^{-1/4}\chi _{4})]^{\diamond }}{2[(\
^{\shortparallel }\zeta _{4}\ ^{\shortparallel }\mathring{g}%
_{4})^{-1/4}]^{\diamond }}+\frac{\int dy^{3}[(~_{2}^{\shortparallel }%
\mathcal{K})(\ ^{\shortparallel }\zeta _{4}\ ^{\shortparallel }\chi _{4}\ \
^{\shortparallel }\mathring{g}_{4})]^{\diamond }}{\int
dy^{3}[(~_{2}^{\shortparallel }\mathcal{K})(\ ^{\shortparallel }\zeta _{4}\
\ ^{\shortparallel }\mathring{g}_{4})]^{\diamond }})}{\ _{1}^{\shortparallel
}n_{k_{1}}+16\ _{2}^{\shortparallel }n_{k_{1}}[\int dy^{3}\frac{\left( [(\
^{\shortparallel }\zeta _{4}\ \ ^{\shortparallel }\mathring{g}%
_{4})^{-1/4}]^{\diamond }\right) ^{2}}{|\int dy^{3}[(~_{2}^{\shortparallel }%
\mathcal{K})(\ ^{\shortparallel }\zeta _{4}\ ^{\shortparallel }\mathring{g}%
_{4})]^{\diamond }|}]}](\ ^{\shortparallel }\mathring{N}_{k_{1}}^{4})dx^{%
\acute{k}_{1}}\}  \notag
\end{eqnarray}%
\begin{eqnarray*}
&&+\ ^{\shortparallel }\zeta ^{5}(1+\kappa ~^{\shortparallel }\chi
^{5})~^{\shortparallel }\mathring{g}^{5}\{d~^{\shortparallel }p_{5}+[\
(~^{\shortparallel }\mathring{N}_{i_{2}5})^{-1}[\ _{1}^{\shortparallel
}n_{i_{2}}+16\ _{2}^{\shortparallel }n_{i_{2}}[\int d~^{\shortparallel
}p_{6}\{\frac{\left( ~^{\shortparallel }\partial ^{6}[(~^{\shortparallel
}\zeta ^{5}\ ~^{\shortparallel }\mathring{g}^{5})^{-1/4}]\right) ^{2}}{|\int
d~^{\shortparallel }p_{6}~^{\shortparallel }\partial
^{6}[(~_{3}^{\shortparallel }\mathcal{K})(~^{\shortparallel }\zeta
^{5}~^{\shortparallel }\mathring{g}^{5})]|}]+ \\
&&\kappa \frac{16\ _{2}^{\shortparallel }n_{i_{2}}\int d~^{\shortparallel
}p_{6}\frac{\left( ~^{\shortparallel }\partial ^{6}[(~^{\shortparallel
}\zeta ^{5}\ ~^{\shortparallel }\mathring{g}^{5})^{-1/4}]\right) ^{2}}{|\int
d~^{\shortparallel }p_{6}(~_{3}^{\shortparallel }\mathcal{K})\
~^{\shortparallel }\partial ^{6}[(~^{\shortparallel }\zeta ^{5}\
~^{\shortparallel }\mathring{g}^{5})]|}(\frac{\ ~^{\shortparallel }\partial
^{6}[(\ ~^{\shortparallel }\zeta ^{5}\ ~^{\shortparallel }\mathring{g}%
^{5})^{-1/4}~^{\shortparallel }\chi ^{5})]}{2\ ~^{\shortparallel }\partial
^{6}[(~^{\shortparallel }\zeta ^{5}~^{\shortparallel }\mathring{g}%
^{5})^{-1/4}]}+\frac{\int d~^{\shortparallel }p_{6}\ (~_{3}^{\shortparallel }%
\mathcal{K})~^{\shortparallel }\partial ^{6}[(~^{\shortparallel }\zeta ^{5}\
~^{\shortparallel }\mathring{g}^{5})~^{\shortparallel }\chi ^{5}]}{\int
d~^{\shortparallel }p_{6}(~_{3}^{\shortparallel }\mathcal{K})\
~^{\shortparallel }\partial ^{6}[(~^{\shortparallel }\zeta ^{5}\
~^{\shortparallel }\mathring{g}^{5})]})}{\ _{1}^{\shortparallel
}n_{i_{2}}+16\ _{2}^{\shortparallel }n_{i_{2}}[\int d~^{\shortparallel }p_{6}%
\frac{\left( \ ~^{\shortparallel }\partial ^{6}[(\ ~^{\shortparallel }\zeta
^{5}\ ~^{\shortparallel }\mathring{g}^{5})^{-1/4}]\right) ^{2}}{|\int
d~^{\shortparallel }p_{6}\ (~_{3}^{\shortparallel }\mathcal{K})\
~^{\shortparallel }\partial ^{6}[(~^{\shortparallel }\zeta ^{5}\
~^{\shortparallel }\mathring{g}^{5})]|}]}]
\end{eqnarray*}%
\begin{eqnarray*}
&&(\ ^{\shortparallel }\mathring{N}_{i_{2}5})dx^{i_{2}}\}-\{\frac{%
4[~^{\shortparallel }\partial ^{6}(|~^{\shortparallel }\zeta
^{5}~^{\shortparallel }\mathring{g}^{5}|^{1/2})]^{2}}{~^{\shortparallel }%
\mathring{g}^{6}|\int d~^{\shortparallel }p_{6}\{(~_{3}^{\shortparallel }%
\mathcal{K})~^{\shortparallel }\partial ^{6}[(~^{\shortparallel }\zeta
^{5}~^{\shortparallel }\mathring{g}^{5})]\}|}-\kappa \lbrack \frac{\partial
_{i_{2}}[\int d~^{\shortparallel }p_{6}(~_{3}^{\shortparallel }\mathcal{K}%
)~^{\shortparallel }\partial ^{6}(~^{\shortparallel }\zeta
^{5}~^{\shortparallel }\mathring{g}^{5})]}{\partial _{i_{2}}\ [\int
d~^{\shortparallel }p_{6}(~_{3}^{\shortparallel }\mathcal{K}%
)~^{\shortparallel }\partial ^{6}(~^{\shortparallel }\zeta ^{5})]}-\frac{%
^{\shortparallel }\partial ^{6}(~^{\shortparallel }\zeta
^{5}~^{\shortparallel }\mathring{g}^{5})}{~^{\shortparallel }\partial
^{6}(~^{\shortparallel }\zeta ^{5})}]\} \\
&&\ ^{\shortparallel }\mathring{g}^{6}\{d~^{\shortparallel }p_{6}+[\frac{%
\partial _{i_{2}}\ \int d~^{\shortparallel }p_{6}(~_{3}^{\shortparallel }%
\mathcal{K})~^{\shortparallel }\partial ^{6}(~^{\shortparallel }\zeta ^{5})}{%
(\ ~^{\shortparallel }\mathring{N}_{i_{2}6})(~_{3}^{\shortparallel }\mathcal{%
K})~^{\shortparallel }\partial ^{6}(~^{\shortparallel }\zeta ^{5})}+\kappa (%
\frac{\partial _{i_{2}}[\int d~^{\shortparallel }p_{6}(~_{3}^{\shortparallel
}\mathcal{K})~^{\shortparallel }\partial ^{6}(~^{\shortparallel }\zeta
^{5}~^{\shortparallel }\mathring{g}^{5})]}{\partial _{i_{2}}\ [\int
d~^{\shortparallel }p_{6}(~_{3}^{\shortparallel }\mathcal{K}%
)~^{\shortparallel }\partial ^{6}(~^{\shortparallel }\zeta ^{5})]}-\frac{%
~^{\shortparallel }\partial ^{6}(~^{\shortparallel }\zeta
^{5}~^{\shortparallel }\mathring{g}^{5})}{~^{\shortparallel }\partial
^{6}(~^{\shortparallel }\zeta ^{5})})](\ ^{\shortparallel }\mathring{N}%
_{i_{2}6})dx^{i_{2}}\}
\end{eqnarray*}%
\begin{eqnarray*}
&&+\ ^{\shortparallel }\zeta ^{7}(1+\kappa ~^{\shortparallel }\chi
^{7})~^{\shortparallel }\mathring{g}^{7}\{d~^{\shortparallel
}p_{7}+[(~^{\shortparallel }\mathring{N}_{i_{3}7})^{-1}[\
_{1}^{\shortparallel }n_{i_{3}}+16\ _{2}^{\shortparallel }n_{i_{3}}[\int
d~^{\shortparallel }E\{\frac{\left( [(~^{\shortparallel }\zeta
^{7}~^{\shortparallel }\mathring{g}^{7})^{-1/4}]^{\ast }\right) ^{2}}{|\int
d~^{\shortparallel }E(~_{4}^{\shortparallel }\mathcal{K})[(~^{\shortparallel
}\zeta ^{7}~^{\shortparallel }\mathring{g}^{7})]^{\ast }|}] \\
&&-\kappa \frac{16\ _{2}^{\shortparallel }n_{i_{3}}\int d~^{\shortparallel }E%
\frac{\left( [(~^{\shortparallel }\zeta ^{7}\ ~^{\shortparallel }\mathring{g}%
^{7})^{-1/4}]^{\ast }\right) ^{2}}{|\int d~^{\shortparallel }E\
(~_{4}^{\shortparallel }\mathcal{K})[(~^{\shortparallel }\zeta
^{7}~^{\shortparallel }\mathring{g}^{7})]^{\ast }|}(\frac{%
[(~^{\shortparallel }\zeta ^{7}~^{\shortparallel }\mathring{g}%
^{7})^{-1/4}~^{\shortparallel }\chi ^{7})]^{\ast }}{2\ [(~^{\shortparallel
}\zeta ^{7}~^{\shortparallel }\mathring{g}^{7})^{-1/4}]^{\ast }}+\frac{\int
d\ ^{\shortparallel }E\ (~_{4}^{\shortparallel }\mathcal{K}%
)[(~^{\shortparallel }\zeta ^{7}\ ~^{\shortparallel }\mathring{g}%
^{7})~^{\shortparallel }\chi ^{7}]^{\ast }}{\int d~^{\shortparallel }E\
(~_{4}^{\shortparallel }\mathcal{K})[(~^{\shortparallel }\zeta ^{7}\
~^{\shortparallel }\mathring{g}^{7})]^{\ast }})}{\ _{1}^{\shortparallel
}n_{i_{3}}+16\ _{2}^{\shortparallel }n_{i_{3}}[\int d~^{\shortparallel }E%
\frac{\left( \ [(~^{\shortparallel }\zeta ^{7}~^{\shortparallel }\mathring{g}%
^{7})^{-1/4}]^{\ast }\right) ^{2}}{|\int d~^{\shortparallel }E\
(~_{4}^{\shortparallel }\mathcal{K})[(~^{\shortparallel }\zeta ^{7}\
~^{\shortparallel }\mathring{g}^{7})]|^{\ast }}]}]\ (\ ^{\shortparallel }%
\mathring{N}_{i_{3}7})dx^{i_{3}}\}
\end{eqnarray*}%
\begin{eqnarray*}
&&-\{4\frac{4[(|~^{\shortparallel }\zeta ^{7}~^{\shortparallel }\mathring{g}%
^{7}|^{1/2})^{\ast }]^{2}}{~^{\shortparallel }\mathring{g}^{8}|\int
d~^{\shortparallel }E\{(~_{4}^{\shortparallel }\mathcal{K}%
)[(~^{\shortparallel }\zeta ^{7}~^{\shortparallel }\mathring{g}^{7})]^{\ast
}\}|}-\kappa \lbrack \frac{(~^{\shortparallel }\chi ^{7}|~^{\shortparallel
}\zeta ^{7}~^{\shortparallel }\mathring{g}^{7}|^{1/2})^{\ast }}{%
4(|~^{\shortparallel }\zeta ^{7}\ ^{\shortparallel }\mathring{g}%
^{7}|^{1/2})^{\ast }}-\frac{\int d~^{\shortparallel
}E\{(~_{4}^{\shortparallel }\mathcal{K})[(~^{\shortparallel }\zeta
^{7}~^{\shortparallel }\mathring{g}^{7})~^{\shortparallel }\chi ^{7}]^{\ast
}\}}{\int d~^{\shortparallel }E\{(~_{4}^{\shortparallel }\mathcal{K}%
)[(~^{\shortparallel }\zeta ^{4}~^{\shortparallel }\mathring{g}^{4})]^{\ast
}\}}]\}~^{\shortparallel }\mathring{g}^{8} \\
&&\{d\ ^{\shortparallel }E+[\frac{~^{\shortparallel }\partial _{i_{3}}\ \int
d\ ^{\shortparallel }E(~_{4}^{\shortparallel }\mathcal{K})\ (\
^{\shortparallel }\zeta ^{7})^{\ast }}{(~^{\shortparallel }\mathring{N}%
_{i_{3}8})(~_{4}^{\shortparallel }\mathcal{K})(~^{\shortparallel }\zeta
^{7})^{\ast }}+\kappa \lbrack \frac{~^{\shortparallel }\partial
_{i_{3}}[\int d~^{\shortparallel }E(~_{4}^{\shortparallel }\mathcal{K}%
)(~^{\shortparallel }\zeta ^{7}~^{\shortparallel }\mathring{g}^{7})^{\ast }]%
}{~^{\shortparallel }\partial _{i_{3}}\ [\int d~^{\shortparallel
}E(~_{4}^{\shortparallel }\mathcal{K})(~^{\shortparallel }\zeta ^{7})^{\ast
}]}-\frac{(~^{\shortparallel }\zeta ^{7}~^{\shortparallel }\mathring{g}%
^{7})^{\ast }}{(~^{\shortparallel }\zeta ^{7})^{\ast }}]](~^{\shortparallel }%
\mathring{N}_{i_{3}8})d~^{\shortparallel }x^{i_{3}}\}.
\end{eqnarray*}

In above formulas, the $\eta $-polarization functions may depend on all
phase space coordinates. For $\kappa $--polarizations, we can work with
variables when respective coefficients depend only on the same shell and
lower ones coordinates for all 4 shells of variables on the 8-d phase space.

We can consider, for instance, that a prime metric $\ _{s}^{\shortparallel}%
\mathbf{\mathring{g}}$ (for $s=1,2)$ defines a (quasi-) stationary solution
in 4-d Einstein gravity or MGT with coefficients $[\ ^{\shortparallel }%
\mathring{g}_{\alpha _{s}},\ ^{\shortparallel }\mathring{N}%
_{i_{s-1}}^{a_{s}}]$ depending only on spacetime variables conventionally
split on shells $1$ and $2.$ In such cases, a target solutions $\
_{s}^{\shortparallel }\mathbf{g}(\kappa )$ (\ref{offdncelepsilon}) describes
possible generalizations, for instance, of black hole solutions on phase
spaces. We can fix such nonholonomic dyadic distributions for the first two
shells when the respective target metrics determine small parametric forms
of quasi-stationary spacetime metrics (\ref{qeltorshv1}) as solutions of (%
\ref{sourc1hv}) with trivial (\ref{sourc1cc}) but with nontrivial R-flux
contributions (\ref{realrflux}) encoded in $_{1}^{\shortparallel }\mathcal{K}
$ and $~_{2}^{\shortparallel }\mathcal{K}.$ Considered as star deformed
s-adapted configurations, such 4-d generic off-diagonal metrics are
symmetric up to 1st order on $\kappa $ as follow from formulas (\ref{nsm1qs}%
) and (\ref{ssm1qs}). For LC-configurations, such s-metrics define (off-)
diagonal quasi-stationary solutions of nonassociative vacuum Einstein
equations introduced in \cite{aschieri17}.

Using parametric generating functions and effective sources, we construct
generic off-diagonal s-metric configurations with respective star
deformations to symmetric and nonsymmetric metrics. Nevertheless, it is
possible to define certain nonholonomic geometric conditions (they may
involve integration functions) on $\eta $- and/or $\kappa $- polarization
functions when the target quadratic line elements are diagonal. For other
classes of phase space, or spacetime quasi-stationary metrics, we can impose
such nonholonomic constraints via prescribed cosmological constants, when
the effective sources are zero and the star deformed canonical Ricci s-tenor
satisfies the conditions (\ref{purestarvacuum}). Sections 5.4 and 5.5 in
\cite{bubuianu20} contain respective details and a number of examples of
diagonal and/or non-diagonal, effective/ pure vacuum solutions of
associative and nonassociative phase space gravitational equations. Those
s-metric can be taken as $\kappa ^{0}$-approximation for solutions which can
be star deformed to nonassociative dyadic gravitational configurations as we
described appendix.

\end{document}